\documentclass[11pt]{article}
\usepackage[margin=2cm,vmargin=2cm]{geometry}
\usepackage{microtype}
\usepackage[T1]{fontenc}
\usepackage[utf8]{inputenc}
\usepackage[affil-it]{authblk}
\usepackage[hidelinks]{hyperref}
\usepackage[usenames,dvipsnames]{color}
\usepackage{pdflscape}
\usepackage{cite}
\usepackage{multicol}
\usepackage{etoolbox}
\usepackage{ragged2e}
\usepackage{amsmath}
\usepackage{amssymb}
\usepackage{mathtools}
\usepackage{stmaryrd}							
\usepackage{accents}
\usepackage{cancel}
\usepackage{tabulary}
\usepackage{booktabs}
\usepackage{caption}
\usepackage{float}
\usepackage{multirow}
\usepackage{tikz}
\usepackage{afterpage}
\usetikzlibrary[matrix,patterns,calc]
\renewcommand*{\thefootnote}{\roman{footnote}}
\newcommand{\ubar}[1]{\underaccent{\bar}{#1}}

\pgfmathtruncatemacro\distance{1}
\title{%
\vspace{-1cm}
\begin{flushright}
\normalsize{QMUL-PH-18-19}
\end{flushright}
$ $\\
\textbf{Exotic Branes in Exceptional Field Theory: \texorpdfstring{$E_{7(7)}$}{E7(7)} and Beyond}}
\date{}
\author{\vspace{-1cm}}
\begin{document}
\maketitle
\begin{center}
\textsc{David S. Berman$^\bullet$}\footnote{\href{mailto:d.s.berman@qmul.ac.uk}{d.s.berman@qmul.ac.uk}},
\textsc{Edvard T. Musaev$^{\dagger\ddagger}$}\footnote{\href{mailto:musaev.et@phystech.edu}{musaev.et@phystech.edu}},
\textsc{ and Ray Otsuki$^\bullet$}\footnote{\href{mailto:r.otsuki@qmul.ac.uk}{r.otsuki@qmul.ac.uk}}\\
\end{center}
\begin{center}
\vspace{1cm}
\textit{$^\bullet$Queen Mary University of London, Centre for Research in String Theory,\\
School of Physics and Astronomy, Mile End Road, London, England, E1 4NS}\\
\vspace{0.3cm}
\textit{$^\dagger$Moscow Institute of Physics and Technology,\\
  Institutskii per. 9, Dolgoprudny, 141700,  Russia}\\
\vspace{0.3cm}
\textit{$^\ddagger$Kazan Federal University, Institute of Physics,\\
 Kremlevskaya st. 16, Kazan, 420008, Russia}
\end{center}
\vspace{2cm}
\abstract{In recent years, it has been widely argued that the duality transformations of string and M-theory naturally imply the existence of so-called `exotic branes'---low codimension objects with highly non-perturbative tensions, scaling as $g_s^{\alpha}$ for $\alpha \leq -3$. We argue that their intimate link with these duality transformations make them an ideal object of study using the general framework of Double Field Theory (DFT) and Exceptional Field Theory (EFT)---collectively referred to as ExFT. Parallel to the theme of dualities, we also stress that these theories unify known solutions in string- and M-theory into a single solution under ExFT. We argue that not only is there a natural unifying description of the lowest codimension objects, many of these exotic states \emph{require} this formalism as a consistent supergravity description does not exist.}
\clearpage
\tableofcontents
\renewcommand*{\thefootnote}{\arabic{footnote}}
\setcounter{footnote}{0}
\section{Introduction}
One of the remarkable aspects of string theory is the presence of non-perturbative branes whose tensions scale as $\frac{1}{g_s}$ (the D-branes) and as $\frac{1}{g_s^2}$ (the Neveu-Schwarz five branes). The study of these branes in string theory over the last 20 years has revealed much about the connection between quantum field theories and gravity and have been a huge part of the construction of M-theory where there are no perturbative brane states.\par
Following the work of \cite{Obers:1998fb,deBoer:2010ud,deBoer:2012ma}, and others, it was realised that string theory also contains so-called `exotic brane' states whose tensions scale as $g_s^\alpha$ with $\alpha<-2$. These objects typically have low codimension\footnote{By low codimension, we mean branes of codimension-2 (`defect branes'), codimension-1 (`domain wall') and codimension-0 (`space-filling branes')} and so potentially suffer from various pathologies. Nevertheless there is now a substantial corpus of work in the area including \cite{Bergshoeff:2011se,Bergshoeff:2012ex,Bergshoeff:2012pm} where such branes have been shown to play an important role in duality symmetries. Of course it is also interesting to speculate what such branes would correspond to in a dual holographic theory with masses scaling as $\frac{N^\alpha}{\lambda^\alpha}$ after taking a 't Hooft limit. These states in the dual field theory with higher $N$ dependences should then be related to multiple traces as in the giant graviton story \cite{Corley:2001zk}\footnote{We thank Sanjaye Ramgoolam for discussions on this.}.\par
Apart from being exotic due to their novel scaling, these branes were also curious objects since it appeared that they were not well defined globally as supergravity solutions; a key part of their construction is to use elements of the duality group to patch together local solutions such that (globally) these branes then have holonomies valued in the duality groups. For the case of U-duality, these produce examples of Hull's U-folds\cite{Hull:2006va} which obviously contain S-folds and T-folds amongst their reductions. If they are not solutions of supergravity then what are they solutions of?\par
For a while Exceptional Field Theory (EFT) appeared to look like a rather nice answer looking for a question. In fact, Exceptional Field Theory (so-called because the exceptional groups are local symmetries of the theory) is the natural setting for exotic branes. Below we shall review EFT in more detail to provide the setting and set up conventions but first let us state some of the ideas behind EFT relevant for this paper. One of the key ideas from M-theory is that branes in IIA and IIB are descendants of a smaller set of branes in the higher dimensional theory of eleven dimensional supergravity. In addition to branes descending to branes, crucially some branes have their origin from the purely geometric solutions in eleven dimensional supergravity; namely the D0 is a null wave solution in eleven dimensions and the D6 brane is an eleven dimensional Kaluza Klein monopole of Gross-Sorkin-Perry type\cite{Sorkin:1983ns,Gross:1983hb}. Ideally we would like a theory with no central charges and no additional external sources. EFT has a chance of being this theory. As was shown in \cite{Berman:2014hna} and based on work in DFT \cite{Berkeley:2014nza,Berman:2014jsa} the membrane, five-brane and their bound states all come from a single EFT solution, namely the EFT version of the superposition of a wave and monopole. Thus the EFT superalgebra doesn't have central charges for these states. Just as the D0 is part of the wave solution in M-theory, and thus its IIA central charge has its origin as an eleven dimensional momentum, so are all the usual M-theory branes in EFT! The next question then is to investigate the role of exotic branes in EFT. This has begun with the works \cite{Sakatani:2014hba,Bergshoeff:2015cba,Bakhmatov:2016kfn,Bergshoeff:2016ncb, Lombardo:2016swq,Bakhmatov:2017les} and others. Ultimately one would wish that all the branes in string and M-theory including the exotic ones come from a single object in EFT. The hope for this is that any object in the same duality orbit must come from a single solution in EFT. So why hasn't this been already achieved?\par
A key problem for EFT is that one picks the exceptional group $E_n$ and works with a particular $n$. For a given $n$, there is a split between so called \textit{internal} and \textit{external} spaces. This split respects the $E_n$ symmetry (by construction) but does not respect the higher $E_{d+n}$ symmetries. Thus there are objects that are connected through higher $E_{d+n}$ symmetries that one sees as separate objects in the $E_n$ theory. Only with the full $E_{11}$ theory \cite{West:2001as,West:2003fc,West:2004kb,West:2004iz,Cook:2008bi,Cook:2009ri,Tumanov:2015yjd} would one expect all the symmetries to be manifest.\par
This paper concentrates on the brane solutions of the $E_{7(7)}$ EFT and constructs the single solution that give rise to the codimension-2 exotic branes in Type IIA, IIB and M-theory. We then look further at what sort of exotic branes may exist beyond those contained in this solution by going through simple duality rotations beyond the explicit solution we give.\par
In the following section, we shall describe these exotic branes in slightly more detail alongside explaining the notation used to denote the branes. In Section \ref{sec:NonGeomSoln}, we demonstrate why ExFT provide an ideal playground in which to probe these exotic branes by explicitly constructing a single solution in $E_{7(7)}$ EFT which unifies many of the exotic branes described to date. This section is perhaps best thought of as complementary to the work described in \cite{Berman:2014hna} as is shown in Figure \ref{fig:Brane}. In Section \ref{sec:Map}, we map out all the exotic branes that one may encounter down to $g_s^{-7}$ using a very simple procedure and we compare with what has been found to date in the literature. We apologise for bombarding the reader with an extensive taxonomy. The purpose is to reveal patterns in the exotic brane spectrum that we will comment in the final section.
\section{Overview of Exotic Branes}\label{sec:ExoticBranes}
Even before the work of \cite{deBoer:2012ma}, low codimension objects were known to possess non-standard features, regardless of the $g_s$ scaling; the D7 (codimension 2 in $D=10$) already modifies the spacetime asymptotics, the D8 (codimension 1 in $D=10$) terminates spacetime at a finite distance due to a fast running of the dilaton and the D9 is space-filling. In addition, the NS7 (the S-dual of the D7 and later reclassified as a $7_3$) possesses a tension scaling as $g_s^{-3}$ and was thus more highly non-perturbative than the other conventional branes. It has since become customary to organise these exotic states in terms of the $g_s$-scaling of their tensions which we discuss now.\par
The embedding of the T-duality group within the U-duality groups
\begin{align}
E_{n(n)} \supset \operatorname{O}(n-1,n-1) \times \mathbb{R}^+,
\end{align}
induces a grading of the tension of the branes, which may be characterised by a single number $\alpha \leq 0$. The highest values of $\alpha$ correspond to the well-known branes, whilst the lower powers form the focus of this paper:
\begin{itemize}
	\item $\alpha=0$: Fundamental $\text{F}1 \equiv 1_0$, P
	\item $\alpha=-1$: Dirichlet $\text{D}p \equiv p_1$
	\item $\alpha=-2$: Solitonic$\text{NS5} \equiv 5_2 \xrightarrow{T} \text{KK5} \equiv 5_2^1 \xrightarrow{T} 5_2^2$
	\item $\alpha\leq - 3$: Exotic e.g. ($p_3^{7-p}$ or $0_4^{(1,6)}$)
\end{itemize}
These exotic branes are generically low-codimension objects that are additionally non-geometric, either globally or locally. Such objects generically require duality transformations, in addition to the conventional diffeomorphisms and gauge transformations, in order to patch correctly and thus realise the T-folds and U-folds of Hull \cite{Hull:2006va}. Further, they may explicitly depend on winding or wrapping coordinates and may thus not even be well-defined locally in conventional supergravity. Since their existence and behaviour is so closely tied to the duality transformations of string- and M-theory, ExFT are an obvious candidate in which to study these objects as they geometrise these pathologies in a way described later. Nevertheless, some of the better-behaved of these exotic branes were explicitly argued to exist in string theory in \cite{deBoer:2012ma} via the supertube effect amongst conventional branes and it follows that a better understanding of these exotic branes is required\footnote{See, for example, \cite{Obers:1998} for early work along these lines; \cite{deBoer:2012ma} for a thorough discussion of codimension-2 exotic branes; \cite{Lombardo:2016swq,Lombardo2017,Bergshoeff:2017gpw,Kleinschmidt:2011vu} for a group-theoretic discussion on classifying mixed-symmetry potentials that these branes couple to; \cite{Kimura2013,Kimura:2018} for a discussion in the GLSM formalism; \cite{Chatzistavrakidis:2013jqa,Kimura:2014upa,Kimura:2016anf} for effective world-volume actions of exotic five-branes and \cite{Lee2016,Sakatani2017,Sakatani:2014hba,Andriot:2014uda} amongst others for a specific discussion of the so-called `non-geometric' parametrisations in the context of ExFT and $\beta$-supergravity.}.\par
As the D7, D8 and D9 show, possessing low codimension (whilst interesting in their own right) is not necessarily indicative of the sort of non-geometry that we seek to study which are characterised by low $g_s^{\alpha}$ scaling. On the other hand, there exists an exotic $5_2^2$-brane (we shall cover the notation denoting these branes shortly) which, whilst being codimension-2, possesses the same $g_s$ scaling as the NS5 and KK5 but manifests the sort of non-geometry that we seek by virtue of the fact that the metric is not single-valued at $\theta =0, 2\pi$ as one traverses around the brane. This form of non-geometry is a realisation of Hull's T-fold which possess the odd property that traversing around the brane does not return it to its original configuration. Such states require more than conventional diffeomorphisms to patch together the spacetime. We thus see that low codimension alone is not sufficient for non-geometry; one appears to require low $g_s$ scaling as well. Yet, as the $5_2^2$ demonstrates, there exists states which are non-geometric but still scale as $g_s^{-2}$.\par
The use of ExFT is that many of the exotic states that one can construct can be better understood or more elegantly unified  when the duality transformations are realised linearly and, indeed, we shall demonstrate in Section \ref{sec:Map} that, not only are codimension$<2$ objects are common, they form the majority of the exotic states and may even require an ExFT description to make sense. In particular, ExFT allow for the construction of non-trivial space-filling branes by allowing for a dependence of the fields on the extended coordinates. This is only possible because of the distinguishing feature of ExFT in that they capture winding mode dependences. We shall give a more detailed argument for this in Section \ref{sec:Map}.\par
For the $E_{7(7)}$ solution presented here, we shall focus on codimension-2 states. However, we shall move onto discussing all the possible exotic states that one should be able to construct. We briefly discuss the notation used in \cite{deBoer:2012ma} for branes, in which they are characterised by the mass-dependence when wrapping an internal torus. For Type II states, the mass of a $b_n^{(\ldots, d, c)}$-brane depends linearly on $b$ radii, quadratically on $c$ radii, cubically on $d$ radii, and so on. Additionally, the subscript denotes the power dependence on the string coupling. Finally, the power of $l_s$ on the denominator is such that the total mass has units of ${(\text{Length})}^{-1}$ as required. For M-theory states, the notation is very similar except for the absence of the string coupling number $n$ and the role of $l_s$ being reprised by the Planck length $l_P$ in eleven dimensions:
\begin{align}
\text{Type II}: && \text{M}(b_n^{(\ldots, d,c)}) & = \frac{\ldots {(R_{k_1} \ldots R_{k_d})}^3 {(R_{j_1} \ldots R_{j_c})}^2 {(R_{i_1} \ldots R_{i_b})}}{g_s^nl_s^{1 + b + 2c + 3d + \ldots}},\\
\text{M-Theory}: && \text{M}(b^{(\ldots, d,c)}) & = \frac{\ldots {(R_{k_1} \ldots R_{k_d})}^3 {(R_{j_1} \ldots R_{j_c})}^2 {(R_{i_1} \ldots R_{i_b})}}{l_p^{1 + b + 2c + 3d + \ldots}}.
\end{align}
For states that appear in both Type IIA and IIB, we may additionally append an A/B suffix to the brane if the theory being discussed is relevant e.g. $0_4^{(1,6)}\text{B}$. These exotic branes couple electrically to the mixed symmetry potentials\footnote{Note that we shall label the type of potential by the power of $g_s$, schematically labelled $E^{(n)}$. Thus, for $n=0,1,2,3,4,\ldots$, we shall denote the potentials $B, C, D, E, F, \ldots$. Thus, the fundamental string ($n=0$) couples to the NS-NS 2-form $B_{2}$, the D$p$-branes ($n=1$) couple to $C_{p+1}$ etc.}
\begin{align}
{(E^{(n)})}_{b+c+d+1, c+d, d} \leftrightarrow b_n^{(d,c)},
\end{align}
where the subscripts on the potentials denote the number of indices in that set. The notation is such that the each set is implicitly antisymmetrised over and contains all the sets of indices to the right of it. For example, the $0_4^{(1,6)}$ couples to
\begin{align}
F_{8,7,1} \sim F_{x y_1 \ldots y_6 z, y_1 \ldots y_6 z, z}.
\end{align}
Note that the we do not consider the Hopf fibre (and more generally, distinguished isometric directions) as a worldvolume direction and so the KK-monopole in eleven dimensions shall be denoted KK6 (or KK6M) and the monopole in ten dimensions as KK5 (or KK5A/B).
\section{The Non-Geometric Solution in \texorpdfstring{$E_{7(7)}$}{E7(7)} EFT}\label{sec:NonGeomSoln}
\subsection{Overview of \texorpdfstring{$E_{7(7)} \times \mathbb{R}^+$}{E7(7)R+} Exceptional Field Theory}
Extended Field Theories (ExFTs) are a class of theories that augment the usual spacetime with a set of novel coordinates that are related to the winding modes of extended objects. Double field theory (DFT) was the first example to be constructed this way. Geometrically, the usual spacetime coordinates (dual to momentum) were supplemented with a second set of coordinates, dual to the winding modes. \cite{Siegel:1993th,Hull:2009mi}\par
Exceptional field theories (EFTs) were the natural progression to this where now the spacetime coordinates are extended to include with coordinates dual to the wrapping modes of the M-branes. These theories possess a natural global $E_{n(n)}$ exceptional symmetry. Note that, the symmetry here is a \emph{continuous} $E_{n(n)}( \mathbb{R})$ action and are thus not themselves the U-duality groups. The arithmetic subgroup associated to U-duality arises in the presence of isometries and relate the ambiguities in the choice of section, as explained in \cite{Berman:2014jsa}.\par
Formalisms to make manifest the various $E_{n(n)}$ symmetry were developed in \cite{Hull:2007zu,Pacheco:2008ps,Hillmann:2009ci,Berman:2010is,Coimbra:2011nw,Coimbra:2011ky,Coimbra:2012af}\footnote{See the reviews \cite{Aldazabal:2013sca,Geissbuhler:2013uka,Berman:2013eva} instead for DFT.}. Later in what became known as EFT these extended spaces with manifest exceptional symmetries were combined with the normal spacetime. These were constructed in the series of papers \cite{Hohm:2014fxa,Hohm:2013uia,Hohm:2013vpa,Abzalov:2015ega,Musaev:2015ces,Hohm:2015xna,Berman:2015rcc} for $n = 8, \ldots, 2$. For a review of the ideas in extended theories and exceptional field theory one may consult the reviews \cite{Aldazabal:2013sca,Berman:2013eva,Hohm:2013bwa}.\par
In this paper we work with the $E_{7(7)}$ theory. What follows is a brief description so as to make the paper self contained and introduce notation. If the reader is not already familiar with EFT then they are urged to consult the original paper in the area \cite{Hohm:2013uia}. The coordinate representation $\rho_1$ of $E_{7(7)}$ EFT is the 56-dimensional fundamental representation which we index with $M = 1, \ldots, 56$. For every EFT, there are exactly two inequivalent (i.e. not related by $E_{n(n)}$ transformations) solutions to the section constraint; the M-theory section and the Type IIB section. For the M-theory section, we decompose the coordinates under $\operatorname{GL}(7)$ as
\begin{align}\label{eq:MSect}
56 \rightarrow 7_{+3} + 21^\prime_{+1} + 21_{-1} + 7^\prime_{-3},
\end{align}
where the subscript denotes the weight under $\operatorname{GL}(1)$. Letting $m,n = 1, \ldots, 7$ denote the vector representation of $\operatorname{GL}(7)$, we thus decompose the 56 coordinates of the internal space $Y^M$ to
\begin{align}
Y^M = ( y^m, y_{mn}, y^{mn}, y_m),
\end{align}
where $y_{mn}$ and $y^{mn}$ are labelled by a pair of antisymmetric indices. The coordinates $y_{mn}$ are dual to the wrapping modes the M2-brane, $y^{mn} \sim \epsilon^{mn k_1 \ldots k_5} y_{k_1 \ldots k_5}$ is dual to the wrapping modes of the M5-brane and $y_{m} \sim y_{n_1 \ldots n_7,m}$ are KK-modes.\par
The other section is the Type IIB which corresponds to decomposition under $\operatorname{GL}(6) \times \operatorname{SL}(2)$:
\begin{align}
56 \rightarrow {(6,1)}_{+2} + {(6^\prime,2)}_{+1} + {(20,1)}_0 + {(6,2)}_{-1} + {(6^\prime,1)}_{-2}.
\end{align}
Denoting $\hat{m} , \hat{n}, \hat{k} = 1, \ldots, 6$ and $\hat{\alpha} = 1,2$ for the $\operatorname{SL}(2)$ index, this corresponds to the decomposition
\begin{align}
{\hat{Y}}^M = ({\hat{y}}^{\hat{m}}, {\hat{y}}_{\hat{m}, \hat{\alpha}}, {\hat{y}}_{\hat{m} \hat{n} \hat{k}}, {\hat{y}}^{\hat{m}, \hat{\alpha}}, {\hat{y}}_{\hat{m}}).
\end{align}
To distinguish between the two sections, we shall adorn all Type IIB quantities with carets as above. Moving on to the field content of the theory, we have the following fields:
\begin{align}
\{ g_{\mu \nu}, \mathcal{M}_{MN}, \mathcal{A}_{\mu}{}^M, {\mathcal{B}}_{\mu \nu, \alpha}, {\mathcal{B}}_{\mu \nu}{}^M \},
\end{align}
where $\alpha = 1, \ldots, 133$ is and adjoint index and $\mu, \nu = 1, \ldots 4$ indexes the external space. The first of these fields is self-explanatory and so we focus on the remainder. The scalar degrees of freedom (from the perspective of the 4-dimensional external space) are held in the generalised metric $\mathcal{M}_{MN}$ which parametrises the coset $(E_{7(7)} \times \mathbb{R}^+)/\operatorname{SU}(8)$. A simple counting reveals that the 70 independent components of the metric, three-form and six-form potentials matches the dimension of the coset, as required. The generalised metric of $E_{7(7)}$ EFT (without the $\mathbb{R}^+$ scaling), which we denote ${\tilde{\mathcal{M}}}_{MN}$, is given in \cite{Hillmann:2009ci,Lee2016} (though note that the latter adopt a slightly modified choice of dualised coordinates to those used here, which follow the conventions of \cite{Hillmann:2009ci,Berman:2011jh}) and is the one used in \cite{Berman:2014hna}. In the absence of internal potentials, it is given by
\begin{align}\label{eq:E7GenMetric}
{\tilde{\mathcal{M}}}_{MN} = \operatorname{diag} [g_{(7)}^{\frac{1}{2}} g_{mn}; g_{(7)}^{\frac{1}{2}} g^{mn,pq}; g_{(7)}^{-\frac{1}{2}} g^{mn}; g_{(7)}^{-\frac{1}{2}} g_{mn,pq}],
\end{align}
where $g_{mn}$ is the internal metric, $g_{(7)}$ is its determinant and $g_{mn,pq} \coloneqq \frac{1}{2} ( g_{mp} g_{qn} - g_{mq} g_{pn})$ (similarly for $g^{mn,pq}$). With this choice, the generalised metric is a true $E_{7(7)}$ element (with determinant 1). However, allowing for the $\mathbb{R}^+$ scaling factor, we consider a generalised metric of the form $\mathcal{M}_{MN} = e^{-\Delta} {\tilde{\mathcal{M}}}_{MN}$ such as in \cite{Berman:2011jh}, which constructed the generalised metric from a non-linear realisation of $E_{11}$, obtaining $e^{-\Delta}= {g_{(7)}}{}^{-1}$ (though note that the external space has been truncated). However, we also require a non-trivial relative scaling between the internal space and external space, in a similar vein to \cite{Malek:2012pw,Bakhmatov:2017les}, and so require a different choice of scaling. In particular, we shall choose
\begin{align}\label{eq:Rescaling}
e^{-\Delta} = g_{(4)}^{-\frac{1}{4}}.
\end{align}
This overall scaling is identified as an extra scalar determining the relative scaling of the internal and external spaces and is the analogue of the $e^\phi = \operatorname{det} g_{\text{ext}}^{\frac{1}{7}}$ introduced first in \cite{Blair:2013gqa,Blair:2014zba}\footnote{See also \cite{Park:2013gaj}} and later used in \cite{Bakhmatov:2017les} for the $\operatorname{SL}(5) \times \mathbb{R}^+$ EFT. We are thus able to induce transformations on (the determinant of) the external metric via transformations of the generalised metric $\mathcal{M}_{MN}$.\par
One way to understand this is as follows. From the perspective of gauged supergravities we consider two distinct symmetries. The first is, of course, the Julia-Cremmer $E_{n(n)}$ duality symmetry which itself contains a natural scaling of the internal torus under the embedding $\operatorname{GL}(1) \subset \operatorname{GL}(n) \subset E_{n(n)}$---essentially a rescaling of the coordinates of the internal torus $y^m$ by $y^m \mapsto \alpha y^m$. The second is the so-called \emph{trombone symmetry} which is a well-known global scaling symmetry acting on the supergravity fields as
\begin{align}
g \mapsto \lambda^2 g, \qquad A_{(3)} \mapsto \lambda^3 A_{(3)}.
\end{align}
This is on-shell for $n\leq 9$, being realised only at the level of the equations of motion (the Lagrangian of 11-dimensional supergravity transforms as $\mathcal{L} \mapsto \lambda^{d-2} \mathcal{L}$ for $\lambda \in \mathbb{R}^+$), but is promoted to an off-shell symmetry for $n=9$ (i.e. a symmetry of both the action and equations of motion \cite{LeDiffon2007}\footnote{The special case of $n=9$ is perhaps best understood in terms of the allowed gaugings of the dimensionally reduced theory. In $n \leq 9$, the standard $E_{n(n)} \subset G$ gaugings (excluding the trombone symmetry) organise themselves into the embedding tensor $\theta_M{}^\alpha$, transforming under a subrepresentation of $\mathcal{R}_{V^\ast} \otimes \mathcal{R}_{\text{adj.}}$ dictated by group theoretic arguments---the so-called algebraic `linear' and `quadratic constraints', related to preserving supersymmetry and requiring $E_{n(n)}$-invariance of $\theta_M{}^\alpha$ respectively. In addition, the gaugings of the trombone symmetry organise themselves into a second object $\theta_M$, transforming under $\mathcal{R}_{V^\ast}$. For $n=9$, one finds that these two objects unify into a single object transforming under a single representation of the affine Kac-Moody algebra $E_{9(9)}$\cite{LeDiffon2008}. Thus, unlike in higher dimensions, generic gaugings in $d=2$ naturally contain trombone gaugings and so one finds that the trombone symmetry is promoted to an off-shell symmetry.}).  The extra $\mathbb{R}^+$ factor that we consider here may thus be considered as a combination of these two symmetries and was first considered in the present context in the closely related exceptional generalised geometry in \cite{Coimbra:2011ky,Coimbra:2012af,Coimbra:2011nw}. Just as the conventional gaugings embedding tensors of higher duality groups seed both the conventional gaugings and trombone gaugings of lower-dimensional gauged supergravities, the extra $\mathbb{R}^+$ factor may be understood as arising from the truncation of a higher duality group and is thus likely indispensable in generating distinct U-duality orbits, related by some \emph{solution-generating transformation} (itself distinct from the global $E_{7(7)}$ group).\par
The remaining fields $(\mathcal{A}_{\mu}{}^M, \mathcal{B}_{\mu \nu, \alpha}, \mathcal{B}_{\mu \nu, M})$ are a set of generalised gauge transformations. The fully gauge-covariant field strength is given by
\begin{align}
\mathcal{F}_{\mu \nu}{}^M = F_{\mu \nu}{}^M - 12 {\left(t^\alpha \right)}^{MN} \partial_N \mathcal{B}_{\mu \nu, \alpha} - \frac{1}{2} \Omega^{MN} \mathcal{B}_{\mu \nu, N},
\end{align}
where $F_{\mu \nu}{}^M \coloneqq 2 \partial_{[\mu} \mathcal{A}_{\nu]}{}^M$ is the na\"{i}ve Abelian field strength. This satisfies the Bianchi identity
\begin{align}\label{eq:Bianchi}
3 \mathcal{D}_{[\mu} \mathcal{F}_{\nu \rho]}{}^M = - 12 {\left(t^\alpha\right)}^{MN} \partial_N \mathcal{H}_{\mu \nu \rho, \alpha} - \frac{1}{2} \Omega^{MN} \mathcal{H}_{\mu \nu \rho, N},
\end{align}
where $\mathcal{D}_\mu \coloneqq \partial_\mu - \mathbb{L}_{\mathcal{A}_\mu}$ denotes the generalised Lie-covariantised derivative.
More relevant to this line of work it also satisfies the twisted self-duality relation
\begin{align}
\mathcal{F}_{\mu \nu}{}^M = - \frac{1}{2} {|g_{(4)}|}^{\frac{1}{2}} \varepsilon_{\mu \nu \rho \sigma} g^{\rho \lambda} g^{\sigma \kappa} \Omega^{MN} \mathcal{M}_{NK} \mathcal{F}_{\lambda \kappa}{}^K,
\end{align}
where $\varepsilon$ is the tensor density. For the purposes of this solution, the $\mathcal{B}_{\mu \nu , \bullet}$ fields (which are only required to close the gauge structure of the algebra will be set to zero).\par
Additionally, the theory possesses two group invariants; a symplectic form $\Omega_{MN}$ and a totally symmetric four-index object $c_{MNPQ}$, though we shall be dealing primarily with only the former. Due to the $\mathbb{R}^+$ factor, this is a weighted symplectic matrix\cite{Aldazabal:2013a}, of weight $\lambda(\Omega) = \frac{1}{2}$, and it is related to ${\tilde{\Omega}}_{MN} \in \operatorname{Sp}(56) \supset E_{7(7)}$ by
\begin{align}\label{eq:WeightedOmega}
\Omega_{MN} = e^{-\Delta} {\tilde{\Omega}}_{MN}.
\end{align}
We adopt the convention that indices are raised and lowered according to
\begin{align}
V^M = \Omega^{MN} V_N, \qquad V_M = V^N \Omega_{NM},
\end{align}
The only non-vanishing components of $\Omega_{MN}$ are
\begin{align}
\Omega_m{}^n & = e^{-\Delta} \delta^m_n = - \Omega^n{}_m,\\
\Omega_{mn}{}^{pq} & = e^{-\Delta} \delta^{mn}_{pq} = - \Omega^{pq}{}_{mn},
\end{align}
and similarly for the inverse, defined by $\Omega^{MK} \Omega_{NK} = \delta^M_N$.\par
In addition to the global $E_{n(n)} \times \mathbb{R}^+$ symmetry, the theory possesses a number of local symmetries---the general coordinate transformations of the metric and the $p$-form gauge transformations. Analogous to how the Lie derivative generates the algebra of infinitesimal diffeomorphisms in GR, we define a \emph{generalised Lie derivative} $\mathbb{L}$ which generates these local symmetries. In component form, the generalised Lie derivative of a generalised vector $V$, of weight $\lambda(V)$, with along $U$ in $E_{7(7)}\times \mathbb{R}^+$ EFT is given by
\begin{align}\label{eq:GenLie}
\mathbb{L}_U V^M = {[U,V]}^M + Y^{MN}{}_{PQ} \partial_N U^P V^Q + \left( \lambda(V) - \frac{1}{2} \right) \partial_N U^N V^M,
\end{align}
where $Y^{MN}{}_{PQ}$ is the \emph{Y-tensor}, given in terms of group invariants as\footnote{
\cite{Berman:2012vc} gives an equivalent form in terms of the quartic invariant $c_{MNPQ}$ of $E_{7(7)}$:
\begin{align}
Y^{MN}{}_{PQ} = 12 c^{MN}{}_{PQ} + \delta^{(M}_P \delta^{N)}_Q + \frac{1}{2} \Omega^{MN} \Omega_{PQ},
\end{align}
where the quartic invariant is defined in terms of the generators ${(t_\alpha)}_{MN} = {(t_\alpha)}_{NM}$ and symplectic form as\cite{LeDiffon2011}
\begin{align}\label{eq:Quartic}
{(t^\alpha)}_{MN} {(t_{\alpha})}_{KL} = \frac{1}{12} \tilde{\Omega}_{M(K} \tilde{\Omega}_{L)N} + c_{MNKL}.
\end{align}
Indeed, this last relation is used to show equivalence between \eqref{eq:P1} and \eqref{eq:P2} by permuting the indices on the two generators using the symmetry of $c_{MNKL}$. 
}
\begin{align}
Y^{MN}{}_{PQ} & = - 12 {\mathbb{P}_{(\text{adj})}}{}^M{}_Q{}^N{}_P + \frac{1}{2} \delta^M_Q \delta^N_P + \delta^M_P \delta^N_Q\\
	& = - 12 {(t^\alpha)}^{MN} {(t_\alpha)}_{PQ} - \frac{1}{2} \Omega^{MN} \Omega_{PQ},
\end{align}
where we have used that the projection onto the adjoint $\mathbb{P}_{(\text{adj})}$ is given by
\begin{align}
\mathbb{P}_{(\text{adj})}{}^M{}_Q{}^N{}_P & = {(t^\alpha)}_Q{}^M {(t_\alpha)}_P{}^N\label{eq:P1}\\
	& = {(t^\alpha)}^{MN} {(t_\alpha)}_{QP} + \frac{1}{24} \delta^M_Q \delta^N_P +\frac{1}{12} \delta^M_P \delta^N_Q - \frac{1}{24} \Omega^{MN} \Omega_{QP}\label{eq:P2},
\end{align}
and normalised according to ${\mathbb{P}_{(\text{adj})}}^M{}_N{}^N{}_M = 133$. One sees that, in the form \eqref{eq:GenLie}, there is a naturally defined \emph{effective} weight defined in the theory given by the bracketed term which naturally singles out $\lambda = \frac{1}{2}$ (indeed, for consistency, one requires that the gauge transformations of the generalised gauge fields must have weight $\lambda = \frac{1}{2}$). We may thus equivalently write the generalised Lie derivative in the form given in \cite{Hohm:2013uia}:
\begin{align}
\mathbb{L}_U V^M & = U^N \partial_N V^M - 12 \mathbb{P}_{(\text{adj})}^M{}_Q{}^N{}_P \partial_N U^P V^Q + \lambda(V) \partial_N U^N V^M.
\end{align}
The strong constraint $Y^{MN}{}_{PQ} \partial_{M} \bullet \partial_N \bullet = 0$ supplements the weak constraint $Y^{MN}{}_{PQ} \partial_M\partial_N \bullet = 0$ which, for this EFT, reduces to
\begin{align}
{(t_\alpha)}^{MN} \partial_{MN} \bullet = 0, \qquad {(t_\alpha)}^{MN} \partial_M \bullet \partial_N \bullet =0, \qquad \Omega^{MN} \partial_M \bullet \partial_N \bullet = 0.
\end{align}
Note that the analogous weak section constraint for the symplectic form is automatically satisfied by the symmetry of partial derivatives.

For this section, we shall denote a particular choice of frame by a superscript such that e.g. $g_{\mu \nu}^{5^3}$ denotes the external metric in the $5^3$ frame (or, more generally, $g_{\mu \nu}^{\text{M}}$ for a generic M-theory solution). Additionally, we shall adorn all objects in the Type IIB section with carets and all objects in the Type IIA reduction with a caron.\par
Here we add to a growing list of solutions in DFT \cite{Berkeley:2014nza,Blair:2016xnn,Berman:2014jsa,Bakhmatov:2016kfn,Kimura:2018} and EFT\cite{Blair:2014zba,Berman:2014hna,Bakhmatov:2017les} in which a single solution in the extended space reduces to a number of distinct solutions upon applying the section constraint. Although apparently unrelated in the reduced section, they are nonetheless related by a duality transformation (or at least a \emph{solution-generating} transformation) acting linearly on the extended space. The solution presented here is a very close analogue of the solution presented in \cite{Berman:2014hna} which described all the conventional branes. We shall henceforth refer to that solution the `geometric solution'. The solution that is presented here covers all the non-geometric branes of de Boer and Shigemori \cite{deBoer:2012ma} and will thus be referred to as the `non-geometric solution'.\par
The solutions that are contained in this solution are indicated in Figure~\ref{fig:Brane}.
\clearpage
\thispagestyle{empty}
\begin{landscape}
\begin{figure}[H]
\centering
\begin{tikzpicture}
\matrix(M)[matrix of math nodes, row sep=3em, column sep=2 em, minimum width=2em]{
\clap{\phantom{a}}&&&&&&&&&\clap{\phantom{a}}&\clap{\phantom{a}}&&&\clap{\phantom{a}}\\
\clap{\text{M:}}&\clap{\phantom{a}}&\clap{\text{PM}}&&\clap{\text{M2}}&&&&&\clap{\text{M5}}&&\clap{\text{KK6}}&\clap{\phantom{a}}&\\
\clap{\text{IIA:}}&\clap{\text{WA}}&\clap{\text{F1A}}&\clap{\text{D0}}&&\clap{\text{D2}}&&\clap{\text{D4}}&&\clap{\text{D6}}&&\clap{\text{NS5A}}&\clap{\text{KK5A}}&\\
\clap{\text{IIB:}}&\clap{\text{WB}}&\clap{\text{F1B}}&&\clap{\text{D1}}&&\clap{\text{D3}}&&\clap{\text{D5}}&\clap{\phantom{a}}&\clap{\text{D7}}&\clap{\text{NS5B}}&\clap{\phantom{a}}&\clap{\phantom{a}}\\
\clap{\phantom{a}}&&&&&&&&&\clap{\phantom{a}}&\clap{\phantom{a}}&\clap{\phantom{a}}&\clap{\text{KK5B}}&\clap{\phantom{a}}\\
\clap{\text{IIB:}}&\mathclap{0_4^{(1,6)}\text{B}}&\mathclap{1_4^6\text{B}}&&\mathclap{1_3^6}&&\mathclap{3_3^4}&&\mathclap{5_3^2}&\mathclap{\phantom{a}}&\mathclap{7_3}&\mathclap{5_2^2\text{B}}&\mathclap{\phantom{a}}&\mathclap{\phantom{a}}\\
\clap{\text{IIA:}}&\mathclap{0_4^{(1,6)}\text{A}}&\mathclap{1_4^6\text{A}}&\mathclap{0_3^7}&&\mathclap{2_3^5}&&\mathclap{4_3^3}&&\mathclap{6_3^1}&\mathclap{\phantom{a}}&\mathclap{5_2^2\text{A}}&\clap{\text{KK5A}}&\\
\clap{\text{M:}}&\clap{\phantom{a}}&\mathclap{0^{(1,7)}}&&\mathclap{2^6}&&&&&\mathclap{5^3}&&\clap{\text{KK6}}&\clap{\phantom{\text{a}}}&\clap{\phantom{a}}\\
\clap{\phantom{a}}&&&&&&&&&\clap{\phantom{a}}&\clap{\phantom{a}}&&\clap{\phantom{a}}&\\
	};
\draw[latex-latex, draw=blue, fill=blue] (M-2-3) -- (M-3-2);
\draw[latex-latex, draw=blue, fill=blue] (M-2-3) -- (M-3-4);
\draw[latex-latex, draw=blue, fill=blue] (M-2-5) -- (M-3-3);
\draw[latex-latex, draw=blue, fill=blue] (M-2-5) -- (M-3-6);
\draw[latex-latex, draw=blue, fill=blue] (M-2-10) -- (M-3-8);
\draw[latex-latex, draw=blue, fill=blue] (M-2-10) -- (M-3-12);
\draw[latex-latex, draw=blue, fill=blue] (M-2-12) -- (M-3-10);
\draw[latex-latex, draw=blue, fill=blue] (M-2-12) -- (M-3-13);
\draw[latex-latex, draw=red, fill=red] (M-3-2) -- (M-4-2);
\draw[latex-latex, draw=red, fill=red] (M-3-3) -- (M-4-3);
\draw[latex-latex, draw=red, fill=red] (M-3-2) -- (M-4-3);
\draw[latex-latex, draw=red, fill=red] (M-3-3) -- (M-4-2);
\draw[latex-latex, draw=red, fill=red] (M-3-4) -- (M-4-5);
\draw[latex-latex, draw=red, fill=red] (M-4-5) -- (M-3-6);
\draw[latex-latex, draw=red, fill=red] (M-3-6) -- (M-4-7);
\draw[latex-latex, draw=red, fill=red] (M-4-7) -- (M-3-8);
\draw[latex-latex, draw=red, fill=red] (M-3-8) -- (M-4-9);
\draw[latex-latex, draw=red, fill=red] (M-4-9) -- (M-3-10);
\draw[latex-latex, draw=red, fill=red] (M-3-10) -- (M-4-11);
\draw[latex-latex, draw=red, fill=red] (M-3-12) -- (M-4-12);
\draw[latex-latex, draw=red, fill=red] (M-3-13) -- (M-5-13);
\draw[latex-latex, draw=red, fill=red] (M-3-12) -- (M-5-13);
\draw[latex-latex, draw=red, fill=red] (M-3-13) -- (M-4-12);
\draw[latex-latex] (M-4-3) to[bend right,min distance=5mm] (M-4-5);
\draw[latex-latex] (M-4-9) to[bend right,min distance=5mm] (M-4-12);
\draw[latex-latex] (M-4-7) to[out=-135,in=-45,min distance=10mm](M-4-7);
\draw[latex-latex] (M-4-11) to[bend left] (M-6-11);
\draw[latex-latex] (M-6-7) to[out=45,in=135, min distance=10mm](M-6-7);
\draw[latex-latex] (M-6-9) to[bend left,min distance=5mm] (M-6-12);
\draw[latex-latex] (M-6-3) to[bend left,min distance=5mm] (M-6-5);
\draw[latex-latex, draw=red, fill=red] (M-6-2) -- (M-7-2);
\draw[latex-latex, draw=red, fill=red] (M-6-3) -- (M-7-3);
\draw[latex-latex, draw=red, fill=red] (M-6-2) -- (M-7-3);
\draw[latex-latex, draw=red, fill=red] (M-6-3) -- (M-7-2);
\draw[latex-latex, draw=red, fill=red] (M-7-4) -- (M-6-5);
\draw[latex-latex, draw=red, fill=red] (M-6-5) -- (M-7-6);
\draw[latex-latex, draw=red, fill=red] (M-7-6) -- (M-6-7);
\draw[latex-latex, draw=red, fill=red] (M-6-7) -- (M-7-8);
\draw[latex-latex, draw=red, fill=red] (M-7-8) -- (M-6-9);
\draw[latex-latex, draw=red, fill=red] (M-6-9) -- (M-7-10);
\draw[latex-latex, draw=red, fill=red] (M-7-10) -- (M-6-11);
\draw[latex-latex, draw=red, fill=red] (M-6-12) -- (M-7-12);
\draw[latex-latex, draw=red, fill=red] (M-5-13) -- (M-7-12);
\draw[latex-latex, draw=red, fill=red] (M-5-13) -- (M-7-13);
\draw[latex-latex, draw=red, fill=red] (M-6-12) -- (M-7-13);
\draw[latex-latex, draw=blue, fill=blue] (M-8-3) -- (M-7-2);
\draw[latex-latex, draw=blue, fill=blue] (M-8-3) -- (M-7-4);
\draw[latex-latex, draw=blue, fill=blue] (M-8-5) -- (M-7-3);
\draw[latex-latex, draw=blue, fill=blue] (M-8-5) -- (M-7-6);
\draw[latex-latex, draw=blue, fill=blue] (M-8-10) -- (M-7-8);
\draw[latex-latex, draw=blue, fill=blue] (M-8-10) -- (M-7-12);
\draw[latex-latex, draw=blue, fill=blue] (M-8-12) -- (M-7-10);
\draw[latex-latex, draw=blue, fill=blue] (M-8-12) -- (M-7-13);
\pattern[pattern=north east lines, pattern color=orange, draw opacity=0.6]
	($(M-1-1.south east)!0.5!(M-2-2.north west)$)
	--($(M-5-1.north east)!0.5!(M-4-2.south west)$)
	--($(M-5-10.north east)!0.5!(M-4-11.south west)$)
	--($(M-3-10.south west)!0.5!(M-4-11.north east)$)
	--($(M-4-11.north east)!0.5!(M-3-12.south west)$)
	--($(M-5-11.north east)!0.5!(M-4-12.south west)$)
	--($(M-5-12.north east)!0.5!(M-4-13.south west)$)
	--($(M-5-13.south west)!0.5!(M-6-12.north east)$)
	--($(M-6-14.north west)!0.5!(M-5-13.south east)$)
	--($(M-2-13.north east)!0.5!(M-1-14.south west)$)
	--($(M-1-1.south east)!0.5!(M-2-2.north west)$);
\pattern[pattern=north west lines, pattern color=green, draw opacity=0.6]
	($(M-5-1.south west)!0.5!(M-6-2.north east)$)
	--($(M-9-1.south west)!0.5!(M-8-2.north east)$)
	--($(M-9-13.north east)!0.5!(M-8-14.south west)$)
	--($(M-4-13.south east)!0.5!(M-5-14.north west)$)
	--($(M-5-12.north east)!0.5!(M-4-13.south west)$)
	--($(M-6-12.north east)!0.5!(M-5-13.south west)$)
	--($(M-6-11.north east)!0.5!(M-5-12.south west)$)
	--($(M-7-12.north west)!0.5!(M-6-11.south east)$)
	--($(M-7-10.north east)!0.5!(M-6-11.south west)$)
	--($(M-5-10.south east)!0.5!(M-6-11.north west)$)
	--($(M-5-1.south west)!0.5!(M-6-2.north east)$);
\end{tikzpicture}
\caption{The branes that we consider. Red lines denote T-duality, blue lines denote lifts/reductions and black lines denote S-dualities. The hashed green area contains all the branes contained within this non-geometric solution whilst the hashed orange area contains all the branes contained in the geometric solution.}
\label{fig:Brane}
\end{figure}
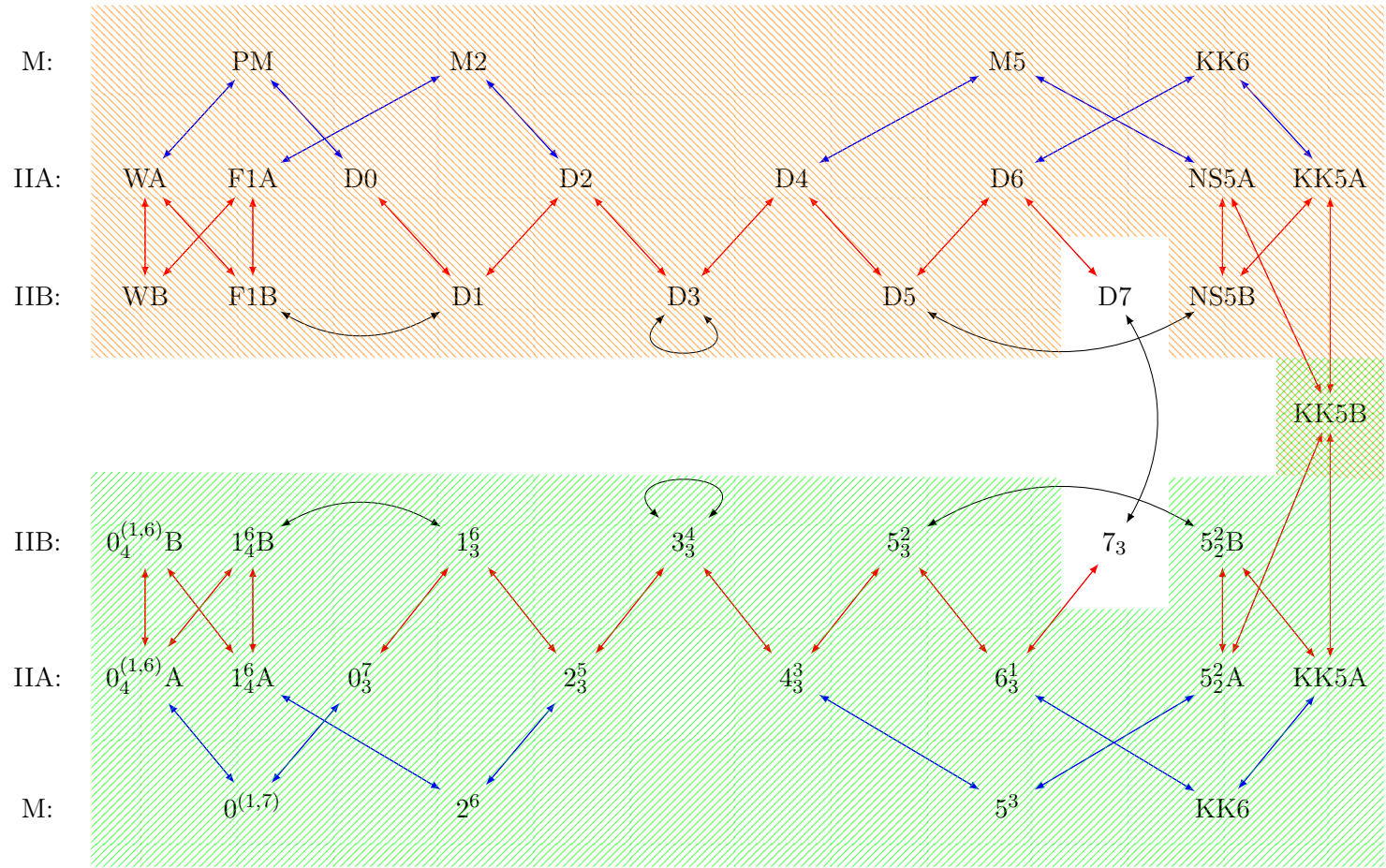
\end{landscape}
The set-up that we choose is as follows. We shall define the four-dimensional external space by the coordinates
\begin{align}
x^\mu = ( t, r, \theta, z),
\end{align} 
and take the external metric to be the same in both M-theory and Type IIB section, given by
\begin{align}\label{eq:Ext}
g_{\mu \nu} = \operatorname{diag} [ - {(HK^{-1})}^{-\frac{1}{2}}, {(HK)}^{\frac{1}{2}}, r^2 {(HK)}^{\frac{1}{2}}, {(HK^{-1})}^{\frac{1}{2}}] = {\hat{g}}_{\hat{\mu} \hat{\nu}},
\end{align}
where $H$ is a harmonic function in the $r$-$\theta$ plane
\begin{align}
H(r) & = h_0 + \sigma \ln \frac{\mu}{r},\\
K & = H^2 + \sigma^2 \theta^2 .
\end{align}
The determinant of the external metric shall be denoted $g_{(4)}$. The external metrics of all the branes shall be proportional to this e.g. $g_{\mu \nu}^{5^3} = {(HK^{-1})}^{\frac{1}{6}} g_{\mu\nu} = {\left| g_{(7)}^{5^3} \right|}^{-\frac{1}{2}} g_{\mu \nu}$.\par
The generalised metric is also chosen to be completely diagonal and thus contains only 56 components all along the diagonal. These are chosen to be 27 pairs of ${(HK^{-1})}^{\pm \frac{1}{2}}$ and one pair of ${(HK^{-1})}^{-1}$. From the form of the generalised vielbein one sees that this diagonal ansatz means that, under the simple coordinate swaps that we consider, every frame has no internal $A_{(3)}$ and $A_{(6)}$ potentials.\par
For the generalised gauge fields we shall take the ${\mathcal{B}}_{\mu \nu, \bullet}$ fields (which are only required to close the gauge structure of the theory and which is broken upon applying the section condition anyway) to vanish and choose the only non-vanishing components of $\mathcal{A}_\mu{}^M$ to be
\begin{align}
\mathcal{A}_t{}^M & = -H^{-1} K a^M,\\
\mathcal{A}_z{}^M & = -K^{-1} \theta \sigma {\tilde{a}}^M,
\end{align}
where $a^M$ and ${\tilde{a}}^M$ determine the direction the vector points in the generalised space. Under these simplifications, the covariantised generalised field strength reduces to the Abelian field strength and one may check that it does indeed satisfy the anti-self duality and Bianchi identities.
\subsection{M-Theory Section}
For this non-geometric solution, we take a subtly different approach to the self-dual solution constructed in \cite{Berman:2014hna}. There, they chose the generalised metric to be a genuine $E_{7(7)}$ element ${\tilde{\mathcal{M}}}_{MN}$, of the form \eqref{eq:E7GenMetric}, and implemented a relative scaling between the internal and external sectors $g_{\mu \nu} \mapsto {\left|g_{(7)}\right|}^{\frac{1}{2}} g_{\mu \nu}$ (reducing to the Einstein frame for the Type II solutions) as a conventional Kaluza-Klein decomposition of the 11-dimensional metric $\textrm{d} s^2 = g_{\mu \nu} \textrm{d} x^\mu \textrm{d} x^\nu + e^{2\alpha} g_{mn}{(\textrm{d} y^m + \mathcal{A}_\mu{}^m \textrm{d} x^\mu)}{(\textrm{d} y^n + \mathcal{A}_\nu{}^n \textrm{d} x^\nu)}$. 
 In particular, fixing a particular frame $\text{M}$ in \eqref{eq:E7GenMetric}, there is no information about the scaling of the external metric contained within ${\tilde{\mathcal{M}}}_{MN}$:
\begin{align}
{\tilde{\mathcal{M}}}_{MN} = \operatorname{diag} [{(g^{\text{M}}_{(7)})}^{\frac{1}{2}} g^{\text{M}}_{mn}; {(g^{\text{M}}_{(7)})}^{\frac{1}{2}} g_{\text{M}}^{mn,pq}; {(g^{\text{M}}_{(7)})}^{-\frac{1}{2}} g_{\text{M}}^{mn}; {(g^{\text{M}}_{(7)})}^{-\frac{1}{2}} g^{\text{M}}_{mn,pq}].
\end{align}
Whilst notionally intuitive, it requires the section condition to have been imposed and has an ad hoc feel in the sense that there appears to be nothing in the EFT framework that compels one to take this particular scaling; it is only imposed to match the known brane metrics. Here we make a small modification. We may use the scaling \eqref{eq:Rescaling} to encode this information directly into the generalised metric such that it forces the required relative scaling of the internal and external sector without imposing the section condition---a perhaps more natural description in terms of EFT. In particular let a superscript/subscript $\text{M}$ denote a given duality frame. Since we wish to impose the scaling of the external metric
\begin{align}\label{eq:ExtScaling}
g_{\mu \nu} = {\left|g_{(7)}^{\text{M}}\right|}^{\frac{1}{2}} g^{\text{M}}_{\mu \nu} \Rightarrow {|g^{\text{M}}_{(4)}|}^{-\frac{1}{4}} = {\left|g_{(7)}^{\text{M}}\right|}^{\frac{1}{2}} {|g_{(4)}|}^{-\frac{1}{4}},
\end{align}
we choose to define the $E_{7(7)} \times \mathbb{R}^+$ element
\begin{align}
{\mathcal{M}}_{MN} & = {|g_{(4)}|}^{-\frac{1}{4}} {\tilde{\mathcal{M}}}_{MN} = {\left|g^{\text{M}}_{(4)}\right|}^{-\frac{1}{4}} \operatorname{diag} [ g^{\text{M}}_{mn}; g_{\text{M}}^{mn,pq}; {(g^{\text{M}}_{(7)})}^{-1} g_{\text{M}}^{mn}; {(g^{\text{M}}_{(7)})}^{-1} g^{\text{M}}_{mn,pq}].
\end{align}
One sees that all terms, including the scaling of the external metric, are in the same duality frame. To be explicit, we are constructing a generalised metric $\mathcal{M}_{MN}$ which reproduces the backgrounds of the exotic branes in the following fashion\footnote{We can easily apply the same trick to the self-dual solution \cite{Berman:2014hna} using the same choice of scaling \eqref{eq:Rescaling} and interpreting $\mathcal{M}_{MN}^{(\text{there})} = {\tilde{\mathcal{M}}}_{MN}^{(\text{here})}$. To be explicit, mirroring \eqref{eq:NonGeom} this means that the generalised metric may now be brought to the similar form for all brane metrics:
\begin{align}
\begin{aligned}
\mathcal{M}_{MN} & = {\left|g^{\text{KK6}}_{(4)}\right|}^{-\frac{1}{4}} [ g^{\text{KK6}}_{mn}; g_{\text{KK6}}^{mn,pq}; {(g^{\text{KK6}}_{(7)})}^{-1} g_{\text{KK6}}^{mn}; {(g^{\text{KK6}}_{(7)})}^{-1} g^{\text{KK6}}_{mn,pq}]\\
	& = {\left|g^{\text{M5}}_{(4)}\right|}^{-\frac{1}{4}} [ g^{\text{M5}}_{mn}; g_{\text{M5}}^{mn,pq}; {(g^{\text{M5}}_{(7)})}^{-1} g_{\text{M5}}^{mn}; {(g^{\text{M5}}_{(7)})}^{-1} g^{\text{M5}}_{mn,pq}]\\
	& \qquad  \qquad \qquad \qquad \vdots
\end{aligned}
\end{align}
and so on for all the branes included in the self-dual solution. Explicitly, in coordinates where the metric of the KK6 takes the form
\begin{align}
\textrm{d} s^2_{\text{KK6}} = - \textrm{d} t^2 + H \textrm{d} {\vec{x}}^2_{(3)} + \textrm{d} {\vec{v}}^2_{(2)} + \textrm{d} {\vec{w}}^2_{(3)} + \textrm{d} y^2 + H^{-1} {(\textrm{d} z + A \cdot \textrm{d} {\vec{x}}_{(3)})}^2,
\end{align}
where the external and internal spaces are spanned by the coordinates $x^\mu = (t,{\vec{x}}_{(3)})$ and $y^m = ({\vec{v}}_{(2)}, {\vec{w}}_{(3)}, \omega, z)$ respectively, the generalised metric is given by
\begin{align}
\mathcal{M}_{MN} & = {|g_{(4)}|}^{-\frac{1}{4}} \operatorname{diag} [ H^{-\frac{1}{2}} \delta_{(3)}, H^{-\frac{1}{2}} \delta_{(3)}, H^{-\frac{1}{2}}, H^{-\frac{3}{2}}; ,\ldots]\\
& = {\left|g^{\text{KK6}}_{(4)}\right|}^{-\frac{1}{4}} \operatorname{diag} [ \delta_{(3)}, \delta_{(3)}, 1, H^{-1}; \ldots],
\end{align}
where $g_{(4)}$ is the determinant of the external metric $g_{\mu \nu} = \operatorname{diag} [H^{-\frac{1}{2}}, H^{\frac{1}{2}} \delta_{(3)}]$. The remaining component of the generalised metric may be reconstructed from the given data as a trivial lift if required. Note that the self-dual solution given there is harmonic in three dimensions, unlike the solution given here which is harmonic in two and, in particular, $\textrm{d} A = \star_3 \textrm{d} H$ with $H = 1 + \frac{1}{\sqrt{\vec{x}_{(3)}^2}}$.}:
\begin{align}\label{eq:NonGeom}
\begin{aligned}
\mathcal{M}_{MN} & = {\left|g^{5^3}_{(4)}\right|}^{-\frac{1}{4}} \operatorname{diag} [ g^{5^3}_{mn}; g_{5^3}^{mn,pq}; {(g^{5^3}_{(7)})}^{-1} g_{5^3}^{mn}; {(g^{5^3}_{(7)})}^{-1} g^{5^3}_{mn,pq}]\\
	& = {\left|g^{2^6}_{(4)}\right|}^{-\frac{1}{4}} \operatorname{diag} [ g^{2^6}_{mn}; g_{2^6}^{mn,pq}; {(g^{2^6}_{(7)})}^{-1} g_{2^6}^{mn}; {(g^{2^6}_{(7)})}^{-1} g^{2^6}_{mn,pq}]\\
	& = {\left|g^{0^{(1,7)}}_{(4)}\right|}^{-\frac{1}{4}} \operatorname{diag} [ g^{0^{(1,7)}}_{mn}; g_{0^{(1,7)}}^{mn,pq}; {(g^{0^{(1,7)}}_{(7)})}^{-1} g_{0^{(1,7)}}^{mn}; {(g^{0^{(1,7)}}_{(7)})}^{-1} g^{0^{(1,7)}}_{mn,pq}]\\
	& = {\left|g^{\text{KK6}}_{(4)}\right|}^{-\frac{1}{4}} \operatorname{diag} [ g^{\text{KK6}}_{mn}; g_{\text{KK6}}^{mn,pq}; {(g^{\text{KK6}}_{(7)})}^{-1} g_{\text{KK6}}^{mn}; {(g^{\text{KK6}}_{(7)})}^{-1} g^{\text{KK6}}_{mn,pq}]\\
	& \qquad \qquad \qquad \qquad \vdots
\end{aligned}
\end{align}
where the vertical dots represent all the Type IIA and Type IIB branes (barring the $7_3$ as discussed above) listed in Tables \ref{tab:NonGeomIIA} and \ref{tab:NonGeomIIB}.\par
In any given frame, the 56 components of the generalised metric split into 27 components of $(HK^{-1})^{\frac{1}{2}}$, 27 components of ${(HK^{-1})}^{-\frac{1}{2}}$ and one component each of ${(HK^{-1})}^{\frac{3}{2}}$ and ${(HK^{-1})}^{-\frac{3}{2}}$. The two components of the EFT vector always point in these distinguished directions.\par
We split the seven internal coordinates into
\begin{align}
y^m = (\xi, \chi, w^a),
\end{align}
where $a = 1, \ldots, 5$. These are promoted to the 56 coordinates of the extended internal space in EFT, indexed by $M=1, \ldots, 56$ which we order according to
\begin{align}\label{eq:IntCoords}
Y^M = ( Y^\xi, Y^\chi, Y^a; Y_{\xi \chi}, Y_{\xi a}, Y_{\chi a}, Y_{ab}; Y_{\xi} Y_{\chi}, Y_a; Y^{\xi a} Y^{\chi a}, Y^{ab}),
\end{align}
where the coordinates with two labels are antisymmetric. Since the generalised metric that we consider here is diagonal, this unambiguously defines the ordering of the components of the generalised matrix. Under this coordinate splitting, the configurations of the branes that we obtain is summarised in Table \ref{tab:NonGeomM}
\begin{table}[!ht]
\centering
\begin{tabulary}{\textwidth}{CLCCCCCCCCC}
\toprule
& & $t$ & $r$ & $\theta$ & $z$ & & $\xi$ & $\chi$ & $w^a$ &\\
\cmidrule{3-6}\cmidrule{8-10}
& $5^3$ & $\ast$ & $\bullet$ & $\bullet$ & $\circ$ & & $\circ$ & $\circ$ & $\ast$\\
& $2^6$ & $\ast$ & $\bullet$ & $\bullet$ & $\circ$ & & $\ast$ & $\ast$ & $\circ$\\
& $0^{(1,7)}$ & $\ast$ & $\bullet$ & $\bullet$ & $\circ$ & & $\odot$ & $\circ$ & $\circ$\\
& KK6 & $\ast$ & $\bullet$ & $\bullet$ & $\odot$ & & $\circ$ & $\ast$ & $\ast$\\
\bottomrule
\end{tabulary}
\caption{The configuration of M-theory branes that we consider. Asterisks $\ast$ denote worldvolume coordinates, empty circles $\circ$ denote smeared transverse coordinates and filled circles $\bullet$ denote coordinates that the harmonic function depends on. Finally, $\odot$ denotes an otherwise distinguished direction; the Hopf fibre for the monopole and the quadratic direction for the $0^{(1,7)}$}
\label{tab:NonGeomM}
\end{table}
Upon taking the M-theory section, the components of the EFT vector take on different roles, depending on the direction in which it points in the generalised space:
\begin{align}
\mathcal{A}_\mu{}^M \rightarrow (\mathcal{A}_\mu{}^m, \mathcal{A}_{\mu, mn}, \mathcal{A}_{\mu m}, \mathcal{A}_{\mu}{}^{mn}).
\end{align}
The first of these sources the conventional Kaluza-Klein cross-sector coupling, of the type seen in the $0^{(1,7)}$, and the third is related to the dual graviton. The remaining two components then source potentials; $\mathcal{A}_{\mu,mn} \sim A_{\mu mn}$ and $\mathcal{A}_\mu{}^{mn} \sim \epsilon^{mn p_1 \ldots p_5} A_{\mu p_1 \ldots p_5}$.\par
The components of the generalised metric are interchanged by a rotation on the internal space and are thus dependent on the frame chosen but for the $5^3$ frame, we have
\begin{align}
\phantom{\mathcal{M}_{MN}} &
\begin{alignedat}{2}\label{eq:53GM}
\mathllap{\mathcal{M}_{MN}} = {|g_{(4)}|}^{-\frac{1}{4}} \operatorname{diag}[
& {(HK^{-1})}^{\frac{1}{2}}, {(HK^{-1})}^{\frac{1}{2}}, {(HK^{-1})}^{-\frac{1}{2}} \delta_{(5)};\\
& {(HK^{-1})}^{-\frac{3}{2}}, {(HK^{-1})}^{-\frac{1}{2}} \delta_{(5)}, {(HK^{-1})}^{-\frac{1}{2}} \delta_{(5)}, {(HK^{-1})}^{\frac{1}{2}} \delta_{(10)};\\
& {(HK^{-1})}^{-\frac{1}{2}}, {(HK^{-1})}^{-\frac{1}{2}}, {(HK^{-1})}^{\frac{1}{2}} \delta_{(5)};\\
& {(HK^{-1})}^{\frac{3}{2}}, {(HK^{-1})}^{\frac{1}{2}} \delta_{(5)}, {(HK^{-1})}^{\frac{1}{2}} \delta_{(5)}, {(HK^{-1})}^{-\frac{1}{2}} \delta_{(10)}]
\end{alignedat}\\
&
\begin{alignedat}{2}\label{eq:53GenMetric}
= \smash{\left[\smash{\underbrace{ {(HK^{-1})}^{-\frac{1}{6}}}_{{\left|g_{(7)}^{5^3}\right|}^{\frac{1}{2}}}} {| g_{(4)}|}^{-\frac{1}{4}} \right] \operatorname{diag} [}
& {(HK^{-1})}^{\frac{2}{3}}, {(HK^{-1})}^{\frac{2}{3}}, {(HK^{-1})}^{-\frac{1}{3}} \delta_{(5)};\\
& {(HK^{-1})}^{-\frac{4}{3}}, {(HK^{-1})}^{-\frac{1}{3}} \delta_{(5)}, {(HK^{-1})}^{-\frac{1}{3}} \delta_{(5)}, {(HK^{-1})}^{\frac{2}{3}} \delta_{(10)};\\
& {(HK^{-1})}^{-\frac{1}{3}}, {(HK^{-1})}^{-\frac{1}{3}}, {(HK^{-1})}^{\frac{2}{3}} \delta_{(5)};\\
& {(HK^{-1})}^{\frac{5}{3}}, {(HK^{-1})}^{\frac{2}{3}} \delta_{(5)}, {(HK^{-1})}^{\frac{2}{3}} \delta_{(5)}, {(HK^{-1})}^{-\frac{1}{3}} \delta_{(10)}],
\end{alignedat}
\end{align}
where we have stressed the factor in the square braces is fixed by the scaling \eqref{eq:ExtScaling}. One may verify that this does indeed give the background of the $5^3$ upon applying the section condition if one identifies
\begin{align}
g_{mn}^{5^3} & = \operatorname{diag} [ {(HK^{-1})}^{\frac{2}{3}}, {(HK^{-1})}^{\frac{2}{3}}, {(HK^{-1})}^{-\frac{1}{3}} \delta_{(5)}],\\
{\left| g_{(4)}^{5^3}\right|}^{-\frac{1}{4}} & \coloneqq \left[ {(HK^{-1})}^{-\frac{1}{6}} {| g_{(4)}|}^{-\frac{1}{4}} \right] \Rightarrow g_{\mu \nu}^{5^3} = {(HK^{-1})}^{\frac{1}{6}} g_{\mu \nu}.
\end{align}
For this frame, we choose the EFT vectors to point out of section such that they do not contribute to the metric, but rather source the potentials of the $5^3$
\begin{align}\label{eq:53EFTVec}
\mathcal{A}_t{}^{\xi \chi} & = - H^{-1}K,\\
\mathcal{A}_{z,\xi \chi} & = - K^{-1} \theta \sigma.
\end{align}
We thus obtain the background of the $5^3$:
\begin{gather}\label{eq:53Metric}
\begin{gathered}
\textrm{d} s^2_{5^3} = {(HK^{-1})}^{-\frac{1}{3}} ( -\textrm{d} t^2 + \textrm{d} {\vec{w}}^2_{(5)} ) + {(HK^{-1})}^{\frac{2}{3}} ( \textrm{d} z^2 + \textrm{d} \xi^2 + \textrm{d} \chi^2) + H^{\frac{2}{3}} K^{\frac{1}{3}} ( \textrm{d} r^2 + r^2 \textrm{d} \theta^2),\\
A_{(3)} = -K^{-1} \theta \sigma \textrm{d} z \wedge \textrm{d} \xi \wedge \textrm{d} \chi, \qquad A_{(6)} = - H^{-1}K \textrm{d} t \wedge \textrm{d} w^1 \ldots \wedge \textrm{d} w^5.\\
\end{gathered}
\end{gather}
After applying the coordinate swap\footnote{This is most easily done on the form multiplied through by the braced prefactor in \eqref{eq:53GenMetric} i.e. apply the rotation on \eqref{eq:53GM} and then factoring out the new scaling ${(HK^{-1})}^{\frac{1}{6}} = {\left| g_{(7)}^{2^6}\right|}^{\frac{1}{2}}$ afterwards.}
\begin{gather}
Y^M \leftrightarrow Y_M,
\end{gather}
one obtains
\begin{align}
\begin{aligned}\label{eq:26GenMetric}
\mathcal{M}_{MN} = \left[ {(HK^{-1})}^{\frac{1}{6}}  {|g_{(4)}|}^{-\frac{1}{4}} \right] \operatorname{diag} [
& {(HK^{-1})}^{-\frac{2}{3}}, {(HK^{-1})}^{-\frac{2}{3}}, {(HK^{-1})}^{\frac{1}{3}} \delta_{(5)};\\
& {(HK^{-1})}^{\frac{4}{3}}, {(HK^{-1})}^{\frac{1}{3}} \delta_{(5)}, {(HK^{-1})}^{\frac{1}{3}} \delta_{(5)}, {(HK^{-1})}^{-\frac{2}{3}} \delta_{(10)};\\
& {(HK^{-1})}^{\frac{1}{3}}, {(HK^{-1})}^{\frac{1}{3}}, {(HK^{-1})}^{-\frac{2}{3}} \delta_{(5)};\\
& {(HK^{-1})}^{-\frac{5}{3}}, {(HK^{-1})}^{-\frac{2}{3}} \delta_{(5)}, {(HK^{-1})}^{-\frac{2}{3}} \delta_{(5)}, {(HK^{-1})}^{\frac{1}{3}} \delta_{(10)}].
\end{aligned}
\end{align}
Identifying
\begin{align}
g_{mn}^{2^6} & = \operatorname{diag}[{(HK^{-1})}^{-\frac{2}{3}}, {(HK^{-1})}^{-\frac{2}{3}}, {(HK^{-1})}^{\frac{1}{3}} \delta_{(5)}],\\
{\left| g_{(4)}^{2^6}\right|}^{-\frac{1}{4}} & \coloneqq \left[ {(HK^{-1})}^{\frac{1}{6}}  {|g_{(4)}|}^{-\frac{1}{4}} \right] \Rightarrow g_{\mu \nu}^{2^6} = {(HK^{-1})}^{-\frac{1}{6}} g_{\mu \nu},
\end{align}
and noting that the EFT vectors get inverted by the rotation, such that they still source potentials
\begin{align}
\mathcal{A}_t{}^{\xi \chi} & \mapsto \mathcal{A}_{t, \xi \chi},\\
\mathcal{A}_{z,\xi \chi} & \mapsto \mathcal{A}_z{}^{\xi \chi},
\end{align}
one obtains the background of the $2^6$:
\begin{gather}
\begin{gathered}
\textrm{d} s^2_{2^6} = {(HK^{-1})}^{-\frac{2}{3}} ( - \textrm{d} t^2 + \textrm{d} \xi^2 + \textrm{d} \chi^2 ) + {(HK^{-1})}^{\frac{1}{3}} ( \textrm{d} z^2 + \textrm{d} {\vec{w}}^2_{(5)} ) + H^{\frac{1}{3}} K^{\frac{2}{3}} ( \textrm{d} r^2 + r^2 \textrm{d} \theta^2 ),\\
A_{(3)} = - H^{-1} K \textrm{d} t \wedge \textrm{d} \xi \wedge \textrm{d} \chi, \qquad A_{(6)} = - K^{-1} \sigma \theta \textrm{d} z \wedge \textrm{d} w^1 \wedge \ldots \wedge \textrm{d} w^5.
\end{gathered}
\end{gather}
Further applying the rotations
\begin{gather}
Y^{ab} \leftrightarrow Y_{ab}\\
Y^{\xi} \leftrightarrow Y_{\xi \chi}, \qquad Y_\xi \leftrightarrow Y^{\xi \chi}\\
Y^\chi \leftrightarrow Y_\chi\\
Y^{\chi a} \leftrightarrow Y_{\chi a},
\end{gather}
one obtains the generalised metric
\begin{align}
\begin{aligned}\label{eq:017GenMetric}
\mathcal{M}_{MN} = \left[ {(HK^{-1})}^{\frac{1}{2}}  {|g_{(4)}|}^{-\frac{1}{4}} \right] \operatorname{diag} [
& HK^{-1}, 1, \delta_{(5)};\\
& H^{-1}K, H^{-1}K \delta_{(5)}, \delta_{(5)}, \delta_{(10)};\\
& {(HK^{-1})}^{-2}, H^{-1}K, H^{-1}K \delta_{(5)};\\
& 1, \delta_{(5)}, H^{-1}K \delta_{(5)}, H^{-1}K \delta_{(10)};],
\end{aligned}
\end{align}
which is consistent with the background of the $0^{(1,7)}$ if one identifies
\begin{align}
g_{mn}^{0^{(1,7)}} & = \operatorname{diag} [ HK^{-1}, 1, \delta_{(5)}],\\
{\left| g_{(4)}^{0^{(1,7)}} \right|}^{-\frac{1}{4}} & \coloneqq \left[ {(HK^{-1})}^{\frac{1}{2}}  {|g_{(4)}|}^{-\frac{1}{4}} \right] \Rightarrow g_{\mu \nu}^{0^{(1,7)}} = {(HK^{-1})}^{-\frac{1}{2}} g_{\mu \nu}.
\end{align}
Combining with the generalised vectors,
\begin{align}
\mathcal{A}_{t,\xi \chi} & \mapsto \mathcal{A}_t{}^\xi,\\
\mathcal{A}_z{}^{\xi \chi} & \mapsto \mathcal{A}_{z, \xi}.
\end{align}
one obtains the background of the $0^{(1,7)}$
\begin{align}
\textrm{d} s^2_{0^{(1,7)}} & = - H^{-1}K \textrm{d} t^2 + {\vec{w}}_{(5)}^2 + \textrm{d} z^2 + \textrm{d} \chi^2 + HK^{-1} {(\textrm{d} \xi - H^{-1}K \textrm{d} t)}^2 + K (\textrm{d} r^2 + r^2 \textrm{d} \theta^2).
\end{align}
Then applying the swap
\begin{align}
Y^{M} \leftrightarrow Y_M,
\end{align}
one obtains
\begin{align}
\begin{aligned}\label{eq:KK6GenMetric}
\mathcal{M}_{MN}= \left[ {(HK^{-1})}^{-\frac{1}{2}}  {|g_{(4)}|}^{-\frac{1}{4}} \right] \operatorname{diag} [
& {(HK^{-1})}^{-1}, 1, \delta_{(5)};\\
& HK^{-1}, HK^{-1} \delta_{(5)}, \delta_{(5)}, \delta_{(10)};\\
& {(HK^{-1})}^2, HK^{-1}, HK^{-1} \delta_{(5)};\\
& 1, \delta_{(5)}, HK^{-1} \delta_{(5)}, HK^{-1} \delta_{(10)}].
\end{aligned}
\end{align}
One may verify that this is sourced by the background
\begin{align}
g_{mn}^{\text{KK6}} & = \operatorname{diag} [ H^{-1}K, 1, \delta_{(5)}],\\
{\left| g_{(4)}^{\text{KK6}} \right|}^{-\frac{1}{4}} & \coloneqq \left[ {(HK^{-1})}^{-\frac{1}{2}} {|g_{(4)}|}^{-\frac{1}{4}} \right] \Rightarrow g_{\mu \nu}^{\text{KK6}} = {(HK^{-1})}^{\frac{1}{2}} g_{\mu \nu}.
\end{align}
The EFT vectors are rotated to
\begin{align}
\mathcal{A}_t{}^\xi & \mapsto \mathcal{A}_{t,\xi},\\
\mathcal{A}_{z, \xi} & \mapsto \mathcal{A}_z{}^\xi.
\end{align}
and so the latter sources a cross-sector coupling. We thus obtain the following background:
\begin{align}
\textrm{d} s^2 = - \textrm{d} t^2 + H \left( \textrm{d} r^2 + r^2 \textrm{d} \theta^2 \right) + HK^{-1} \textrm{d} z + H^{-1}K{\left( \textrm{d} \xi - K^{-1} \theta \sigma \textrm{d} z \right)}^2 + \textrm{d} \chi^2 + \textrm{d} {\vec{w}}^2_{(5)}.
\end{align}
However, focusing on the $\textrm{d}z$ and $\textrm{d} \xi$ terms, one finds that this is a disguised KK6 metric:
\begin{align}\label{eq:DisguisedKK}
(HK^{-1} + H^{-1}K^{-1} \theta^2 \sigma^2 )\textrm{d} z^2 + H^{-1}K  \textrm{d} \xi^2 + H^{-1}\sigma \theta \textrm{d} \xi \textrm{d} z & = H^{-1} \textrm{d} z^2 + ( H + H^{-1} \theta^2 \sigma^2) \textrm{d} \xi^2 + H^{-1} \theta \sigma \textrm{d} \xi \textrm{d} z\\
	& = H \textrm{d} \xi^2 + H^{-1} {(\textrm{d} z + \theta \sigma \textrm{d} \xi)}^2.
\end{align}
We thus obtain the metric of the KK6:
\begin{align}
\textrm{d} s^2_{\text{KK6}} & = - \textrm{d} t^2 + \textrm{d} \chi^2 + \textrm{d} {\vec{w}}^2_{(5)} + H (\textrm{d} r^2 + r^2 \textrm{d} \theta^2 + \textrm{d} \xi^2 ) + H^{-1} {(\textrm{d} z + \theta \sigma \textrm{d} \xi)}^2.
\end{align}
Note that the harmonic function is smeared in $\xi$ (one of the transverse directions).
\subsubsection{Type IIA Reduction}
The reduction to Type IIA solutions is not an independent solution to the section condition but rather a simple re-identification of degrees of freedom in terms of the Type IIA fields. We exploit the fact that the M-theory background, reduced on an isometry $\eta$, is equivalent to a Type IIA background (in the Einstein frame) under the identification
\begin{align}
\textrm{d} s^2_{\text{M}} & = e^{-\frac{\phi}{6}} \textrm{d} s^2_{\text{IIA},\text{E}} + e^{\frac{4\phi}{3}} {(\textrm{d} \eta + A_{(1)})}^2,\\
A_{(3)} & = B_{(2)} \wedge \textrm{d} \eta + C_{(3)}\label{eq:MPot},
\end{align}
where $A_{(p)}$ are the M-theory potentials and $B_{(q)}, C_{(r)}$ are the Type IIA NS-NS and R-R potentials. Grouping the surviving six coordinates into ${\check{y}}^{\check{m}}$, the reduction induces a decomposition of the generalised coordinates according to
\begin{align}\label{eq:IIACoords}
Y^M = (Y^{\check{m}}, Y^\eta; Y_{\check{m} \check{n}}, Y_{\check{m} \eta}; Y_{\check{m}}, Y_\eta; Y^{\check{m} \check{n}}; Y^{\check{m} \eta}).
\end{align}
One thus sees that the $g_{\eta \eta}$ component gives the ten-dimensional dilaton. Since the reduction is always along an internal direction, the external metric is rescaled by $e^{-\frac{\phi}{6}}$ as follows:
\begin{align}\label{eq:IIAExtFrame}
g^{\text{M}}_{\mu \nu} = e^{-\frac{\phi}{6}} {\check{g}}^{\text{A}}_{\check{\mu} \check{\nu}}, \qquad {\left|g^{\text{M}}_{(4)}\right|}^{-\frac{1}{4}} = e^{\frac{\phi}{6}} {\left|{\check{g}}^{\text{A}}_{(4)} \right|}^{-\frac{1}{4}},
\end{align}
with ${\check{y}}^{\check{\mu}} = (t,r,\theta z)$ still indexing the same coordinates on the external space as the M-theory section and ${\check{g}}_{\check{\mu} \check{\nu}}$ taking on the same numerical values as in the M-theory frame:
\begin{align}\label{eq:IIAExt}
{\check{g}}_{\check{\mu}\check{\nu}} = \operatorname{diag} [ - {(HK^{-1})}^{\frac{1}{2}}, {(HK)}^{\frac{1}{2}}, r^2 {(HK)}^{\frac{1}{2}}, {(HK^{-1})}^{\frac{1}{2}}].
\end{align}
The internal metric decomposes according to
\begin{align}
g^{\text{M}}_{mn} = \operatorname{diag} [ e^{-\frac{\phi}{6}} {\check{g}}_{\check{m} \check{n}}^{\text{A}}, e^{\frac{4\phi}{3}} ], \qquad g^{\text{M}}_{(7)} = e^{\frac{\phi}{3}} {\check{g}}^{\text{A}}_{(6)},
\end{align}
and so the generalised metric decomposes to\footnote{If one wishes to work in the string frame, the analogous decomposition is
\begin{align}
\mathcal{M}_{MN} = {\left| \check{g}_{(4)}\right|}^{-\frac{1}{4}}  \operatorname{diag} [ & {\check{g}}_{\check{m}\check{n}}, e^{2\phi}; e^{2\phi} {\check{g}}^{\check{m}\check{n}, \check{p}\check{q}}, {\check{g}}^{\check{m}\check{n}}; e^{4\phi} {\check{g}}_{(6)}^{-1} {\check{g}}^{\check{m} \check{n}}, e^{2\phi} {\check{g}}_{(6)}^{-1}; e^{2\phi} {\check{g}}_{(6)}^{-1} {\check{g}}_{\check{m}\check{n}, \check{p} \check{q}}, e^{4\phi} {\check{g}}_{(6)}^{-1}  {\check{g}}_{\check{m}\check{n}} ].
\end{align}
which follows simply from
\begin{align}
\textrm{d} s^2_{\text{M}} = e^{-\frac{2\phi}{3}} \textrm{d} s^2_{\text{IIA,s}} + e^{\frac{4\phi}{3}} {(\textrm{d} \eta + A_{(1})}^2.
\end{align}
}
\begin{align}\label{eq:IIAGenMetric}
\phantom{\mathcal{M}_{MN}} &
\begin{alignedat}{2}
\mathllap{\mathcal{M}_{MN}} = \smash{\underbrace{e^{\frac{\phi}{6}} {\left|{\check{g}}^{\text{A}}_{(4)} \right|}^{-\frac{1}{4}}}_{\text{from } {| g_{(4)}^{\text{M}}|}^{-\frac{1}{4}}} \operatorname{diag} [}
& e^{-\frac{\phi}{6}} {\check{g}}_{\check{m}\check{n}}^{\text{A}}, e^{\frac{4\phi}{3}}; e^{\frac{\phi}{3}} g^{\check{m} \check{n}, \check{p} \check{q}}_{\text{A}}, e^{-\frac{7\phi}{6}} g^{\check{m} \check{n}}_{\text{A}};\\
& e^{-\frac{\phi}{6}} {\check{g}}^{\text{A}}_{(6)}{}^{-1} {\check{g}}^{\check{m} \check{n}}_{\text{A}}, e^{-\frac{5\phi}{3}} {\check{g}}^{\text{A}}_{(6)}{}^{-1}; e^{-\frac{2\phi}{3}} {\check{g}}^{\text{A}}_{(6)}{}^{-1} {\check{g}}^{\text{A}}_{\check{m} \check{n}, \check{p} \check{q}}, e^{\frac{5\phi}{6}} {\check{g}}^{\text{A}}_{(6)}{}^{-1} {\check{g}}^{\text{A}}_{\check{m} \check{n}} ],
\end{alignedat}
\end{align}
where the ordering of components follows that of the coordinates \eqref{eq:IIACoords}. The EFT vector likewise decomposes to 
\begin{align}
\mathcal{A}_\mu{}^M \rightarrow  (\mathcal{A}_{\check{\mu}}{}^{\check{m}}, \mathcal{A}_{\check{\mu}}{}^{\eta}; \mathcal{A}_{\check{\mu}, \check{m} \check{n}}, \mathcal{A}_{\check{\mu}, \check{m} \eta}; \mathcal{A}_{\check{\mu}, \check{m}}, \mathcal{A}_{\check{\mu}, \eta}; \mathcal{A}_{\check{\mu}}{}^{\check{m}\check{n}}, \mathcal{A}_{\check{\mu}}{}^{\check{m} \eta}).
\end{align}
As before, the $\mathcal{A}_{\check{\mu}}{}^{\check{m}}$ components sources the KK-vector of the 4+6 split and the $\mathcal{A}_{\check{\mu}, \check{m}}$ is related to the dual graviton. Of the remaining components, the R-R potentials $C_{(1)}, C_{(3)}, C_{(5)}$ and $C_{(7)}$ are encoded in the components $\mathcal{A}_{\check{\mu}}{}^{\eta}, \mathcal{A}_{\check{\mu}, \check{m} \check{n}}, \mathcal{A}_{\check{\mu}}{}^{\check{m}\check{n}}$ and $\mathcal{A}_{\check{\mu} \eta}$ respectively (where the latter two are to be dualised on the internal space) and the NS-NS potentials $B_{(2)}$ and $B_{(6)}$ are held in $\mathcal{A}_{\check{\mu}, \eta \check{m}}$ and $\mathcal{A}_{\check{\mu}}{}^{\eta \check{m}}$ respectively.\par
In order to tabulate the brane configurations that we obtain, it will be convenient to split the five $w^a$ coordinates into $w^a = (u^{\textrm{a}}, v)$ with $\textrm{a} = 1, \ldots, 4$. The results are summarised in Table \ref{tab:NonGeomIIA}.\par
\begin{table}[!ht]
\centering
\begin{tabulary}{\textwidth}{CLLCCCCCCCCCC}
\toprule
& & & & & & & & & & \multicolumn{2}{c}{$w^a$} & \\
\cmidrule{11-12}
& Parent & & $t$ & $r$ & $\theta$ & $z$ & & $\xi$ & $\chi$ & $u^{\textrm{a}}$ & $v$ & \\
\cmidrule{4-7}\cmidrule{9-12}
& \multirow{2}{*}{$5^3$} & $5_2^2\text{A}$ & $\ast$ & $\bullet$ & $\bullet$ & $\circ$ & & $\times$ & $\circ$ & $\ast$ & $\ast$\\
& & $4_3^3$ & $\ast$ & $\bullet$ & $\bullet$ & $\circ$ & & $\circ$ & $\circ$ & $\ast$ & $\times$\\
& \multirow{2}{*}{$2^6$} & $2_3^5$ & $\ast$ & $\bullet$ & $\bullet$ & $\circ$ & & $\ast$ & $\ast$ & $\circ$ & $\times$\\
& & $1_4^6\text{A}$ & $\ast$ & $\bullet$ & $\bullet$ & $\circ$ & & $\times$ & $\ast$ & $\circ$ & $\circ$\\
& \multirow{2}{*}{$0^{(1,7)}$} & $0_4^{(1,6)}\text{A}$ & $\ast$ & $\bullet$ & $\bullet$ & $\circ$ & & $\odot$ & $\circ$ & $\circ$ & $\times$\\
& & $0_3^7$ & $\ast$ & $\bullet$ & $\bullet$ & $\circ$ & & $\times$ & $\circ$ & $\circ$ & $\circ$\\
& \multirow{2}{*}{KK6} & $6_3^1$ & $\ast$ & $\bullet$ & $\bullet$ & $\circ$ & & $\times$ & $\ast$ & $\ast$ & $\ast$\\
& & KK5A & $\ast$ & $\bullet$ & $\bullet$ & $\odot$ & & $\circ$ & $\ast$ & $\ast$ & $\times$\\
\bottomrule
\end{tabulary}
\caption{The configuration of the Type IIA branes that we consider. Note that the coordinates heading the columns are those of the M-theory section. A cross $\times$ denotes the direction that is being reduced on.}
\label{tab:NonGeomIIA}
\end{table}
Note that since some of the M-theory solutions are symmetric under certain coordinate transformations, these are not the only reductions that we could have done to obtain the Type IIA branes. For example, the generalised metrics of the $5^3$ and $2^6$ are invariant under the exchange $\xi \leftrightarrow \chi$ and so we could have obtained the $5_2^2\text{A}$ and $1_4^6\text{A}$ by reducing along $\chi$ instead of $\xi$ (although this further requires a re-identification of $\mathcal{A}_{\check{\mu}}{}^{\eta \check{m}} \sim - B_{(6)}$ and $\mathcal{A}_{\check{\mu}, \eta \check{m}} \sim - B_{(2)}$). Likewise, the role of $\chi$ is indistinguishable from any of the $w^a$ in the generalised metrics of the $0^{(1,7)}$ and KK6 and so we may have equally swapped $\chi$ with any one of the $w^a$ coordinates (which we nominally called $v$ in Table \ref{tab:NonGeomIIA}) and obtained a valid reduction to the KK6A and $0_4^{(1,6)}\text{A}$ by reducing on $\chi$ instead of $v$ (again with a suitable re-identification of the potentials to $\mathcal{A}_{\check{\mu}}{}^{\eta} \sim - C_{(1)}$ and $\mathcal{A}_{\check{\mu},\eta} \sim - C_{(7)}$). Nonetheless, the choice given in Table \ref{tab:NonGeomIIA} is the most symmetric choice of reductions.\par
Since all the reductions given above are along $\xi$ or $v$, we work through two examples in detail to illustrate the two reduction. The first is the reduction of the $5^3$ generalised metric along $\eta = \xi$. We begin by noting that the external metric scales as
\begin{align}
{\left| g_{(4)}^{5^3}\right|}^{-\frac{1}{4}} & = {(HK^{-1})}^{-\frac{1}{6}} {|g_{(4)}|}^{-\frac{1}{4}}\\
	& = {(HK^{-1})}^{-\frac{1}{6}} {|{\check{g}}_{(4)}|}^{-\frac{1}{4}}\\
	& \coloneqq e^{\frac{\phi}{6}} {\left| g_{(4)}^{5_2^2\text{A}} \right|}^{-\frac{1}{4}},
\end{align}
where we have used \eqref{eq:IIAExtFrame} and \eqref{eq:IIAExt}. The dilaton is obtained from the $\mathcal{M}_{\eta \eta} = \mathcal{M}_{\xi\xi}$ component of the generalised metric in the $5^3$ frame \eqref{eq:53GenMetric}:
\begin{align}
e^{\frac{4\phi}{3}} = {(HK^{-1})}^{\frac{2}{3}}
\end{align}
We thus obtain the scaling of the external metric as
\begin{align}
{\left|{\check{g}}^{5^2_2\text{A}}_{(4)}\right|}^{-\frac{1}{4}} & = {(HK^{-1})}^{-\frac{1}{4}} {| {\check{g}}_{(4)} |}^{-\frac{1}{4}} \Rightarrow {\check{g}}_{\check{\mu} \check{\nu}} = {(HK^{-1})}^{\frac{1}{4}} {\check{g}}_{\check{\mu}\check{\nu}}.
\end{align}
One may verify that the rest of the $5^3$ generalised metric \eqref{eq:53GenMetric} is sourced by the background
\begin{align}
{\check{g}}_{\check{m} \check{n}}^{5^2_2\text{A}} & = \operatorname{diag} \left[{(HK^{-1})}^{\frac{3}{4}}, {(HK^{-1})}^{-\frac{1}{4}} \delta_{(5)} \right],
\end{align}
according to \eqref{eq:IIAGenMetric}. Here, the reduced internal index spans $\check{m} =(\chi, w^a)$. Since the non-vanishing components of the EFT vector in this frame $\mathcal{A}_t{}^{\xi \chi}$ and $\mathcal{A}_{z,\xi \chi}$ lie along the reduction direction $\eta = \xi$, they must source the NS-NS potentials, giving the background of the $5^2_2\text{A}$ in the Einstein frame:
\begin{gather}
\begin{gathered}
\textrm{d} s^2_{5^2_2\text{A},\text{E}} = {(HK^{-1})}^{-\frac{1}{4}} \left( -\textrm{d} t^2 + \textrm{d} {\vec{w}}^2_{(5)} \right) + {(HK^{-1})}^{\frac{3}{4}} \left( \textrm{d} z^2 + \textrm{d} \chi^2 \right) + H^{\frac{3}{4}} K^{\frac{1}{4}} \left(\textrm{d} r^2 + r^2 \textrm{d} \theta^2 \right),\\
B_{(2)} = - K^{-1} \theta \sigma \textrm{d} z \wedge \textrm{d} \chi, \qquad B_{(6)} = - H^{-1} K \textrm{d} t^2 \wedge \textrm{d} w^1 \wedge \ldots \wedge \textrm{d} w^5,\\
e^{2(\phi -\phi_0)} = HK^{-1}.
\end{gathered}
\end{gather}
The second reduction of the $5^3$ is along $\eta = v \equiv w^5$ and so the coordinates of the Type IIA internal space are $y^{\check{m}} = (\xi, \chi, u^{\text{a}})$. We begin by rewriting the generalised metric of the $5^3$, given in \eqref{eq:53GenMetric}, adapted to this coordinate splitting: 
\begin{align}\label{eq:SplitGenMetric}
\begin{aligned}
\mathcal{M}_{MN} = {\left|g_{(4)}^{5^3}\right|}^{-\frac{1}{4}} \operatorname{diag}[
& {(HK^{-1})}^{\frac{2}{3}}, {(HK^{-1})}^{\frac{2}{3}}, {(HK^{-1})}^{-\frac{1}{3}} \delta_{(4)}, {(HK^{-1})}^{-\frac{1}{3}};\\
& {(HK^{-1})}^{-\frac{4}{3}}, {(HK^{-1})}^{-\frac{1}{3}} \delta_{(4)}, {(HK^{-1})}^{-\frac{1}{3}}, {(HK^{-1})}^{-\frac{1}{3}} \delta_{(4)}, {(HK^{-1})}^{-\frac{1}{3}},  {(HK^{-1})}^{\frac{2}{3}} \delta_{(6)}, {(HK^{-1})}^{\frac{2}{3}} \delta_{(4)};\\
& {(HK^{-1})}^{-\frac{1}{3}}, {(HK^{-1})}^{-\frac{1}{3}}, {(HK^{-1})}^{\frac{2}{3}} \delta_{(4)}, {(HK^{-1})}^{\frac{2}{3}};\\
& {(HK^{-1})}^{\frac{5}{3}}, {(HK^{-1})}^{\frac{2}{3}} \delta_{(4)}, {(HK^{-1})}^{\frac{2}{3}}, {(HK^{-1})}^{\frac{2}{3}} \delta_{(4)}, {(HK^{-1})}^{\frac{2}{3}},  {(HK^{-1})}^{-\frac{1}{3}} \delta_{(6)}, {(HK^{-1})}^{-\frac{1}{3}} \delta_{(4)}].
\end{aligned}
\end{align}
We now proceed as before and examine the prefactor and $\mathcal{M}_{vv}$ component to obtain the dilaton and relative scaling:
\begin{align}
{\left|g_{(4)}^{5^3} \right|}^{-\frac{1}{4}} & = {(HK^{-1})}^{-\frac{1}{6}} {|{\check{g}}_{(4)}|}^{-\frac{1}{4}} \coloneqq e^{\frac{\phi}{6}} {\left|{\check{g}}_{(4)}^{4_3^3} \right|}^{-\frac{1}{4}},\\
{(HK^{-1})}^{-\frac{1}{3}} & = e^{\frac{4\phi}{3}},
\end{align}
to obtain
\begin{align}
e^{2(\phi- \phi_0)} & = {(HK^{-1})}^{-\frac{1}{2}},\\
{\left| {\check{g}}_{(4)}^{4_3^3} \right|}^{-\frac{1}{4}} & = {(HK^{-1})}^{-\frac{1}{8}} {|{\check{g}}_{(4)}|}^{-\frac{1}{4}} \Rightarrow {\check{g}}_{\check{\mu}\check{\nu}}^{4_3^3} = {(HK^{-1})}^{\frac{1}{8}} {\check{g}}_{\check{\mu} \check{\nu}}.
\end{align}
One may verify that the rest of the generalised background is sourced, conforming to the Type IIA decomposition \eqref{eq:IIAGenMetric}, by
\begin{align}
{\check{g}}^{4^3_3}_{\check{m}\check{n}} & = \operatorname{diag} [ {(HK^{-1})}^{\frac{5}{8}}, {(HK^{-1})}^{\frac{5}{8}}, {(HK^{-1})}^{-\frac{3}{8}} \delta_{(4)}].
\end{align}
Since the direction being reduced on is not contained in the EFT vector, it reduces trivially to source the 5-form and 3-form R-R potentials. We thus obtain the background of the $4_3^3$ in the Einstein frame:
\begin{gather}
\begin{gathered}
\textrm{d}s^2_{4^3_3,\text{E}} = {(HK^{-1})}^{-\frac{3}{8}} \left( -\textrm{d} t^2 + \textrm{d} {\vec{w}}^2_{(4)} \right) + {(HK^{-1})}^{\frac{5}{8}} \left( \textrm{d} z^2 + \textrm{d} \xi^2 + \textrm{d} \chi^2 \right) + H^{\frac{5}{8}} K^{\frac{3}{8}} ( \textrm{d} r^2 + r^2 \textrm{d} \theta^2),\\
C_{(3)} = -K^{-1} \theta \sigma \textrm{d} z \wedge \textrm{d} \xi \wedge \textrm{d} \chi, \qquad C_{(5)} = - H^{-1} K \textrm{d} t^2 \wedge \textrm{d} w^1 \wedge \ldots \wedge \textrm{d} w^4,\\
e^{2(\phi - \phi_0)} = {(HK^{-1})}^{-\frac{1}{2}}.
\end{gathered}
\end{gather}
The remaining reductions of the M-theory solutions are done in exactly the same fashion as described above. The only complication is that the reduction of the KK6 reduces to the KK5A, along $\eta=v$, to give the non-canonical form
\begin{gather}
\textrm{d} s^2_{\text{KK5A}} = - \textrm{d} t^2 + \textrm{d} \chi^2 + \textrm{d} {\vec{u}}^2_{(4)} + H (\textrm{d} r^2 + r^2 \textrm{d} \theta^2) + HK^{-1} \textrm{d}z^2 + H^{-1}K {(\textrm{d} \xi - K^{-1} \theta \sigma) \textrm{d} z}^2,
\end{gather}
and one needs to apply the same trick \eqref{eq:DisguisedKK} to obtain the canonical form of the KK-monopole.
\subsection{Type IIB Section}
The generalised metric in the Type IIB section, in the absence of internal potentials, is given by
\begin{align}
\mathcal{M}_{MN} = {|{\hat{g}}_{(4)}|}^{-\frac{1}{4}} \operatorname{diag} [ {\hat{g}}_{\hat{m} \hat{n}}; {\hat{g}}^{\hat{m}\hat{n}} \hat{\gamma}^{\hat{\alpha} \hat{\beta}}; {\hat{g}}_{(6)}^{-1} {\hat{g}}_{\hat{m} \hat{k} \hat{p}, \hat{n} \hat{k} \hat{q}}; {\hat{g}}^{-1}_{(6)} {\hat{g}}_{\hat{m} \hat{n}} {\hat{\gamma}}_{\hat{\alpha} \hat{\beta}}; {\hat{g}}^{-1}_{(6)} {\hat{g}}^{\hat{m} \hat{n}} ],
\end{align}
where $\hat{m}, \hat{n} = 1, \ldots, 6$ index the Type IIB section and $\hat{\alpha}, \hat{\beta} =1,2$ are $\operatorname{SL}(2)$ indices. Accordingly, ${\hat{g}}_{\hat{m} \hat{n}}$ and ${\hat{\gamma}}_{\hat{\alpha} \hat{\beta}}$ are the metric on the internal space and torus respectively. The former we take to be of the same form as the M-theory section
\begin{align}
{\hat{g}}_{\hat{\mu} \hat{\nu}} = \operatorname{diag} [ - {(HK^{-1})}^{-\frac{1}{2}}, {(HK)}^{\frac{1}{2}}, r^2 {(HK)}^{\frac{1}{2}}, {(HK^{-1})}^{\frac{1}{2}} ],
\end{align}
and the latter parametrised by the axio-dilaton $\tau = A_{(0)} + i e^{-\phi}$ by
\begin{align}
{\hat{\gamma}}_{\hat{\alpha} \hat{\beta}} = \frac{1}{\operatorname{Im} \tau} \begin{pmatrix} {|\tau|}^2 & \operatorname{Re} \tau\\ \operatorname{Re} \tau & 1 \end{pmatrix}.
\end{align}
Additionally, we have defined ${\hat{g}}_{\hat{m} \hat{k} \hat{p}, \hat{n} \hat{l} \hat{q}} \coloneqq {\hat{g}}_{\hat{m} [ \hat{n}|} {\hat{g}}_{\hat{k}| \hat{l}|} {\hat{g}}_{\hat{p}|\hat{q}]}$. The EFT vector splits according to this coordinate decomposition to 
\begin{align}
\mathcal{A}_{\mu}{}^M \rightarrow ( \mathcal{A}_{\hat{\mu}}{}^{\hat{m}}, \mathcal{A}_{\hat{\mu}, \hat{m} \hat{\alpha}}, \mathcal{A}_{\hat{\mu}}{}^{\hat{m} \hat{k} \hat{p}}, \mathcal{A}_{\hat{\mu}}{}^{\hat{m} \hat{\alpha}}, \mathcal{A}_{\hat{\mu}, \hat{m}}).
\end{align}
As always, the $\mathcal{A}_{\hat{\mu}}{}^{\hat{m}}$ component is the KK-vector and $\mathcal{A}_{\hat{\mu} \hat{m}}$ is the dual graviton vector. The $\operatorname{SL}(2)$ index $\hat{\alpha}$ distinguishes between $C_{(2)}/B_{(2)}$ in $\mathcal{A}_{\mu \hat{m} \hat{\alpha}}$ and $C_{(6)}/ B_{(6)}$ in $\mathcal{A}_{\mu}{}^{\hat{m} \hat{\alpha}}$. In particular, when $\hat{\alpha} = 1$, the potential is of R-R type and when $\hat{\alpha} =2$, the potentials is of NS-NS type. Finally, the $\mathcal{A}_{\hat{\mu}}{}^{\hat{m} \hat{k} \hat{p}}$ component, once dualised on the internal space, sources the self-dual 4-form potential $A_{(4)}$.\par
In order to tabulate the brane configurations that we obtain, it will be necessary to split the five $w^a$ coordinates into $w^a = (\omega, {\bar{w}}^{\bar{a}}, {\ubar{w}}_{\ubar{a}})$ with $\bar{a} = 1,2$ and $\ubar{a} = 1,2$. The results are conveniently summarised in Table \ref{tab:NonGeomIIB}. 
\begin{table}[!ht]
\centering
\begin{tabulary}{\textwidth}{LCCCCCCCCC}
\toprule
& & & & & & & & $w^a$\\
\cmidrule{8-10}
& $t$ & $r$ & $\theta$ & $z$ & & $\zeta$ & $\omega$ & ${\bar{w}}^{\bar{a}}$ & ${\ubar{w}}^{\ubar{a}}$\\
\cmidrule{2-5}\cmidrule{7-10}
$5_2^2\text{B}$ & $\ast$ & $\bullet$ & $\bullet$ & $\circ$ & & $\circ$ & $\ast$ & $\ast$ & $\ast$\\
$5_3^2$ & $\ast$ & $\bullet$ & $\bullet$ & $\circ$ & & $\circ$ & $\ast$ & $\ast$ & $\ast$\\
$3_3^4$ & $\ast$ & $\bullet$ & $\bullet$ & $\circ$ & & $\circ$ & $\ast$ & $\circ$ & $\ast$\\
$1_3^6$ & $\ast$ & $\bullet$ & $\bullet$ & $\circ$ & & $\circ$ & $\ast$ & $\circ$ & $\circ$\\
$1_4^6\text{B}$ & $\ast$ & $\bullet$ & $\bullet$ & $\circ$ & & $\circ$ & $\ast$ & $\circ$ & $\circ$\\
$0_4^{(1,6)}\text{B}$ & $\ast$ & $\bullet$ & $\bullet$ & $\circ$ & & $\circ$ & $\circ$ & $\circ$ & $\circ$\\
KK5B & $\ast$ & $\bullet$ & $\bullet$ & $\odot$ & & $\ast$ & $\circ$ & $\ast$ & $\ast$\\
\bottomrule
\end{tabulary}
\caption{The configuration of the Type IIB branes that we consider.}
\label{tab:NonGeomIIB}
\end{table}
In order to identify the relation between the M-theory section and Type IIB section, we examine the $5^3$ generalised metric. Recall the internal coordinates of the M-theory section were $(\xi, \chi, w^a)$. Of these, the five $w^a$ coordinates enter directly into the Type IIB section but the remaining two coordinates $(\xi, \chi)$ become, loosely speaking, the $\operatorname{SL}(2)$ indices. In particular, denoting the six coordinates of the Type IIB section as ${\hat{y}}^{\hat{m}} = ( \zeta, w^a)$ and generating the remaining generalised coordinates ${\hat{Y}}^M$, the correspondence between the M-theory coordinates $Y^M$ and the Type IIB coordinates ${\hat{Y}}^M$ are given by
\begin{gather}
Y^{\xi} \equiv {\hat{Y}}^{\zeta 1}, \qquad Y_{\xi} \equiv {\hat{Y}}_{\zeta 1},\\
Y^{\chi} \equiv {\hat{Y}}_{\zeta}, \qquad Y^{\chi} \equiv {\hat{Y}}^{\zeta},\\
Y^a \equiv {\hat{Y}}^a,\\
Y^{ab} \equiv {\hat{Y}}^{abc}, \qquad Y_{ab} \equiv {\hat{Y}}^{\zeta a b},\\
Y^{\xi \chi} \equiv {\hat{Y}}^{\zeta 2}, \qquad Y_{\xi \chi} \equiv {\hat{Y}}_{\zeta 2}\label{eq:MIIB}.\\
Y^{\xi a} \equiv {\hat{Y}}_{a 1}, \qquad Y_{\xi a} \equiv {\hat{Y}}^{a 1},\\
Y^{\chi a} \equiv {\hat{Y}}^{a 2}, \qquad Y_{\chi a} \equiv {\hat{Y}}_{a 2}.
\end{gather}
This gives the generalised metric
\begin{align}
\phantom{\mathcal{M}_{MN}}&
\begin{alignedat}{2}
\mathllap{\mathcal{M}_{MN}} = {\left|g_{(4)} \right|}^{-\frac{1}{4}} \operatorname{diag} [
& {(HK^{-1})}^{\frac{1}{2}}, {(HK^{-1})}^{-\frac{1}{2}} \delta_{(5)};\\
& {(HK^{-1})}^{-\frac{1}{2}}, {(HK^{-1})}^{\frac{1}{2}} \delta_{(5)}, {(HK^{-1})}^{-\frac{3}{2}}, {(HK^{-1})}^{-\frac{1}{2}} \delta_{(5)};\\
& {(HK^{-1})}^{\frac{1}{2}} \delta_{(10)}, {(HK^{-1})}^{-\frac{1}{2}} \delta_{(10)};\\
& {(HK^{-1})}^{\frac{1}{2}}, {(HK^{-1})}^{-\frac{1}{2}} \delta_{(5)}, {(HK^{-1})}^{\frac{3}{2}}, {(HK^{-1})}^{\frac{1}{2}} \delta_{(5)};\\
& {(HK^{-1})}^{-\frac{1}{2}}, {(HK^{-1})}^{\frac{1}{2}} \delta_{(5)}]
\end{alignedat}\\
&
\begin{alignedat}{2}\label{eq:522BGenMetric}
= \left[ {(HK^{-1})}^{-\frac{1}{4}}  {|g_{(4)}|}^{-\frac{1}{4}} \right] \operatorname{diag} [
& {(HK^{-1})}^{\frac{3}{4}}, {(HK^{-1})}^{-\frac{1}{4}} \delta_{(5)};\\
& {(HK^{-1})}^{-\frac{1}{4}}, {(HK^{-1})}^{\frac{3}{4}} \delta_{(5)}, {(HK^{-1})}^{-\frac{5}{4}}, {(HK^{-1})}^{-\frac{1}{4}} \delta_{(5)};\\
& {(HK^{-1})}^{\frac{3}{4}} \delta_{(10)}, {(HK^{-1})}^{-\frac{1}{4}} \delta_{(10)};\\
& {(HK^{-1})}^{\frac{3}{4}}, {(HK^{-1})}^{-\frac{1}{4}} \delta_{(5)}, {(HK^{-1})}^{\frac{7}{4}}, {(HK^{-1})}^{\frac{3}{4}} \delta_{(5)};\\
& {(HK^{-1})}^{-\frac{1}{4}}, {(HK^{-1})}^{\frac{3}{4}} \delta_{(5)}],
\end{alignedat}
\end{align}
which is consistent with the background of the $5_2^2\text{B}$ if one identifies
\begin{align}
{\hat{g}}_{\hat{m}\hat{n}}^{5_2^2\text{B}} & = \operatorname{diag} [ {(HK^{-1})}^{\frac{3}{4}}, {(HK^{-1})}^{-\frac{1}{4}} \delta_{(5)}],\\
{\left| {\hat{g}}_{(4)}^{5^2_2\text{B}} \right|}^{-\frac{1}{4}} & \coloneqq \left[ {(HK^{-1})}^{-\frac{1}{4}}  {|{\hat{g}}_{(4)}|}^{-\frac{1}{4}} \right] \Rightarrow {\hat{g}}_{\hat{\mu} \hat{\nu}}^{5^2_2\text{B}} = {(HK^{-1})}^{\frac{1}{4}} {\hat{g}}_{\hat{\mu} \hat{\nu}},\\
\hat{\gamma}_{\hat{\alpha} \hat{\beta}} & = \operatorname{diag} [ {(HK^{-1})}^{- \frac{1}{2}}, {(HK^{-1})}^{\frac{1}{2}}], \Rightarrow \tau = i {(HK^{-1})}^{-\frac{1}{2}}.
\end{align}
Applying \eqref{eq:MIIB} on \eqref{eq:53EFTVec} to identify the direction that the EFT vector points in the Type IIB frame, we have
\begin{align}
\mathcal{A}_t{}^{\xi \chi} \rightarrow \mathcal{A}_t{}^{\zeta 2} = - H^{-1} K,\\
\mathcal{A}_{z, \xi \chi} \rightarrow \mathcal{A}_{z, \zeta 2} = - K^{-1} \theta \sigma,
\end{align}
of which the first sources $B_{(6)}$ (upon being dualised on the internal space) and the latter sources $B_{(2)}$. We thus obtain the background of the $5^2_2\text{B}$ in the Einstein frame:
\begin{gather}
\begin{gathered}
\textrm{d} s^2_{5^2_2\text{B},\text{E}} = {(HK^{-1})}^{-\frac{1}{4}} \left( -\textrm{d} t^2 + \textrm{d} {\vec{w}}^2_{(5)} \right) + {(HK^{-1})}^{\frac{3}{4}} \left( \textrm{d} z^2 + \textrm{d} \zeta^2 \right) + H^{\frac{3}{4}} K^{\frac{1}{4}} \left(\textrm{d} r^2 + r^2 \textrm{d} \theta^2 \right),\\
B_{(2)} = - K^{-1} \theta \sigma \textrm{d} z \wedge \textrm{d} \zeta, \qquad B_{(6)} = - H^{-1} K \textrm{d} t^2 \wedge \textrm{d} w^1 \wedge \ldots \wedge \textrm{d} w^5,\\ 
e^{2(\phi -\phi_0)} = HK^{-1}.
\end{gathered}
\end{gather}
Alternatively, we could have reduced the $5^3$ to the $5_3^2$---the S-dual of the $5_2^2\text{B}$. The coordinate identifications that we make between the M-theory and Type IIB coordinates are essentially the same as for the $5_2^2\text{B}$ except for with the $\operatorname{SL}(2)$ indices 1 and 2 exchanged, as expected. The generalised metric that one obtains is then
\begin{align}
\phantom{\mathcal{M}_{MN}}&
\begin{alignedat}{2}
\mathllap{\mathcal{M}_{MN}} = {\left|g_{(4)} \right|}^{-\frac{1}{4}} \operatorname{diag} [
& {(HK^{-1})}^{\frac{1}{2}}, {(HK^{-1})}^{-\frac{1}{2}} \delta_{(5)};\\
& {(HK^{-1})}^{-\frac{3}{2}}, {(HK^{-1})}^{-\frac{1}{2}} \delta_{(5)}, {(HK^{-1})}^{-\frac{1}{2}}, {(HK^{-1})}^{\frac{1}{2}} \delta_{(5)};\\
& {(HK^{-1})}^{\frac{1}{2}} \delta_{(10)}, {(HK^{-1})}^{-\frac{1}{2}} \delta_{(10)};\\
& {(HK^{-1})}^{\frac{3}{2}}, {(HK^{-1})}^{\frac{1}{2}} \delta_{(5)}, {(HK^{-1})}^{\frac{1}{2}}, {(HK^{-1})}^{-\frac{1}{2}} \delta_{(5)};\\
& {(HK^{-1})}^{-\frac{1}{2}}, {(HK^{-1})}^{\frac{1}{2}} \delta_{(5)}]
\end{alignedat}\\
&
\begin{alignedat}{2}\label{eq:523GenMetric}
= \left[ {(HK^{-1})}^{-\frac{1}{4}}  {|g_{(4)}|}^{-\frac{1}{4}} \right] \operatorname{diag} [
& {(HK^{-1})}^{\frac{3}{4}}, {(HK^{-1})}^{-\frac{1}{4}} \delta_{(5)};\\
& {(HK^{-1})}^{-\frac{5}{4}}, {(HK^{-1})}^{-\frac{1}{4}} \delta_{(5)}, {(HK^{-1})}^{-\frac{1}{4}}, {(HK^{-1})}^{\frac{3}{4}} \delta_{(5)};\\
& {(HK^{-1})}^{\frac{3}{4}} \delta_{(10)}, {(HK^{-1})}^{-\frac{1}{4}} \delta_{(10)};\\
& {(HK^{-1})}^{\frac{7}{4}}, {(HK^{-1})}^{\frac{3}{4}} \delta_{(5)}, {(HK^{-1})}^{\frac{3}{4}}, {(HK^{-1})}^{-\frac{1}{4}} \delta_{(5)};\\
& {(HK^{-1})}^{-\frac{1}{4}}, {(HK^{-1})}^{\frac{3}{4}} \delta_{(5)}],
\end{alignedat}
\end{align}
which is consistent with the background of the $5_3^2$ if one identifies
\begin{align}
{\hat{g}}_{\hat{m}\hat{n}}^{5_3^2} & = \operatorname{diag} [ {(HK^{-1})}^{\frac{3}{4}}, {(HK^{-1})}^{-\frac{1}{4}}\delta_{(5)}],\\
{\left| {\hat{g}}_{(4)}^{5^2_3} \right|}^{-\frac{1}{4}} & \coloneqq \left[ {(HK^{-1})}^{-\frac{1}{4}}  {|{\hat{g}}_{(4)}|}^{-\frac{1}{4}} \right] \Rightarrow {\hat{g}}_{\hat{\mu} \hat{\nu}}^{5^2_3} = {(HK^{-1})}^{\frac{1}{4}} {\hat{g}}_{\hat{\mu} \hat{\nu}},\\
\hat{\gamma}_{\hat{\alpha} \hat{\beta}} & = \operatorname{diag} [ {(HK^{-1})}^{\frac{1}{2}}, {(HK^{-1})}^{-\frac{1}{2}}] \Rightarrow \tau = i {(HK^{-1})}^{\frac{1}{2}}.
\end{align}
Additionally, in the identification of coordinates used here, we have
\begin{align}
\mathcal{A}_t{}^{\xi \chi} \rightarrow \mathcal{A}_t{}^{\zeta 1} = - H^{-1} K,\\
\mathcal{A}_{i, \xi \chi} \rightarrow \mathcal{A}_{i, \zeta 1} = - K^{-1} \theta \sigma,
\end{align}
and so these source $C_{(2)}$ and $C_{(6)}$ potentials respectively. We thus obtain the background of the $5_3^2$ with the inverted dilaton relative to the $5_2^2\text{B}$:
\begin{gather}
\begin{gathered}
\textrm{d} s^2_{5^2_3,\text{E}} = {(HK^{-1})}^{-\frac{1}{4}} \left( -\textrm{d} t^2 + \textrm{d} {\vec{w}}^2_{(5)} \right) + {(HK^{-1})}^{\frac{3}{4}} \left( \textrm{d} z^2 + \textrm{d} \zeta^2 \right) + H^{\frac{3}{4}} K^{\frac{1}{4}} \left(\textrm{d} r^2 + r^2 \textrm{d} \theta^2 \right),\\
C_{(2)} = - K^{-1} \theta \sigma \textrm{d} z \wedge \textrm{d} \zeta, \qquad C_{(6)} = - H^{-1} K \textrm{d} t^2 \wedge \textrm{d} w^1 \wedge \ldots \wedge \textrm{d} w^5,\\ 
e^{2(\phi -\phi_0)} = H^{-1}K.
\end{gathered}
\end{gather}
Equivalently, this is obtainable from the $5_2^2\text{B}$ generalised metric by the rotations
\begin{align}
{\hat{Y}}^{\hat{m}, 1} \leftrightarrow {\hat{Y}}^{\hat{m},2},\\
{\hat{Y}}_{\hat{m}, 1} \leftrightarrow {\hat{Y}}_{\hat{m},2}.
\end{align}
Once we are in a Type IIB frame, we are now free to apply rotations to the generalised metric as before. In order to rotate to the $3_3^4$, we must first split the five $w^a$ coordinates to a further 1+2+2 splitting. As such, let $w^a \rightarrow (\omega, {\bar{w}}^{\bar{a}},{\ubar{w}}^{\ubar{a}})$ where $\bar{a}$ and $\ubar{a}$ can each take on values 1 or 2. The generalised metric of the $5_2^2\text{B}$ in this coordinate splitting becomes
\begin{align}
\begin{aligned}
\mathcal{M}_{MN} = {\left|g_{(4)} \right|}^{-\frac{1}{4}} \operatorname{diag} [
& {(HK^{-1})}^{\frac{1}{2}}, {(HK^{-1})}^{-\frac{1}{2}}, {(HK^{-1})}^{-\frac{1}{2}} \delta_{(2)}, {(HK^{-1})}^{-\frac{1}{2}} \delta_{(2)};\\
& {(HK^{-1})}^{-\frac{1}{2}}, {(HK^{-1})}^{\frac{1}{2}}, {(HK^{-1})}^{\frac{1}{2}} \delta_{(2)}, {(HK^{-1})}^{\frac{1}{2}} \delta_{(2)},\\
	& \qquad \qquad {(HK^{-1})}^{-\frac{3}{2}}, {(HK^{-1})}^{-\frac{1}{2}}, {(HK^{-1})}^{-\frac{1}{2}} \delta_{(2)}, {(HK^{-1})}^{-\frac{1}{2}} \delta_{(2)};\\
& {(HK^{-1})}^{\frac{1}{2}} \delta_{(2)}, {(HK^{-1})}^{\frac{1}{2}} \delta_{(2)}, {(HK^{-1})}^{\frac{1}{2}}, {(HK^{-1})}^{\frac{1}{2}} \delta_{(4)}, {(HK^{-1})}^{\frac{1}{2}},\\
	& \qquad \qquad {(HK^{-1})}^{-\frac{1}{2}}, {(HK^{-1})}^{-\frac{1}{2}} \delta_{(4)}, {(HK^{-1})}^{-\frac{1}{2}}, {(HK^{-1})}^{-\frac{1}{2}} \delta_{(2)}, {(HK^{-1})}^{-\frac{1}{2}} \delta_{(2)};\\
& {(HK^{-1})}^{\frac{1}{2}}, {(HK^{-1})}^{-\frac{1}{2}}, {(HK^{-1})}^{-\frac{1}{2}} \delta_{(2)}, {(HK^{-1})}^{-\frac{1}{2}} \delta_{(2)},\\
	& \qquad \qquad {(HK^{-1})}^{\frac{3}{2}}, {(HK^{-1})}^{\frac{1}{2}}, {(HK^{-1})}^{\frac{1}{2}} \delta_{(2)}, {(HK^{-1})}^{\frac{1}{2}} \delta_{(2)};\\
& {(HK^{-1})}^{-\frac{1}{2}}, {(HK^{-1})}^{\frac{1}{2}}, {(HK^{-1})}^{\frac{1}{2}} \delta_{(2)}, {(HK^{-1})}^{\frac{1}{2}} \delta_{(2)}].
\end{aligned}
\end{align}
Note the positions of the semicolons that delimit each part of the generalised metric. Then, applying the rotations
\begin{gather}
{\hat{Y}}^{\zeta \ubar{a}\ubar{b}} \leftrightarrow {\hat{Y}}^{\zeta 2}, \qquad {\hat{Y}}^{\omega \ubar{a} \ubar{b}} \leftrightarrow {\hat{Y}}^{\zeta \bar{a} \bar{b}},\\
{\hat{Y}}^{\omega \bar{a} \bar{b}} \leftrightarrow {\hat{Y}}_{\zeta 2},\\
{\hat{Y}}^{\bar{a} \ubar{b}\ubar{c}} \leftrightarrow {\hat{Y}}^{\zeta \omega \bar{a}},\\
{\hat{Y}}_{\omega 2} \leftrightarrow {\hat{Y}}^{\omega 2}, \qquad {\hat{Y}}_{\bar{a} 2} \leftrightarrow {\hat{Y}}^{\bar{a} 2}, \qquad {\hat{Y}}^{\ubar{a}} \leftrightarrow {\hat{Y}}_{\ubar{a}}\label{eq:522B433},\\
{\hat{Y}}^{\ubar{a}1} \leftrightarrow {\hat{Y}}^{\ubar{a}2}.
\end{gather}
The generalised metric that one obtains is
\begin{align}
\begin{aligned}
\mathcal{M}_{MN} = {\left|g_{(4)} \right|}^{-\frac{1}{4}} \operatorname{diag} [
& {(HK^{-1})}^{\frac{1}{2}}, {(HK^{-1})}^{-\frac{1}{2}}, {(HK^{-1})}^{-\frac{1}{2}} \delta_{(2)}, {(HK^{-1})}^{\frac{1}{2}} \delta_{(2)};\\
& {(HK^{-1})}^{-\frac{1}{2}}, {(HK^{-1})}^{\frac{1}{2}}, {(HK^{-1})}^{\frac{1}{2}} \delta_{(2)}, {(HK^{-1})}^{-\frac{1}{2}} \delta_{(2)},\\
	& \qquad \qquad {(HK^{-1})}^{-\frac{1}{2}}, {(HK^{-1})}^{\frac{1}{2}}, {(HK^{-1})}^{\frac{1}{2}} \delta_{(2)}, {(HK^{-1})}^{-\frac{1}{2}} \delta_{(2)};\\
& {(HK^{-1})}^{-\frac{1}{2}} \delta_{(2)}, {(HK^{-1})}^{\frac{1}{2}} \delta_{(2)}, {(HK^{-1})}^{-\frac{1}{2}}, {(HK^{-1})}^{\frac{1}{2}} \delta_{(4)}, {(HK^{-1})}^{\frac{3}{2}},\\
	& \qquad \qquad {(HK^{-1})}^{-\frac{3}{2}}, {(HK^{-1})}^{-\frac{1}{2}} \delta_{(4)}, {(HK^{-1})}^{\frac{1}{2}}, {(HK^{-1})}^{-\frac{1}{2}} \delta_{(2)}, {(HK^{-1})}^{\frac{1}{2}} \delta_{(2)};\\
& {(HK^{-1})}^{\frac{1}{2}}, {(HK^{-1})}^{-\frac{1}{2}}, {(HK^{-1})}^{-\frac{1}{2}} \delta_{(2)}, {(HK^{-1})}^{\frac{1}{2}} \delta_{(2)},\\
	& \qquad \qquad {(HK^{-1})}^{\frac{1}{2}}, {(HK^{-1})}^{-\frac{1}{2}}, {(HK^{-1})}^{-\frac{1}{2}} \delta_{(2)}, {(HK^{-1})}^{\frac{1}{2}} \delta_{(2)};\\
& {(HK^{-1})}^{-\frac{1}{2}}, {(HK^{-1})}^{\frac{1}{2}}, {(HK^{-1})}^{\frac{1}{2}} \delta_{(2)}, {(HK^{-1})}^{-\frac{1}{2}} \delta_{(2)}],
\end{aligned}
\end{align}
which one may verify is sourced by the background
\begin{align}
{\hat{g}}^{3_3^4}_{\hat{m}\hat{n}} & = \operatorname{diag} [ {(HK^{-1})}^{\frac{1}{2}}, {(HK^{-1})}^{-\frac{1}{2}}, {(HK^{-1})}^{-\frac{1}{2}} \delta_{(2)}, {(HK^{-1})}^{\frac{1}{2}} \delta_{(2)}],\\
{\left| {\hat{g}}_{(4)}^{3_3^4} \right|}^{-\frac{1}{4}} & = {\left| {\hat{g}}_{(4)}\right|}^{-\frac{1}{4}} \Rightarrow {\hat{g}}_{\hat{\mu} \hat{\nu}}^{3_3^4} = {\hat{g}}_{\hat{\mu} \hat{\nu}},\\
{\hat{\gamma}}_{\hat{\alpha}\hat{\beta}} & = \operatorname{diag} [1,1] \Rightarrow \tau = i.
\end{align}
Since the dilaton is trivial, there is no distinction between string and Einstein frames and, indeed, the $3_3^4$ is self-dual under S-duality, much like the D3-brane. The EFT vector is rotated to
\begin{align}
\mathcal{A}_t{}^{\zeta 2} \mapsto \mathcal{A}_t{}^{\zeta \ubar{a} \ubar{b}},\\
\mathcal{A}_{z, \zeta 2} \mapsto \mathcal{A}_{z}{}^{\omega \bar{a}\bar{b}},
\end{align}
and these both source the self-dual 4-form potential $C_{(4)}$. Thus, the background that one obtains is that of the $3_3^4$:
\begin{gather}
\begin{gathered}
\textrm{d} s^2_{3_3^4} = {(HK^{-1})}^{-\frac{1}{2}} \left( - \textrm{d} t^2 + \textrm{d} \omega^2 + \textrm{d} {\vec{\bar{w}}}^2_{(2)} \right) + {(HK^{-1})}^{\frac{1}{2}} \left( \textrm{d} z^2 + \textrm{d} \zeta^2 + \textrm{d} {\vec{\ubar{w}}}^2_{(2)} \right) + {(HK)}^{\frac{1}{2}} \left( \textrm{d} r^2 + r^2 \textrm{d} \theta^2 \right),\\
C_{(4)} = - K^{-1} \theta \sigma \textrm{d} t \wedge \textrm{d} \omega \wedge \textrm{d} {\bar{w}}^{\bar{1}} \wedge \textrm{d} {\bar{w}}^{\bar{2}}, \qquad C_{(4)} = - H^{-1} K \textrm{d} z \wedge \textrm{d} \zeta \wedge \textrm{d} {\ubar{w}}^{\ubar{1}} \wedge \textrm{d} {\ubar{w}}^{\ubar{2}},\\
e^{2(\phi - \phi_0)} = 1.
\end{gathered}
\end{gather}
Note that the apparent distinction of the $\hat{\alpha}=2$ index in, for example, \eqref{eq:522B433} is a consequence of rotating from the $5_2^2\text{B}$; were we to rotate from the $5_3^2$ instead, they would be replaced with $\hat{\alpha}=1$.\par
Using the same coordinate splitting as for the $3_3^4$, we rotate the generalised internal coordinates according to
\begin{gather}
{\hat{Y}}^{\bar{a}} \leftrightarrow {\hat{Y}}_{\bar{a}}, \qquad {\hat{Y}}_{\bar{a} 1} \leftrightarrow {\hat{Y}}^{\bar{a}1}, \qquad {\hat{Y}}_{\ubar{a}2} \leftrightarrow {\hat{Y}}^{\ubar{a}2}, \qquad {\hat{Y}}_{\zeta 2} \leftrightarrow {\hat{Y}}^{\zeta 2},\\
{\hat{Y}}^{\zeta \ubar{a} \ubar{b}} \leftrightarrow {\hat{Y}}_{\omega 2}, \qquad {\hat{Y}}^{\omega \bar{a} \bar{b}} \leftrightarrow {\hat{Y}}^{\omega 2},\\
{\hat{Y}}^{\zeta \bar{a} \bar{b}} \leftrightarrow {\hat{Y}}^{\omega \ubar{a} \ubar{b}}, \qquad {\hat{Y}}^{\bar{a} \bar{b} \ubar{a|}} \leftrightarrow {\hat{Y}}^{\zeta \omega \ubar{a}},
\end{gather}
to obtain the generalised metric of the $1_4^6\text{B}$:
\begin{align}
\phantom{\mathcal{M}_{MN}}&
\begin{alignedat}{2}
\mathllap{\mathcal{M}_{MN}} = {\left|{\hat{g}}_{(4)} \right|}^{-\frac{1}{4}} \operatorname{diag} [
& {(HK^{-1})}^{\frac{1}{2}}, {(HK^{-1})}^{-\frac{1}{2}}, {(HK^{-1})}^{\frac{1}{2}} \delta_{(2)}, {(HK^{-1})}^{\frac{1}{2}} \delta_{(2)};\\
& {(HK^{-1})}^{-\frac{1}{2}}, {(HK^{-1})}^{\frac{1}{2}}, {(HK^{-1})}^{-\frac{1}{2}} \delta_{(2)}, {(HK^{-1})}^{-\frac{1}{2}} \delta_{(2)},\\
	& \qquad \qquad {(HK^{-1})}^{\frac{1}{2}}, {(HK^{-1})}^{\frac{3}{2}}, {(HK^{-1})}^{\frac{1}{2}} \delta_{(2)}, {(HK^{-1})}^{\frac{1}{2}} \delta_{(2)};\\
& {(HK^{-1})}^{-\frac{1}{2}} \delta_{(2)}, {(HK^{-1})}^{-\frac{1}{2}} \delta_{(2)}, {(HK^{-1})}^{\frac{1}{2}}, {(HK^{-1})}^{\frac{1}{2}} \delta_{(4)}, {(HK^{-1})}^{\frac{1}{2}},\\
	& \qquad \qquad {(HK^{-1})}^{-\frac{1}{2}}, {(HK^{-1})}^{-\frac{1}{2}} \delta_{(4)}, {(HK^{-1})}^{-\frac{1}{2}}, {(HK^{-1})}^{\frac{1}{2}} \delta_{(2)}, {(HK^{-1})}^{\frac{1}{2}} \delta_{(2)};\\
& {(HK^{-1})}^{\frac{1}{2}}, {(HK^{-1})}^{-\frac{1}{2}}, {(HK^{-1})}^{\frac{1}{2}} \delta_{(2)}, {(HK^{-1})}^{\frac{1}{2}} \delta_{(2)},\\
	& \qquad \qquad {(HK^{-1})}^{-\frac{1}{2}}, {(HK^{-1})}^{-\frac{3}{2}}, {(HK^{-1})}^{-\frac{1}{2}} \delta_{(2)}, {(HK^{-1})}^{-\frac{1}{2}} \delta_{(2)};\\
& {(HK^{-1})}^{-\frac{1}{2}}, {(HK^{-1})}^{\frac{1}{2}}, {(HK^{-1})}^{-\frac{1}{2}} \delta_{(2)}, {(HK^{-1})}^{-\frac{1}{2}} \delta_{(2)}]
\end{alignedat}\\
&
\begin{alignedat}{2}
= \left[ {(HK^{-1})}^{\frac{1}{4}} \cdot {\left|{\hat{g}}_{(4)} \right|}^{-\frac{1}{4}} \right] \operatorname{diag} [
& {(HK^{-1})}^{\frac{1}{4}}, {(HK^{-1})}^{-\frac{3}{4}}, {(HK^{-1})}^{\frac{1}{4}} \delta_{(2)}, {(HK^{-1})}^{\frac{1}{4}} \delta_{(2)};\\
& {(HK^{-1})}^{-\frac{3}{4}}, {(HK^{-1})}^{\frac{1}{4}}, {(HK^{-1})}^{-\frac{3}{4}} \delta_{(2)}, {(HK^{-1})}^{-\frac{3}{4}} \delta_{(2)},\\
	& \qquad \qquad {(HK^{-1})}^{\frac{1}{4}}, {(HK^{-1})}^{\frac{5}{4}}, {(HK^{-1})}^{\frac{1}{4}} \delta_{(2)}, {(HK^{-1})}^{\frac{1}{4}} \delta_{(2)};\\
& {(HK^{-1})}^{-\frac{3}{4}} \delta_{(2)}, {(HK^{-1})}^{-\frac{3}{4}} \delta_{(2)}, {(HK^{-1})}^{\frac{1}{4}}, {(HK^{-1})}^{\frac{1}{4}} \delta_{(4)}, {(HK^{-1})}^{\frac{1}{4}},\\
	& \qquad \qquad {(HK^{-1})}^{-\frac{3}{4}}, {(HK^{-1})}^{-\frac{3}{4}} \delta_{(4)}, {(HK^{-1})}^{-\frac{3}{4}}, {(HK^{-1})}^{\frac{1}{4}} \delta_{(2)}, {(HK^{-1})}^{\frac{1}{4}} \delta_{(2)};\\
& {(HK^{-1})}^{\frac{1}{4}}, {(HK^{-1})}^{-\frac{3}{4}}, {(HK^{-1})}^{\frac{1}{4}} \delta_{(2)}, {(HK^{-1})}^{\frac{1}{4}} \delta_{(2)},\\
	& \qquad \qquad {(HK^{-1})}^{-\frac{3}{4}}, {(HK^{-1})}^{-\frac{7}{4}}, {(HK^{-1})}^{-\frac{3}{4}} \delta_{(2)}, {(HK^{-1})}^{-\frac{3}{4}} \delta_{(2)};\\
& {(HK^{-1})}^{-\frac{3}{4}}, {(HK^{-1})}^{\frac{1}{4}}, {(HK^{-1})}^{-\frac{3}{4}} \delta_{(2)}, {(HK^{-1})}^{-\frac{3}{4}} \delta_{(2)}].
\end{alignedat}
\end{align}
This is sourced by the background
\begin{align}
{\hat{g}}^{1_4^6\text{B}}_{\hat{m}\hat{n}} & = \operatorname{diag} [ {(HK^{-1})}^{\frac{1}{4}}, {(HK^{-1})}^{-\frac{3}{4}},  {(HK^{-1})}^{\frac{1}{4}} \delta_{(2)},  {(HK^{-1})}^{\frac{1}{4}} \delta_{(2)}],\\
{\left| {\hat{g}}_{(4)}^{1_4^6\text{B}} \right|}^{-\frac{1}{4}} & = {(HK^{-1})}^{\frac{1}{4}} {\left| {\hat{g}}_{(4)}\right|}^{-\frac{1}{4}} \Rightarrow {\hat{g}}^{1_4^6 \text{B}}_{\hat{\mu}\hat{\nu}} = {(HK^{-1})}^{-\frac{1}{4}} {\hat{g}}_{\hat{\mu} \hat{\nu}},\\
{\hat{\gamma}}_{\hat{\alpha} \hat{\beta}} & = \operatorname{diag} [ {(HK^{-1})}^{\frac{1}{2}}, {(HK^{-1})}^{-\frac{1}{2}}] \Rightarrow \tau = i {(HK^{-1})}^{\frac{1}{2}}.
\end{align}
The EFT vector is rotated to
\begin{align}
\mathcal{A}_t{}^{\zeta\ubar{a} \ubar{b}} \mapsto \mathcal{A}_{t \omega 2},\\
\mathcal{A}_z{}^{\omega \bar{a} \bar{b}} \mapsto \mathcal{A}_z{}^{\omega 2},
\end{align}
and so must source the 2-form and 6-form NS-NS potentials. We thus obtain the background of the $1_4^6\text{B}$:
\begin{gather}
\begin{gathered}
\textrm{d} s^2_{1_4^6\text{B}} = {(HK^{-1})}^{-\frac{3}{4}} \left( - \textrm{d} t^2 + \textrm{d} \omega^2 \right) + {(HK^{-1})}^{\frac{1}{4}} \left( \textrm{d} z^2 + \textrm{d} \zeta^2 + \textrm{d} {\vec{\bar{w}}}_{(2)}^2 + \textrm{d} {\vec{\ubar{w}}}_{(2)}^2 \right) + H^{\frac{1}{4}} K^{\frac{3}{4}} \left( \textrm{d} r^2 + r^2 \textrm{d} \theta^2 \right),\\
B_{(2)} = -H^{-1} K \textrm{d} t \wedge \textrm{d} \omega, \qquad B_{(6)} = -K^{-1} \sigma \theta \textrm{d} z \wedge \textrm{d} \zeta \wedge \textrm{d} {\bar{w}}^{\bar{1}} \wedge \textrm{d} {\bar{w}}^{\bar{2}} \wedge \textrm{d} {\ubar{w}}^{\ubar{1}} \wedge \textrm{d} {\ubar{w}}^{\ubar{2}},\\
e^{2(\phi - \phi_0)} = {(HK^{-1})}^{-1}.
\end{gathered}
\end{gather}
The rotation of the $3_3^4$ to the dual $1_3^6$ is very similar, except with the $\operatorname{SL}(2)$ indices exchanged in the rotations above.\par
We begin by noting that we may clean up the generalised metric of the $1_4^6\text{B}$ by defining the set of coordinates $v^a = (\zeta, {\bar{w}}^{\bar{a}}, {\ubar{w}}_{\ubar{a}})$. Note that this is essentially the similar grouping to $w^a$ before, but with $\omega$ exchanged for $\zeta$. Then, the six coordinates of the Type IIB section are $(\omega, v^a)$ and the generalised metric becomes
\begin{align}
\begin{aligned}
\mathcal{M}_{MN} = {\left| {\hat{g}}_{(4)} \right|}^{-\frac{1}{4}}\operatorname{diag} [
& {(HK^{-1})}^{-\frac{1}{2}}, {(HK^{-1})}^{\frac{1}{2}} \delta_{(5)};\\
& {(HK^{-1})}^{\frac{1}{2}}, {(HK^{-1})}^{-\frac{1}{2}} \delta_{(5)}, {(HK^{-1})}^{\frac{3}{2}}, {(HK^{-1})}^{\frac{1}{2}} \delta_{(5)};\\
& {(HK^{-1})}^{-\frac{1}{2}} \delta_{(10)}, {(HK^{-1})}^{\frac{1}{2}} \delta_{(10)};\\
& {(HK^{-1})}^{-\frac{1}{2}}, {(HK^{-1})}^{\frac{1}{2}} \delta_{(5)}, {(HK^{-1})}^{-\frac{3}{2}}, {(HK^{-1})}^{-\frac{1}{2}} \delta_{(5)};\\
& {(HK^{-1})}^{\frac{1}{2}}, {(HK^{-1})}^{-\frac{1}{2}} \delta_{(5)}].
\end{aligned}
\end{align}
If we apply the rotations
\begin{gather}
{\hat{Y}}^{\omega} \leftrightarrow {\hat{Y}}_{\omega 2}, \qquad {\hat{Y}}_{\omega} \leftrightarrow {\hat{Y}}^{\omega 2},\\
{\hat{Y}}_{\omega 1} \leftrightarrow {\hat{Y}}^{\omega 1}, \qquad {\hat{Y}}_{a1} \leftrightarrow {\hat{Y}}^{a1},\\
{\hat{Y}}^{\omega a b} \leftrightarrow {\hat{Y}}^{c d e},
\end{gather}
we obtain the generalised metric of the $0_4^{(1,6)}\text{B}$:
\begin{align}
\phantom{\mathcal{M}_{MN}}&
\begin{alignedat}{2}
\mathllap{\mathcal{M}_{MN}} = {\left|{\hat{g}}_{(4)} \right|}^{-\frac{1}{4}} \operatorname{diag} [
& {(HK^{-1})}^{\frac{3}{2}},  {(HK^{-1})}^{\frac{1}{2}} \delta_{(5)};\\
& {(HK^{-1})}^{-\frac{1}{2}}, {(HK^{-1})}^{\frac{1}{2}} \delta_{(5)}, {(HK^{-1})}^{-\frac{1}{2}}, {(HK^{-1})}^{\frac{1}{2}}\delta_{(5)};\\
& {(HK^{-1})}^{\frac{1}{2}} \delta_{(10)}, {(HK^{-1})}^{-\frac{1}{2}} \delta_{(10)};\\
& {(HK^{-1})}^{\frac{1}{2}}, {(HK^{-1})}^{-\frac{1}{2}} \delta_{(5)}, {(HK^{-1})}^{\frac{1}{2}}, {(HK^{-1})}^{-\frac{1}{2}}\delta_{(5)};\\
& {(HK^{-1})}^{-\frac{3}{2}}, {(HK^{-1})}^{-\frac{1}{2}} \delta_{(5)}]
\end{alignedat}\\
&
\begin{alignedat}{2}
= \left[ {(HK^{-1})}^{\frac{1}{2}} \cdot {\left|{\hat{g}}_{(4)} \right|}^{-\frac{1}{4}} \right] \operatorname{diag} [
& HK^{-1}, \delta_{(5)};\\
& {(HK^{-1})}^{-1}, \delta_{(5)}, {(HK^{-1})}^{-1}, \delta_{(5)};\\
& \delta_{(10)}, {(HK^{-1})}^{-1} \delta_{(10)};\\
& 1, {(HK^{-1})}^{-1}\delta_{(5)}, 1, {(HK^{-1})}^{-1} \delta_{(5)};\\
& {(HK^{-1})}^{-2}, {(HK^{-1})}^{-1} \delta_{(5)}].
\end{alignedat}
\end{align}
One may verify that this is sourced by the background
\begin{align}
{\hat{g}}^{0_4^{(1,6)}\text{B}}_{\hat{m}\hat{n}} & = \operatorname{diag}[HK^{-1}, \delta_{(5)}],\\
{\left| {\hat{g}}_{(4)}^{0_4^{(1,6)}\text{B}}\right|}^{-\frac{1}{4}} & = {(HK^{-1})}^{\frac{1}{2}} {\left| {\hat{g}}_{(4)} \right|}^{-\frac{1}{4}} \Rightarrow {\hat{g}}_{\hat{\mu} \hat{\nu}}^{0_4^{(1,6)}\text{B}} = {(HK^{-1})}^{-\frac{1}{2}} {\hat{g}}_{\hat{\mu} \hat{\nu}},\\
{\hat{\gamma}}_{\hat{\alpha} \hat{\beta}} & = \operatorname{diag} [ 1,1] \Rightarrow \tau = i.
\end{align}
The EFT vector is rotated to source the Kaluza-Klein vector:
\begin{align}
\mathcal{A}_{t,\omega 2} & \mapsto \mathcal{A}_t{}^{\omega},\\
\mathcal{A}_{z}{}^{\omega 2} & \mapsto \mathcal{A}_{z \omega},
\end{align}
and so one obtains the background of the $0_4^{(1,6)}\text{B}$:
\begin{gather}
\begin{gathered}
\textrm{d} s^2_{0_4^{(1,6)}\text{B}} = - {(HK^{-1})}^{-1} \textrm{d} t^2 + HK^{-1} {\left( \textrm{d} \omega - H^{-1}K \textrm{d} t\right)}^2 + \textrm{d} z^2 + \textrm{d} {\vec{v}}^2_{(5)} + K \left( \textrm{d} r^2 + r^2 \textrm{d} \theta^2\right),\\
e^{2(\phi - \phi_0)} = 1.
\end{gathered}
\end{gather}
Applying the rotation
\begin{gather}
{\hat{Y}}^{\hat{m}} \leftrightarrow {\hat{Y}}_{\hat{m}}, \qquad {\hat{Y}}^{\hat{m} \hat{\alpha}} \leftrightarrow {\hat{Y}}_{\hat{m} \hat{\alpha}},\\
{\hat{Y}}^{\zeta a b} \leftrightarrow {\hat{Y}}^{cde},
\end{gather}
one obtains
\begin{align}
\phantom{\mathcal{M}_{MN}}&
\begin{alignedat}{2}
\mathllap{\mathcal{M}_{MN}}
= {\left|{\hat{g}}_{(4)} \right|}^{-\frac{1}{4}} \operatorname{diag} [
& {(HK^{-1})}^{-\frac{3}{2}}, {(HK^{-1})}^{-\frac{1}{2}} \delta_{(5)};\\
& {(HK^{-1})}^{\frac{1}{2}}, {(HK^{-1})}^{-\frac{1}{2}} \delta_{(5)}, {(HK^{-1})}^{\frac{1}{2}}, {(HK^{-1})}^{-\frac{1}{2}}\delta_{(5)};\\
& {(HK^{-1})}^{-\frac{1}{2}} \delta_{(10)}, {(HK^{-1})}^{\frac{1}{2}} \delta_{(10)};\\
& {(HK^{-1})}^{-\frac{1}{2}}, {(HK^{-1})}^{\frac{1}{2}} \delta_{(5)}, {(HK^{-1})}^{-\frac{1}{2}}, {(HK^{-1})}^{\frac{1}{2}}\delta_{(5)};\\
& {(HK^{-1})}^{\frac{3}{2}},  {(HK^{-1})}^{\frac{1}{2}} \delta_{(5)}]
\end{alignedat}\\
&
\begin{alignedat}{2}
= \left[ {(HK^{-1})}^{-\frac{1}{2}} \cdot {\left|{\hat{g}}_{(4)} \right|}^{-\frac{1}{4}} \right] \operatorname{diag} [
& H^{-1}K, \delta_{(5)};\\
& HK^{-1}, \delta_{(5)}, HK^{-1}, \delta_{(5)};\\
& \delta_{(10)}, HK^{-1} \delta_{(10)};\\
& 1, HK^{-1}\delta_{(5)}, 1, HK^{-1} \delta_{(5)};\\
& {(HK^{-1})}^2, HK^{-1}\delta_{(5)}].
\end{alignedat}
\end{align}
This is sourced by the background
\begin{align}
{\hat{g}}^{\text{KK5B}}_{\hat{m}\hat{n}} & = \operatorname{diag} [ H^{-1}K, \delta_{(5)}],\\
{\left| {\hat{g}}_{(4)}^{\text{KK5B}}\right|}^{-\frac{1}{4}} & = {(HK^{-1})}^{-\frac{1}{2}} {\left| {\hat{g}}_{(4)} \right|}^{-\frac{1}{4}} \Rightarrow {\hat{g}}_{\hat{\mu}\hat{\nu}}^{\text{KK5B}} = {(HK^{-1})}^{\frac{1}{2}} {\hat{g}}_{\hat{\mu}\hat{\nu}},\\
{\hat{\gamma}}_{\hat{\alpha}\hat{\beta}} & = \operatorname{diag} [1,1]\Rightarrow \tau = i.
\end{align}
The EFT vector is rotated to
\begin{align}
\mathcal{A}_t{}^\omega & \mapsto \mathcal{A}_{t,\omega},\\
\mathcal{A}_{z \omega} & \mapsto \mathcal{A}_z{}^\omega,\\
\end{align}
and so we obtain the background
\begin{gather}
\begin{aligned}
\textrm{d}s^2_{\text{KK5B}} & = -\textrm{d} t^2 + \textrm{d} {\vec{v}}^2_{(5)} + H (\textrm{d} r^2 + r^2 \textrm{d} \theta^2) + HK^{-1} \textrm{d} z + H^{-1}K {(\textrm{d} \omega - K^{-1} \theta \sigma \textrm{d} z)}^2\\
	& = - \textrm{d} t^2 + \textrm{d} {\vec{v}}^2_{(5)} + H( \textrm{d} r^2 + r^2 \textrm{d} \theta^2 + \textrm{d} \omega^2) + H^{-1}{( \textrm{d} z + \theta \sigma \textrm{d} \omega)}^2,
\end{aligned}\\
e^{2(\phi - \phi_0)} = 1.
\end{gather}
Recall that we switched coordinates in the $0_4^{(1,6)}\text{B}$ frame such that it has coordinates $(\omega, v^a)$ with $\zeta$ in $v^a$. Thus, one sees that the transverse three-space of the KK5B is spanned by $(r,\theta, \omega)$ whereas the three-transverse space in the KK6A was spanned by $(r,\theta, \zeta)$.
\subsection{Discussion}
The rotations amongst the branes presented here need not, in principle, be $E_7$ rotations (which would need to preserve $\Omega_{[MN]}$ and $c_{(MNPQ)}$); they may be part of the larger \emph{solution-generating} transformations, despite acting on an $E_7$ index.\par
Note that there the following hold:
\begin{align}
Y^M \leftrightarrow Y_M : \begin{cases}
\text{M2} \leftrightarrow \text{M5},\\
0^{(1,7)} \leftrightarrow \text{KK6},
\end{cases}
\end{align}
where $Y^M \leftrightarrow Y_M$ corresponds to $Y^m \leftrightarrow Y_m$ and $Y^{mn} \leftrightarrow Y_{mn}$. For the Type IIB section, we have
\begin{align}
{\hat{Y}}^M \leftrightarrow {\hat{Y}}_M: \begin{cases}
5_2^2\text{B} \leftrightarrow 1_4^6\text{B},\\
5_3^2 \leftrightarrow 1_3^6,\\
3_3^4 \leftrightarrow 3_3^4,\\
0_4^{(1,6)} \leftrightarrow \text{KK5B},
\end{cases}
\end{align}
where ${\hat{Y}}^M \leftrightarrow {\hat{Y}}_M$ corresponds to ${\hat{Y}}^{\hat{m}} \leftrightarrow {\hat{Y}}_{\hat{m}}, {\hat{Y}}^{\hat{m} \hat{\alpha}} \leftrightarrow {\hat{Y}}_{\hat{m} \hat{\alpha}}, {\hat{Y}}^{\zeta a b} \leftrightarrow {\hat{Y}}^{cde}$. Under this transformation, the $3_3^4$ is self-dual, but with the roles of $(\omega, {\bar{w}}_{(2)}) \leftrightarrow (\zeta, {\ubar{w}}_{(2)})$ in the resulting metric.\par
Note that this solution shows a strong resemblance to the self-dual solution if one notes that the combination $HK^{-1}$ is itself harmonic in two dimensions. More concretely, recall that the geometric solution was constructed with a three-dimensional transverse space. Denoting the coordinates of this transverse space as $(r,\theta, z)$ and the harmonic function as $\tilde{H}$, the potential $\tilde{A}$ sourcing this is obtained by solving $\textrm{d} \tilde{A} = \star_3 \textrm{d} \tilde{H}$. The non-geometric solution as presented above is obtained by smearing this solution over $z$ to give the $(H,A)$ used in the solution as presented above. However, noting that $HK^{-1}$ is itself harmonic in two dimensions, we obtain the following identification:
\begin{table}[H]
\centering
\begin{tabulary}{\textwidth}{LCLCL}
Geometric & & \multicolumn{3}{c}{Non-Geometric}\\
\cmidrule{1-1} \cmidrule{3-5}
3-dimensional on $\mathbb{R}^3$ & & Effective 2-dimensional on $\mathbb{R}^2 \times S^1$ & & 2-dimensional on $\mathbb{R}^2$\\
$\mathcal{A}_t{}^M = (1 - H^{-1}) a^M$ & & $\mathcal{A}_t{}^M = - H^{-1} K a^M$ & & $\mathcal{A}_t{}^M = - \hat{H}^{-1} a^M$\\
$\mathcal{A}_i{}^M = A_{i} {\tilde{a}}^M$ & & $\mathcal{A}_z{}^M = - K^{-1} \theta \sigma {\tilde{a}}^M$ & & $\mathcal{A}_z{}^M = \hat{A} {\tilde{a}}^M$\\
$\mathcal{M}_{MN} = \{ {\tilde{H}}^{\pm \frac{3}{2}}, {\tilde{H}}^{\pm \frac{1}{2}} \delta_{(27)} \}$ & & $\mathcal{M}_{MN} = \{ {(HK^{-1})}^{\pm \frac{3}{2}}, {(HK^{-1})}^{\pm \frac{1}{2}} \delta_{(27)} \}$ & & $\mathcal{M}_{MN} = \{ {\hat{H}}^{\pm \frac{3}{2}}, {\hat{H}}^{\pm \frac{1}{2}} \delta_{(27)} \}$\\
\end{tabulary}
\end{table}
where
\begin{align}
\textrm{d} {\tilde{A}} & = \star_3 \textrm{d} \tilde{H}, & \textrm{d} A & = \star_3 \textrm{d}H, & \textrm{d} \hat{A} & = \star_2 \textrm{d} \hat{H},\\
\tilde{H} & = 1 + \frac{{\tilde{h}}_0}{\sqrt{r^2 + z^2}}, & H & = h_0 + \sigma \ln \frac{\mu}{r}, & \hat{H} & = HK^{-1},\\
{\tilde{A}} & = \frac{{\tilde{h}}_0 z}{2r^2 \sqrt{r^2 + z^2}} \textrm{d} \theta, & A & = - \sigma \theta \textrm{d} z, & \hat{A} & = - \sigma \theta K^{-1},
\end{align}
and so we see a clear parallel between the geometric solution with transverse space $(r,\theta,z)$ and the non-geometric solution with transverse space $(r,\theta)$. The fact that the	 $\mathcal{A}_t{}^M \sim - \hat{H} a^M$ rather than $\mathcal{A}_t{}^M \sim (1-\hat{H}) a^M$ is mostly irrelevant considering the fact that the asymptotics of the harmonic function require some method of regularising the divergence (e.g. some anti-brane configuration around the exotic branes to absorb any flux along similar lines to what happens with the D8). More generally, we may construct the solutions as a genuine codimension-2 solution with an arbitrary harmonic function $\hat{H}(r,\theta)$ rather than the smearing a codimension-3 solution.\par
\section{Mapping out the Exotic States}\label{sec:Map}
Whilst we have given an EFT parent to a modest number of exotic branes in the previous sections, it has become increasingly clear that exotic branes are far more common than were previously thought. Indeed, the branes discussed in the solution above are only a tiny fraction of exotic branes that one may find in ExFT-like theories (specifically, those studying $E_9$ and larger). Here, we discuss a very simple algorithm for mapping out all dual branes based on the transformations of the masses under the various dualities. We then compare what we obtain with what has previously appeared in the literature where work has primarily focused on the mixed-symmetry potentials that these branes couple to. We warn the reader that after the following sections on notation, duality transformations and a worked example, we will simply list the duality orbits, stratified by their $g_s$-dependence. As $\alpha$ decreases, the size of the orbits generally grows, leaving us with a vast taxonomy. The hope is that this taxonomy will allow us to find patterns in the exotic brane structure and point to the existence of unifying solutions in EFT.
\subsection{Introduction}
\subsubsection{Notation}
In order to describe all the possible branes that one obtains, we require an extension of the notation introduced in Section \ref{sec:ExoticBranes}. We also define branes with a prefixed superscript to denote branes with an inverse dependence on given radii as follows:
\begin{align}
\text{Type II}: && \text{M}({}^ab_n^{(\ldots, d,c)}) & = \frac{\ldots {(R_{k_1} \ldots R_{k_d})}^3 {(R_{j_1} \ldots R_{j_c})}^2 {(R_{i_1} \ldots R_{i_b})}}{{(R_{l_1} \ldots R_{l_a})}g_s^nl_s^{1 - a + b + 2c + 3d + \ldots}},\\
\text{M-Theory}: && \text{M}({}^ab^{(\ldots, d,c)}) & = \frac{\ldots {(R_{k_1} \ldots R_{k_d})}^3 {(R_{j_1} \ldots R_{j_c})}^2 {(R_{i_1} \ldots R_{i_b})}}{ {(R_{l_1} \ldots R_{l_a})} l_p^{1 - a + b + 2c + 3d + \ldots}}.
\end{align}
For the work presented here, only the momentum mode in Type II is of this form.
\begin{align}
\text{M}(\text{P} = {}^10_0) = \frac{1}{R_1}
\end{align}
\subsubsection{Duality Transformations}\label{sec:Mapping}
In the following sections, we map out all the allowed exotic branes down to $\alpha=7$---the lowest power of $g_s^{-\alpha}$ admissible in $E_{7(7)}$ EFT\footnote{We shall work in the string frame throughout.}. The general scheme is to map out all allowed S- and T-duality transformations and lifts from a given brane. The result of a given transformation is identified by the mass of the resulting object. By noting that T-duality does not change the scaling of $g_s$, these are then organised by their mass scaling into T-duality orbits. Each figure in the proceeding pages correspond to a single T-duality orbit i.e. every brane in each figure may be reached from any other brane in the same figure by judicious T-dualities alone. The T-duality transformation along the direction $y$ is given by
\begin{align}\label{eq:TTrans}
T_y: R_y \mapsto \frac{l_s^2}{R_y}, \qquad g_s \mapsto \frac{l_s}{R_y} g_s.
\end{align}
We stress that this process only has a natural description in ExFT wherein the duality transformations correspond to different choices of section condition that allow winding mode dependences. It is a well-known fact that  the T-duality rules encoded in the Buscher rules or the reduction from M-theory to Type IIA both require an isometry but the ExFT description of this is simply the rotation of coordinates in and out of section which does not require an isometry. These extended theories thus afford us a much richer spectrum of branes since one can take duality transformation in directions which classically would not be allowed.\par
For example, in supergravity, whilst a codimension-1 brane may be T-dualised along the transverse direction after smearing the harmonic function in that direction, this removes any dependence of the harmonic function on any of the coordinates and thus become a simple constant which renders it equivalent to the trivial D9. The DFT description of this, however, still allows for a meaningful duality transformation since the dependence of the harmonic function is simply shifted to a dependence on a winding coordinate, rather than being lost entirely. Thus, one may still construct space-filling branes in DFT that remain non-trivial by virtue of this winding mode dependence. A similar story holds for reductions of M-theory branes; a codimension-1 brane in M-theory may be `reduced' along the transverse direction to yield a non-trivial codimension-0 solution in ten dimensions simply because the coordinate dependence is simply shifted out of section.\par
The dependence on winding modes pre-dates DFT and has been well-studied in the context of Gauged Linear Sigma Models (GLSM). By comparing their interpretations on either sides of the T-dual pair $(\text{NS5},\text{KK5})$, it was shown that such a winding mode dependence may be understood as worldsheet instanton corrections \cite{Tong:2002rq,Harvey:2005ab}. More specifically, the worldsheet instanton corrections of an H-monopole break the isometry in the $S^1$, localising it to an NS5 and this transfers over to the T-dual picture as the breaking of the isometry in the dual circle. Thus, one concludes that the information encoded in a dependence on dual coordinates is equivalent to that of worldsheet instanton corrections. More recently, the GLSM analysis was extended to include the $5_2^2$ \cite{Kimura:2013fda,Kimura2013,Lust:2017jox} and further studied in \cite{Kimura:2018} in the context of DFT with similar conclusions that winding mode dependences may be interpreted as worldsheet instanton corrections to the geometry.\par
It is easy to see that the T-duality rules given in \eqref{eq:TTrans} are, together equivalent to the general rule proposed in \cite{Lombardo:2016swq}
\begin{align}
\alpha = -n: \qquad \underbrace{a,a,\ldots, a}_p \xleftrightarrow{T_a}\underbrace{a, a, \ldots, a}_{n-p}.
\end{align}
The S-duals of each of the Type IIB branes, which map between the orbits/figures, are also given and are determined by the following transformations:
\begin{align}
S: g_s \mapsto \frac{1}{g_s}, \qquad l_s \mapsto g_s^{\frac{1}{2}} l_s.
\end{align}
Note that this does not affect the wrapping structure of the brane and only touches the $g_s$ scaling i.e. a $b_{n}^{(\dots, d,c)}$-brane is mapped to some $b_{n^\prime}^{(\ldots, d,c)}$-brane. Finally, the lift of each Type IIA brane is determined by using the relations between the ten- and eleven-dimensional constants
\begin{align}
\begin{rcases}
l_s & = \frac{l_p^{\frac{3}{2}}}{R_\natural^{\frac{1}{2}}}\\
g_s & = {\left( \frac{R_\natural}{l_p} \right)}^{\frac{3}{2}}
\end{rcases} \leftrightarrow
\begin{cases}
R_\natural & = l_s g_s\\
l_p & = g_s^{\frac{1}{3}} l_s
\end{cases}
\end{align}
This, in turn, indicates the existence of other Type IIA branes. The above procedure is repeated iteratively until all possible duality transformations and lifts have been saturated.\par
All the figures presented here were generated by saturating all possible S- and T-duality transformations as well as lifts/reductions. We have chosen to display any even-$\alpha$ branes that appear in both Type II theories as separate nodes such that the number of lines coming out of each brane is always equal to the number of T-dual partners that the brane possesses---this provides a simple verification that all possible T-duality transformations have been accounted for. Specifically, representing each $b_n^{(\ldots, d, c)}$-brane as a single node, one must always have $t$ lines emanating from it where
\begin{align}
t = \begin{cases} (b + c + d + \ldots ) + 1  & \text{ if codimension}\neq 0,\\
(b + c + d +  \ldots ) & \text{ if codimension}=0.
\end{cases}
\end{align}
Since T-dualising along a transverse direction produces a brane of 1 lower codimension, the special case of codimension-0 branes in eleven dimensions is precisely why these T-duality orbits close. For example, the $0_4^{(2,1,6)}$ (obtained from a double T-duality along the two transverse coordinates of the $0_4^{(1,6)}$) has only three T-dual partners, not four.\par
The branes presented here are `complete' to $g_s^{-7}$ in so far as all branes down to there whose existence is implied by the above rules are included. The missing figure references are all for branes of $\alpha \leq -8$ but these are also expected to fall into T-duality orbits. For example, at $g_s^{-8}$, the branes of higher $g_s$ scaling imply the existence of 64 branes in Type IIA and 26 branes in Type IIB. Another 190 further branes are required to organise these into eight complete T-duality orbits. The proliferation of branes is evident and it is not clear whether the process will terminate at finite $g_s^\alpha$. Already at $g_s^{-7}$, one finds the implied existence of branes down to $g_s^{-15}$ and at $g_s^{-8}$ there is an implied existence of branes down to $g_s^{-17}$ (the lift of an $0_8^{(6,1,2,0,0)}$ will give rise to a $0_{17}^{(1,0,0,0,0,0,0,6,1,1,0,0)}$ as one of its descendants). 
\subsubsection{A Partial Example}
To illustrate the procedure, we give a partial example below. Consider the $0_4^{(1,6)}$-brane in Type IIA. Its mass is given by
\begin{align}
\text{M}(0_4^{(1,6)}) = \frac{R_7^3 {(R_6 \ldots R_1)}^2}{g_s^4 l_s^{16}}.
\end{align}
We have three possible distinct T-duality transformations that we may apply (up to renaming of coordinates); a duality transformation along the cubic direction, the quadratic direction or a direction entirely transverse to the brane:
\begin{align}
\text{M}(0_4^{(1,6)}\text{A}) = \frac{R_7^3 {(R_6 \ldots R_1)}^2}{g_s^4 l_s^{16}} \rightarrow
\begin{cases}
\xrightarrow{T_8} \frac{R_7^3 {(R_6 \ldots R_1)}^2}{{\left(\frac{l_s}{R_8} g_s\right)}^4 l_s^{16}} = \frac{R_8^4 R_7^3 {(R_6 \ldots R_1)}^2}{g_s^4 l_s^{20}} = \text{M} (0^{(1,1,6)}_4\text{B}),\\
\xrightarrow{T_7} \frac{{\left(\frac{l_s^2}{R_7} \right)}^3 {(R_6 \ldots R_1)}^2}{{\left( \frac{l_s}{R_7} g_s \right)}^4 l_s^{16}} = \frac{R_7 {(R_6 \ldots R_1)}^2}{g_s^4 l_s^{14}} = \text{M} ( 1_4^6\text{B}),\\
\xrightarrow{T_6} \frac{R_7^3 {\left( \frac{l_s^2} R_6 \right)}^2 {(R_5 \ldots R_1)}^2}{{\left( \frac{l_s}{R_6} g_s \right)}^4 l_s^{16}} = \frac{R_7^3 {(R_6 \ldots R_1)}^2}{g_s^4 l_s^{16}} = \text{M} ( 0_4^{(1,6)}\text{B}).\\
\end{cases}
\end{align}
Of these, the first is a novel codimension-1 object that appears only because we are allowing transformations along non-isometric directions (if one is more careful, one should be able to obtain these in the standard supergravity picture by appropriate arraying and smearing of the $0_4^{(1,6)}$). Finally, note that the appearance of the $0_4^{(1,6)}$ in the Type IIB theory also means that one must have $0_4^{(1,1,6)}\text{A}$- and $1_4^6\text{A}$-branes as well. This is a manifestation of the even-$\alpha$ effect that was discussed earlier.\par
We now proceed with the example. The respective S-duals of these branes are given by
\begin{align}
\text{M} (0_4^{(1,1,6)}\text{B}) & \xrightarrow{S} \frac{R_8^4 R_7^3 {(R_6 \ldots R_1)}^2}{{\left( \frac{1}{g_s} \right)}^4 {\left( g_s^{\frac{1}{2}} l_s\right)}^{20}} = \frac{R_8^4 R_7^3 {(R_6 \ldots R_1)}^2}{g_s^{6} l_s^{10}} = \text{M}(0_6^{(1,1,6)}\text{B}),\\
\text{M} (1_4^6\text{B}) & \xrightarrow{S} \frac{R_7 {(R_6 \ldots R_1)}^2}{{\left( \frac{1}{g_s} \right)}^4 {\left( g_s^{\frac{1}{2}} l_s\right)}^{14}} = \frac{R_7 {(R_6 \ldots R_1)}^2}{g_s^3 l_s^{14}} = \text{M}(1_3^6\text{B}),\\
\text{M} ( 0_4^{(1,6)}\text{B}) & \xrightarrow{S} \frac{R_7 {(R_6 \ldots R_1)}^2}{{\left( \frac{1}{g_s} \right)}^4 {\left( g_s^{\frac{1}{2}} l_s\right)}^{16}} = \frac{R_7 {(R_6 \ldots R_1)}^2}{{\left( \frac{1}{g_s} \right)}^4 g_s^8 l_s^{16}} = \text{M} (0_4^{(1,6)}\text{B}).
\end{align}
Again, the last two are known results and it is only the $0_6^{(1,1,6)}\text{B}$ which is novel. Its existence means that there is at least one T-duality orbit at $g_s^{-6}$ which must be fleshed out. One must then map out all allowed S- and T-duals of these objects. Finally, we may lift the $0_4^{(1,6)}\text{A}$ to M-theory by rewriting it in terms of M-theory constants:
\begin{align}
\text{M}(0_4^{(1,6)}\text{A}) = \frac{R_7^3 {(R_6 \ldots R_1)}^2}{g_s^4 l_s^{16}} = \frac{R_7^3{(R_6 \ldots R_1)}^2 R_\natural^2}{l_p^{18}},
\end{align}
where $R_\natural$ is the M-theory circle. Thus, we may deduce that the $0_4^{(1,6)}\text{A}$ is obtained from the $0^{(1,7)}$ by choosing the M-theory circle to correspond to one of the quadratic directions. The existence of the parent brane in M-theory then requires the introduction of other branes in Type IIA. In particular, we have three distinct choices for the reduction of the $0^{(1,7)}$: the M-theory circle may lie along a direction entirely transverse to the brane, along the cubic direction or along one of the quadratic directions. Relabelling coordinates, we have
\begin{align}
\text{M}(0^{(1,7)}) = \frac{R_8^3 {(R_7 \ldots R_1)}^2}{l_p^{18}} \begin{cases}
R_\natural = R_9: \frac{R_8^3 {(R_7^2 \ldots R_1)}^2}{l_p^{18}} = \frac{R_8^3{(R_7 \ldots R_1)}^2}{g_s^6 l_s^{18}} = \text{M} (0_6^{(1,7)}\text{A}),\\
R_\natural = R_8: \frac{R_\natural^3 {(R_7 \dots R_1)}^2}{l_p^{18}} = \frac{{(R_7 \ldots R_1)}^2}{g_s^3 l_s^{18}} = \text{M}(0_3^7\text{A}),\\
R_\natural = R_7: \frac{R_8^3 R_\natural^2 {(R_6 \ldots R_1)}^2}{l_p^{18}} = \frac{R_8^3 {(R_6 \ldots R_1)}^2}{g_s^4 l_s^{16}} = \text{M}(0_4^{(1,6)}\text{A}).
\end{cases}
\end{align}
The last two are in agreement with the de Boer-Shigemori classification. Indeed the $0_3^7$ obtained in this way happens to be in the same $p_3^{7-p}$ T-duality orbit (the only $g_s^{-3}$ orbit) as the $1_3^6\text{B}$ found above. Likewise, the $0_6^{(1,7)}\text{A}$ obtained here happens to be in the same $g_s^{-6}$ orbit (of which there are multiple) as the $0_6^{(1,1,6)}\text{B}$ found above. We thus see the beginnings of a heavily intertwined, complex structure in these dualities and lifts/reductions. The novel branes appear only because we are allowing for dependence on winding modes. The number of such branes is seen to quickly proliferate once one starts to apply this procedure iteratively, until exhaustion of all possible duality transformations and lifts/reductions.
\clearpage
\newgeometry{vmargin=2cm, hmargin=2cm}
\begin{landscape}
\subsection{\texorpdfstring{$g_s^0$}{gs0} Duality Orbits}
\begin{figure}[H]
\centering
\begin{tikzpicture}
\matrix(M)[matrix of math nodes, row sep=3em, column sep=2 em, minimum width=2em]{
\clap{\text{A}} & \text{P}/{}^10_0 & \text{F1}/1_0\\
\clap{\text{B}} & \text{P}/{}^10_0 & \text{F1}/1_0\\
};
\draw[latex-latex] (M-1-2) -- (M-2-2);
\draw[latex-latex] (M-1-3) -- (M-2-3);
\draw[latex-latex] (M-1-2) -- (M-2-3);
\draw[latex-latex] (M-1-3) -- (M-2-2);
\end{tikzpicture}
\caption{The T-duality orbit of the $\text{F1}=1_0$.}
\label{fig:10Orbit}
\end{figure}
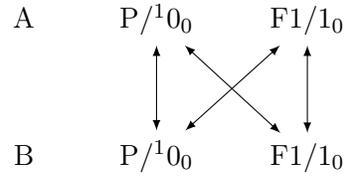
\begin{minipage}{0.5\linewidth}
S-dualities:
\begin{itemize}
	\item $\text{P}={}^10_0 \leftrightarrow \text{P}={}^10_0$
	\item $\text{F1}=1_0 \leftrightarrow \text{D1}=1_1$ See Figure \ref{fig:11Orbit}
\end{itemize}
\end{minipage}
\begin{minipage}{0.5\linewidth}
M-theory origins:
\begin{itemize}
	\item $\text{P}={}^10_0 \rightarrow \text{WM} = 0$
	\item $\text{F1}=1_0 \rightarrow \text{M2} = 2$
\end{itemize}
\end{minipage}
$ $\par
Note that the massless WM must be treated separately from the remaining branes; one instead uses $P^2=0$ such that the masses of the PA and D0 are obtained from the radii of the 11th direction. 
\subsection{\texorpdfstring{$g_s^{-1}$}{gs-1} Duality Orbits}
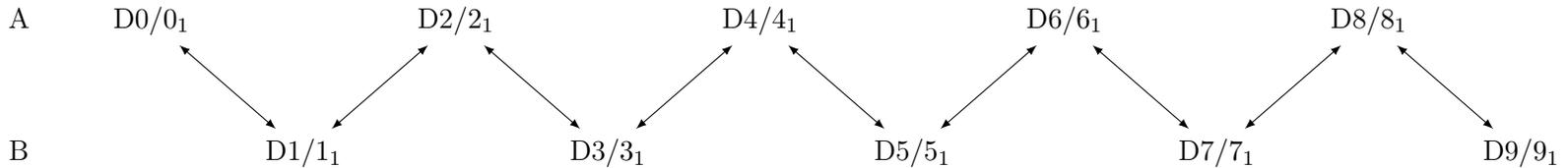
\begin{figure}[H]
\centering
\begin{tikzpicture}
\matrix(M)[matrix of math nodes, row sep=3em, column sep=2 em, minimum width=2em]{
\clap{\text{A}} & \text{D0}/0_1 & & \text{D2}/2_1 & & \text{D4}/4_1 && \text{D6}/6_1 && \text{D8}/8_1\\
\clap{\text{B}} & & \text{D1}/1_1 & & \text{D3}/3_1 & & \text{D5}/5_1 && \text{D7}/7_1 && \text{D9}/9_1\\
};
\draw[latex-latex] (M-1-2) -- (M-2-3);
\draw[latex-latex] (M-2-3) -- (M-1-4);
\draw[latex-latex] (M-1-4) -- (M-2-5);
\draw[latex-latex] (M-2-5) -- (M-1-6);
\draw[latex-latex] (M-1-6) -- (M-2-7);
\draw[latex-latex] (M-2-7) -- (M-1-8);
\draw[latex-latex] (M-1-8) -- (M-2-9);
\draw[latex-latex] (M-2-9) -- (M-1-10);
\draw[latex-latex] (M-1-10) -- (M-2-11);
\end{tikzpicture}
\caption{The T-duality orbit of the $\text{D1}=1_1$.}
\label{fig:11Orbit}
\end{figure}
\begin{minipage}{0.5\linewidth}
S-dualities:
\begin{itemize}
	\item $\text{D1}=1_1 \leftrightarrow \text{F1}=1_0$ See Figure \ref{fig:10Orbit}
	\item $\text{D3}=3_1 \leftrightarrow \text{D3}=3_1$ Self-dual
	\item $\text{D5}=5_1 \leftrightarrow \text{NS5}=5_2$ See Figure \ref{fig:52Orbit}
	\item $\text{D7}=7_1 \leftrightarrow \text{NS7}=7_3$ See Figure \ref{fig:532Orbit}
	\item $\text{D9}=9_1 \leftrightarrow 9_4$ See Figure \ref{fig:7420Orbit}
\end{itemize}
\end{minipage}
\begin{minipage}{0.5\linewidth}
M-theory origins:
\begin{itemize}
	\item $\text{D0}=0_1 \rightarrow \text{WM}=0$
	\item $\text{D2}=2_1 \rightarrow \text{M2}=2$
	\item $\text{D4}=4_1 \rightarrow \text{M5}=5$
	\item $\text{D6}=6_1 \rightarrow \text{KK6M}=6^1$
	\item $\text{D8}=8_1 \rightarrow \text{KK8M}=8^{(1,0)}$
\end{itemize}
\end{minipage}
\clearpage
\end{landscape}
\subsection{\texorpdfstring{$g_s^{-2}$}{gs-2} Duality Orbits}
\begin{figure}[H]
\centering
\begin{tikzpicture}
\matrix(M)[matrix of math nodes, row sep=3em, column sep=4 em, minimum width=2em]{
\clap{\text{A}} & \text{NS5}/5_2 & \text{KK5A}/5_2^1 & 5_2^2 & 5_2^3 & 5_2^4\\
\clap{\text{B}} & \text{NS5}/5_2 & \text{KK5A}/5_2^1 & 5_2^2 & 5_2^3 & 5_2^4\\
};
\draw[latex-latex] (M-1-2) -- (M-2-3);
\draw[latex-latex] (M-1-2) -- (M-2-2);
\draw[latex-latex] (M-2-3) -- (M-1-4);
\draw[latex-latex] (M-2-3) -- (M-1-3);
\draw[latex-latex] (M-1-4) -- (M-2-5);
\draw[latex-latex] (M-1-4) -- (M-2-4);
\draw[latex-latex] (M-2-5) -- (M-1-6);
\draw[latex-latex] (M-2-5) -- (M-1-5);
\draw[latex-latex] (M-2-6) -- (M-1-6);
\draw[latex-latex] (M-2-2) -- (M-1-3);
\draw[latex-latex] (M-1-3) -- (M-2-4);
\draw[latex-latex] (M-2-4) -- (M-1-5);
\draw[latex-latex] (M-1-5) -- (M-2-6);
\end{tikzpicture}
\caption{The T-duality orbit of the $5_2^3$.}
\label{fig:52Orbit}
\end{figure}
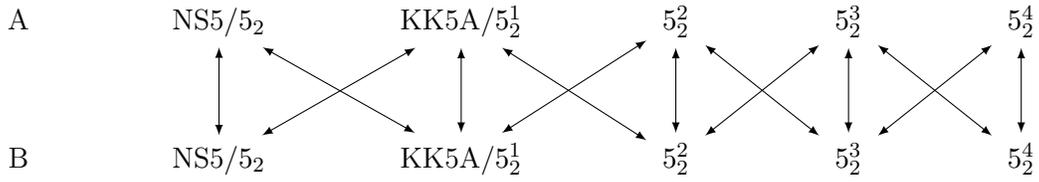
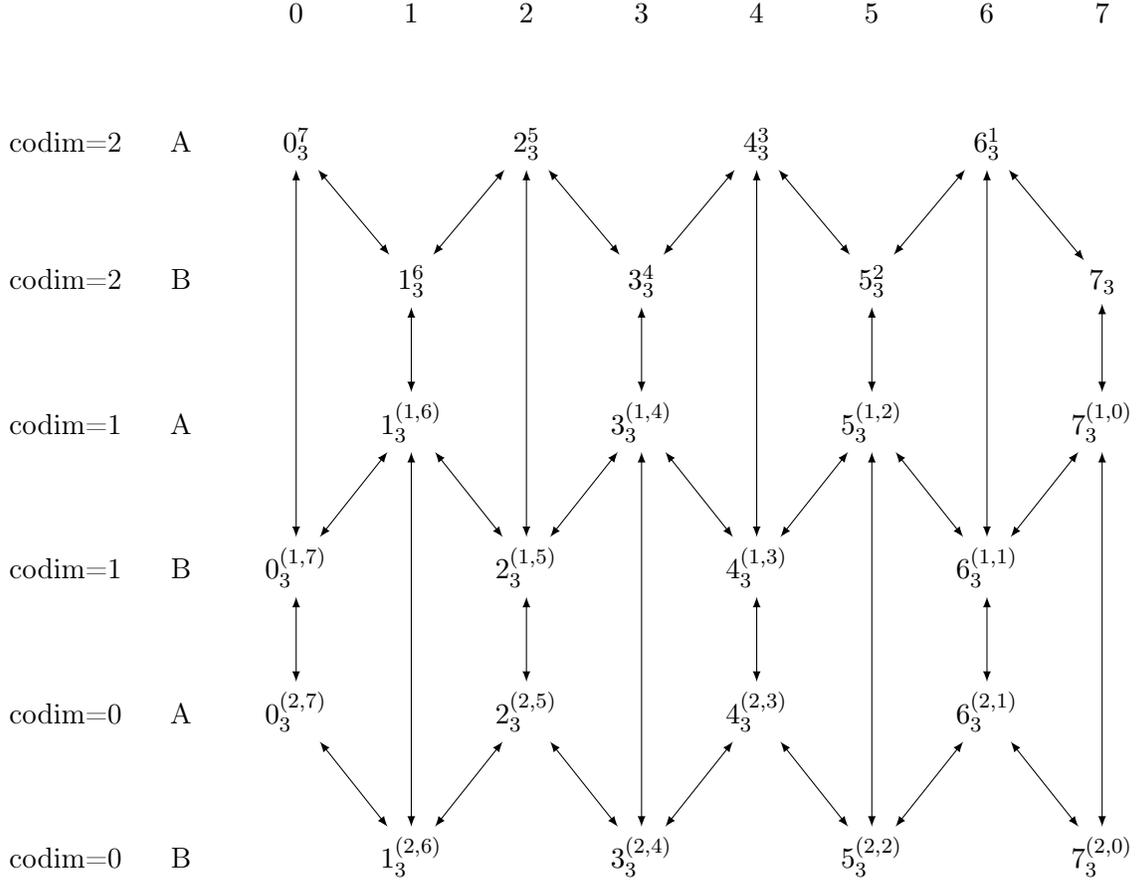
\begin{figure}[H]
\centering
\begin{tikzpicture}
\matrix(M)[matrix of math nodes, row sep=3em, column sep=2 em, minimum width=2em]{
& & \mathclap{0} & \mathclap{1} & \mathclap{2} & \mathclap{3} & \mathclap{4} & \mathclap{5} & \mathclap{6} & \mathclap{7}\\
\clap{\text{codim=2}} & \clap{\text{A}} & \mathclap{0_3^7} & & \mathclap{2_3^5} & & \mathclap{4_3^3} & & \mathclap{6_3^1} &\\
\clap{\text{codim=2}} & \clap{\text{B}} & & \mathclap{1_3^6} & & \mathclap{3_3^4} & & \mathclap{5_3^2} & & \mathclap{7_3}\\
\clap{\text{codim=1}} & \clap{\text{A}} & & \mathclap{1_3^{(1,6)}} & & \mathclap{3_3^{(1,4)}} & & \mathclap{5_3^{(1,2)}} & & \mathclap{7_3^{(1,0)}}\\
\clap{\text{codim=1}} & \clap{\text{B}} & \mathclap{0_3^{(1,7)}} & & \mathclap{2_3^{(1,5)}} & & \mathclap{4_3^{(1,3)}} & & \mathclap{6_3^{(1,1)}} &\\
\clap{\text{codim=0}} & \clap{\text{A}} & \mathclap{0_3^{(2,7)}} & & \mathclap{2_3^{(2,5)}} & & \mathclap{4_3^{(2,3)}} & & \mathclap{6_3^{(2,1)}} &\\
\clap{\text{codim=0}} & \clap{\text{B}} & & \mathclap{1_3^{(2,6)}} & & \mathclap{3_3^{(2,4)}} & & \mathclap{5_3^{(2,2)}} & & \mathclap{7_3^{(2,0)}}\\
};
\draw[latex-latex] (M-2-3) -- (M-3-4);
\draw[latex-latex] (M-3-4) -- (M-2-5);
\draw[latex-latex] (M-2-5) -- (M-3-6);
\draw[latex-latex] (M-3-6) -- (M-2-7);
\draw[latex-latex] (M-2-7) -- (M-3-8);
\draw[latex-latex] (M-3-8) -- (M-2-9);
\draw[latex-latex] (M-2-9) -- (M-3-10);
\draw[latex-latex] (M-5-3) -- (M-4-4);
\draw[latex-latex] (M-4-4) -- (M-5-5);
\draw[latex-latex] (M-5-5) -- (M-4-6);
\draw[latex-latex] (M-4-6) -- (M-5-7);
\draw[latex-latex] (M-5-7) -- (M-4-8);
\draw[latex-latex] (M-4-8) -- (M-5-9);
\draw[latex-latex] (M-5-9) -- (M-4-10);
\draw[latex-latex] (M-6-3) -- (M-7-4);
\draw[latex-latex] (M-7-4) -- (M-6-5);
\draw[latex-latex] (M-6-5) -- (M-7-6);
\draw[latex-latex] (M-7-6) -- (M-6-7);
\draw[latex-latex] (M-6-7) -- (M-7-8);
\draw[latex-latex] (M-7-8) -- (M-6-9);
\draw[latex-latex] (M-6-9) -- (M-7-10);
\draw[latex-latex] (M-2-3) -- (M-5-3);
\draw[latex-latex] (M-5-3) -- (M-6-3);
\draw[latex-latex] (M-2-5) -- (M-5-5);
\draw[latex-latex] (M-5-5) -- (M-6-5);
\draw[latex-latex] (M-2-7) -- (M-5-7);
\draw[latex-latex] (M-5-7) -- (M-6-7);
\draw[latex-latex] (M-2-9) -- (M-5-9);
\draw[latex-latex] (M-5-9) -- (M-6-9);
\draw[latex-latex] (M-3-4) -- (M-4-4);
\draw[latex-latex] (M-4-4) -- (M-7-4);
\draw[latex-latex] (M-3-6) -- (M-4-6);
\draw[latex-latex] (M-4-6) -- (M-7-6);
\draw[latex-latex] (M-3-8) -- (M-4-8);
\draw[latex-latex] (M-4-8) -- (M-7-8);
\draw[latex-latex] (M-3-10) -- (M-4-10);
\draw[latex-latex] (M-4-10) -- (M-7-10);
\end{tikzpicture}
\caption{The T-duality orbit of the $5_3^2$.}
\label{fig:532Orbit}
\end{figure}

\begin{minipage}{0.5\linewidth}
S-dualities:
\begin{itemize}
	\item $\text{NS5}/5_2 \leftrightarrow \text{D5}/5_1$ See Figure \ref{fig:11Orbit}
	\item $\text{KK5A}/5_2^1 \leftrightarrow \text{KK5A}/5_2^1$ Self-dual
	\item $5_2^2 \leftrightarrow 5_3^2$ See Figure \ref{fig:532Orbit}
	\item $5_2^3 \leftrightarrow 5_4^3$ See Figure \ref{fig:4413Orbit}
	\item $5_2^4 \leftrightarrow 5_5^4$ See Figure \ref{fig:5522Orbit}
\end{itemize}
\end{minipage}
\begin{minipage}{0.5\linewidth}
M-theory origins:
\begin{itemize}
	\item $\text{NS5}/5_2 \rightarrow \text{M5}/5$
	\item $\text{KK5A}/5_2^1 \rightarrow \text{KK6M}/6^1$
	\item $5_2^2 \rightarrow 5^3$
	\item $5_2^3 \rightarrow 5^{(1,3)}$
	\item $5_2^4 \rightarrow 5^{(1,0,4)}$
\end{itemize}
\end{minipage}
\clearpage
\subsection{\texorpdfstring{$g_s^{-3}$}{gs-3} Duality Orbits}
\begin{minipage}{0.5\linewidth}
S-dualities:
\begin{itemize}
 	\item $1_3^6 \leftrightarrow 1_4^6$ See Figure \ref{fig:146Orbit}
	\item $3_3^4 \leftrightarrow 3_3^4$ Self-dual
	\item $5_3^2 \leftrightarrow 5_2^2$ See Figure \ref{fig:52Orbit}
	\item $7_3 \leftrightarrow 7_1$ See Figure \ref{fig:11Orbit}
	\item $0_3^{(1,7)} \leftrightarrow 0_6^{(1,7)}$ See Figure \ref{fig:0617Orbit}
	\item $2_3^{(1,5)} \leftrightarrow 2_5^{(1,5)}$ See Figure \ref{fig:2515Orbit}
	\item $4_3^{(1,3)} \leftrightarrow 4_4^{(1,3)}$ See Figure \ref{fig:4413Orbit}
	\item $6_3^{(1,1)} \leftrightarrow 6_3^{(1,1)}$ Self-dual
	\item $1_3^{(2,6)} \leftrightarrow 1_7^{(2,6)}$ See Figure \ref{fig:1726Orbit}
	\item $3_3^{(2,4)} \leftrightarrow 3_6^{(2,4)}$ See Figure \ref{fig:3624Orbit}
	\item $5_3^{(2,2)} \leftrightarrow 5_5^{(2,2)}$ See Figure \ref{fig:5522Orbit}
	\item $7_3^{(2,0)} \leftrightarrow 7_4^{(2,0)}$ See Figure \ref{fig:7420Orbit}
\end{itemize}
\end{minipage}
\begin{minipage}{0.5\linewidth}
M-theory origins:
\begin{itemize}
	\item $0_3^7 \rightarrow 0^{(1,7)}$
	\item $2_3^5 \rightarrow 2^6$
	\item $4_3^3 \rightarrow 5^3$
	\item $6_3^1 \rightarrow 6^1=\text{KK6M}$
	\item $1_3^{(1,6)} \rightarrow 1^{(1,1,6)}$
	\item $3_3^{(1,4)} \rightarrow 3^{(2,4)}$
	\item $5_3^{(1,2)} \rightarrow 5^{(1,3)}$
	\item $7_3^{(1,0)}=\text{KK7A} \rightarrow 8^{(1,0)}=\text{KK8M}$
	\item $0_3^{(2,7)} \rightarrow 0^{(1,0,0,2,7)}$
	\item $2_3^{(2,5)} \rightarrow 2^{(1,0,2,5)}$
	\item $4_3^{(2,3)} \rightarrow 4^{(1,2,3)}$
	\item $6_3^{(2,1)} \rightarrow 6^{(3,1)}$
\end{itemize}
\end{minipage}
\clearpage
\subsection{\texorpdfstring{$g_s^{-4}$}{gs-4} Duality Orbits}
\begin{figure}[H]
\centering
\begin{tikzpicture}
\matrix(M)[matrix of math nodes, row sep=3em, column sep=4 em, minimum width=2em]{
& & \mathclap{0} & \mathclap{1}\\
\clap{\text{codim=2}} & \clap{\text{A}} & \mathclap{0_4^{(1,6)}} & \mathclap{1_4^6}\\
\clap{\text{codim=2}} & \clap{\text{B}} & \mathclap{0_4^{(1,6)}} & \mathclap{1_4^6}\\
\clap{\text{codim=1}} & \clap{\text{A}} & \mathclap{0_4^{(1,1,6)}} & \mathclap{1_4^{(1,0,6)}}\\
\clap{\text{codim=1}} & \clap{\text{B}} & \mathclap{0_4^{(1,1,6)}} & \mathclap{1_4^{(1,0,6)}}\\
\clap{\text{codim=0}} & \clap{\text{A}} & \mathclap{0_4^{(2,1,6)}} & \mathclap{1_4^{(2,0,6)}}\\
\clap{\text{codim=0}} & \clap{\text{B}} & \mathclap{0_4^{(2,1,6)}} & \mathclap{1_4^{(2,0,6)}}\\
};
\draw[latex-latex] (M-2-3) -- (M-3-3);
\draw[latex-latex] (M-3-3) -- (M-4-3);
\draw[latex-latex] (M-4-3) -- (M-5-3);
\draw[latex-latex] (M-5-3) -- (M-6-3);
\draw[latex-latex] (M-6-3) -- (M-7-3);
\draw[latex-latex] (M-2-4) -- (M-3-4);
\draw[latex-latex] (M-3-4) -- (M-4-4);
\draw[latex-latex] (M-4-4) -- (M-5-4);
\draw[latex-latex] (M-5-4) -- (M-6-4);
\draw[latex-latex] (M-6-4) -- (M-7-4);
\draw[latex-latex] (M-2-3) to[out=225, in=135] (M-5-3);
\draw[latex-latex] (M-4-3) to[out=225, in=135] (M-7-3);
\draw[latex-latex] (M-2-4) to[out=315, in=45] (M-5-4);
\draw[latex-latex] (M-4-4) to[out=315, in=45] (M-7-4);
\draw[latex-latex] (M-2-3) -- (M-3-4);
\draw[latex-latex] (M-3-3) -- (M-2-4);
\draw[latex-latex] (M-4-3) -- (M-5-4);
\draw[latex-latex] (M-5-3) -- (M-4-4);
\draw[latex-latex] (M-6-3) -- (M-7-4);
\draw[latex-latex] (M-7-3) -- (M-6-4);
\end{tikzpicture}
\caption{The T-duality orbit of the $1_4^6$.}
\label{fig:146Orbit}
\end{figure}
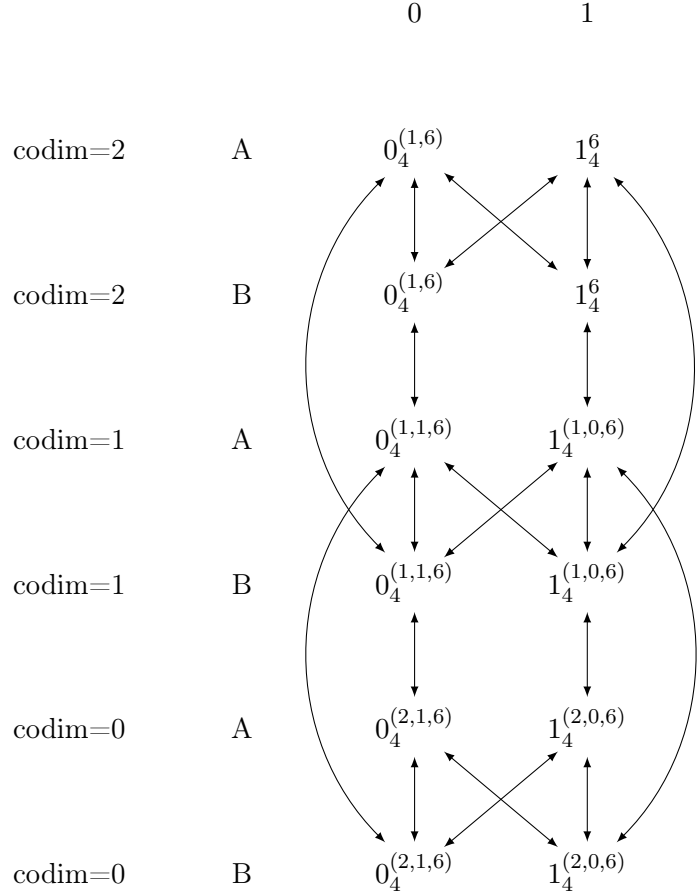
\begin{minipage}{0.5\linewidth}
S-dualties:
\begin{itemize}
	\item $0_4^{(1,6)} \leftrightarrow 0_4^{(1,6)}$ Self-dual
	\item $0_4^{(1,1,6)} \leftrightarrow 0_6^{(1,1,6)}$ See Figure \ref{fig:0617Orbit}
	\item $0_4^{(2,1,6)} \leftrightarrow 0_8^{(2,1,6)}$
	\item $1_4^6 \leftrightarrow 1_3^6$ See Figure \ref{fig:532Orbit}
	\item $1_4^{(1,0,6)} \leftrightarrow 1_5^{(1,0,6)}$ See Figure \ref{fig:2515Orbit}
	\item $1_4^{(2,0,6)} \leftrightarrow 1_7^{(2,0,6)}$ See Figure \ref{fig:1726Orbit}
\end{itemize}
\end{minipage}
\begin{minipage}{0.5\linewidth}
M-theory origins:
\begin{itemize}
	\item $0_4^{(1,6)} \rightarrow 0^{(1,7)}$
	\item $0_4^{(1,1,6)} \rightarrow 0^{(2,1,6)}$
	\item $0_4^{(2,1,6)} \rightarrow 0^{(1,0,2,1,6)}$
	\item $1_4^6 \rightarrow 2^6$
	\item $1_4^{(1,0,6)} \rightarrow 1^{(1,1,6)}$
	\item $1_4^{(2,0,6)} \rightarrow 1^{(1,2,0,6)}$
\end{itemize}
\end{minipage}
\clearpage
\begin{landscape}

\begin{figure}[H]
\centering
\begin{tikzpicture}
\matrix(M)[matrix of math nodes, row sep=3em, column sep=4 em, minimum width=2em]{
& & \mathclap{0} & \mathclap{1} & \mathclap{2} & \mathclap{3} & \mathclap{4} & \mathclap{5}\\
\clap{\text{codim=1}} & \clap{\text{A}} & \mathclap{0_4^{(5,3)}} & \mathclap{1_4^{(4,3)}} & \mathclap{2_4^{(3,3)}} & \mathclap{3_4^{(2,3)}} & \mathclap{4_4^{(1,3)}} & \mathclap{5_4^3}\\
\clap{\text{codim=1}} & \clap{\text{B}} & \mathclap{0_4^{(5,3)}} & \mathclap{1_4^{(4,3)}} & \mathclap{2_4^{(3,3)}} & \mathclap{3_4^{(2,3)}} & \mathclap{4_4^{(1,3)}} & \mathclap{5_4^3}\\
\clap{\text{codim=0}} & \clap{\text{A}} & \mathclap{0_4^{(1,5,3)}} & \mathclap{1_4^{(1,4,3)}} & \mathclap{2_4^{(1,3,3)}} & \mathclap{3_4^{(1,2,3)}} & \mathclap{4_4^{(1,1,3)}} & \mathclap{5_4^{(1,0,3)}}\\
\clap{\text{codim=0}} & \clap{\text{B}} & \mathclap{0_4^{(1,5,3)}} & \mathclap{1_4^{(1,4,3)}} & \mathclap{2_4^{(1,3,3)}} & \mathclap{3_4^{(1,2,3)}} & \mathclap{4_4^{(1,1,3)}} & \mathclap{5_4^{(1,0,3)}}\\
};
\draw[latex-latex] (M-2-3) -- (M-3-3);
\draw[latex-latex] (M-3-3) -- (M-4-3);
\draw[latex-latex] (M-4-3) -- (M-5-3);
\draw[latex-latex] (M-2-3) to[out=225, in=135] (M-5-3);
\draw[latex-latex] (M-2-4) -- (M-3-4);
\draw[latex-latex] (M-3-4) -- (M-4-4);
\draw[latex-latex] (M-4-4) -- (M-5-4);
\draw[latex-latex] (M-2-4) to[out=225, in=135] (M-5-4);
\draw[latex-latex] (M-2-5) -- (M-3-5);
\draw[latex-latex] (M-3-5) -- (M-4-5);
\draw[latex-latex] (M-4-5) -- (M-5-5);
\draw[latex-latex] (M-2-5) to[out=225, in=135] (M-5-5);
\draw[latex-latex] (M-2-6) -- (M-3-6);
\draw[latex-latex] (M-3-6) -- (M-4-6);
\draw[latex-latex] (M-4-6) -- (M-5-6);
\draw[latex-latex] (M-2-6) to[out=315, in=45] (M-5-6);
\draw[latex-latex] (M-2-7) -- (M-3-7);
\draw[latex-latex] (M-3-7) -- (M-4-7);
\draw[latex-latex] (M-4-7) -- (M-5-7);
\draw[latex-latex] (M-2-7) to[out=315, in=45] (M-5-7);
\draw[latex-latex] (M-2-8) -- (M-3-8);
\draw[latex-latex] (M-3-8) -- (M-4-8);
\draw[latex-latex] (M-4-8) -- (M-5-8);
\draw[latex-latex] (M-2-8) to[out=315, in=45] (M-5-8);
\draw[latex-latex] (M-2-3) -- (M-3-4);
\draw[latex-latex] (M-3-3) -- (M-2-4);
\draw[latex-latex] (M-2-4) -- (M-3-5);
\draw[latex-latex] (M-3-4) -- (M-2-5);
\draw[latex-latex] (M-2-5) -- (M-3-6);
\draw[latex-latex] (M-3-5) -- (M-2-6);
\draw[latex-latex] (M-2-6) -- (M-3-7);
\draw[latex-latex] (M-3-6) -- (M-2-7);
\draw[latex-latex] (M-2-7) -- (M-3-8);
\draw[latex-latex] (M-3-7) -- (M-2-8);
\draw[latex-latex] (M-4-3) -- (M-5-4);
\draw[latex-latex] (M-5-3) -- (M-4-4);
\draw[latex-latex] (M-4-4) -- (M-5-5);
\draw[latex-latex] (M-5-4) -- (M-4-5);
\draw[latex-latex] (M-4-5) -- (M-5-6);
\draw[latex-latex] (M-5-5) -- (M-4-6);
\draw[latex-latex] (M-4-6) -- (M-5-7);
\draw[latex-latex] (M-5-6) -- (M-4-7);
\draw[latex-latex] (M-4-7) -- (M-5-8);
\draw[latex-latex] (M-5-7) -- (M-4-8);
\end{tikzpicture}
\caption{The T-duality orbit of the $4_4^{(1,3)}$.}
\label{fig:4413Orbit}
\end{figure}
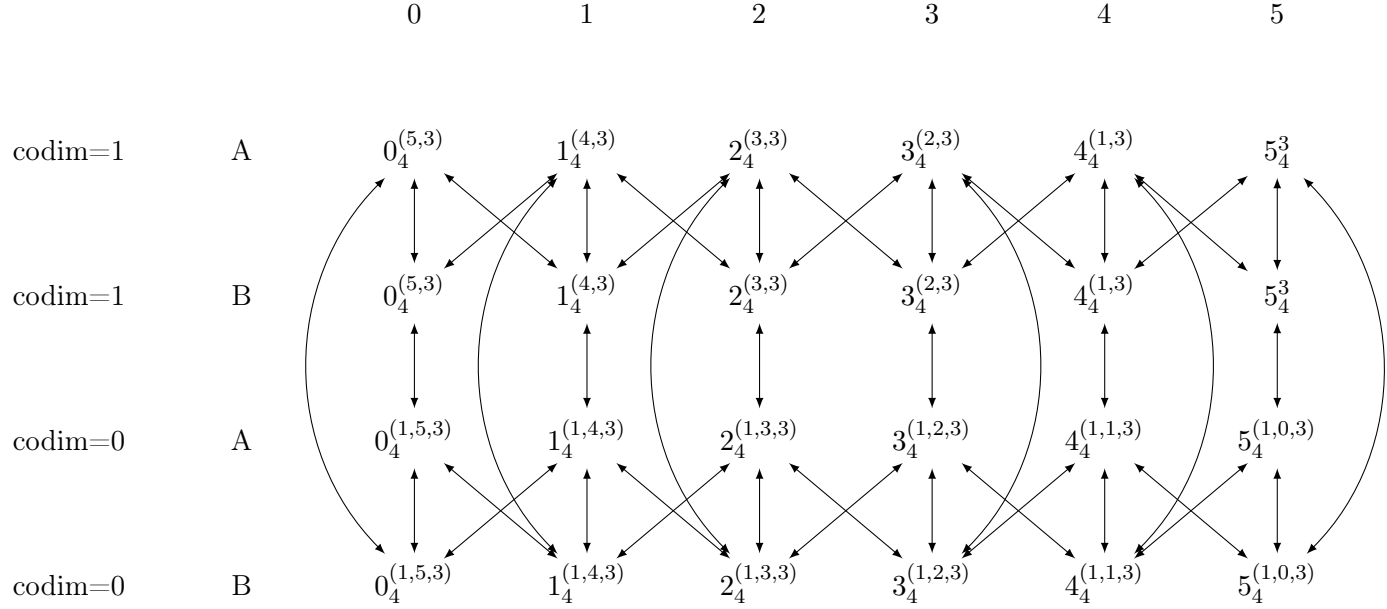
\begin{minipage}{0.5\linewidth}
S-dualities:
\begin{multicols}{2}
\begin{itemize}
	\item $0_4^{(5,3)} \leftrightarrow 0_7^{(5,3)}$ See Figure \ref{fig:0753Orbit}
	\item $1_4^{(4,3)} \leftrightarrow 1_6^{(4,3)}$ See Figure \ref{fig:1643Orbit}
	\item $2_4^{(3,3)} \leftrightarrow 2_5^{(3,3)}$ See Figure \ref{fig:2515Orbit}
	\item $3_4^{(2,3)} \leftrightarrow 3_4^{(2,3)}$ Self-dual
	\item $4_4^{(1,3)} \leftrightarrow 4_3^{(1,3)}$ See Figure \ref{fig:532Orbit}
	\item $5_4^3 \leftrightarrow 5_2^3$ See Figure \ref{fig:52Orbit}
	\item $0_4^{(1,5,3)} \leftrightarrow 0_9^{(1,5,3)}$
	\item $1_4^{(1,4,3)} \leftrightarrow 1_8^{(1,4,3)}$
	\item $2_4^{(1,3,3)} \leftrightarrow 2_7^{(1,3,3)}$ See Figure \ref{fig:27133Orbit}
	\item $3_4^{(1,2,3)} \leftrightarrow 3_6^{(1,2,3)}$ See Figure \ref{fig:3624Orbit}
	\item $4_4^{(1,1,3)} \leftrightarrow 4_5^{(1,1,3)}$ See Figure \ref{fig:5522Orbit}
	\item $5_4^{(1,0,3)} \leftrightarrow 5_4^{(1,0,3)}$ Self-dual
\end{itemize}
\end{multicols}
\end{minipage}
\begin{minipage}{0.5\linewidth}
M-theory origins:
\begin{multicols}{2}
\begin{itemize}
	\item $0_4^{(5,3)} \rightarrow 0^{(1,0,5,3)}$
	\item $1_4^{(4,3)} \rightarrow 1^{(1,4,3)}$
	\item $2_4^{(3,3)} \rightarrow 2^{(4,3)}$
	\item $3_4^{(2,3)} \rightarrow 3^{(2,4)}$
	\item $4_4^{(1,3)} \rightarrow 5^{(1,3)}$
	\item $5_4^3 \rightarrow 5^3$
	\item $0_4^{(1,5,3)} \rightarrow 0^{(1,0,0,1,5,3)}$
	\item $1_4^{(1,4,3)} \rightarrow 1^{(1,0,1,4,3)}$
	\item $2_4^{(1,3,3)} \rightarrow 2^{(1,1,3,3)}$
	\item $3_4^{(1,2,3)} \rightarrow 3^{(2,2,3)}$
	\item $4_4^{(1,1,3)} \rightarrow 4^{(1,2,3)}$
	\item $5_4^{(1,0,3)} \rightarrow 5^{(1,0,4)}$
\end{itemize}
\end{multicols}
\end{minipage}
\clearpage

\begin{figure}[H]
\centering
\begin{tikzpicture}
\matrix(M)[matrix of math nodes, row sep=3em, column sep=4 em, minimum width=2em]{
\clap{\text{A}} & \mathclap{0_4^{(9,0)}} & & \mathclap{2_4^{(7,0)}} & & \mathclap{4_4^{(5,0)}} & & \mathclap{6_4^{(3,0)}} & & \mathclap{8_4^{(1,0)}} &\\
\clap{\text{B}} & & \mathclap{1_4^{(8,0)}} & & \mathclap{3_4^{(6,0)}} & & \mathclap{5_4^{(4,0)}} & & \mathclap{7_4^{(2,0)}} & & \mathclap{\text{NS9B}/9_4}\\
};
\draw[latex-latex] (M-1-2) -- (M-2-3);
\draw[latex-latex] (M-2-3) -- (M-1-4);
\draw[latex-latex] (M-1-4) -- (M-2-5);
\draw[latex-latex] (M-2-5) -- (M-1-6);
\draw[latex-latex] (M-1-6) -- (M-2-7);
\draw[latex-latex] (M-2-7) -- (M-1-8);
\draw[latex-latex] (M-1-8) -- (M-2-9);
\draw[latex-latex] (M-2-9) -- (M-1-10);
\draw[latex-latex] (M-1-10) -- (M-2-11);
\end{tikzpicture}
\caption{T-duality orbit of the $7_4^{(2,0)}$. Note that this is formed from two distinct, though mirrored, orbits. See the discussion in Section \ref{sec:Mapping}}
\label{fig:7420Orbit}
\end{figure}
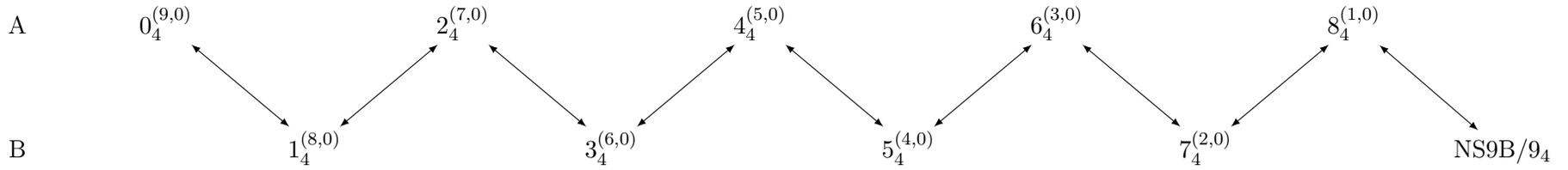
\begin{minipage}{0.5\linewidth}
S-dualities:
\begin{itemize}
	\item $0_4^{(9,0)} \leftrightarrow 0_{10}^{(9,0)}$
	\item $1_4^{(8,0)} \leftrightarrow 1_9^{(8,0)}$
	\item $3_4^{(6,0)} \leftrightarrow 3_7^{(6,0)}$ See Figure \ref{fig:3760Orbit}
	\item $5_4^{(4,0)} \leftrightarrow 5_5^{(4,0)}$ See Figure \ref{fig:5522Orbit}
	\item $7_4^{(2,0)} \leftrightarrow 7_3^{(2,0)}$ See Figure \ref{fig:532Orbit}
	\item $\text{NS9B}/9_4 \leftrightarrow \text{D9}/9_1$ See Figure \ref{fig:11Orbit}
\end{itemize}
\end{minipage}
\begin{minipage}{0.5\linewidth}
M-theory origins:
\begin{itemize}
	\item $0_4^{(9,0)} \rightarrow 0^{(1,0,0,0,0,9,0)}$
	\item $2_4^{(7,0)} \rightarrow 2^{(1,0,0,7,0)}$
	\item $4_4^{(5,0)} \rightarrow 4^{(1,5,0)}$
	\item $6_4^{(3,0)} \rightarrow 6^{(3,1)}$
	\item $8_4^{(1,0)}=\text{KK8A} \rightarrow 8^{(1,0)}=\text{KK8M}$
\end{itemize}
\end{minipage}
\clearpage
\end{landscape}
\subsection{\texorpdfstring{$g_s^{-5}$}{gs-5} Duality Orbits}
\begin{figure}[H]
\centering
\begin{tikzpicture}
\matrix(M)[matrix of math nodes, row sep=2.92em, column sep=4 em, minimum width=2em]{
\clap{\text{A}} & \mathclap{0_5^{(2,6,0)}} & & \mathclap{0_5^{(2,4,2)}} & & \mathclap{0_5^{(2,2,4)}} & & \mathclap{0_5^{(2,0,6)}}\\
\clap{\text{B}} & & \mathclap{0_5^{(2,5,1)}} & & \mathclap{0_5^{(2,3,3)}} & & \mathclap{0_5^{(2,1,5)}} &\\
\clap{\text{A}} & & \mathclap{1_5^{(1,5,1)}} & & \mathclap{1_5^{(1,3,3)}} & & \mathclap{1_5^{(1,1,5)}} &\\
\clap{\text{B}} & \mathclap{1_5^{(1,6,0)}} & & \mathclap{1_5^{(1,4,2)}} & & \mathclap{1_5^{(1,2,4)}} & & \mathclap{1_5^{(1,0,6)}}\\
\clap{\text{A}} & \mathclap{2_5^{(6,0)}} & & \mathclap{2_5^{(4,2)}} & & \mathclap{2_5^{(2,4)}} & & \mathclap{2_5^{6}}\\
\clap{\text{B}} & & \mathclap{2_5^{(5,1)}} & & \mathclap{2_5^{(3,3)}} & & \mathclap{2_5^{(1,5)}} &\\
\clap{\text{A}} & & \mathclap{0_5^{(1,2,5,1)}} & & \mathclap{0_5^{(1,2,3,3)}} & & \mathclap{0_5^{(1,2,1,5)}} &\\
\clap{\text{B}} & \mathclap{0_5^{(1,2,6,0)}} & & \mathclap{0_5^{(1,2,4,2)}} & & \mathclap{0_5^{(1,2,2,4)}} & & \mathclap{0_5^{(1,2,0,6)}}\\
\clap{\text{A}} & \mathclap{1_5^{(1,1,6,0)}} & & \mathclap{1_5^{(1,1,4,2)}} & & \mathclap{1_5^{(1,1,2,4)}} & & \mathclap{1_5^{(1,1,0,6)}}\\
\clap{\text{B}} & & \mathclap{1_5^{(1,1,5,1)}} & & \mathclap{1_5^{(1,1,3,3)}} & & \mathclap{1_5^{(1,1,1,5)}} &\\
\clap{\text{A}} & & \mathclap{2_5^{(1,0,5,1)}} & & \mathclap{2_5^{(1,0,3,3)}} & & \mathclap{2_5^{(1,0,1,5)}} &\\
\clap{\text{B}} & \mathclap{2_5^{(1,0,6,0)}} & & \mathclap{2_5^{(1,0,4,2)}} & & \mathclap{2_5^{(1,0,2,4)}} & & \mathclap{2_5^{(1,0,0,6)}}\\
};
\draw[latex-latex] (M-1-2) -- (M-4-2);
\draw[latex-latex] (M-4-2) -- (M-5-2);
\draw[latex-latex] (M-8-2) -- (M-9-2);
\draw[latex-latex] (M-9-2) -- (M-12-2);
\draw[latex-latex] (M-1-4) -- (M-4-4);
\draw[latex-latex] (M-4-4) -- (M-5-4);
\draw[latex-latex] (M-8-4) -- (M-9-4);
\draw[latex-latex] (M-9-4) -- (M-12-4);
\draw[latex-latex] (M-1-6) -- (M-4-6);
\draw[latex-latex] (M-4-6) -- (M-5-6);
\draw[latex-latex] (M-8-6) -- (M-9-6);
\draw[latex-latex] (M-9-6) -- (M-12-6);
\draw[latex-latex] (M-1-8) -- (M-4-8);
\draw[latex-latex] (M-4-8) -- (M-5-8);
\draw[latex-latex] (M-5-8) -- (M-8-8);
\draw[latex-latex] (M-8-8) -- (M-9-8);
\draw[latex-latex] (M-9-8) -- (M-12-8);
\draw[latex-latex] (M-2-3) -- (M-3-3);
\draw[latex-latex] (M-3-3) -- (M-6-3);
\draw[latex-latex] (M-7-3) -- (M-10-3);
\draw[latex-latex] (M-10-3) -- (M-11-3);
\draw[latex-latex] (M-2-5) -- (M-3-5);
\draw[latex-latex] (M-3-5) -- (M-6-5);
\draw[latex-latex] (M-7-5) -- (M-10-5);
\draw[latex-latex] (M-10-5) -- (M-11-5);
\draw[latex-latex] (M-2-7) -- (M-3-7);
\draw[latex-latex] (M-3-7) -- (M-6-7);
\draw[latex-latex] (M-7-7) -- (M-10-7);
\draw[latex-latex] (M-10-7) -- (M-11-7);
\draw[latex-latex] (M-1-2) -- (M-2-3);
\draw[latex-latex] (M-2-3) -- (M-1-4);
\draw[latex-latex] (M-1-4) -- (M-2-5);
\draw[latex-latex] (M-2-5) -- (M-1-6);
\draw[latex-latex] (M-1-6) -- (M-2-7);
\draw[latex-latex] (M-2-7) -- (M-1-8);
\draw[latex-latex] (M-5-2) -- (M-6-3);
\draw[latex-latex] (M-6-3) -- (M-5-4);
\draw[latex-latex] (M-5-4) -- (M-6-5);
\draw[latex-latex] (M-6-5) -- (M-5-6);
\draw[latex-latex] (M-5-6) -- (M-6-7);
\draw[latex-latex] (M-6-7) -- (M-5-8);
\draw[latex-latex] (M-9-2) -- (M-10-3);
\draw[latex-latex] (M-10-3) -- (M-9-4);
\draw[latex-latex] (M-9-4) -- (M-10-5);
\draw[latex-latex] (M-10-5) -- (M-9-6);
\draw[latex-latex] (M-9-6) -- (M-10-7);
\draw[latex-latex] (M-10-7) -- (M-9-8);
\draw[latex-latex] (M-4-2) -- (M-3-3);
\draw[latex-latex] (M-3-3) -- (M-4-4);
\draw[latex-latex] (M-4-4) -- (M-3-5);
\draw[latex-latex] (M-3-5) -- (M-4-6);
\draw[latex-latex] (M-4-6) -- (M-3-7);
\draw[latex-latex] (M-3-7) -- (M-4-8);
\draw[latex-latex] (M-8-2) -- (M-7-3);
\draw[latex-latex] (M-7-3) -- (M-8-4);
\draw[latex-latex] (M-8-4) -- (M-7-5);
\draw[latex-latex] (M-7-5) -- (M-8-6);
\draw[latex-latex] (M-8-6) -- (M-7-7);
\draw[latex-latex] (M-7-7) -- (M-8-8);
\draw[latex-latex] (M-12-2) -- (M-11-3);
\draw[latex-latex] (M-11-3) -- (M-12-4);
\draw[latex-latex] (M-12-4) -- (M-11-5);
\draw[latex-latex] (M-11-5) -- (M-12-6);
\draw[latex-latex] (M-12-6) -- (M-11-7);
\draw[latex-latex] (M-11-7) -- (M-12-8);
\draw[latex-latex] (M-1-8) to[out=280, in=80] (M-8-8);
\draw[latex-latex] (M-4-8) to[out=292, in=67] (M-9-8);
\draw[latex-latex] (M-5-8) to[out=280, in=80] (M-12-8);
\draw[latex-latex] (M-1-6) to[out=280, in=80] (M-8-6);
\draw[latex-latex] (M-4-6) to[out=292, in=67] (M-9-6);
\draw[latex-latex] (M-5-6) to[out=280, in=80] (M-12-6);
\draw[latex-latex] (M-1-4) to[out=260, in=100] (M-8-4);
\draw[latex-latex] (M-4-4) to[out=247, in=112] (M-9-4);
\draw[latex-latex] (M-5-4) to[out=260, in=100] (M-12-4);
\draw[latex-latex] (M-1-2) to[out=260, in=100] (M-8-2);
\draw[latex-latex] (M-4-2) to[out=247, in=112] (M-9-2);
\draw[latex-latex] (M-5-2) to[out=260, in=100] (M-12-2);
\draw[latex-latex] (M-2-7) to[out=280, in=80] (M-7-7);
\draw[latex-latex] (M-3-7) to[out=292, in=67] (M-10-7);
\draw[latex-latex] (M-6-7) to[out=280, in=80] (M-11-7);
\draw[latex-latex] (M-2-3) to[out=260, in=100] (M-7-3);
\draw[latex-latex] (M-3-3) to[out=247, in=112] (M-10-3);
\draw[latex-latex] (M-6-3) to[out=260, in=100] (M-11-3);
\draw[latex-latex] (M-2-5) to[out=247, in=112] (M-7-5);
\draw[latex-latex] (M-3-5) to[out=292, in=67] (M-10-5);
\draw[latex-latex] (M-6-5) to[out=247, in=112] (M-11-5);
\end{tikzpicture}
\caption{T-duality orbit of $2_5^{(1,5)}$.}
\label{fig:2515Orbit}
\end{figure}
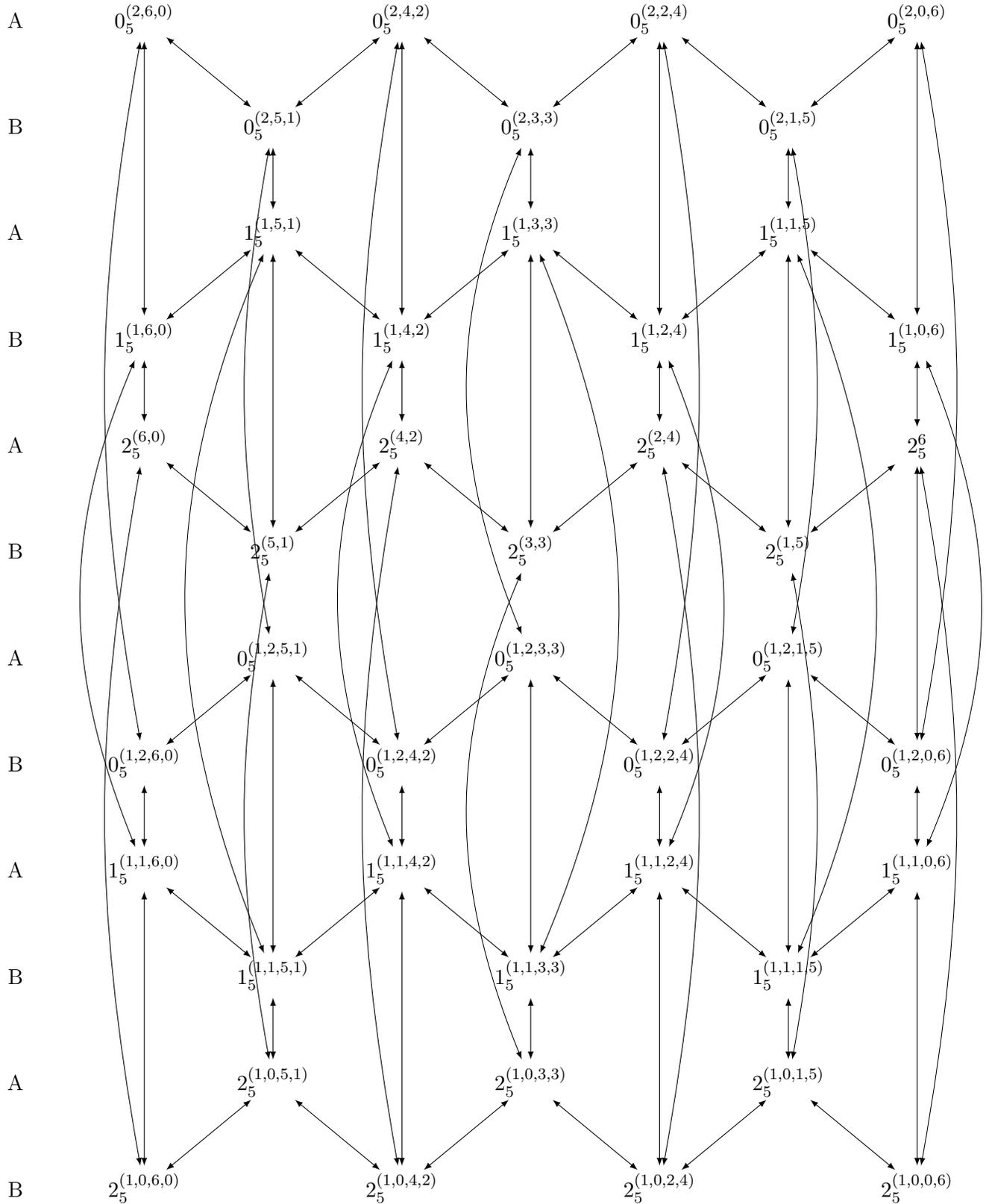
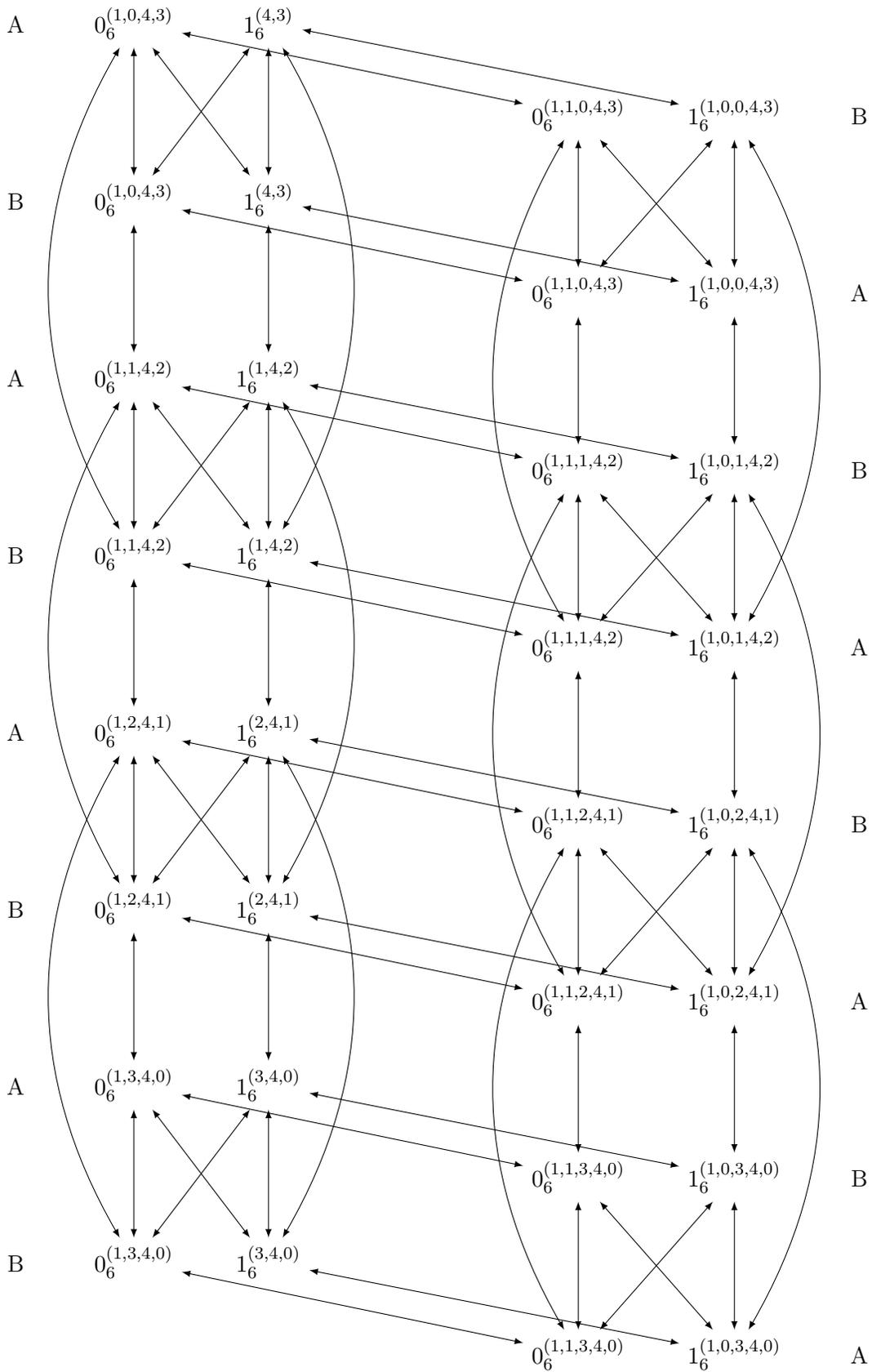
\begin{figure}[H]
\centering
\begin{tikzpicture}
\begin{scope}
\matrix(M)[matrix of math nodes, row sep=5.5em, column sep=2em, minimum width=2em]{
\clap{\text{A}} & 0_6^{(1,0,4,3)} & 1_6^{(4,3)}\\
\clap{\text{B}} & 0_6^{(1,0,4,3)} & 1_6^{(4,3)}\\
\clap{\text{A}} & 0_6^{(1,1,4,2)} & 1_6^{(1,4,2)}\\
\clap{\text{B}} & 0_6^{(1,1,4,2)} & 1_6^{(1,4,2)}\\
\clap{\text{A}} & 0_6^{(1,2,4,1)} & 1_6^{(2,4,1)}\\
\clap{\text{B}} & 0_6^{(1,2,4,1)} & 1_6^{(2,4,1)}\\
\clap{\text{A}} & 0_6^{(1,3,4,0)} & 1_6^{(3,4,0)}\\
\clap{\text{B}} & 0_6^{(1,3,4,0)} & 1_6^{(3,4,0)}\\
};
\end{scope}
\begin{scope}[xshift=9cm, yshift=-1.5cm]
\matrix(N)[matrix of math nodes, row sep=5.5em, column sep=2em, minimum width=2em]{
0_6^{(1,1,0,4,3)} & 1_6^{(1,0,0,4,3)} & \clap{\text{B}}\\
0_6^{(1,1,0,4,3)} & 1_6^{(1,0,0,4,3)} & \clap{\text{A}}\\
0_6^{(1,1,1,4,2)} & 1_6^{(1,0,1,4,2)} & \clap{\text{B}}\\
0_6^{(1,1,1,4,2)} & 1_6^{(1,0,1,4,2)} & \clap{\text{A}}\\
0_6^{(1,1,2,4,1)} & 1_6^{(1,0,2,4,1)} & \clap{\text{B}}\\
0_6^{(1,1,2,4,1)} & 1_6^{(1,0,2,4,1)} & \clap{\text{A}}\\
0_6^{(1,1,3,4,0)} & 1_6^{(1,0,3,4,0)} & \clap{\text{B}}\\
0_6^{(1,1,3,4,0)} & 1_6^{(1,0,3,4,0)} & \clap{\text{A}}\\
};
\end{scope}
\draw[latex-latex] (M-1-2) -- (M-2-2);
\draw[latex-latex] (M-1-3) -- (M-2-3);
\draw[latex-latex] (M-2-2) -- (M-3-2);
\draw[latex-latex] (M-2-3) -- (M-3-3);
\draw[latex-latex] (M-3-2) -- (M-4-2);
\draw[latex-latex] (M-3-3) -- (M-4-3);
\draw[latex-latex] (M-4-2) -- (M-5-2);
\draw[latex-latex] (M-4-3) -- (M-5-3);
\draw[latex-latex] (M-5-2) -- (M-6-2);
\draw[latex-latex] (M-5-3) -- (M-6-3);
\draw[latex-latex] (M-6-2) -- (M-7-2);
\draw[latex-latex] (M-6-3) -- (M-7-3);
\draw[latex-latex] (M-7-2) -- (M-8-2);
\draw[latex-latex] (M-7-3) -- (M-8-3);
\draw[latex-latex] (N-1-1) -- (N-2-1);
\draw[latex-latex] (N-1-2) -- (N-2-2);
\draw[latex-latex] (N-2-1) -- (N-3-1);
\draw[latex-latex] (N-2-2) -- (N-3-2);
\draw[latex-latex] (N-3-1) -- (N-4-1);
\draw[latex-latex] (N-3-2) -- (N-4-2);
\draw[latex-latex] (N-4-1) -- (N-5-1);
\draw[latex-latex] (N-4-2) -- (N-5-2);
\draw[latex-latex] (N-5-1) -- (N-6-1);
\draw[latex-latex] (N-5-2) -- (N-6-2);
\draw[latex-latex] (N-6-1) -- (N-7-1);
\draw[latex-latex] (N-6-2) -- (N-7-2);
\draw[latex-latex] (N-7-1) -- (N-8-1);
\draw[latex-latex] (N-7-2) -- (N-8-2);
\draw[latex-latex] (M-1-2) to[out=240, in=120] (M-4-2);
\draw[latex-latex] (M-1-3) to[out=300, in=60] (M-4-3);
\draw[latex-latex] (M-3-2) to[out=240, in=120] (M-6-2);
\draw[latex-latex] (M-3-3) to[out=300, in=60] (M-6-3);
\draw[latex-latex] (M-5-2) to[out=240, in=120] (M-8-2);
\draw[latex-latex] (M-5-3) to[out=300, in=60] (M-8-3);
\draw[latex-latex] (N-1-1) to[out=240, in=120] (N-4-1);
\draw[latex-latex] (N-1-2) to[out=300, in=60] (N-4-2);
\draw[latex-latex] (N-3-1) to[out=240, in=120] (N-6-1);
\draw[latex-latex] (N-3-2) to[out=300, in=60] (N-6-2);
\draw[latex-latex] (N-5-1) to[out=240, in=120] (N-8-1);
\draw[latex-latex] (N-5-2) to[out=300, in=60] (N-8-2);
\draw[latex-latex] (M-1-2) -- (M-2-3);
\draw[latex-latex] (M-2-2) -- (M-1-3);
\draw[latex-latex] (M-3-2) -- (M-4-3);
\draw[latex-latex] (M-4-2) -- (M-3-3);
\draw[latex-latex] (M-5-2) -- (M-6-3);
\draw[latex-latex] (M-6-2) -- (M-5-3);
\draw[latex-latex] (M-7-2) -- (M-8-3);
\draw[latex-latex] (M-8-2) -- (M-7-3);
\draw[latex-latex] (N-1-1) -- (N-2-2);
\draw[latex-latex] (N-2-1) -- (N-1-2);
\draw[latex-latex] (N-3-1) -- (N-4-2);
\draw[latex-latex] (N-4-1) -- (N-3-2);
\draw[latex-latex] (N-5-1) -- (N-6-2);
\draw[latex-latex] (N-6-1) -- (N-5-2);
\draw[latex-latex] (N-7-1) -- (N-8-2);
\draw[latex-latex] (N-8-1) -- (N-7-2);
\draw[latex-latex] (M-1-2) --(N-1-1);
\draw[latex-latex] (M-1-3) --(N-1-2);
\draw[latex-latex] (M-2-2) --(N-2-1);
\draw[latex-latex] (M-2-3) --(N-2-2);
\draw[latex-latex] (M-3-2) --(N-3-1);
\draw[latex-latex] (M-3-3) --(N-3-2);
\draw[latex-latex] (M-4-2) --(N-4-1);
\draw[latex-latex] (M-4-3) --(N-4-2);
\draw[latex-latex] (M-5-2) --(N-5-1);
\draw[latex-latex] (M-5-3) --(N-5-2);
\draw[latex-latex] (M-6-2) --(N-6-1);
\draw[latex-latex] (M-6-3) --(N-6-2);
\draw[latex-latex] (M-7-2) --(N-7-1);
\draw[latex-latex] (M-7-3) --(N-7-2);
\draw[latex-latex] (M-8-2) --(N-8-1);
\draw[latex-latex] (M-8-3) --(N-8-2);
\end{tikzpicture}
\caption{The T-duality orbit of the $1_6^{(4,3)}$}
\label{fig:1643Orbit}
\end{figure}
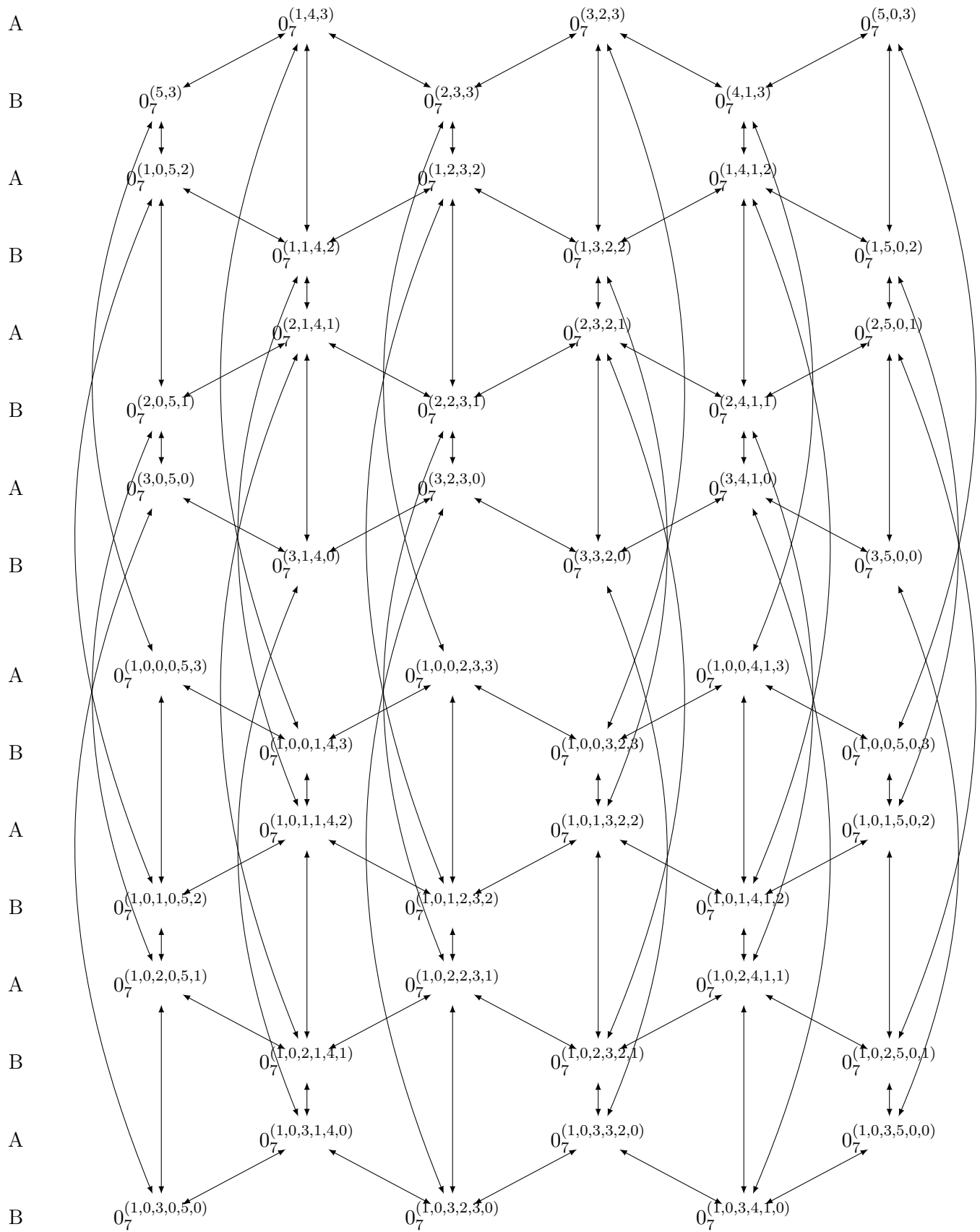
\begin{figure}[H]
\centering
\begin{tikzpicture}
\begin{scope}
\matrix(M)[matrix of math nodes, row sep=1.69em, column sep=5em, minimum width=2em]{
\clap{\text{A}} & & \mathclap{0_7^{(1,4,3)}} & & \mathclap{0_7^{(3,2,3)}} & & \mathclap{0_7^{(5,0,3)}}\\
\clap{\text{B}} & \mathclap{0_7^{(5,3)}} & & \mathclap{0_7^{(2,3,3)}} & & \mathclap{0_7^{(4,1,3)}}\\
\clap{\text{A}} & \mathclap{0_7^{(1,0,5,2)}} & & \mathclap{0_7^{(1,2,3,2)}} & & \mathclap{0_7^{(1,4,1,2)}}\\
\clap{\text{B}} & & \mathclap{0_7^{(1,1,4,2)}} & & \mathclap{0_7^{(1,3,2,2)}} & & \mathclap{0_7^{(1,5,0,2)}}\\
\clap{\text{A}} & &\mathclap{0_7^{(2,1,4,1)}} & & \mathclap{0_7^{(2,3,2,1)}} & & \mathclap{0_7^{(2,5,0,1)}}\\
\clap{\text{B}} & \mathclap{0_7^{(2,0,5,1)}} & & \mathclap{0_7^{(2,2,3,1)}} & & \mathclap{0_7^{(2,4,1,1)}}\\
\clap{\text{A}} & \mathclap{0_7^{(3,0,5,0)}} & & \mathclap{0_7^{(3,2,3,0)}} & & \mathclap{0_7^{(3,4,1,0)}}\\
\clap{\text{B}} & & \mathclap{0_7^{(3,1,4,0)}} & & \mathclap{0_7^{(3,3,2,0)}} & & \mathclap{0_7^{(3,5,0,0)}}\\
};
\end{scope}
\begin{scope}[yshift=-12cm]
\matrix(N)[matrix of math nodes, row sep=1.69em, column sep=5em, minimum width=2em]{
\clap{\text{A}} & \mathclap{0_7^{(1,0,0,0,5,3)}} & & \mathclap{0_7^{(1,0,0,2,3,3)}} & & \mathclap{0_7^{(1,0,0,4,1,3)}}\\
\clap{\text{B}} & & \mathclap{0_7^{(1,0,0,1,4,3)}} & & \mathclap{0_7^{(1,0,0,3,2,3)}} & & \mathclap{0_7^{(1,0,0,5,0,3)}}\\
\clap{\text{A}} & & \mathclap{0_7^{(1,0,1,1,4,2)}} & & \mathclap{0_7^{(1,0,1,3,2,2)}} & & \mathclap{0_7^{(1,0,1,5,0,2)}}\\
\clap{\text{B}} &  \mathclap{0_7^{(1,0,1,0,5,2)}} & & \mathclap{0_7^{(1,0,1,2,3,2)}} & & \mathclap{0_7^{(1,0,1,4,1,2)}}\\
\clap{\text{A}} & \mathclap{0_7^{(1,0,2,0,5,1)}} & & \mathclap{0_7^{(1,0,2,2,3,1)}} & & \mathclap{0_7^{(1,0,2,4,1,1)}}\\
\clap{\text{B}} & &\mathclap{0_7^{(1,0,2,1,4,1)}} & & \mathclap{0_7^{(1,0,2,3,2,1)}} & & \mathclap{0_7^{(1,0,2,5,0,1)}}\\
\clap{\text{A}} & & \mathclap{0_7^{(1,0,3,1,4,0)}} & & \mathclap{0_7^{(1,0,3,3,2,0)}} & & \mathclap{0_7^{(1,0,3,5,0,0)}}\\
\clap{\text{B}} &  \mathclap{0_7^{(1,0,3,0,5,0)}} & & \mathclap{0_7^{(1,0,3,2,3,0)}} & & \mathclap{0_7^{(1,0,3,4,1,0)}}\\
};
\draw[latex-latex] (M-2-2) -- (M-1-3);
\draw[latex-latex] (M-1-3) -- (M-2-4);
\draw[latex-latex] (M-2-4) -- (M-1-5);
\draw[latex-latex] (M-1-5) -- (M-2-6);
\draw[latex-latex] (M-2-6) -- (M-1-7);
\draw[latex-latex] (M-3-2) -- (M-4-3);
\draw[latex-latex] (M-4-3) -- (M-3-4);
\draw[latex-latex] (M-3-4) -- (M-4-5);
\draw[latex-latex] (M-4-5) -- (M-3-6);
\draw[latex-latex] (M-3-6) -- (M-4-7);
\draw[latex-latex] (M-6-2) -- (M-5-3);
\draw[latex-latex] (M-5-3) -- (M-6-4);
\draw[latex-latex] (M-6-4) -- (M-5-5);
\draw[latex-latex] (M-5-5) -- (M-6-6);
\draw[latex-latex] (M-6-6) -- (M-5-7);
\draw[latex-latex] (M-7-2) -- (M-8-3);
\draw[latex-latex] (M-8-3) -- (M-7-4);
\draw[latex-latex] (M-7-4) -- (M-8-5);
\draw[latex-latex] (M-8-5) -- (M-7-6);
\draw[latex-latex] (M-7-6) -- (M-8-7);
\draw[latex-latex] (M-2-2) -- (M-3-2);
\draw[latex-latex] (M-3-2) -- (M-6-2);
\draw[latex-latex] (M-6-2) -- (M-7-2);
\draw[latex-latex] (M-1-3) -- (M-4-3);
\draw[latex-latex] (M-4-3) -- (M-5-3);
\draw[latex-latex] (M-5-3) -- (M-8-3);
\draw[latex-latex] (M-2-4) -- (M-3-4);
\draw[latex-latex] (M-3-4) -- (M-6-4);
\draw[latex-latex] (M-6-4) -- (M-7-4);
\draw[latex-latex] (M-1-5) -- (M-4-5);
\draw[latex-latex] (M-4-5) -- (M-5-5);
\draw[latex-latex] (M-5-5) -- (M-8-5);
\draw[latex-latex] (M-2-6) -- (M-3-6);
\draw[latex-latex] (M-3-6) -- (M-6-6);
\draw[latex-latex] (M-6-6) -- (M-7-6);
\draw[latex-latex] (M-1-7) -- (M-4-7);
\draw[latex-latex] (M-4-7) -- (M-5-7);
\draw[latex-latex] (M-5-7) -- (M-8-7);
\draw[latex-latex] (N-1-2) -- (N-2-3);
\draw[latex-latex] (N-2-3) -- (N-1-4);
\draw[latex-latex] (N-1-4) -- (N-2-5);
\draw[latex-latex] (N-2-5) -- (N-1-6);
\draw[latex-latex] (N-1-6) -- (N-2-7);
\draw[latex-latex] (N-4-2) -- (N-3-3);
\draw[latex-latex] (N-3-3) -- (N-4-4);
\draw[latex-latex] (N-4-4) -- (N-3-5);
\draw[latex-latex] (N-3-5) -- (N-4-6);
\draw[latex-latex] (N-4-6) -- (N-3-7);
\draw[latex-latex] (N-5-2) -- (N-6-3);
\draw[latex-latex] (N-6-3) -- (N-5-4);
\draw[latex-latex] (N-5-4) -- (N-6-5);
\draw[latex-latex] (N-6-5) -- (N-5-6);
\draw[latex-latex] (N-5-6) -- (N-6-7);
\draw[latex-latex] (N-8-2) -- (N-7-3);
\draw[latex-latex] (N-7-3) -- (N-8-4);
\draw[latex-latex] (N-8-4) -- (N-7-5);
\draw[latex-latex] (N-7-5) -- (N-8-6);
\draw[latex-latex] (N-8-6) -- (N-7-7);
\draw[latex-latex] (N-1-2) -- (N-4-2);
\draw[latex-latex] (N-4-2) -- (N-5-2);
\draw[latex-latex] (N-5-2) -- (N-8-2);
\draw[latex-latex] (N-2-3) -- (N-3-3);
\draw[latex-latex] (N-3-3) -- (N-6-3);
\draw[latex-latex] (N-6-3) -- (N-7-3);
\draw[latex-latex] (N-1-4) -- (N-4-4);
\draw[latex-latex] (N-4-4) -- (N-5-4);
\draw[latex-latex] (N-5-4) -- (N-8-4);
\draw[latex-latex] (N-2-5) -- (N-3-5);
\draw[latex-latex] (N-3-5) -- (N-6-5);
\draw[latex-latex] (N-6-5) -- (N-7-5);
\draw[latex-latex] (N-1-6) -- (N-4-6);
\draw[latex-latex] (N-4-6) -- (N-5-6);
\draw[latex-latex] (N-5-6) -- (N-8-6);
\draw[latex-latex] (N-2-7) -- (N-3-7);
\draw[latex-latex] (N-3-7) -- (N-6-7);
\draw[latex-latex] (N-6-7) -- (N-7-7);
\draw[latex-latex] (M-2-2) to[out=247, in=112] (N-1-2);
\draw[latex-latex] (M-3-2) to[out=247, in=112] (N-4-2);
\draw[latex-latex] (M-6-2) to[out=247, in=112] (N-5-2);
\draw[latex-latex] (M-7-2) to[out=247, in=112] (N-8-2);
\draw[latex-latex] (M-1-3) to[out=247, in=112] (N-2-3);
\draw[latex-latex] (M-4-3) to[out=247, in=112] (N-3-3);
\draw[latex-latex] (M-5-3) to[out=247, in=112] (N-6-3);
\draw[latex-latex] (M-8-3) to[out=247, in=112] (N-7-3);
\draw[latex-latex] (M-2-4) to[out=247, in=112] (N-1-4);
\draw[latex-latex] (M-3-4) to[out=247, in=112] (N-4-4);
\draw[latex-latex] (M-6-4) to[out=247, in=112] (N-5-4);
\draw[latex-latex] (M-7-4) to[out=247, in=112] (N-8-4);
\draw[latex-latex] (M-1-5) to[out=292, in=67] (N-2-5);
\draw[latex-latex] (M-4-5) to[out=292, in=67] (N-3-5);
\draw[latex-latex] (M-5-5) to[out=292, in=67] (N-6-5);
\draw[latex-latex] (M-8-5) to[out=292, in=67] (N-7-5);
\draw[latex-latex] (M-2-6) to[out=292, in=67] (N-1-6);
\draw[latex-latex] (M-3-6) to[out=292, in=67] (N-4-6);
\draw[latex-latex] (M-6-6) to[out=292, in=67] (N-5-6);
\draw[latex-latex] (M-7-6) to[out=292, in=67] (N-8-6);
\draw[latex-latex] (M-1-7) to[out=292, in=67] (N-2-7);
\draw[latex-latex] (M-4-7) to[out=292, in=67] (N-3-7);
\draw[latex-latex] (M-5-7) to[out=292, in=67] (N-6-7);
\draw[latex-latex] (M-8-7) to[out=292, in=67] (N-7-7);
\end{scope}
\end{tikzpicture}
\caption{The T-duality orbit of the $0_7^{(5,3)}$.}
\label{fig:0753Orbit}
\end{figure}

\begin{minipage}{0.5\linewidth}
S-dualities:
\begin{itemize}
	\item $0_5^{(2,5,1)} \leftrightarrow 0_8^{(2,5,1)}$
	\item $0_5^{(2,3,3)} \leftrightarrow 0_7^{(2,3,3)}$ See Figure \ref{fig:0753Orbit}
	\item $0_5^{(2,1,5)} \leftrightarrow 0_6^{(2,1,5)}$ See Figure \ref{fig:0617Orbit}
	\item $1_5^{(1,6,0)} \leftrightarrow 1_7^{(1,6,0)}$ See Figure \ref{fig:17160Orbit}
	\item $1_5^{(1,4,2)} \leftrightarrow 1_6^{(1,4,2)}$ See Figure \ref{fig:1643Orbit}
	\item $1_5^{(1,2,4)} \leftrightarrow 1_5^{(1,2,4)}$ Self-dual
	\item $1_5^{(1,0,6)} \leftrightarrow 1_4^{(1,0,6)}$ See Figure \ref{fig:146Orbit}
	\item $2_5^{(5,1)} \leftrightarrow 2_5^{(5,1)}$ Self-dual
	\item $2_5^{(3,3)} \leftrightarrow 2_4^{(3,3)}$ See Figure \ref{fig:4413Orbit}
	\item $2_5^{(1,5)} \leftrightarrow 2_3^{(1,5)}$ See Figure \ref{fig:532Orbit}
	\item $0_5^{(1,2,6,0)} \leftrightarrow 0_{11}^{(1,2,6,0)}$
	\item $0_5^{(1,2,4,2)} \leftrightarrow 0_{10}^{(1,2,4,2)}$
	\item $0_5^{(1,2,2,4)} \leftrightarrow 0_9^{(1,2,2,4)}$
	\item $0_5^{(1,2,0,6)} \leftrightarrow 0_8^{(1,2,0,6)}$
	\item $1_5^{(1,1,5,1)} \leftrightarrow 1_9^{(1,1,5,1)}$
	\item $1_5^{(1,1,3,3)} \leftrightarrow 1_8^{(1,1,3,3)}$
	\item $1_5^{(1,1,1,5)} \leftrightarrow 1_7^{(1,1,1,5)}$ See Figure \ref{fig:1726Orbit}
	\item $2_5^{(1,0,6,0)} \leftrightarrow 2_8^{(1,0,6,0)}$
	\item $2_5^{(1,0,4,2)} \leftrightarrow 2_7^{(1,0,4,2)}$ See Figure \ref{fig:27133Orbit}
	\item $2_5^{(1,0,2,4)} \leftrightarrow 2_6^{(1,0,2,4)}$ See Figure \ref{fig:3624Orbit}
	\item $2_5^{(1,0,0,6)} \leftrightarrow 2_5^{(1,0,0,6)}$ Self-dual
\end{itemize}
\end{minipage}
\begin{minipage}{0.5\linewidth}
M-theory origins:
\begin{itemize}
	\item $0_5^{(2,6,0)} \rightarrow 0^{(1,0,2,6,0)}$
	\item $0_5^{(2,4,2)} \rightarrow 0^{(1,2,4,2)}$
	\item $0_5^{(2,2,4)} \rightarrow 0^{(3,2,4)}$
	\item $0_5^{(2,0,6)} \rightarrow 0^{(2,1,6)}$
	\item $1_5^{(1,5,1)} \rightarrow 1^{(2,5,1)}$
	\item $1_5^{(1,3,3)} \rightarrow 1^{(1,4,3)}$
	\item $1_5^{(1,1,5)} \rightarrow 1^{(1,1,6)}$
	\item $2_5^{(6,0)} \rightarrow 2^{(1,0,0,6,0)}$
	\item $2_5^{(4,2)} \rightarrow 2^{(4,3)}$
	\item $2_5^{(2,4)} \rightarrow 3^{(2,4)}$
	\item $2_5^6 \rightarrow 2^6$
	\item $0_5^{(1,2,5,1)} \rightarrow 0^{(1,0,1,2,5,1)}$
	\item $0_5^{(1,2,3,3)} \rightarrow 0^{(1,0,1,2,3,3)}$
	\item $0_5^{(1,2,1,5)} \rightarrow 0^{(1,1,2,1,5)}$
	\item $1_5^{(1,1,6,0)} \rightarrow 1^{(1,0,1,1,6,0)}$
	\item $1_5^{(1,1,4,2)} \rightarrow 1^{(1,1,1,4,2)} $
	\item $1_5^{(1,1,2,4)} \rightarrow 1^{(2,1,2,4)}$
	\item $1_5^{(1,1,0,6)} \rightarrow 1^{(1,2,0,6)}$
	\item $2_5^{(1,0,5,1)} \rightarrow 2^{(2,0,5,1)}$
	\item $2_5^{(1,0,3,3)} \rightarrow 2^{(1,1,3,3)}$
	\item $2_5^{(1,0,1,5)} \rightarrow 2^{(1,0,2,5)}$
\end{itemize}
\end{minipage}
\clearpage

\begin{figure}[H]
\centering
\begin{tikzpicture}
\matrix(M)[matrix of math nodes, row sep=3em, column sep=4 em, minimum width=2em]{
\clap{\text{A}} & \mathclap{0_5^{(5,0,4)}} & & \mathclap{0_5^{(5,2,2)}} & & \mathclap{0_5^{(5,4,0)}}\\
\clap{\text{B}} & & \mathclap{0_5^{(5,1,3)}} & & \mathclap{0_5^{(5,3,1)}} &\\
\clap{\text{A}} & & \mathclap{1_5^{(4,1,3)}} & & \mathclap{1_5^{(4,3,1)}} &\\
\clap{\text{B}} & \mathclap{1_5^{(4,0,4)}} & & \mathclap{1_5^{(4,2,2)}} & & \mathclap{1_5^{(4,4,0)}}\\
\clap{\text{A}} & \mathclap{2_5^{(3,0,4)}} & & \mathclap{2_5^{(3,2,2)}} & & \mathclap{2_5^{(3,4,0)}} &\\
\clap{\text{B}} & & \mathclap{2_5^{(3,1,3)}} & & \mathclap{2_5^{(3,3,1)}} &\\
\clap{\text{A}} & & \mathclap{3_5^{(2,1,3)}} & & \mathclap{3_5^{(2,3,1)}} &\\
\clap{\text{B}} & \mathclap{3_5^{(2,0,4)}} & & \mathclap{3_5^{(2,2,2)}} & & \mathclap{3_5^{(2,4,0)}}\\
\clap{\text{A}} & \mathclap{4_5^{(1,0,4)}} & & \mathclap{4_5^{(1,2,2)}} & & \mathclap{4_5^{(1,4,0)}}\\
\clap{\text{B}} & & \mathclap{4_5^{(1,1,3)}} & & \mathclap{4_5^{(1,3,1)}} &\\
\clap{\text{A}} & & \mathclap{5_5^{(1,3)}} & & \mathclap{5_5^{(3,1)}} &\\
\clap{\text{B}} & \mathclap{5_5^4} & & \mathclap{5_5^{(2,2)}} & & \mathclap{5_5^{(4,0)}} & \\
};
\draw[latex-latex] (M-1-2) -- (M-4-2);
\draw[latex-latex] (M-4-2) -- (M-5-2);
\draw[latex-latex] (M-5-2) -- (M-8-2);
\draw[latex-latex] (M-8-2) -- (M-9-2);
\draw[latex-latex] (M-9-2) -- (M-12-2);
\draw[latex-latex] (M-1-4) -- (M-4-4);
\draw[latex-latex] (M-4-4) -- (M-5-4);
\draw[latex-latex] (M-5-4) -- (M-8-4);
\draw[latex-latex] (M-8-4) -- (M-9-4);
\draw[latex-latex] (M-9-4) -- (M-12-4);
\draw[latex-latex] (M-1-6) -- (M-4-6);
\draw[latex-latex] (M-4-6) -- (M-5-6);
\draw[latex-latex] (M-5-6) -- (M-8-6);
\draw[latex-latex] (M-8-6) -- (M-9-6);
\draw[latex-latex] (M-9-6) -- (M-12-6);
\draw[latex-latex] (M-2-3) -- (M-3-3);
\draw[latex-latex] (M-3-3) -- (M-6-3);
\draw[latex-latex] (M-6-3) -- (M-7-3);
\draw[latex-latex] (M-7-3) -- (M-10-3);
\draw[latex-latex] (M-10-3) -- (M-11-3);
\draw[latex-latex] (M-2-5) -- (M-3-5);
\draw[latex-latex] (M-3-5) -- (M-6-5);
\draw[latex-latex] (M-6-5) -- (M-7-5);
\draw[latex-latex] (M-7-5) -- (M-10-5);
\draw[latex-latex] (M-10-5) -- (M-11-5);
\draw[latex-latex] (M-1-2) -- (M-2-3);
\draw[latex-latex] (M-2-3) -- (M-1-4);
\draw[latex-latex] (M-1-4) -- (M-2-5);
\draw[latex-latex] (M-2-5) -- (M-1-6);
\draw[latex-latex] (M-5-2) -- (M-6-3);
\draw[latex-latex] (M-6-3) -- (M-5-4);
\draw[latex-latex] (M-5-4) -- (M-6-5);
\draw[latex-latex] (M-6-5) -- (M-5-6);
\draw[latex-latex] (M-9-2) -- (M-10-3);
\draw[latex-latex] (M-10-3) -- (M-9-4);
\draw[latex-latex] (M-9-4) -- (M-10-5);
\draw[latex-latex] (M-10-5) -- (M-9-6);
\draw[latex-latex] (M-4-2) -- (M-3-3);
\draw[latex-latex] (M-3-3) -- (M-4-4);
\draw[latex-latex] (M-4-4) -- (M-3-5);
\draw[latex-latex] (M-3-5) -- (M-4-6);
\draw[latex-latex] (M-8-2) -- (M-7-3);
\draw[latex-latex] (M-7-3) -- (M-8-4);
\draw[latex-latex] (M-8-4) -- (M-7-5);
\draw[latex-latex] (M-7-5) -- (M-8-6);
\draw[latex-latex] (M-12-2) -- (M-11-3);
\draw[latex-latex] (M-11-3) -- (M-12-4);
\draw[latex-latex] (M-12-4) -- (M-11-5);
\draw[latex-latex] (M-11-5) -- (M-12-6);
\end{tikzpicture}
\caption{T-duality orbit of $5_5^{(2,2)}$.}
\label{fig:5522Orbit}
\end{figure}
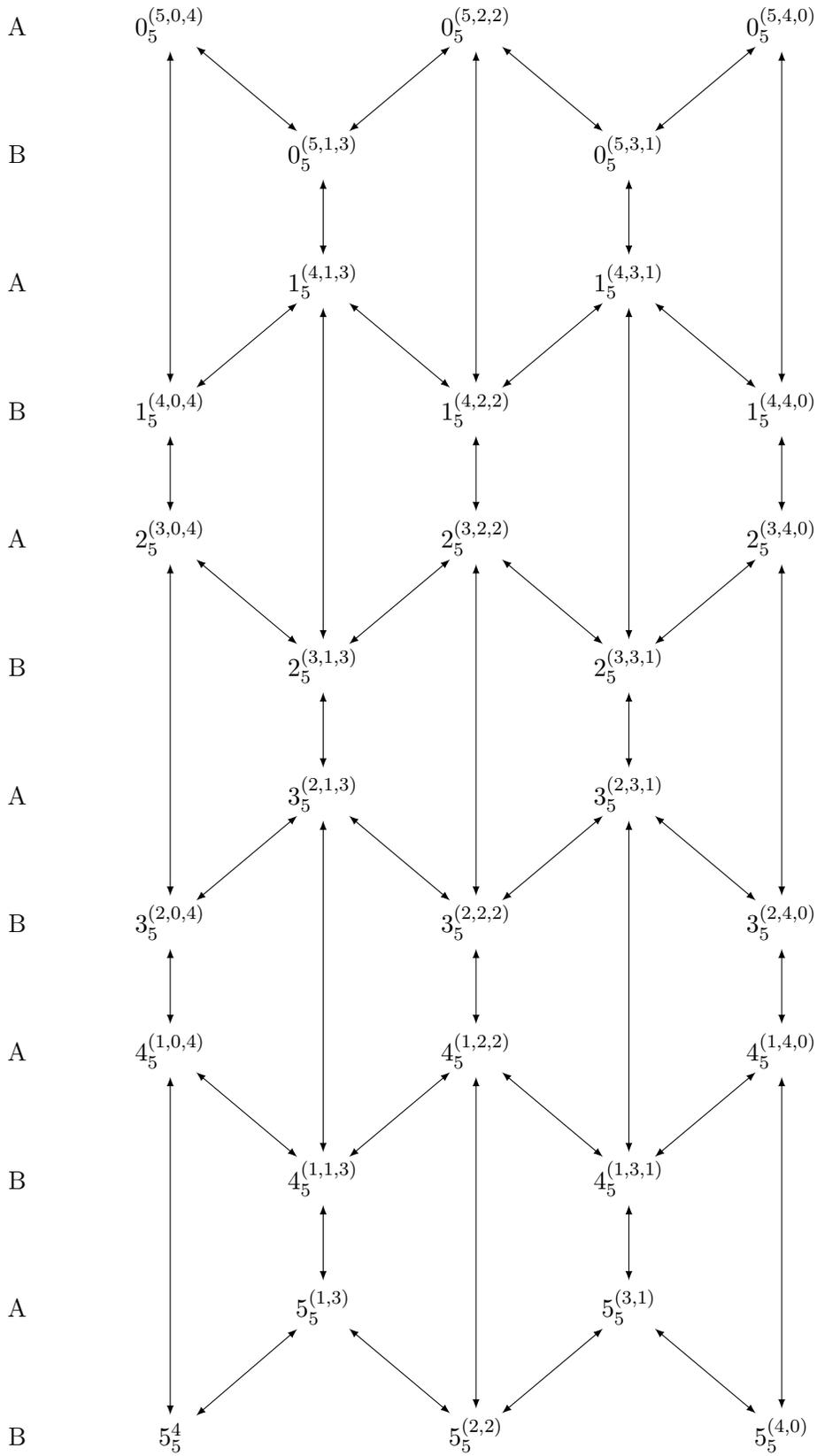
\begin{minipage}{0.5\linewidth}
S-dualities:
\begin{itemize}
	\item $0_5^{(5,1,3)} \leftrightarrow 0_{10}^{(5,1,3)}$
	\item $0_5^{(5,3,1)} \leftrightarrow 0_{11}^{(5,3,1)}$
	\item $1_5^{(4,0,4)} \leftrightarrow 1_8^{(4,0,4)}$
	\item $1_5^{(4,2,2)} \leftrightarrow 1_9^{(4,2,2)}$
	\item $1_5^{(4,4,0)} \leftrightarrow 1_{10}^{(4,4,0)}$
	\item $2_5^{(3,1,3)} \leftrightarrow 2_7^{(3,1,3)}$ See Figure \ref{fig:27133Orbit}
	\item $2_5^{(3,3,1)} \leftrightarrow 2_8^{(3,3,1)}$
	\item $3_5^{(2,0,4)} \leftrightarrow 3_5^{(2,0,4)}$ Self-dual
	\item $3_5^{(2,2,2)} \leftrightarrow 3_6^{(2,2,2)}$ See Figure \ref{fig:3624Orbit}
	\item $3_5^{(2,4,0)} \leftrightarrow 3_7^{(2,4,0)}$ See Figure \ref{fig:3760Orbit}
	\item $4_5^{(1,1,3)} \leftrightarrow 4_4^{(1,1,3)}$ See Figure \ref{fig:4413Orbit}
	\item $4_5^{(1,3,1)} \leftrightarrow 4_5^{(1,3,1)}$ Self-dual
	\item $5_5^4 \leftrightarrow 5_2^4$ See Figure \ref{fig:52Orbit}
	\item $5_5^{(2,2)} \leftrightarrow 5_3^{(2,2)}$ See Figure \ref{fig:532Orbit}
	\item $5_5^{(4,0)} \leftrightarrow 5_4^{(4,0)}$ See Figure \ref{fig:7420Orbit}
\end{itemize}
\end{minipage}
\begin{minipage}{0.5\linewidth}
M-theory origins:
\begin{itemize}
	\item $0_5^{(5,0,4)} \rightarrow 0^{(1,0,0,5,0,4)}$
	\item $0_5^{(5,2,2)} \rightarrow 0^{(1,0,0,0,5,2,2)}$
	\item $0_5^{(5,4,0)} \rightarrow 0^{(1,0,0,0,0,5,4,0)}$
	\item $1_5^{(4,1,3)} \rightarrow 1^{(1,0,4,1,3)}$
	\item $1_5^{(4,3,1)} \rightarrow 1^{(1,0,0,4,3,1)}$
	\item $2_5^{(3,0,4)} \rightarrow 2^{(4,0,4)}$
	\item $2_5^{(3,2,2)} \rightarrow 2^{(1,3,2,2)}$
	\item $2_5^{(3,4,0)} \rightarrow 2^{(1,0,3,4,0)}$
	\item $3_5^{(2,1,3)} \rightarrow 3^{(2,2,3)}$
	\item $3_5^{(2,3,1)} \rightarrow 3^{(3,3,1)}$
	\item $4_5^{(1,0,4)} \rightarrow 5^{(1,0,4)}$
	\item $4_5^{(1,2,2)} \rightarrow 4^{(1,2,3)}$
	\item $4_5^{(1,4,0)} \rightarrow 4^{(1,5,0)}$
	\item $5_5^{(1,3)} \rightarrow 5^{(1,3)}$
	\item $5_5^{(3,1)} \rightarrow 6^{(3,1)}$
\end{itemize}
\end{minipage}
\begin{landscape}
\thispagestyle{empty}
\subsection{\texorpdfstring{$g_s^{-6}$}{gs-6} Duality Orbits}
\begin{figure}[H]
\centering
\begin{tikzpicture}
\matrix(M)[matrix of math nodes, row sep=3em, column sep=4 em, minimum width=2em]{
\clap{\text{A}} & \mathclap{0_6^{(1,7)}} & \mathclap{0_6^{(1,1,6)}} & \mathclap{0_6^{(2,1,5)}} & \mathclap{0_6^{(3,1,4)}} & \mathclap{0_6^{(4,1,3)}} & \mathclap{0_6^{(5,1,2)}} & \mathclap{0_6^{(6,1,1)}} & \mathclap{0_6^{(7,1,0)}}\\
\clap{\text{B}} & \mathclap{0_6^{(1,7)}} & \mathclap{0_6^{(1,1,6)}} & \mathclap{0_6^{(2,1,5)}} & \mathclap{0_6^{(3,1,4)}} & \mathclap{0_6^{(4,1,3)}} & \mathclap{0_6^{(5,1,2)}} & \mathclap{0_6^{(6,1,1)}} & \mathclap{0_6^{(7,1,0)}}\\
\clap{\text{A}} & \mathclap{0_6^{(1,0,0,1,7)}} & \mathclap{0_6^{(1,0,1,1,6)}} & \mathclap{0_6^{(1,0,2,1,5)}} & \mathclap{0_6^{(1,0,3,1,4)}} & \mathclap{0_6^{(1,0,4,1,3)}} & \mathclap{0_6^{(1,0,5,1,2)}} & \mathclap{0_6^{(1,0,6,1,1)}} & \mathclap{0_6^{(1,0,7,1,0)}}\\
\clap{\text{B}} & \mathclap{0_6^{(1,0,0,1,7)}} & \mathclap{0_6^{(1,0,1,1,6)}} & \mathclap{0_6^{(1,0,2,1,5)}} & \mathclap{0_6^{(1,0,3,1,4)}} & \mathclap{0_6^{(1,0,4,1,3)}} & \mathclap{0_6^{(1,0,5,1,2)}} & \mathclap{0_6^{(1,0,6,1,1)}} & \mathclap{0_6^{(1,0,7,1,0)}}\\
};
\draw[latex-latex] (M-1-2) -- (M-2-2);
\draw[latex-latex] (M-2-2) -- (M-3-2);
\draw[latex-latex] (M-3-2) -- (M-4-2);
\draw[latex-latex] (M-1-3) -- (M-2-3);
\draw[latex-latex] (M-2-3) -- (M-3-3);
\draw[latex-latex] (M-3-3) -- (M-4-3);
\draw[latex-latex] (M-1-4) -- (M-2-4);
\draw[latex-latex] (M-2-4) -- (M-3-4);
\draw[latex-latex] (M-3-4) -- (M-4-4);
\draw[latex-latex] (M-1-5) -- (M-2-5);
\draw[latex-latex] (M-2-5) -- (M-3-5);
\draw[latex-latex] (M-3-5) -- (M-4-5);
\draw[latex-latex] (M-1-6) -- (M-2-6);
\draw[latex-latex] (M-2-6) -- (M-3-6);
\draw[latex-latex] (M-3-6) -- (M-4-6);
\draw[latex-latex] (M-1-7) -- (M-2-7);
\draw[latex-latex] (M-2-7) -- (M-3-7);
\draw[latex-latex] (M-3-7) -- (M-4-7);
\draw[latex-latex] (M-1-8) -- (M-2-8);
\draw[latex-latex] (M-2-8) -- (M-3-8);
\draw[latex-latex] (M-3-8) -- (M-4-8);
\draw[latex-latex] (M-1-9) -- (M-2-9);
\draw[latex-latex] (M-2-9) -- (M-3-9);
\draw[latex-latex] (M-3-9) -- (M-4-9);
\draw[latex-latex] (M-1-2) -- (M-2-3);
\draw[latex-latex] (M-2-2) -- (M-1-3);
\draw[latex-latex] (M-1-3) -- (M-2-4);
\draw[latex-latex] (M-2-3) -- (M-1-4);
\draw[latex-latex] (M-1-4) -- (M-2-5);
\draw[latex-latex] (M-2-4) -- (M-1-5);
\draw[latex-latex] (M-1-5) -- (M-2-6);
\draw[latex-latex] (M-2-5) -- (M-1-6);
\draw[latex-latex] (M-1-6) -- (M-2-7);
\draw[latex-latex] (M-2-6) -- (M-1-7);
\draw[latex-latex] (M-1-7) -- (M-2-8);
\draw[latex-latex] (M-2-7) -- (M-1-8);
\draw[latex-latex] (M-1-8) -- (M-2-9);
\draw[latex-latex] (M-2-8) -- (M-1-9);
\draw[latex-latex] (M-3-2) -- (M-4-3);
\draw[latex-latex] (M-4-2) -- (M-3-3);
\draw[latex-latex] (M-3-3) -- (M-4-4);
\draw[latex-latex] (M-4-3) -- (M-3-4);
\draw[latex-latex] (M-3-4) -- (M-4-5);
\draw[latex-latex] (M-4-4) -- (M-3-5);
\draw[latex-latex] (M-3-5) -- (M-4-6);
\draw[latex-latex] (M-4-5) -- (M-3-6);
\draw[latex-latex] (M-3-6) -- (M-4-7);
\draw[latex-latex] (M-4-6) -- (M-3-7);
\draw[latex-latex] (M-3-7) -- (M-4-8);
\draw[latex-latex] (M-4-7) -- (M-3-8);
\draw[latex-latex] (M-3-8) -- (M-4-9);
\draw[latex-latex] (M-4-8) -- (M-3-9);
\draw[latex-latex] (M-1-2) to[out=225, in=135] (M-4-2);
\draw[latex-latex] (M-1-3) to[out=225, in=135] (M-4-3);
\draw[latex-latex] (M-1-4) to[out=225, in=135] (M-4-4);
\draw[latex-latex] (M-1-5) to[out=225, in=135] (M-4-5);
\draw[latex-latex] (M-1-6) to[out=315, in=45] (M-4-6);
\draw[latex-latex] (M-1-7) to[out=315, in=45] (M-4-7);
\draw[latex-latex] (M-1-8) to[out=315, in=45] (M-4-8);
\draw[latex-latex] (M-1-9) to[out=315, in=45] (M-4-9);
\end{tikzpicture}
\caption{The T-duality orbit of the $0_6^{(1,7)}$}
\label{fig:0617Orbit}
\end{figure}
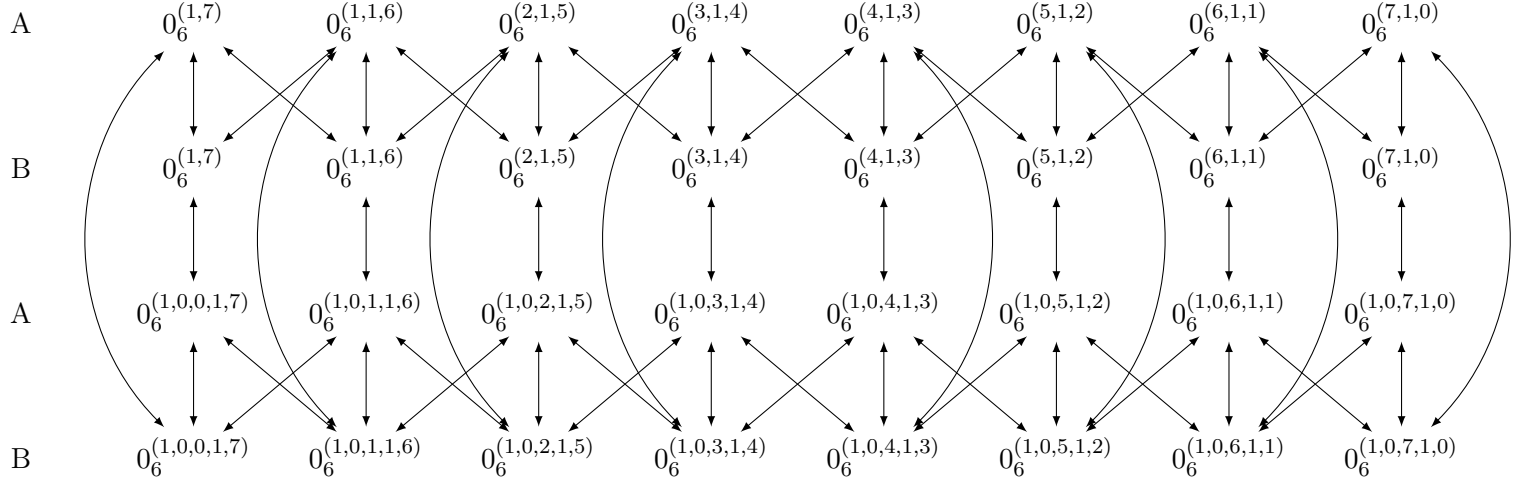
\begin{minipage}{0.52\linewidth}
S-dualities:
\begin{multicols}{2}
\begin{itemize}
	\item $0_6^{(1,7)} \leftrightarrow 0_3^{(1,7)}$ See Figure \ref{fig:532Orbit}
	\item $0_6^{(1,1,6)} \leftrightarrow 0_4^{(1,1,6)}$ See Figure \ref{fig:146Orbit}
	\item $0_6^{(2,1,5)} \leftrightarrow 0_5^{(2,1,5)}$ See Figure \ref{fig:2515Orbit}
	\item $0_6^{(3,1,4)} \leftrightarrow 0_6^{(3,1,4)}$ Self-dual
	\item $0_6^{(4,1,3)} \leftrightarrow 0_7^{(4,1,3)}$ See Figure \ref{fig:0753Orbit}
	\item $0_6^{(5,1,2)} \leftrightarrow 0_8^{(5,1,2)}$
	\item $0_6^{(6,1,1)} \leftrightarrow 0_9^{(6,1,1)}$
	\item $0_6^{(7,1,0)} \leftrightarrow 0_{10}^{(7,1,0)}$
	\item $0_6^{(1,0,0,1,7)} \leftrightarrow 0_6^{(1,0,0,1,7)}$ Self-dual
	\item $0_6^{(1,0,1,1,6)} \leftrightarrow 0_7^{(1,0,1,1,6)}$ See Figure \ref{fig:1726Orbit}
	\item $0_6^{(1,0,2,1,5)} \leftrightarrow 0_8^{(1,0,2,1,5)}$
	\item $0_6^{(1,0,3,1,4)} \leftrightarrow 0_9^{(1,0,3,1,4)}$
	\item $0_6^{(1,0,4,1,3)} \leftrightarrow 0_{10}^{(1,0,4,1,3)}$
	\item $0_6^{(1,0,5,1,2)} \leftrightarrow 0_{11}^{(1,0,5,1,2)}$
	\item $0_6^{(1,0,6,1,1)} \leftrightarrow 0_{12}^{(1,0,6,1,1)}$
	\item $0_6^{(1,0,7,1,0)} \leftrightarrow 0_{13}^{(1,0,7,1,0)}$
\end{itemize}
\end{multicols}
\end{minipage}
\begin{minipage}{0.48\linewidth}
M-theory origins:
\begin{multicols}{2}
\begin{itemize}
	\item $0_6^{(1,7)} \rightarrow 0^{(1,7)}$
	\item $0_6^{(1,1,6)} \rightarrow 1^{(1,1,6)} $
	\item $0_6^{(2,1,5)} \rightarrow 0^{(2,1,6)}$
	\item $0_6^{(3,1,4)} \rightarrow 0^{(3,2,4)}$
	\item $0_6^{(4,1,3)} \rightarrow 0^{(5,1,3)}$
	\item $0_6^{(5,1,2)} \rightarrow 0^{(1,5,1,2)}$
	\item $0_6^{(6,1,1)} \rightarrow 0^{(1,0,6,1,1)}$
	\item $0_6^{(7,1,0)} \rightarrow 0^{(1,0,0,7,1,0)}$
	\item $0_6^{(1,0,0,1,7)} \rightarrow 0^{(1,0,0,2,7)}$
	\item $0_6^{(1,0,1,1,6)} \rightarrow 0^{(1,0,2,1,6)}$
	\item $0_6^{(1,0,2,1,5)} \rightarrow 0^{(1,1,2,1,5)}$
	\item $0_6^{(1,0,3,1,4)} \rightarrow 0^{(2,0,3,1,4)}$
	\item $0_6^{(1,0,4,1,3)} \rightarrow 0^{(1,1,0,4,1,3)}$
	\item $0_6^{(1,0,5,1,2)} \rightarrow 0^{(1,0,1,0,5,1,2)}$
	\item $0_6^{(1,0,6,1,1)} \rightarrow 0^{(1,0,0,1,0,6,1,1)}$
	\item $0_6^{(1,0,7,1,0)} \rightarrow 0^{(1,0,0,0,1,0,7,1,0)}$
\end{itemize}
\end{multicols}
\end{minipage}
\clearpage
\end{landscape}
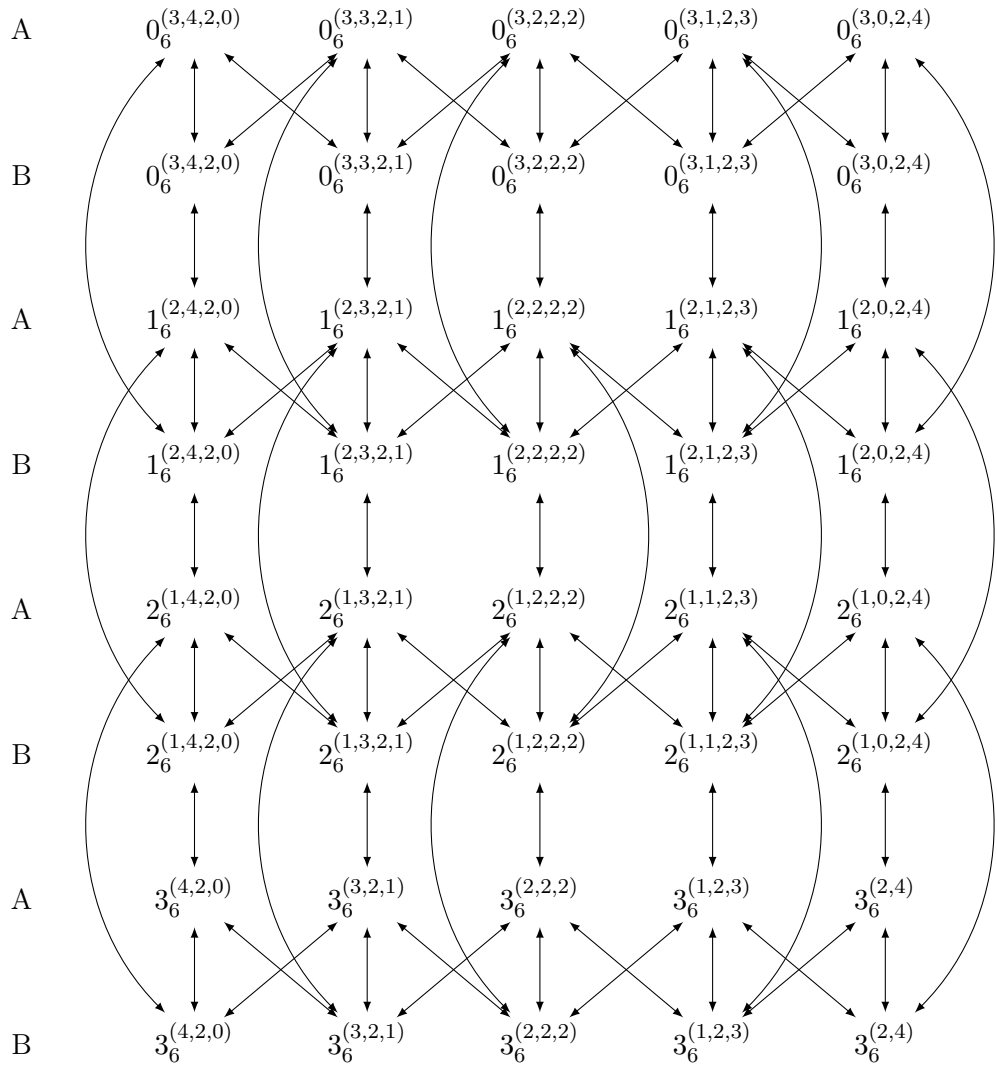
\begin{figure}[H]
\centering
\begin{tikzpicture}
\matrix(M)[matrix of math nodes, row sep=3em, column sep=4 em, minimum width=2em]{
\clap{\text{A}} & \mathclap{0_6^{(3,4,2,0)}} & \mathclap{0_6^{(3,3,2,1)}} & \mathclap{0_6^{(3,2,2,2)}} & \mathclap{0_6^{(3,1,2,3)}} & \mathclap{0_6^{(3,0,2,4)}}\\
\clap{\text{B}} & \mathclap{0_6^{(3,4,2,0)}} & \mathclap{0_6^{(3,3,2,1)}} & \mathclap{0_6^{(3,2,2,2)}} & \mathclap{0_6^{(3,1,2,3)}} & \mathclap{0_6^{(3,0,2,4)}}\\
\clap{\text{A}} & \mathclap{1_6^{(2,4,2,0)}} & \mathclap{1_6^{(2,3,2,1)}} & \mathclap{1_6^{(2,2,2,2)}} & \mathclap{1_6^{(2,1,2,3)}} & \mathclap{1_6^{(2,0,2,4)}}\\
\clap{\text{B}} & \mathclap{1_6^{(2,4,2,0)}} & \mathclap{1_6^{(2,3,2,1)}} & \mathclap{1_6^{(2,2,2,2)}} & \mathclap{1_6^{(2,1,2,3)}} & \mathclap{1_6^{(2,0,2,4)}}\\
\clap{\text{A}} & \mathclap{2_6^{(1,4,2,0)}} & \mathclap{2_6^{(1,3,2,1)}} & \mathclap{2_6^{(1,2,2,2)}} & \mathclap{2_6^{(1,1,2,3)}} & \mathclap{2_6^{(1,0,2,4)}}\\
\clap{\text{B}} & \mathclap{2_6^{(1,4,2,0)}} & \mathclap{2_6^{(1,3,2,1)}} & \mathclap{2_6^{(1,2,2,2)}} & \mathclap{2_6^{(1,1,2,3)}} & \mathclap{2_6^{(1,0,2,4)}}\\
\clap{\text{A}} & \mathclap{3_6^{(4,2,0)}} & \mathclap{3_6^{(3,2,1)}} & \mathclap{3_6^{(2,2,2)}} & \mathclap{3_6^{(1,2,3)}} & \mathclap{3_6^{(2,4)}}\\
\clap{\text{B}} & \mathclap{3_6^{(4,2,0)}} & \mathclap{3_6^{(3,2,1)}} & \mathclap{3_6^{(2,2,2)}} & \mathclap{3_6^{(1,2,3)}} & \mathclap{3_6^{(2,4)}}\\
};
\draw[latex-latex] (M-1-2) -- (M-2-2);
\draw[latex-latex] (M-2-2) -- (M-3-2);
\draw[latex-latex] (M-3-2) -- (M-4-2);
\draw[latex-latex] (M-4-2) -- (M-5-2);
\draw[latex-latex] (M-5-2) -- (M-6-2);
\draw[latex-latex] (M-6-2) -- (M-7-2);
\draw[latex-latex] (M-7-2) -- (M-8-2);
\draw[latex-latex] (M-1-3) -- (M-2-3);
\draw[latex-latex] (M-2-3) -- (M-3-3);
\draw[latex-latex] (M-3-3) -- (M-4-3);
\draw[latex-latex] (M-4-3) -- (M-5-3);
\draw[latex-latex] (M-5-3) -- (M-6-3);
\draw[latex-latex] (M-6-3) -- (M-7-3);
\draw[latex-latex] (M-7-3) -- (M-8-3);
\draw[latex-latex] (M-1-4) -- (M-2-4);
\draw[latex-latex] (M-2-4) -- (M-3-4);
\draw[latex-latex] (M-3-4) -- (M-4-4);
\draw[latex-latex] (M-4-4) -- (M-5-4);
\draw[latex-latex] (M-5-4) -- (M-6-4);
\draw[latex-latex] (M-6-4) -- (M-7-4);
\draw[latex-latex] (M-7-4) -- (M-8-4);
\draw[latex-latex] (M-1-5) -- (M-2-5);
\draw[latex-latex] (M-2-5) -- (M-3-5);
\draw[latex-latex] (M-3-5) -- (M-4-5);
\draw[latex-latex] (M-4-5) -- (M-5-5);
\draw[latex-latex] (M-5-5) -- (M-6-5);
\draw[latex-latex] (M-6-5) -- (M-7-5);
\draw[latex-latex] (M-7-5) -- (M-8-5);
\draw[latex-latex] (M-1-6) -- (M-2-6);
\draw[latex-latex] (M-2-6) -- (M-3-6);
\draw[latex-latex] (M-3-6) -- (M-4-6);
\draw[latex-latex] (M-4-6) -- (M-5-6);
\draw[latex-latex] (M-5-6) -- (M-6-6);
\draw[latex-latex] (M-6-6) -- (M-7-6);
\draw[latex-latex] (M-7-6) -- (M-8-6);
\draw[latex-latex] (M-1-2) -- (M-2-3);
\draw[latex-latex] (M-2-2) -- (M-1-3);
\draw[latex-latex] (M-1-3) -- (M-2-4);
\draw[latex-latex] (M-2-3) -- (M-1-4);
\draw[latex-latex] (M-1-4) -- (M-2-5);
\draw[latex-latex] (M-2-4) -- (M-1-5);
\draw[latex-latex] (M-1-5) -- (M-2-6);
\draw[latex-latex] (M-2-5) -- (M-1-6);
\draw[latex-latex] (M-3-2) -- (M-4-3);
\draw[latex-latex] (M-4-2) -- (M-3-3);
\draw[latex-latex] (M-3-3) -- (M-4-4);
\draw[latex-latex] (M-4-3) -- (M-3-4);
\draw[latex-latex] (M-3-4) -- (M-4-5);
\draw[latex-latex] (M-4-4) -- (M-3-5);
\draw[latex-latex] (M-3-5) -- (M-4-6);
\draw[latex-latex] (M-4-5) -- (M-3-6);
\draw[latex-latex] (M-5-2) -- (M-6-3);
\draw[latex-latex] (M-6-2) -- (M-5-3);
\draw[latex-latex] (M-5-3) -- (M-6-4);
\draw[latex-latex] (M-6-3) -- (M-5-4);
\draw[latex-latex] (M-5-4) -- (M-6-5);
\draw[latex-latex] (M-6-4) -- (M-5-5);
\draw[latex-latex] (M-5-5) -- (M-6-6);
\draw[latex-latex] (M-6-5) -- (M-5-6);
\draw[latex-latex] (M-7-2) -- (M-8-3);
\draw[latex-latex] (M-8-2) -- (M-7-3);
\draw[latex-latex] (M-7-3) -- (M-8-4);
\draw[latex-latex] (M-8-3) -- (M-7-4);
\draw[latex-latex] (M-7-4) -- (M-8-5);
\draw[latex-latex] (M-8-4) -- (M-7-5);
\draw[latex-latex] (M-7-5) -- (M-8-6);
\draw[latex-latex] (M-8-5) -- (M-7-6);
\draw[latex-latex] (M-1-2) to[out=225, in=135] (M-4-2);
\draw[latex-latex] (M-3-2) to[out=225, in=135] (M-6-2);
\draw[latex-latex] (M-5-2) to[out=225, in=135] (M-8-2);
\draw[latex-latex] (M-1-3) to[out=225, in=135] (M-4-3);
\draw[latex-latex] (M-3-3) to[out=225, in=135] (M-6-3);
\draw[latex-latex] (M-5-3) to[out=225, in=135] (M-8-3);
\draw[latex-latex] (M-1-4) to[out=225, in=135] (M-4-4);
\draw[latex-latex] (M-3-4) to[out=315, in=45] (M-6-4);
\draw[latex-latex] (M-5-4) to[out=225, in=135] (M-8-4);
\draw[latex-latex] (M-1-5) to[out=315, in=45] (M-4-5);
\draw[latex-latex] (M-3-5) to[out=315, in=45] (M-6-5);
\draw[latex-latex] (M-5-5) to[out=315, in=45] (M-8-5);
\draw[latex-latex] (M-1-6) to[out=315, in=45] (M-4-6);
\draw[latex-latex] (M-3-6) to[out=315, in=45] (M-6-6);
\draw[latex-latex] (M-5-6) to[out=315, in=45] (M-8-6);
\end{tikzpicture}
\caption{The T-duality orbit of the $3_6^{(2,4)}$}
\label{fig:3624Orbit}
\end{figure}
\clearpage
\begin{multicols}{2}
S-dualities:
\begin{itemize}
	\item $0_6^{(3,4,2,0)} \leftrightarrow 0_{13}^{(3,4,2,0)}$
	\item $0_6^{(3,3,2,1)} \leftrightarrow 0_{12}^{(3,3,2,1)}$
	\item $0_6^{(3,2,2,2)} \leftrightarrow 0_{11}^{(3,2,2,2)}$
	\item $0_6^{(3,1,2,3)} \leftrightarrow 0_{10}^{(3,1,2,3)}$
	\item $0_6^{(3,0,2,4)} \leftrightarrow 0_9^{(3,0,2,4)}$
	\item $1_6^{(2,4,2,0)} \leftrightarrow 1_{11}^{(2,4,2,0)}$
	\item $1_6^{(2,3,2,1)} \leftrightarrow 1_{10}^{(2,3,2,1)}$
	\item $1_6^{(2,2,2,2)} \leftrightarrow 1_9^{(2,2,2,2)}$
	\item $1_6^{(2,1,2,3)} \leftrightarrow 1_8^{(2,1,2,3)}$
	\item $1_6^{(2,0,2,4)} \leftrightarrow 1_7^{(2,0,2,4)}$ See Figure \ref{fig:1726Orbit}
	\item $2_6^{(1,4,2,0)} \leftrightarrow 2_9^{(1,4,2,0)}$
	\item $2_6^{(1,3,2,1)} \leftrightarrow 2_8^{(1,3,2,1)}$
	\item $2_6^{(1,2,2,2)} \leftrightarrow 2_7^{(1,2,2,2)}$ See Figure \ref{fig:27133Orbit}
	\item $2_6^{(1,1,2,3)} \leftrightarrow 2_6^{(1,1,2,3)}$ Self-dual
	\item $2_6^{(1,0,2,4)} \leftrightarrow 2_5^{(1,0,2,4)}$ See Figure \ref{fig:2515Orbit}
	\item $3_6^{(4,2,0)} \leftrightarrow 3_7^{(4,2,0)}$ See Figure \ref{fig:3760Orbit}
	\item $3_6^{(3,2,1)} \leftrightarrow 3_6^{(3,2,1)}$ Self-dual
	\item $3_6^{(2,2,2)} \leftrightarrow 3_5^{(2,2,2)}$ See Figure \ref{fig:5522Orbit}
	\item $3_6^{(1,2,3)} \leftrightarrow 3_4^{(1,2,3)}$ See Figure \ref{fig:4413Orbit}
	\item $3_6^{(2,4)} \leftrightarrow 3_3^{(2,4)}$ See Figure \ref{fig:532Orbit}
\end{itemize}
M-theory origins:
\begin{itemize}
	\item $0_6^{(3,4,2,0)} \rightarrow 0^{(1,0,0,0,0,3,4,2,0)}$
	\item $0_6^{(3,3,2,1)} \rightarrow 0^{(1,0,0,0,3,3,2,1)}$
	\item $0_6^{(3,2,2,2)} \rightarrow 0^{(1,0,0,3,2,2,2)}$
	\item $0_6^{(3,1,2,3)} \rightarrow 0^{(1,0,3,1,2,3)}$
	\item $0_6^{(3,0,2,4)} \rightarrow 0^{(1,3,0,2,4)}$
	\item $1_6^{(2,4,2,0)} \rightarrow 1^{(1,0,0,2,4,2,0)}$
	\item $1_6^{(2,3,2,1)} \rightarrow 1^{(1,0,2,3,2,1)}$
	\item $1_6^{(2,2,2,2)} \rightarrow 1^{(1,2,2,2,2)}$
	\item $1_6^{(2,1,2,3)} \rightarrow 1^{(3,1,2,3)}$
	\item $1_6^{(2,0,2,4)} \rightarrow 1^{(2,1,2,4)}$
	\item $2_6^{(1,4,2,0)} \rightarrow 2^{(1,1,4,2,0)}$
	\item $2_6^{(1,3,2,1)} \rightarrow 2^{(2,3,2,1)}$
	\item $2_6^{(1,2,2,2)} \rightarrow 2^{(1,3,2,2)}$
	\item $2_6^{(1,1,2,3)} \rightarrow 2^{(1,1,3,3)}$
	\item $2_6^{(1,0,2,4)} \rightarrow 2^{(1,0,2,5)}$
	\item $3_6^{(4,2,0)} \rightarrow 3^{(5,2,0)}$
	\item $3_6^{(3,2,1)} \rightarrow 3^{(3,3,1)}$
	\item $3_6^{(2,2,2)} \rightarrow 3^{(2,2,3)}$
	\item $3_6^{(1,2,3)} \rightarrow 4^{(1,2,3)}$
	\item $3_6^{(2,4)} \rightarrow 3^{(2,4)}$
\end{itemize}
\end{multicols}
\clearpage
\begin{minipage}{0.5\linewidth}
S-dualities:
\begin{itemize}
	\item $0_6^{(1,0,4,3)} \leftrightarrow 0_6^{(1,0,4,3)}$ Self-dual
	\item $0_6^{(1,1,4,2)} \leftrightarrow 0_7^{(1,1,4,2)}$ See Figure \ref{fig:0753Orbit}
	\item $0_6^{(1,2,4,1)} \leftrightarrow 0_8^{(1,2,4,1)}$
	\item $0_6^{(1,3,4,0)} \leftrightarrow 0_9^{(1,3,4,0)}$
	\item $0_6^{(1,1,0,4,3)} \leftrightarrow 0_9^{(1,1,0,4,3)}$
	\item $0_6^{(1,1,1,4,2)} \leftrightarrow 0_{10}^{(1,1,1,4,2)}$
	\item $0_6^{(1,1,2,4,1)} \leftrightarrow 0_{11}^{(1,1,2,4,1)}$
	\item $0_6^{(1,1,3,4,0)} \leftrightarrow 0_{12}^{(1,1,3,4,0)}$
	\item $1_6^{(4,3)} \leftrightarrow 1_4^{(4,3)}$ See Figure \ref{fig:4413Orbit}
	\item $1_6^{(1,4,2)} \leftrightarrow 1_5^{(1,4,2)}$ See Figure \ref{fig:2515Orbit}
	\item $1_6^{(2,4,1)} \leftrightarrow 1_6^{(2,4,1)}$ Self-dual
	\item $1_6^{(3,4,0)} \leftrightarrow 1_7^{(3,4,0)}$ See Figure \ref{fig:17160Orbit}
	\item $1_6^{(1,0,0,4,3)} \leftrightarrow 1_7^{(1,0,0,4,3)}$ See Figure \ref{fig:27133Orbit}
	\item $1_6^{(1,0,1,4,2)} \leftrightarrow 1_8^{(1,0,1,4,2)}$
	\item $1_6^{(1,0,2,4,1)} \leftrightarrow 1_9^{(1,0,2,4,1)}$
	\item $1_6^{(1,0,3,4,0)} \leftrightarrow 1_{10}^{(1,0,3,4,0)}$
\end{itemize}
\end{minipage}
\begin{minipage}{0.5\linewidth}
M-theory origins:
\begin{itemize}
	\item $0_6^{(1,0,4,3)} \rightarrow 0^{(1,0,5,3)}$
	\item $0_6^{(1,1,4,2)} \rightarrow 0^{(1,2,4,2)}$
	\item $0_6^{(1,2,4,1)} \rightarrow 0^{(2,2,4,1)}$
	\item $0_6^{(1,3,4,0)} \rightarrow 0^{(1,1,3,4,0)}$
	\item $0_6^{(1,1,0,4,3)} \rightarrow 0^{(2,1,0,4,3)}$
	\item $0_6^{(1,1,1,4,2)} \rightarrow 0^{(1,1,1,1,4,2)}$
	\item $0_6^{(1,1,2,4,1)} \rightarrow 0^{(1,0,1,1,2,4,1)}$
	\item $0_6^{(1,1,3,4,0)} \rightarrow 0^{(1,0,0,1,1,3,4,0)}$
	\item $1_6^{(4,3)} \rightarrow 2^{(4,3)}$
	\item $1_6^{(1,4,2)} \rightarrow 1^{(1,4,3)} $
	\item $1_6^{(2,4,1)} \rightarrow 1^{(2,5,1)}$
	\item $1_6^{(3,4,0)} \rightarrow 1^{(4,4,0)}$
	\item $1_6^{(1,0,0,4,3)} \rightarrow 1^{(1,0,1,4,3)}$
	\item $1_6^{(1,0,1,4,2)} \rightarrow 1^{(1,1,1,4,2)}$
	\item $1_6^{(1,0,2,4,1)} \rightarrow 1^{(2,0,2,4,1)}$
	\item $1_6^{(1,0,3,4,0)} \rightarrow 1^{(1,1,0,3,4,0)}$
\end{itemize}
\end{minipage}
\subsection{\texorpdfstring{$g_s^{-7}$}{gs-7} Duality Orbits}
\thispagestyle{empty}
\begin{figure}[H]
\centering
\begin{tikzpicture}
\begin{scope}
\matrix(M)[matrix of math nodes, row sep=1.85em, column sep=2em, minimum width=2em]{
\clap{\text{A}} & 0_7^{(1,0,0,2,6)} & & 0_7^{(1,0,2,0,6)}\\
\clap{\text{B}} & & 0_7^{(1,0,1,1,6)} &\\
\clap{\text{A}} & & 0_7^{(1,1,1,1,5)} &\\
\clap{\text{B}} & 0_7^{(1,1,0,2,5)} & & 0_7^{(1,1,2,0,5)}\\
\clap{\text{A}} & 0_7^{(1,2,0,2,4)} & & 0_7^{(1,2,2,0,4)}\\
\clap{\text{B}} & & 0_7^{(1,2,1,1,4)} &\\
\clap{\text{A}} & & 0_7^{(1,3,1,1,3)} &\\
\clap{\text{B}} & 0_7^{(1,3,0,2,3)} & & 0_7^{(1,3,2,0,3)}\\
\clap{\text{A}} & 0_7^{(1,4,0,2,2)} & & 0_7^{(1,4,2,0,2)}\\
\clap{\text{B}} & & 0_7^{(1,4,1,1,2)} &\\
\clap{\text{A}} & & 0_7^{(1,5,1,1,1)} & \\
\clap{\text{B}} & 0_7^{(1,5,0,2,1)} & & 0_7^{(1,5,2,0,1)}\\
\clap{\text{A}} & 0_7^{(1,6,0,2,0)} & & 0_7^{(1,6,2,0,0)}\\
\clap{\text{B}} & & 0_7^{(1,6,1,1,0)} & \\
};
\end{scope}
\begin{scope}[xshift=10cm, yshift=-1.5cm]
\matrix(N)[matrix of math nodes, row sep=1.85em, column sep=2em, minimum width=2em]{
1_7^{(2,6)} & & 1_7^{(2,0,6)} & \clap{\text{B}}\\
& 1_7^{(1,1,6)} & & \clap{\text{A}}\\
& 1_7^{(1,1,1,5)} & & \clap{\text{B}}\\
1_7^{(1,0,2,5)} & & 1_7^{(1,2,0,5)} & \clap{\text{A}}\\
1_7^{(2,0,2,4)} & & 1_7^{(2,2,0,4)} & \clap{\text{B}}\\
& 1_7^{(2,1,1,4)} & & \clap{\text{A}}\\
& 1_7^{(3,1,1,3)} & & \clap{\text{B}}\\
1_7^{(3,0,2,3)} & & 1_7^{(3,2,0,3)} & \clap{\text{A}}\\
1_7^{(4,0,2,2)} & & 1_7^{(4,2,0,2)} & \clap{\text{B}}\\
& 1_7^{(4,1,1,2)} & & \clap{\text{A}}\\
& 1_7^{(5,1,1,1)} & & \clap{\text{B}}\\
1_7^{(5,0,2,1)} & & 1_7^{(5,2,0,1)} & \clap{\text{A}}\\
1_7^{(6,0,2,0)} & & 1_7^{(6,2,0,0)} & \clap{\text{B}}\\
& 1_7^{(6,1,1,0)} & & \clap{\text{A}}\\
};
\end{scope}
\draw[latex-latex] (M-1-2) -- (M-4-2);
\draw[latex-latex] (M-4-2) -- (M-5-2);
\draw[latex-latex] (M-5-2) -- (M-8-2);
\draw[latex-latex] (M-8-2) -- (M-9-2);
\draw[latex-latex] (M-9-2) -- (M-12-2);
\draw[latex-latex] (M-12-2) -- (M-13-2);
\draw[latex-latex] (M-2-3) -- (M-3-3);
\draw[latex-latex] (M-3-3) -- (M-6-3);
\draw[latex-latex] (M-6-3) -- (M-7-3);
\draw[latex-latex] (M-7-3) -- (M-10-3);
\draw[latex-latex] (M-10-3) -- (M-11-3);
\draw[latex-latex] (M-11-3) -- (M-14-3);
\draw[latex-latex] (M-1-4) -- (M-4-4);
\draw[latex-latex] (M-4-4) -- (M-5-4);
\draw[latex-latex] (M-5-4) -- (M-8-4);
\draw[latex-latex] (M-8-4) -- (M-9-4);
\draw[latex-latex] (M-9-4) -- (M-12-4);
\draw[latex-latex] (M-12-4) -- (M-13-4);
\draw[latex-latex] (N-1-1) -- (N-4-1);
\draw[latex-latex] (N-4-1) -- (N-5-1);
\draw[latex-latex] (N-5-1) -- (N-8-1);
\draw[latex-latex] (N-8-1) -- (N-9-1);
\draw[latex-latex] (N-9-1) -- (N-12-1);
\draw[latex-latex] (N-12-1) -- (N-13-1);
\draw[latex-latex] (N-2-2) -- (N-3-2);
\draw[latex-latex] (N-3-2) -- (N-6-2);
\draw[latex-latex] (N-6-2) -- (N-7-2);
\draw[latex-latex] (N-7-2) -- (N-10-2);
\draw[latex-latex] (N-10-2) -- (N-11-2);
\draw[latex-latex] (N-11-2) -- (N-14-2);
\draw[latex-latex] (N-1-3) -- (N-4-3);
\draw[latex-latex] (N-4-3) -- (N-5-3);
\draw[latex-latex] (N-5-3) -- (N-8-3);
\draw[latex-latex] (N-8-3) -- (N-9-3);
\draw[latex-latex] (N-9-3) -- (N-12-3);
\draw[latex-latex] (N-12-3) -- (N-13-3);
\draw[latex-latex] (M-1-2) -- (M-2-3);
\draw[latex-latex] (M-2-3) -- (M-1-4);
\draw[latex-latex] (M-4-2) -- (M-3-3);
\draw[latex-latex] (M-3-3) -- (M-4-4);
\draw[latex-latex] (M-5-2) -- (M-6-3);
\draw[latex-latex] (M-6-3) -- (M-5-4);
\draw[latex-latex] (M-8-2) -- (M-7-3);
\draw[latex-latex] (M-7-3) -- (M-8-4);
\draw[latex-latex] (M-9-2) -- (M-10-3);
\draw[latex-latex] (M-10-3) -- (M-9-4);
\draw[latex-latex] (M-12-2) -- (M-11-3);
\draw[latex-latex] (M-11-3) -- (M-12-4);
\draw[latex-latex] (M-13-2) -- (M-14-3);
\draw[latex-latex] (M-14-3) -- (M-13-4);
\draw[latex-latex] (N-1-1) -- (N-2-2);
\draw[latex-latex] (N-2-2) -- (N-1-3);
\draw[latex-latex] (N-4-1) -- (N-3-2);
\draw[latex-latex] (N-3-2) -- (N-4-3);
\draw[latex-latex] (N-5-1) -- (N-6-2);
\draw[latex-latex] (N-6-2) -- (N-5-3);
\draw[latex-latex] (N-8-1) -- (N-7-2);
\draw[latex-latex] (N-7-2) -- (N-8-3);
\draw[latex-latex] (N-9-1) -- (N-10-2);
\draw[latex-latex] (N-10-2) -- (N-9-3);
\draw[latex-latex] (N-12-1) -- (N-11-2);
\draw[latex-latex] (N-11-2) -- (N-12-3);
\draw[latex-latex] (N-13-1) -- (N-14-2);
\draw[latex-latex] (N-14-2) -- (N-13-3);
\draw[latex-latex] (M-1-2) --(N-1-1);
\draw[latex-latex] (M-2-3) --(N-2-2);
\draw[latex-latex] (M-1-4) --(N-1-3);
\draw[latex-latex] (M-4-2) --(N-4-1);
\draw[latex-latex] (M-3-3) --(N-3-2);
\draw[latex-latex] (M-4-4) --(N-4-3);
\draw[latex-latex] (M-5-2) --(N-5-1);
\draw[latex-latex] (M-6-3) --(N-6-2);
\draw[latex-latex] (M-5-4) --(N-5-3);
\draw[latex-latex] (M-8-2) --(N-8-1);
\draw[latex-latex] (M-7-3) --(N-7-2);
\draw[latex-latex] (M-8-4) --(N-8-3);
\draw[latex-latex] (M-9-2) --(N-9-1);
\draw[latex-latex] (M-10-3) --(N-10-2);
\draw[latex-latex] (M-9-4) --(N-9-3);
\draw[latex-latex] (M-12-2) --(N-12-1);
\draw[latex-latex] (M-11-3) --(N-11-2);
\draw[latex-latex] (M-12-4) --(N-12-3);
\draw[latex-latex] (M-13-2) --(N-13-1);
\draw[latex-latex] (M-14-3) --(N-14-2);
\draw[latex-latex] (M-13-4) --(N-13-3);
\end{tikzpicture}
\caption{The T-duality orbit of the $1_7^{(2,6)}$}
\label{fig:1726Orbit}
\end{figure}
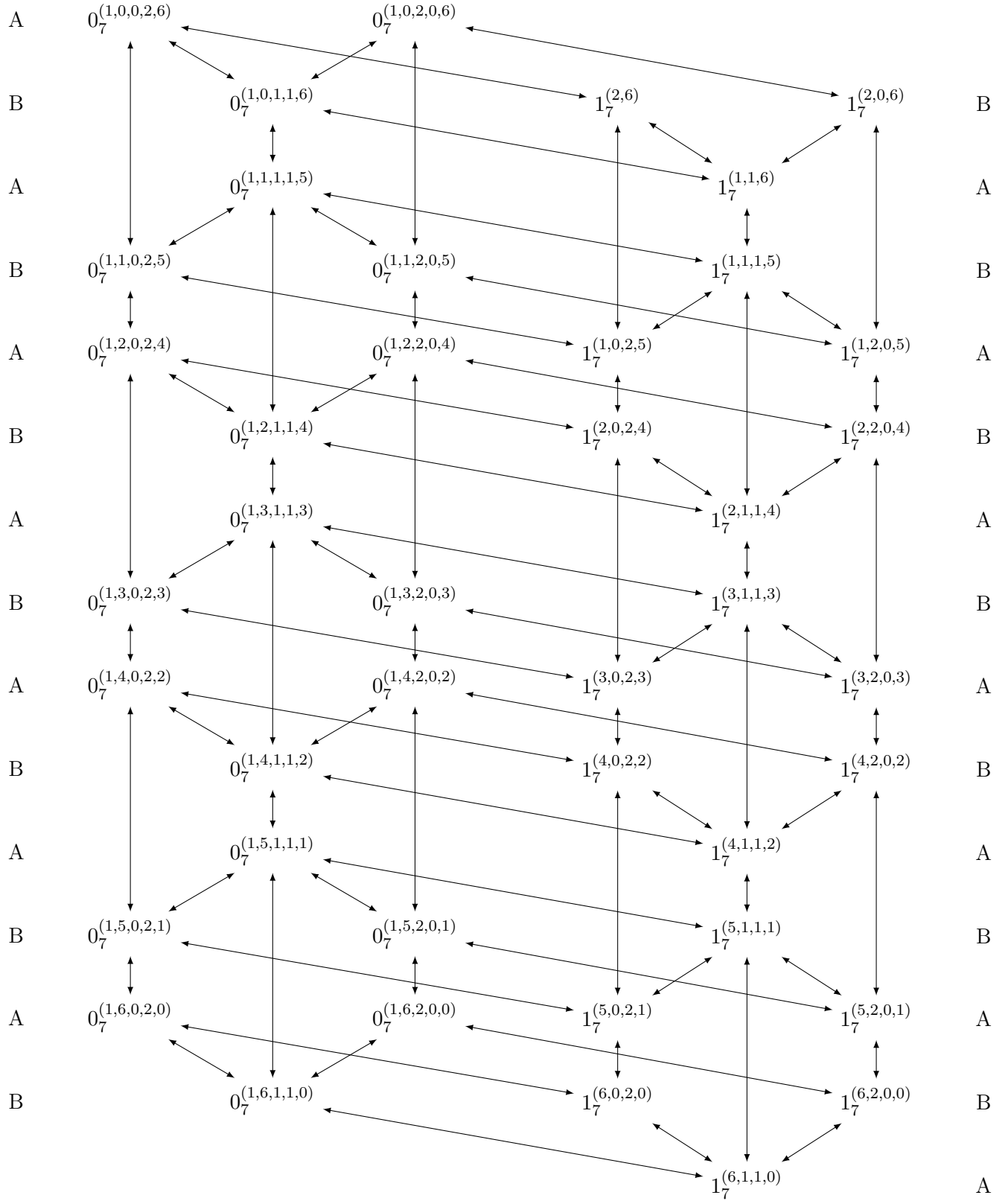
\begin{minipage}{0.5\linewidth}
S-dualities: 
\begin{itemize}
	\item $0_7^{(1,0,1,1,6)} \leftrightarrow 0_6^{(1,0,1,1,6)}$ See Figure \ref{fig:0617Orbit}
	\item $0_7^{(1,1,0,2,5)} \leftrightarrow 0_7^{(1,1,0,2,5)}$ Self-dual
	\item $0_7^{(1,1,2,0,5)} \leftrightarrow 0_8^{(1,1,2,0,5)}$
	\item $0_7^{(1,2,1,1,4)} \leftrightarrow 0_9^{(1,2,1,1,4)}$
	\item $0_7^{(1,3,0,2,3)} \leftrightarrow 0_{10}^{(1,3,0,2,3)}$
	\item $0_7^{(1,3,2,0,3)} \leftrightarrow 0_{11}^{(1,3,2,0,3)}$
	\item $0_7^{(1,4,1,1,2)} \leftrightarrow 0_{12}^{(1,4,1,1,2)}$
	\item $0_7^{(1,5,0,2,1)} \leftrightarrow 0_{13}^{(1,5,0,2,1)}$
	\item $0_7^{(1,5,2,0,1)} \leftrightarrow 0_{14}^{(1,5,2,0,1)}$
	\item $0_7^{(1,6,1,1,0)} \leftrightarrow 0_{15}^{(1,6,1,1,0)}$
	\item $1_7^{(2,6)} \leftrightarrow 1_3^{(2,6)}$ See Figure \ref{fig:532Orbit}
	\item $1_7^{(2,0,6)} \leftrightarrow 1_4^{(2,0,6)}$ See Figure \ref{fig:146Orbit}
	\item $1_7^{(1,1,1,5)} \leftrightarrow 1_5^{(1,1,1,5)}$ See Figure \ref{fig:2515Orbit}
	\item $1_7^{(2,0,2,4)} \leftrightarrow 1_6^{(2,0,2,4)}$ See Figure \ref{fig:3624Orbit}
	\item $1_7^{(2,2,0,4)} \leftrightarrow 1_7^{(2,2,0,4)}$ Self-dual
	\item $1_7^{(3,1,1,3)} \leftrightarrow 1_8^{(3,1,1,3)}$
	\item $1_7^{(4,0,2,2)} \leftrightarrow 1_9^{(4,0,2,2)}$
	\item $1_7^{(4,2,0,2)} \leftrightarrow 1_{10}^{(4,2,0,2)}$
	\item $1_7^{(5,1,1,1)} \leftrightarrow 1_{11}^{(5,1,1,1)}$
	\item $1_7^{(6,0,2,0)} \leftrightarrow 1_{12}^{(6,0,2,0)}$
	\item $1_7^{(6,2,0,0)} \leftrightarrow 1_{13}^{(6,2,0,0)}$
\end{itemize}
\end{minipage}
\begin{minipage}{0.5\linewidth}
M-theory origins:
\begin{itemize}
	\item $0_7^{(1,0,0,2,6)} \rightarrow 0^{(1,0,0,2,7)}$
	\item $0_7^{(1,0,2,0,6)} \rightarrow 0^{(1,0,2,1,6)}$
	\item $0_7^{(1,1,1,1,5)} \rightarrow 0^{(1,1,2,1,5)}$
	\item $0_7^{(1,2,0,2,4)} \rightarrow 0^{(1,3,0,2,4)}$
	\item $0_7^{(1,2,2,0,4)} \rightarrow 0^{(2,2,2,0,4)}$
	\item $0_7^{(1,3,1,1,3)} \rightarrow 0^{(1,1,3,1,1,3)}$
	\item $0_7^{(1,4,0,2,2)} \rightarrow 0^{(1,0,1,4,0,2,2)}$
	\item $0_7^{(1,4,2,0,2)} \rightarrow 0^{(1,0,0,1,4,2,0,2)}$
	\item $0_7^{(1,5,1,1,1)} \rightarrow 0^{(1,0,0,0,1,5,1,1,1)}$
	\item $0_7^{(1,6,0,2,0)} \rightarrow 0^{(1,0,0,0,0,1,6,0,2,0)}$
	\item $0_7^{(1,6,2,0,0)} \rightarrow 0^{(1,0,0,0,0,0,1,6,2,0,0)}$
	\item $1_7^{(1,1,6)} \rightarrow 1^{(1,1,6)}$
	\item $1_7^{(1,0,2,5)} \rightarrow 2^{(1,0,2,5)}$
	\item $1_7^{(1,2,0,5)} \rightarrow 1^{(1,2,0,6)}$
	\item $1_7^{(2,1,1,4)} \rightarrow 1^{(2,1,2,4)}$
	\item $1_7^{(3,0,2,3)} \rightarrow 1^{(3,1,2,3)}$
	\item $1_7^{(3,2,0,3)} \rightarrow 1^{(4,2,0,3)}$
	\item $1_7^{(4,1,1,2)} \rightarrow 1^{(1,4,1,1,2)}$
	\item $1_7^{(5,0,2,1)} \rightarrow 1^{(1,0,5,0,2,1)}$
	\item $1_7^{(5,2,0,1)} \rightarrow 1^{(1,0,0,5,2,0,1)}$
	\item $1_7^{(6,1,1,0)} \rightarrow 1^{(1,0,0,0,6,1,1,0)}$
\end{itemize}
\end{minipage}
\clearpage
\thispagestyle{empty}
\begin{minipage}{0.5\linewidth}
S-dualities:
\begin{itemize}
	\item $0_7^{(5,3)} \leftrightarrow 0_4^{(5,3)}$ See Figure \ref{fig:4413Orbit}
	\item $0_7^{(2,3,3)} \leftrightarrow 0_5^{(2,3,3)}$ See Figure \ref{fig:2515Orbit}
	\item $0_7^{(4,1,3)} \leftrightarrow 0_6^{(4,1,3)}$ See Figure \ref{fig:0617Orbit}
	\item $0_7^{(1,1,4,2)} \leftrightarrow 0_6^{(1,1,4,2)}$ See Figure \ref{fig:1643Orbit}
	\item $0_7^{(1,3,2,2)} \leftrightarrow 0_7^{(1,3,2,2)}$ Self-dual
	\item $0_7^{(1,5,0,2)} \leftrightarrow 0_8^{(1,5,0,2)}$
	\item $0_7^{(2,0,5,1)} \leftrightarrow 0_7^{(2,0,5,1)}$ Self-dual
	\item $0_7^{(2,2,3,1)} \leftrightarrow 0_8^{(2,2,3,1)}$
	\item $0_7^{(2,4,1,1)} \leftrightarrow 0_9^{(2,4,1,1)}$
	\item $0_7^{(3,1,4,0)} \leftrightarrow 0_9^{(3,1,4,0)}$ 
	\item $0_7^{(3,3,2,0)} \leftrightarrow 0_{10}^{(3,3,2,0)}$ 
	\item $0_7^{(3,5,0,0)} \leftrightarrow 0_{11}^{(3,5,0,0)}$ 
	\item $0_7^{(1,0,0,1,4,3)} \leftrightarrow 0_8^{(1,0,0,1,4,3)}$ 
	\item $0_7^{(1,0,0,3,2,3)} \leftrightarrow 0_9^{(1,0,0,3,2,3)}$ 
	\item $0_7^{(1,0,0,5,0,3)} \leftrightarrow 0_{10}^{(1,0,0,5,0,3)}$ 
	\item $0_7^{(1,0,1,0,5,2)} \leftrightarrow 0_9^{(1,0,1,0,5,2)}$ 
	\item $0_7^{(1,0,1,2,3,2)} \leftrightarrow 0_{10}^{(1,0,1,2,3,2)}$ 
	\item $0_7^{(1,0,1,4,1,2)} \leftrightarrow 0_{11}^{(1,0,1,4,1,2)}$ 
	\item $0_7^{(1,0,2,1,4,1)} \leftrightarrow 0_{11}^{(1,0,2,1,4,1)}$ 
	\item $0_7^{(1,0,2,3,2,1)} \leftrightarrow 0_{12}^{(1,0,2,3,2,1)}$ 
	\item $0_7^{(1,0,2,5,0,1)} \leftrightarrow 0_{13}^{(1,0,2,5,0,1)}$ 
	\item $0_7^{(1,0,3,0,5,0)} \leftrightarrow 0_{12}^{(1,0,3,0,5,0)}$ 
	\item $0_7^{(1,0,3,2,3,0)} \leftrightarrow 0_{13}^{(1,0,3,2,3,0)}$ 
	\item $0_7^{(1,0,3,4,1,0)} \leftrightarrow 0_{14}^{(1,0,3,4,1,0)}$ 
\end{itemize}
\end{minipage}
\begin{minipage}{0.5\linewidth}
M-theory origins:
\begin{itemize}
	\item $0_7^{(1,4,3)} \rightarrow 1^{(1,4,3)}$
	\item $0_7^{(3,2,3)} \rightarrow 0^{(3,2,4)}$
	\item $0_7^{(5,0,3)} \rightarrow 0^{(5,1,3)}$
	\item $0_7^{(1,0,5,2)} \rightarrow 0^{(1,0,5,3)}$
	\item $0_7^{(1,2,3,2)} \rightarrow 0^{(1,2,4,2)}$
	\item $0_7^{(1,4,1,2)} \rightarrow 0^{(1,5,1,2)}$
	\item $0_7^{(2,1,4,1)} \rightarrow 0^{(2,2,4,1)}$
	\item $0_7^{(2,3,2,1)} \rightarrow 0^{(3,3,2,1)}$
	\item $0_7^{(2,5,0,1)} \rightarrow 0^{(1,2,5,0,1)}$
	\item $0_7^{(3,0,5,0)} \rightarrow 0^{(4,0,5,0)}$
	\item $0_7^{(3,2,3,0)} \rightarrow 0^{(1,3,2,3,0)}$
	\item $0_7^{(3,4,1,0)} \rightarrow 0^{(1,0,3,4,1,0)}$
	\item $0_7^{(1,0,0,0,5,3)} \rightarrow 0^{(1,0,0,1,5,3)}$
	\item $0_7^{(1,0,0,2,3,3)} \rightarrow 0^{(1,0,1,2,3,3)}$
	\item $0_7^{(1,0,0,4,1,3)} \rightarrow 0^{(1,1,0,4,1,3)}$
	\item $0_7^{(1,0,1,1,4,2)} \rightarrow 0^{(1,1,1,1,4,2)}$
	\item $0_7^{(1,0,1,3,2,2)} \rightarrow 0^{(2,0,1,3,2,2)}$
	\item $0_7^{(1,0,1,5,0,2)} \rightarrow 0^{(1,1,0,1,5,0,2)}$
	\item $0_7^{(1,0,2,0,5,1)} \rightarrow 0^{(2,0,2,0,5,1)}$
	\item $0_7^{(1,0,2,2,3,1)} \rightarrow 0^{(1,1,0,2,2,3,1)}$
	\item $0_7^{(1,0,2,4,1,1)} \rightarrow 0^{(1,0,1,0,2,4,1,1)}$
	\item $0_7^{(1,0,3,1,4,0)} \rightarrow 0^{(1,0,1,0,3,1,4,0)}$
	\item $0_7^{(1,0,3,3,2,0)} \rightarrow 0^{(1,0,0,1,0,3,3,2,0)}$
	\item $0_7^{(1,0,3,5,0,0)} \rightarrow 0^{(1,0,0,0,1,0,3,5,0,0)}$
\end{itemize}
\end{minipage}
\clearpage
\thispagestyle{empty}
\begin{landscape}
\begin{figure}[H]
\centering
\begin{tikzpicture}
\begin{scope}
\matrix(M)[matrix of math nodes, row sep=3em, column sep=2em, minimum width=0em]{
\text{A} & 1_7^{(7,0)} & & 1_7^{(2,5,0)} & & 1_7^{(4,3,0)} & & 1_7^{(6,1,0)} &\\
\text{B} & & 1_7^{(1,6,0)} & & 1_7^{(3,4,0)} & & 1_7^{(5,2,0)} & & 1_7^{(7,0,0)} \\
\text{A} & & 1_7^{(1,0,0,1,6,0)} & & 1_7^{(1,0,0,3,4,0)} & & 1_7^{(1,0,0,5,2,0)} & & 1_7^{(1,0,0,7,0,0)} \\
\text{B} & 1_7^{(1,0,0,0,7,0)} & & 1_7^{(1,0,0,2,5,0)} & & 1_7^{(1,0,0,4,3,0)} & & 1_7^{(1,0,0,6,1,0)} &\\
};
\end{scope}
\begin{scope}[yshift=-8cm,xshift=2cm]
\matrix(N)[matrix of math nodes, row sep=3em, column sep=2em, minimum width=0em]{
\text{B} & 0_7^{(1,0,0,7,0)} & & 0_7^{(1,0,2,5,0)} & & 0_7^{(1,0,4,3,0)} & & 0_7^{(1,0,6,1,0)} &\\
\text{A} & & 0_7^{(1,0,1,6,0)} & & 0_7^{(1,0,3,4,0)} & & 0_7^{(1,0,5,2,0)} & & 0_7^{(1,0,7,0,0)} \\
\text{B} & & 0_7^{(1,1,0,1,6,0)} & & 0_7^{(1,1,0,3,4,0)} & & 0_7^{(1,1,0,5,2,0)} & & 0_7^{(1,1,0,7,0,0)} \\
\text{A} & 0_7^{(1,1,0,0,7,0)} & & 0_7^{(1,1,0,2,5,0)} & & 0_7^{(1,1,0,4,3,0)} & & 0_7^{(1,1,0,6,1,0)} &\\
};
\end{scope}
\draw[latex-latex] (M-1-2) -- (M-2-3);
\draw[latex-latex] (M-2-3) -- (M-1-4);
\draw[latex-latex] (M-1-4) -- (M-2-5);
\draw[latex-latex] (M-2-5) -- (M-1-6);
\draw[latex-latex] (M-1-6) -- (M-2-7);
\draw[latex-latex] (M-2-7) -- (M-1-8);
\draw[latex-latex] (M-1-8) -- (M-2-9);
\draw[latex-latex] (M-4-2) -- (M-3-3);
\draw[latex-latex] (M-3-3) -- (M-4-4);
\draw[latex-latex] (M-4-4) -- (M-3-5);
\draw[latex-latex] (M-3-5) -- (M-4-6);
\draw[latex-latex] (M-4-6) -- (M-3-7);
\draw[latex-latex] (M-3-7) -- (M-4-8);
\draw[latex-latex] (M-4-8) -- (M-3-9);
\draw[latex-latex] (M-1-2) -- (M-4-2);
\draw[latex-latex] (M-2-3) -- (M-3-3);
\draw[latex-latex] (M-1-4) -- (M-4-4);
\draw[latex-latex] (M-2-5) -- (M-3-5);
\draw[latex-latex] (M-1-6) -- (M-4-6);
\draw[latex-latex] (M-2-7) -- (M-3-7);
\draw[latex-latex] (M-1-8) -- (M-4-8);
\draw[latex-latex] (M-2-9) -- (M-3-9);
\draw[latex-latex] (N-1-2) -- (N-2-3);
\draw[latex-latex] (N-2-3) -- (N-1-4);
\draw[latex-latex] (N-1-4) -- (N-2-5);
\draw[latex-latex] (N-2-5) -- (N-1-6);
\draw[latex-latex] (N-1-6) -- (N-2-7);
\draw[latex-latex] (N-2-7) -- (N-1-8);
\draw[latex-latex] (N-1-8) -- (N-2-9);
\draw[latex-latex] (N-4-2) -- (N-3-3);
\draw[latex-latex] (N-3-3) -- (N-4-4);
\draw[latex-latex] (N-4-4) -- (N-3-5);
\draw[latex-latex] (N-3-5) -- (N-4-6);
\draw[latex-latex] (N-4-6) -- (N-3-7);
\draw[latex-latex] (N-3-7) -- (N-4-8);
\draw[latex-latex] (N-4-8) -- (N-3-9);
\draw[latex-latex] (N-1-2) -- (N-4-2);
\draw[latex-latex] (N-2-3) -- (N-3-3);
\draw[latex-latex] (N-1-4) -- (N-4-4);
\draw[latex-latex] (N-2-5) -- (N-3-5);
\draw[latex-latex] (N-1-6) -- (N-4-6);
\draw[latex-latex] (N-2-7) -- (N-3-7);
\draw[latex-latex] (N-1-8) -- (N-4-8);
\draw[latex-latex] (N-2-9) -- (N-3-9);
\draw[latex-latex] (M-1-2) -- (N-1-2);
\draw[latex-latex] (M-4-2) -- (N-4-2);
\draw[latex-latex] (M-2-3) -- (N-2-3);
\draw[latex-latex] (M-3-3) -- (N-3-3);
\draw[latex-latex] (M-1-4) -- (N-1-4);
\draw[latex-latex] (M-4-4) -- (N-4-4);
\draw[latex-latex] (M-2-5) -- (N-2-5);
\draw[latex-latex] (M-3-5) -- (N-3-5);
\draw[latex-latex] (M-1-6) -- (N-1-6);
\draw[latex-latex] (M-4-6) -- (N-4-6);
\draw[latex-latex] (M-2-7) -- (N-2-7);
\draw[latex-latex] (M-3-7) -- (N-3-7);
\draw[latex-latex] (M-1-8) -- (N-1-8);
\draw[latex-latex] (M-4-8) -- (N-4-8);
\draw[latex-latex] (M-2-9) -- (N-2-9);
\draw[latex-latex] (M-3-9) -- (N-3-9);
\end{tikzpicture}
\caption{The T-duality orbit of the $1_7^{(1,6,0)}$.}
\label{fig:17160Orbit}
\end{figure}
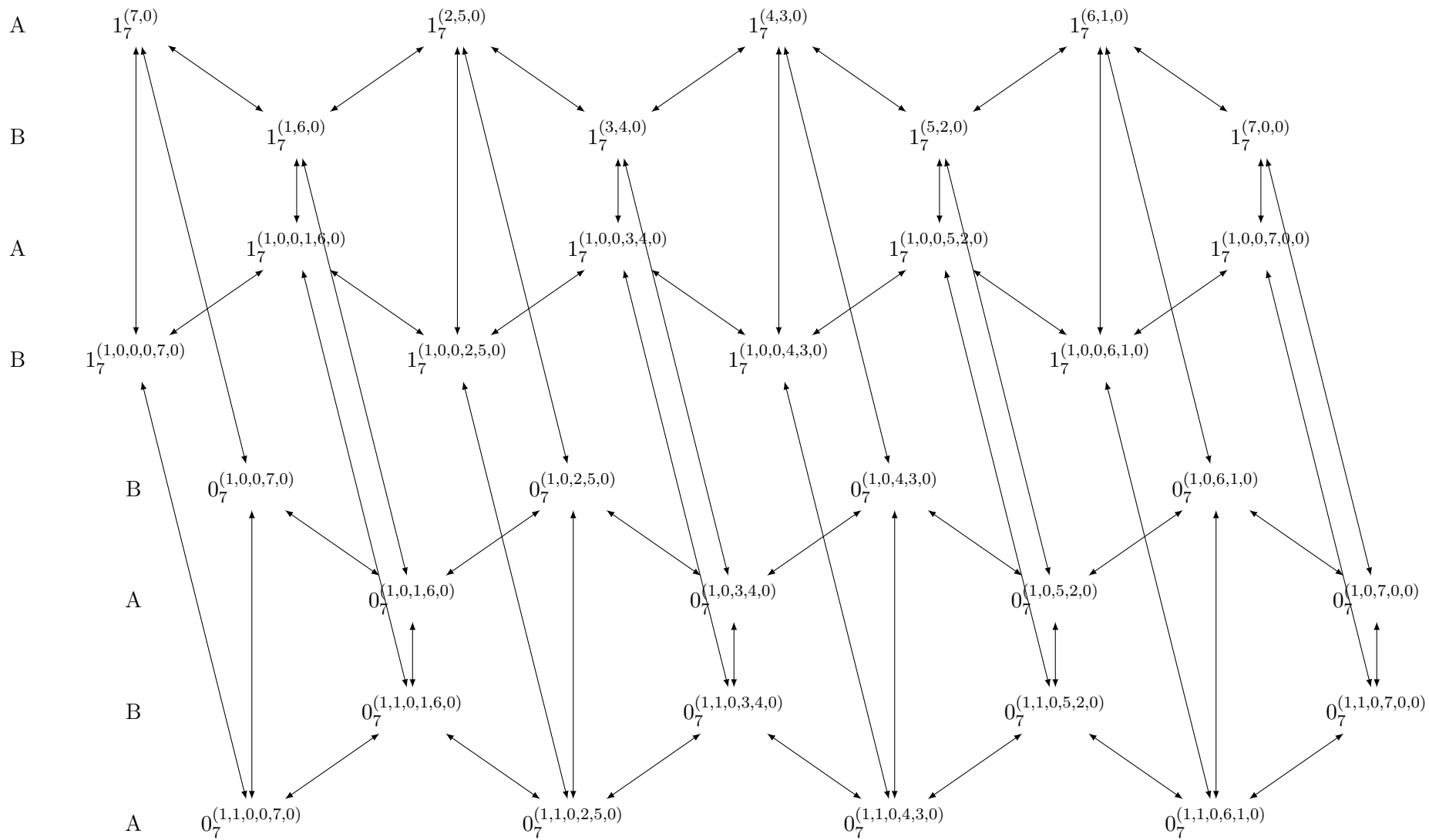
\end{landscape}
\clearpage
\begin{minipage}{0.5\linewidth}
S-dualities:
\begin{itemize}
	\item $1_7^{(1,6,0)} \leftrightarrow 1_5^{(1,6,0)}$ See Figure \ref{fig:2515Orbit}
	\item $1_7^{(3,4,0)} \leftrightarrow 1_6^{(3,4,0)}$ See Figure \ref{fig:1643Orbit}
	\item $1_7^{(5,2,0)} \leftrightarrow 1_7^{(5,2,0)}$ Self-dual
	\item $1_7^{(7,0,0)} \leftrightarrow 1_8^{(7,0,0)}$
	\item $1_7^{(1,0,0,0,7,0)} \leftrightarrow 1_8^{(1,0,0,0,7,0)}$
	\item $1_7^{(1,0,0,2,5,0)} \leftrightarrow 1_9^{(1,0,0,2,5,0)}$
	\item $1_7^{(1,0,0,4,3,0)} \leftrightarrow 1_{10}^{(1,0,0,4,3,0)}$
	\item $1_7^{(1,0,0,6,1,0)} \leftrightarrow 1_{11}^{(1,0,0,6,1,0)}$
	\item $0_7^{(1,0,0,7,0)} \leftrightarrow 0_7^{(1,0,0,7,0)}$ Self-dual
	\item $0_7^{(1,0,2,5,0)} \leftrightarrow 0_8^{(1,0,2,5,0)}$
	\item $0_7^{(1,0,4,3,0)} \leftrightarrow 0_9^{(1,0,4,3,0)}$
	\item $0_7^{(1,0,6,1,0)} \leftrightarrow 0_{10}^{(1,0,6,1,0)}$
	\item $0_7^{(1,1,0,1,6,0)} \leftrightarrow 0_{11}^{(1,1,0,1,6,0)}$
	\item $0_7^{(1,1,0,3,4,0)} \leftrightarrow 0_{12}^{(1,1,0,3,4,0)}$
	\item $0_7^{(1,1,0,5,2,0)} \leftrightarrow 0_{13}^{(1,1,0,5,2,0)}$
	\item $0_7^{(1,1,0,7,0,0)} \leftrightarrow 0_{14}^{(1,1,0,7,0,0)}$
\end{itemize}
\end{minipage}
\begin{minipage}{0.5\linewidth}
M-theory origins:
\begin{itemize}
	\item $1_7^{(7,0)} \rightarrow 2^{(7,0)}$
	\item $1_7^{(2,5,0)} \rightarrow 1^{(2,5,1)}$
	\item $1_7^{(4,3,0)} \rightarrow 1^{(4,4,0)}$
	\item $1_7^{(6,1,0)} \rightarrow 1^{(7,1,0)}$
	\item $1_7^{(1,0,0,1,6,0)} \rightarrow 1^{(1,0,1,1,6,0)}$
	\item $1_7^{(1,0,0,3,4,0)} \rightarrow 1^{(1,1,0,3,4,0)}$
	\item $1_7^{(1,0,0,5,2,0)} \rightarrow 1^{(2,0,0,5,2,0)}$
	\item $1_7^{(1,0,0,7,0,0)} \rightarrow 1^{(1,1,0,0,7,0,0)}$
	\item $0_7^{(1,0,1,6,0)} \rightarrow 0^{(1,0,2,6,0)}$
	\item $0_7^{(1,0,3,4,0)} \rightarrow 0^{(1,1,3,4,0)}$
	\item $0_7^{(1,0,5,2,0)} \rightarrow 0^{(2,0,5,2,0)}$
	\item $0_7^{(1,0,7,0,0)} \rightarrow 0^{(1,1,0,7,0,0)}$
	\item $0_7^{(1,1,0,0,7,0)} \rightarrow 0^{(2,1,0,0,7,0)}$
	\item $0_7^{(1,1,0,2,5,0)} \rightarrow 0^{(1,1,1,0,2,5,0)}$
	\item $0_7^{(1,1,0,4,3,0)} \rightarrow 0^{(1,0,1,1,0,4,3,0)}$
	\item $0_7^{(1,1,0,6,1,0)} \rightarrow 0^{(1,0,0,1,1,0,6,1,0)}$
\end{itemize}
\end{minipage}
\clearpage
\thispagestyle{empty}
\begin{landscape}
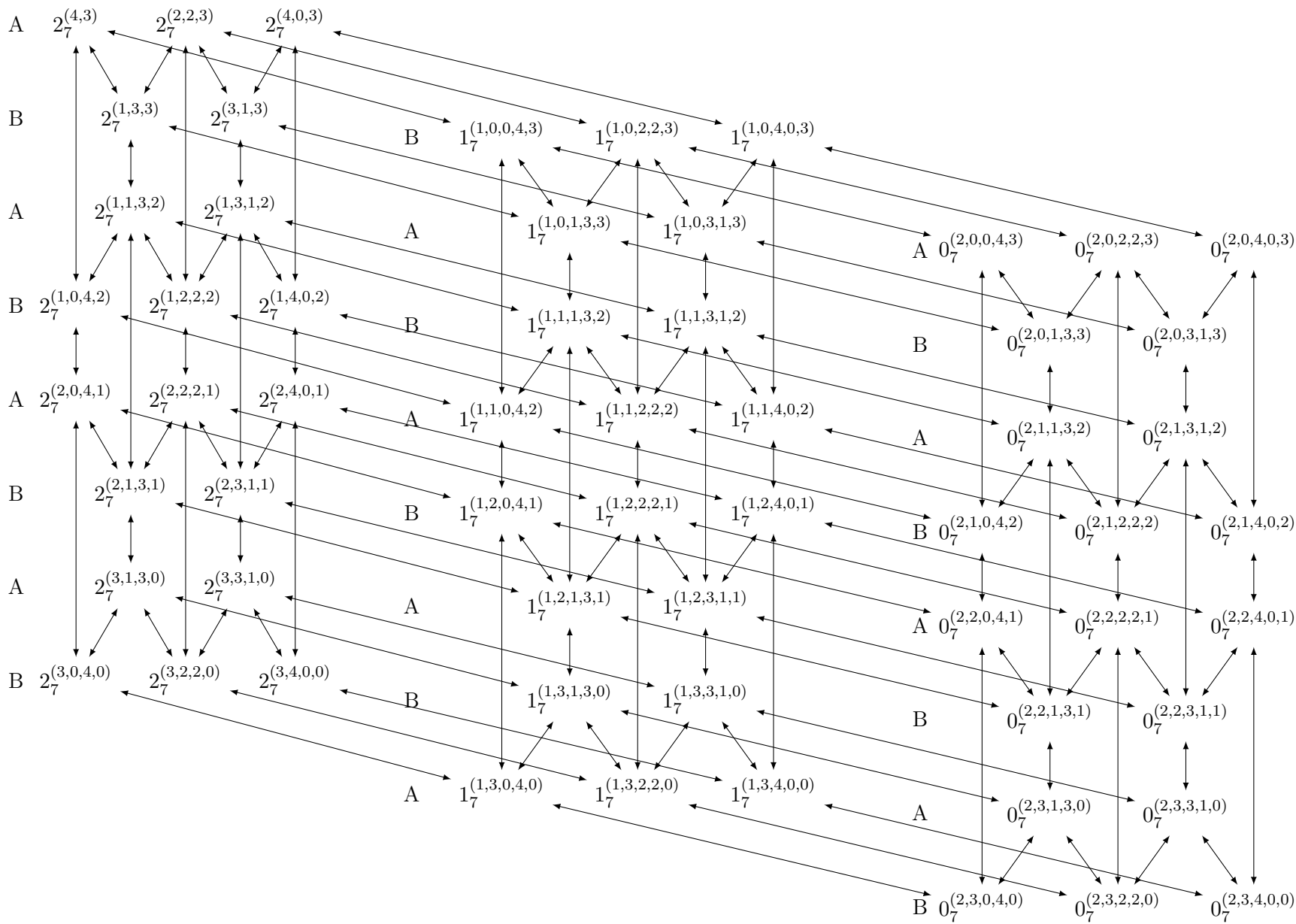
\begin{figure}[H]
\centering
\begin{tikzpicture}
\begin{scope}[xshift=-2.5cm]
\matrix(m)[matrix of math nodes, row sep=2.3em, column sep=-1em]{
\text{A}\phantom{\mathclap{2_7^{(4,3)}}}\\
\text{B}\phantom{\mathclap{2_7^{(1,3,3)}}}\\
\text{A}\phantom{\mathclap{2_7^{(1,1,3,2)}}}\\
\text{B}\phantom{\mathclap{2_7^{(1,0,4,2)}}}\\
\text{A}\phantom{\mathclap{2_7^{(2,0,4,1)}}}\\
\text{B}\phantom{\mathclap{2_7^{(2,1,3,1)}}}\\
\text{A}\phantom{\mathclap{2_7^{(3,1,3,0)}}}\\
\text{B}\phantom{\mathclap{2_7^{(3,0,4,0)}}}\\
};
\end{scope}
\begin{scope}[xshift=0.5cm]
\matrix(M)[matrix of math nodes, row sep=2.3em, column sep=-1.5em, minimum width=2em]{
2_7^{(4,3)} & & 2_7^{(2,2,3)} & & 2_7^{(4,0,3)}\\
& 2_7^{(1,3,3)} & & 2_7^{(3,1,3)} & \\
& 2_7^{(1,1,3,2)} & & 2_7^{(1,3,1,2)} & \\
2_7^{(1,0,4,2)} & & 2_7^{(1,2,2,2)} & & 2_7^{(1,4,0,2)}\\
2_7^{(2,0,4,1)} & & 2_7^{(2,2,2,1)} & & 2_7^{(2,4,0,1)}\\
& 2_7^{(2,1,3,1)} & & 2_7^{(2,3,1,1)} & \\
& 2_7^{(3,1,3,0)} & & 2_7^{(3,3,1,0)} & \\
2_7^{(3,0,4,0)} & & 2_7^{(3,2,2,0)} & & 2_7^{(3,4,0,0)}\\
};
\end{scope}
\begin{scope}[xshift=4.5cm, yshift=-2cm]
\matrix(n)[matrix of math nodes, row sep=2.3em]{
\text{B}\phantom{\mathclap{1_7^{(1,0,0,4,3)}}}\\
\text{A}\phantom{\mathclap{1_7^{(1,0,1,3,3)}}}\\
\text{B}\phantom{\mathclap{1_7^{(1,1,1,3,2)}}}\\
\text{A}\phantom{\mathclap{1_7^{(1,1,0,4,2)}}}\\
\text{B}\phantom{\mathclap{1_7^{(1,2,0,4,1)}}}\\
\text{A}\phantom{\mathclap{1_7^{(1,2,1,3,1)}}}\\
\text{B}\phantom{\mathclap{1_7^{(1,3,1,3,0)}}}\\
\text{A}\phantom{\mathclap{1_7^{(1,3,0,4,0)}}}\\
};
\end{scope}
\begin{scope}[xshift=8.5cm,yshift=-2cm]
\matrix(N)[matrix of math nodes, row sep=2.3em, column sep=-1.5em, minimum width=2em]{
1_7^{(1,0,0,4,3)} & & 1_7^{(1,0,2,2,3)} & & 1_7^{(1,0,4,0,3)}\\
& 1_7^{(1,0,1,3,3)} & & 1_7^{(1,0,3,1,3)} & \\
& 1_7^{(1,1,1,3,2)} & & 1_7^{(1,1,3,1,2)} & \\
1_7^{(1,1,0,4,2)} & & 1_7^{(1,1,2,2,2)} & & 1_7^{(1,1,4,0,2)}\\
1_7^{(1,2,0,4,1)} & & 1_7^{(1,2,2,2,1)} & & 1_7^{(1,2,4,0,1)}\\
& 1_7^{(1,2,1,3,1)} & & 1_7^{(1,2,3,1,1)} & \\
& 1_7^{(1,3,1,3,0)} & & 1_7^{(1,3,3,1,0)} & \\
1_7^{(1,3,0,4,0)} & & 1_7^{(1,3,2,2,0)} & & 1_7^{(1,3,4,0,0)}\\
};
\end{scope}
\begin{scope}[xshift=13.5cm,yshift=-4cm]
\matrix(p)[matrix of math nodes, row sep=2.3em]{
\text{A}\phantom{\mathclap{0_7^{(2,0,0,4,3)}}}\\
\text{B}\phantom{\mathclap{0_7^{(2,0,1,3,3)}}}\\
\text{A}\phantom{\mathclap{0_7^{(2,1,1,3,2)}}}\\
\text{B}\phantom{\mathclap{0_7^{(2,1,0,4,2)}}}\\
\text{A}\phantom{\mathclap{0_7^{(2,2,0,4,1)}}}\\
\text{B}\phantom{\mathclap{0_7^{(2,2,1,3,1)}}}\\
\text{A}\phantom{\mathclap{0_7^{(2,3,1,3,0)}}}\\
\text{B}\phantom{\mathclap{0_7^{(2,3,0,4,0)}}}\\
};
\end{scope}
\begin{scope}[xshift=17cm,yshift=-4cm]
\matrix(P)[matrix of math nodes, row sep=2.32em, column sep=-1.5em, minimum width=2em]{
0_7^{(2,0,0,4,3)} & & 0_7^{(2,0,2,2,3)} & & 0_7^{(2,0,4,0,3)}\\
& 0_7^{(2,0,1,3,3)} & & 0_7^{(2,0,3,1,3)} & \\
& 0_7^{(2,1,1,3,2)} & & 0_7^{(2,1,3,1,2)} & \\
0_7^{(2,1,0,4,2)} & & 0_7^{(2,1,2,2,2)} & & 0_7^{(2,1,4,0,2)}\\
0_7^{(2,2,0,4,1)} & & 0_7^{(2,2,2,2,1)} & & 0_7^{(2,2,4,0,1)}\\
& 0_7^{(2,2,1,3,1)} & & 0_7^{(2,2,3,1,1)} & \\
& 0_7^{(2,3,1,3,0)} & & 0_7^{(2,3,3,1,0)} & \\
0_7^{(2,3,0,4,0)} & & 0_7^{(2,3,2,2,0)} & & 0_7^{(2,3,4,0,0)}\\
};
\end{scope}
\draw[latex-latex] (M-1-1) -- (M-2-2);
\draw[latex-latex] (M-2-2) -- (M-1-3);
\draw[latex-latex] (M-1-3) -- (M-2-4);
\draw[latex-latex] (M-2-4) -- (M-1-5);
\draw[latex-latex] (M-4-1) -- (M-3-2);
\draw[latex-latex] (M-3-2) -- (M-4-3);
\draw[latex-latex] (M-4-3) -- (M-3-4);
\draw[latex-latex] (M-3-4) -- (M-4-5);
\draw[latex-latex] (M-5-1) -- (M-6-2);
\draw[latex-latex] (M-6-2) -- (M-5-3);
\draw[latex-latex] (M-5-3) -- (M-6-4);
\draw[latex-latex] (M-6-4) -- (M-5-5);
\draw[latex-latex] (M-8-1) -- (M-7-2);
\draw[latex-latex] (M-7-2) -- (M-8-3);
\draw[latex-latex] (M-8-3) -- (M-7-4);
\draw[latex-latex] (M-7-4) -- (M-8-5);
\draw[latex-latex] (M-1-1) -- (M-4-1);
\draw[latex-latex] (M-4-1) -- (M-5-1);
\draw[latex-latex] (M-5-1) -- (M-8-1);
\draw[latex-latex] (M-2-2) -- (M-3-2);
\draw[latex-latex] (M-3-2) -- (M-6-2);
\draw[latex-latex] (M-6-2) -- (M-7-2);
\draw[latex-latex] (M-1-3) -- (M-4-3);
\draw[latex-latex] (M-4-3) -- (M-5-3);
\draw[latex-latex] (M-5-3) -- (M-8-3);
\draw[latex-latex] (M-2-4) -- (M-3-4);
\draw[latex-latex] (M-3-4) -- (M-6-4);
\draw[latex-latex] (M-6-4) -- (M-7-4);
\draw[latex-latex] (M-1-5) -- (M-4-5);
\draw[latex-latex] (M-4-5) -- (M-5-5);
\draw[latex-latex] (M-5-5) -- (M-8-5);
\draw[latex-latex] (N-1-1) -- (N-2-2);
\draw[latex-latex] (N-2-2) -- (N-1-3);
\draw[latex-latex] (N-1-3) -- (N-2-4);
\draw[latex-latex] (N-2-4) -- (N-1-5);
\draw[latex-latex] (N-4-1) -- (N-3-2);
\draw[latex-latex] (N-3-2) -- (N-4-3);
\draw[latex-latex] (N-4-3) -- (N-3-4);
\draw[latex-latex] (N-3-4) -- (N-4-5);
\draw[latex-latex] (N-5-1) -- (N-6-2);
\draw[latex-latex] (N-6-2) -- (N-5-3);
\draw[latex-latex] (N-5-3) -- (N-6-4);
\draw[latex-latex] (N-6-4) -- (N-5-5);
\draw[latex-latex] (N-8-1) -- (N-7-2);
\draw[latex-latex] (N-7-2) -- (N-8-3);
\draw[latex-latex] (N-8-3) -- (N-7-4);
\draw[latex-latex] (N-7-4) -- (N-8-5);
\draw[latex-latex] (N-1-1) -- (N-4-1);
\draw[latex-latex] (N-4-1) -- (N-5-1);
\draw[latex-latex] (N-5-1) -- (N-8-1);
\draw[latex-latex] (N-2-2) -- (N-3-2);
\draw[latex-latex] (N-3-2) -- (N-6-2);
\draw[latex-latex] (N-6-2) -- (N-7-2);
\draw[latex-latex] (N-1-3) -- (N-4-3);
\draw[latex-latex] (N-4-3) -- (N-5-3);
\draw[latex-latex] (N-5-3) -- (N-8-3);
\draw[latex-latex] (N-2-4) -- (N-3-4);
\draw[latex-latex] (N-3-4) -- (N-6-4);
\draw[latex-latex] (N-6-4) -- (N-7-4);
\draw[latex-latex] (N-1-5) -- (N-4-5);
\draw[latex-latex] (N-4-5) -- (N-5-5);
\draw[latex-latex] (N-5-5) -- (N-8-5);
\draw[latex-latex] (P-1-1) -- (P-2-2);
\draw[latex-latex] (P-2-2) -- (P-1-3);
\draw[latex-latex] (P-1-3) -- (P-2-4);
\draw[latex-latex] (P-2-4) -- (P-1-5);
\draw[latex-latex] (P-4-1) -- (P-3-2);
\draw[latex-latex] (P-3-2) -- (P-4-3);
\draw[latex-latex] (P-4-3) -- (P-3-4);
\draw[latex-latex] (P-3-4) -- (P-4-5);
\draw[latex-latex] (P-5-1) -- (P-6-2);
\draw[latex-latex] (P-6-2) -- (P-5-3);
\draw[latex-latex] (P-5-3) -- (P-6-4);
\draw[latex-latex] (P-6-4) -- (P-5-5);
\draw[latex-latex] (P-8-1) -- (P-7-2);
\draw[latex-latex] (P-7-2) -- (P-8-3);
\draw[latex-latex] (P-8-3) -- (P-7-4);
\draw[latex-latex] (P-7-4) -- (P-8-5);
\draw[latex-latex] (P-1-1) -- (P-4-1);
\draw[latex-latex] (P-4-1) -- (P-5-1);
\draw[latex-latex] (P-5-1) -- (P-8-1);
\draw[latex-latex] (P-2-2) -- (P-3-2);
\draw[latex-latex] (P-3-2) -- (P-6-2);
\draw[latex-latex] (P-6-2) -- (P-7-2);
\draw[latex-latex] (P-1-3) -- (P-4-3);
\draw[latex-latex] (P-4-3) -- (P-5-3);
\draw[latex-latex] (P-5-3) -- (P-8-3);
\draw[latex-latex] (P-2-4) -- (P-3-4);
\draw[latex-latex] (P-3-4) -- (P-6-4);
\draw[latex-latex] (P-6-4) -- (P-7-4);
\draw[latex-latex] (P-1-5) -- (P-4-5);
\draw[latex-latex] (P-4-5) -- (P-5-5);
\draw[latex-latex] (P-5-5) -- (P-8-5);
\draw[latex-latex] (M-1-1) -- (N-1-1);
\draw[latex-latex] (M-1-3) -- (N-1-3);
\draw[latex-latex] (M-1-5) -- (N-1-5);
\draw[latex-latex] (M-2-2) -- (N-2-2);
\draw[latex-latex] (M-2-4) -- (N-2-4);
\draw[latex-latex] (M-3-2) -- (N-3-2);
\draw[latex-latex] (M-3-4) -- (N-3-4);
\draw[latex-latex] (M-4-1) -- (N-4-1);
\draw[latex-latex] (M-4-3) -- (N-4-3);
\draw[latex-latex] (M-4-5) -- (N-4-5);
\draw[latex-latex] (M-5-1) -- (N-5-1);
\draw[latex-latex] (M-5-3) -- (N-5-3);
\draw[latex-latex] (M-5-5) -- (N-5-5);
\draw[latex-latex] (M-6-2) -- (N-6-2);
\draw[latex-latex] (M-6-4) -- (N-6-4);
\draw[latex-latex] (M-7-2) -- (N-7-2);
\draw[latex-latex] (M-7-4) -- (N-7-4);
\draw[latex-latex] (M-8-1) -- (N-8-1);
\draw[latex-latex] (M-8-3) -- (N-8-3);
\draw[latex-latex] (M-8-5) -- (N-8-5);
\draw[latex-latex] (N-1-1) -- (P-1-1);
\draw[latex-latex] (N-1-3) -- (P-1-3);
\draw[latex-latex] (N-1-5) -- (P-1-5);
\draw[latex-latex] (N-2-2) -- (P-2-2);
\draw[latex-latex] (N-2-4) -- (P-2-4);
\draw[latex-latex] (N-3-2) -- (P-3-2);
\draw[latex-latex] (N-3-4) -- (P-3-4);
\draw[latex-latex] (N-4-1) -- (P-4-1);
\draw[latex-latex] (N-4-3) -- (P-4-3);
\draw[latex-latex] (N-4-5) -- (P-4-5);
\draw[latex-latex] (N-5-1) -- (P-5-1);
\draw[latex-latex] (N-5-3) -- (P-5-3);
\draw[latex-latex] (N-5-5) -- (P-5-5);
\draw[latex-latex] (N-6-2) -- (P-6-2);
\draw[latex-latex] (N-6-4) -- (P-6-4);
\draw[latex-latex] (N-7-2) -- (P-7-2);
\draw[latex-latex] (N-7-4) -- (P-7-4);
\draw[latex-latex] (N-8-1) -- (P-8-1);
\draw[latex-latex] (N-8-3) -- (P-8-3);
\draw[latex-latex] (N-8-5) -- (P-8-5);
\end{tikzpicture}
\caption{The T-duality orbit of the $2_7^{(1,3,3)}$.}
\label{fig:27133Orbit}
\end{figure}
\end{landscape}
\clearpage
\begin{minipage}{0.5\linewidth}
S-dualities:
\begin{itemize}
	\item $2_7^{(1,3,3)} \leftrightarrow 2_4^{(1,3,3)}$ See Figure \ref{fig:4413Orbit}
	\item $2_7^{(3,1,3)} \leftrightarrow 2_5^{(3,1,3)}$ See Figure \ref{fig:5522Orbit}
	\item $2_7^{(1,0,4,2)} \leftrightarrow 2_5^{(1,0,4,2)}$ See Figure \ref{fig:2515Orbit}
	\item $2_7^{(1,2,2,2)} \leftrightarrow 2_6^{(1,2,2,2)}$ See Figure \ref{fig:3624Orbit}
	\item $2_7^{(1,4,0,2)} \leftrightarrow 2_7^{(1,4,0,2)}$ Self-dual
	\item $2_7^{(2,1,3,1)} \leftrightarrow 2_7^{(2,1,3,1)}$ Self-dual
	\item $2_7^{(2,3,1,1)} \leftrightarrow 2_8^{(2,3,1,1)}$
	\item $2_7^{(3,0,4,0)} \leftrightarrow 2_8^{(3,0,4,0)}$
	\item $2_7^{(3,2,2,0)} \leftrightarrow 2_9^{(3,2,2,0)}$
	\item $2_7^{(3,4,0,0)} \leftrightarrow 2_{10}^{(3,4,0,0)}$
	\item $1_7^{(1,0,0,4,3)} \leftrightarrow 1_6^{(1,0,0,4,3)}$ See Figure \ref{fig:1643Orbit}
	\item $1_7^{(1,0,2,2,3)} \leftrightarrow 1_7^{(1,0,2,2,3)}$ Self-dual
	\item $1_7^{(1,0,4,0,3)} \leftrightarrow 1_8^{(1,0,4,0,3)}$
	\item $1_7^{(1,1,1,3,2)} \leftrightarrow 1_9^{(1,1,1,3,2)}$
	\item $1_7^{(1,1,3,1,2)} \leftrightarrow 1_{10}^{(1,1,3,1,2)}$
	\item $1_7^{(1,2,0,4,1)} \leftrightarrow 1_9^{(1,2,0,4,1)}$
	\item $1_7^{(1,2,2,2,1)} \leftrightarrow 1_{10}^{(1,2,2,2,1)}$
	\item $1_7^{(1,2,4,0,1)} \leftrightarrow 1_{11}^{(1,2,4,0,1)}$
	\item $1_7^{(1,3,1,3,0)} \leftrightarrow 1_{11}^{(1,3,1,3,0)}$
	\item $1_7^{(1,3,3,1,0)} \leftrightarrow 1_{12}^{(1,3,3,1,0)}$
	\item $0_7^{(2,0,1,3,3)} \leftrightarrow 0_9^{(2,0,1,3,3)}$
	\item $0_7^{(2,0,3,1,3)} \leftrightarrow 0_{10}^{(2,0,3,1,3)}$
	\item $0_7^{(2,1,0,4,2)} \leftrightarrow 0_{10}^{(2,1,0,4,2)}$
	\item $0_7^{(2,1,2,2,2)} \leftrightarrow 0_{11}^{(2,1,2,2,2)}$
	\item $0_7^{(2,1,4,0,2)} \leftrightarrow 0_{12}^{(2,1,4,0,2)}$
	\item $0_7^{(2,2,1,3,1)} \leftrightarrow 0_{12}^{(2,2,1,3,1)}$
	\item $0_7^{(2,2,3,1,1)} \leftrightarrow 0_{13}^{(2,2,3,1,1)}$
	\item $0_7^{(2,3,0,4,0)} \leftrightarrow 0_{13}^{(2,3,0,4,0)}$
	\item $0_7^{(2,3,2,2,0)} \leftrightarrow 0_{14}^{(2,3,2,2,0)}$
	\item $0_7^{(2,3,4,0,0)} \leftrightarrow 0_{15}^{(2,3,4,0,0)}$
\end{itemize}
\end{minipage}
\begin{minipage}{0.5\linewidth}
M-theory origins:
\begin{itemize}
	\item $2_7^{(4,3)} \rightarrow 2^{(4,3)}$
	\item $2_7^{(2,2,3)} \rightarrow 3^{(2,2,3)}$
	\item $2_7^{(4,0,3)} \rightarrow 2^{(4,0,4)}$
	\item $2_7^{(1,1,3,2)} \rightarrow 2^{(1,1,3,3)}$
	\item $2_7^{(1,3,1,2)} \rightarrow 2^{(1,3,2,2)}$
	\item $2_7^{(2,0,4,1)} \rightarrow 2^{(2,0,5,1)}$
	\item $2_7^{(2,2,2,1)} \rightarrow 2^{(2,3,2,1)}$
	\item $2_7^{(2,4,0,1)} \rightarrow 2^{(3,4,0,1)}$
	\item $2_7^{(3,1,3,0)} \rightarrow 2^{(4,1,3,0)}$
	\item $2_7^{(3,3,1,0)} \rightarrow 2^{1,3,3,1,0)}$
	\item $1_7^{(1,0,1,3,3)} \rightarrow 1^{(1,0,1,4,3)}$
	\item $1_7^{(1,0,3,1,3)} \rightarrow 1^{(1,0,4,1,3)}$
	\item $1_7^{(1,1,0,4,2)} \rightarrow 1^{(1,1,1,4,2)}$
	\item $1_7^{(1,1,2,2,2)} \rightarrow 1^{(1,2,2,2,2)}$
	\item $1_7^{(1,1,4,0,2)} \rightarrow 1^{(1,1,1,4,0,2)}$
	\item $1_7^{(1,2,1,3,1)} \rightarrow 1^{(2,2,1,3,1)}$
	\item $1_7^{(1,2,3,1,1)} \rightarrow 1^{(1,1,2,3,1,1)}$
	\item $1_7^{(1,3,0,4,0)} \rightarrow 1^{(1,1,3,0,4,0)}$
	\item $1_7^{(1,3,2,2,0)} \rightarrow 1^{(1,0,1,3,2,2,0)}$
	\item $1_7^{(1,3,4,0,0)} \rightarrow 1^{(1,0,0,1,3,4,0,0)}$
	\item $0_7^{(2,0,0,4,3)} \rightarrow 0^{(2,1,0,4,3)}$
	\item $0_7^{(2,0,2,2,3)} \rightarrow 0^{(3,0,2,2,3)}$
	\item $0_7^{(2,0,4,0,3)} \rightarrow 0^{(1,2,0,4,0,3)}$
	\item $0_7^{(2,1,1,3,2)} \rightarrow 0^{(1,2,1,1,3,2)}$
	\item $0_7^{(2,1,3,1,2)} \rightarrow 0^{(1,0,2,1,3,1,2)}$
	\item $0_7^{(2,2,0,4,1)} \rightarrow 0^{(1,0,2,2,0,4,1)}$
	\item $0_7^{(2,2,2,2,1)} \rightarrow 0^{(1,0,0,2,2,2,2,1)}$
	\item $0_7^{(2,2,4,0,1)} \rightarrow 0^{(1,0,0,0,2,2,4,0,1)}$
	\item $0_7^{(2,3,1,3,0)} \rightarrow 0^{(1,0,0,0,2,3,1,3,0)}$
	\item $0_7^{(2,3,3,1,0)} \rightarrow 0^{(1,0,0,0,0,2,3,3,1,0)}$
\end{itemize}
\end{minipage}
\clearpage
\begin{landscape}
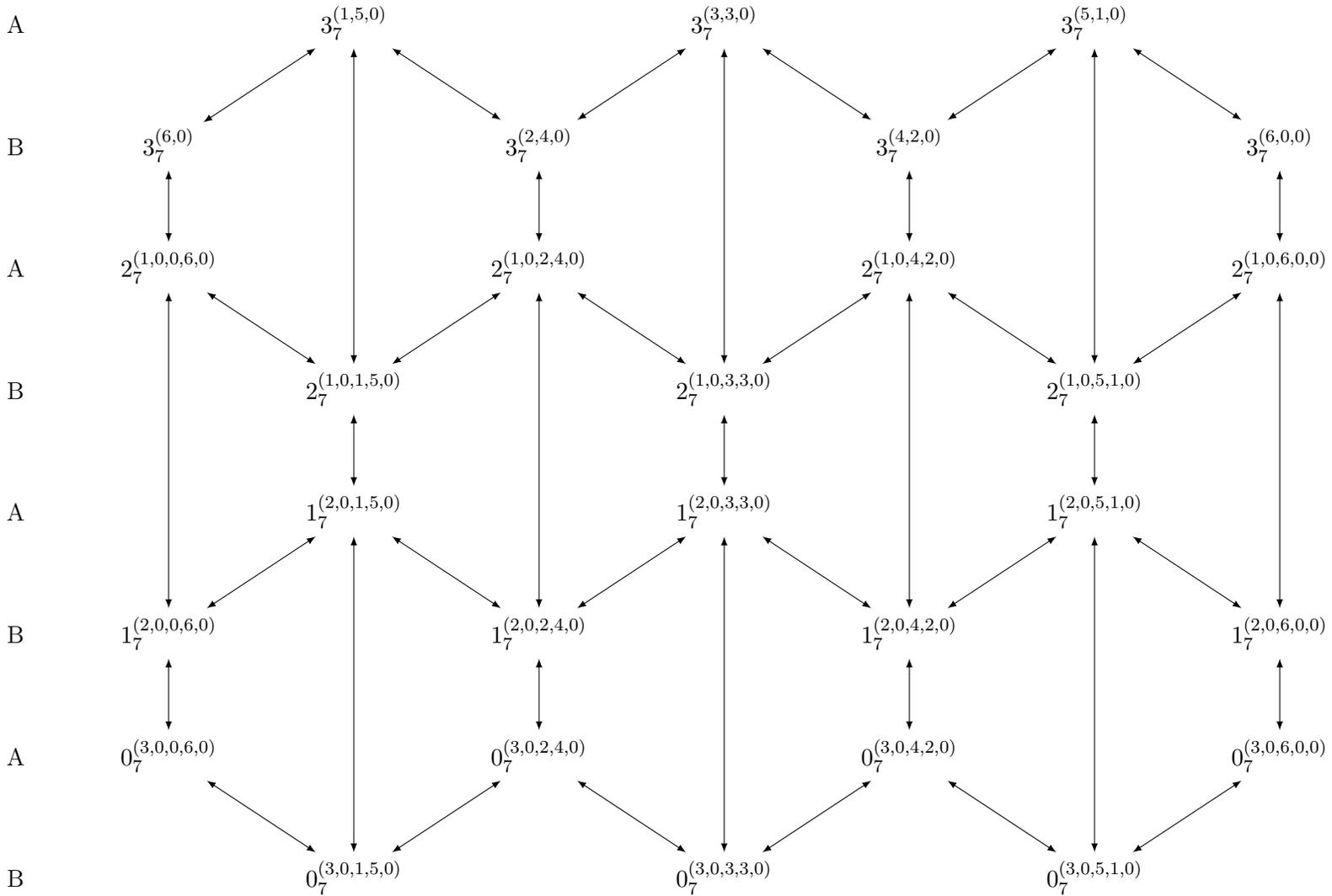
\begin{figure}[H]
\centering
\begin{tikzpicture}
\matrix(M)[matrix of math nodes, row sep=3em, column sep=3em, minimum width=2em]{
\text{A} & & 3_7^{(1,5,0)} & & 3_7^{(3,3,0)} & & 3_7^{(5,1,0)} &\\
\text{B} & 3_7^{(6,0)} & & 3_7^{(2,4,0)} & & 3_7^{(4,2,0)} & & 3_7^{(6,0,0)}\\
\text{A} & 2_7^{(1,0,0,6,0)} & & 2_7^{(1,0,2,4,0)} & & 2_7^{(1,0,4,2,0)} & & 2_7^{(1,0,6,0,0)}\\
\text{B} & & 2_7^{(1,0,1,5,0)} & & 2_7^{(1,0,3,3,0)} & & 2_7^{(1,0,5,1,0)} &\\
\text{A} & & 1_7^{(2,0,1,5,0)} & & 1_7^{(2,0,3,3,0)} & & 1_7^{(2,0,5,1,0)} &\\
\text{B} & 1_7^{(2,0,0,6,0)} & & 1_7^{(2,0,2,4,0)} & & 1_7^{(2,0,4,2,0)} & & 1_7^{(2,0,6,0,0)}\\
\text{A} & 0_7^{(3,0,0,6,0)} & & 0_7^{(3,0,2,4,0)} & & 0_7^{(3,0,4,2,0)} & & 0_7^{(3,0,6,0,0)}\\
\text{B} & & 0_7^{(3,0,1,5,0)} & & 0_7^{(3,0,3,3,0)} & & 0_7^{(3,0,5,1,0)} &\\
};
\draw[latex-latex] (M-2-2) -- (M-1-3);
\draw[latex-latex] (M-1-3) -- (M-2-4);
\draw[latex-latex] (M-2-4) -- (M-1-5);
\draw[latex-latex] (M-1-5) -- (M-2-6);
\draw[latex-latex] (M-2-6) -- (M-1-7);
\draw[latex-latex] (M-1-7) -- (M-2-8);
\draw[latex-latex] (M-3-2) -- (M-4-3);
\draw[latex-latex] (M-4-3) -- (M-3-4);
\draw[latex-latex] (M-3-4) -- (M-4-5);
\draw[latex-latex] (M-4-5) -- (M-3-6);
\draw[latex-latex] (M-3-6) -- (M-4-7);
\draw[latex-latex] (M-4-7) -- (M-3-8);
\draw[latex-latex] (M-6-2) -- (M-5-3);
\draw[latex-latex] (M-5-3) -- (M-6-4);
\draw[latex-latex] (M-6-4) -- (M-5-5);
\draw[latex-latex] (M-5-5) -- (M-6-6);
\draw[latex-latex] (M-6-6) -- (M-5-7);
\draw[latex-latex] (M-5-7) -- (M-6-8);
\draw[latex-latex] (M-7-2) -- (M-8-3);
\draw[latex-latex] (M-8-3) -- (M-7-4);
\draw[latex-latex] (M-7-4) -- (M-8-5);
\draw[latex-latex] (M-8-5) -- (M-7-6);
\draw[latex-latex] (M-7-6) -- (M-8-7);
\draw[latex-latex] (M-8-7) -- (M-7-8);
\draw[latex-latex] (M-2-2) -- (M-3-2);
\draw[latex-latex] (M-3-2) -- (M-6-2);
\draw[latex-latex] (M-6-2) -- (M-7-2);
\draw[latex-latex] (M-2-4) -- (M-3-4);
\draw[latex-latex] (M-3-4) -- (M-6-4);
\draw[latex-latex] (M-6-4) -- (M-7-4);
\draw[latex-latex] (M-2-6) -- (M-3-6);
\draw[latex-latex] (M-3-6) -- (M-6-6);
\draw[latex-latex] (M-6-6) -- (M-7-6);
\draw[latex-latex] (M-2-8) -- (M-3-8);
\draw[latex-latex] (M-3-8) -- (M-6-8);
\draw[latex-latex] (M-6-8) -- (M-7-8);
\draw[latex-latex] (M-1-3) -- (M-4-3);
\draw[latex-latex] (M-4-3) -- (M-5-3);
\draw[latex-latex] (M-5-3) -- (M-8-3);
\draw[latex-latex] (M-1-5) -- (M-4-5);
\draw[latex-latex] (M-4-5) -- (M-5-5);
\draw[latex-latex] (M-5-5) -- (M-8-5);
\draw[latex-latex] (M-1-7) -- (M-4-7);
\draw[latex-latex] (M-4-7) -- (M-5-7);
\draw[latex-latex] (M-5-7) -- (M-8-7);
\end{tikzpicture}
\caption{The T-duality orbit of the $3_7^{(6,0)}$.}
\label{fig:3760Orbit}
\end{figure}
\end{landscape}
\begin{minipage}{0.5\linewidth}
S-dualities:
\begin{itemize}
	\item $3_7^{(6,0)} \leftrightarrow 3_4^{(6,0)}$ See Figure \ref{fig:7420Orbit}
	\item $3_7^{(2,4,0)} \leftrightarrow 3_5^{(2,4,0)}$ See Figure \ref{fig:5522Orbit}
	\item $3_7^{(4,2,0)} \leftrightarrow 3_6^{(4,2,0)}$ See Figure \ref{fig:3624Orbit}
	\item $3_7^{(6,0,0)} \leftrightarrow 3_7^{(6,0,0)}$ Self-dual
	\item $2_7^{(1,0,1,5,0)} \leftrightarrow 2_7^{(1,0,1,5,0)}$ Self-dual
	\item $2_7^{(1,0,3,3,0)} \leftrightarrow 2_8^{(1,0,3,3,0)}$ 
	\item $2_7^{(1,0,5,1,0)} \leftrightarrow 2_9^{(1,0,5,1,0)}$
	\item $1_7^{(2,0,0,6,0)} \leftrightarrow 1_9^{(2,0,0,6,0)}$
	\item $1_7^{(2,0,2,4,0)} \leftrightarrow 1_{10}^{(2,0,2,4,0)}$
	\item $1_7^{(2,0,4,2,0)} \leftrightarrow 1_{11}^{(2,0,4,2,0)}$
	\item $1_7^{(2,0,6,0,0)} \leftrightarrow 1_{12}^{(2,0,6,0,0)}$
	\item $0_7^{(3,0,1,5,0)} \leftrightarrow 0_{12}^{(3,0,1,5,0)}$
	\item $0_7^{(3,0,3,3,0)} \leftrightarrow 0_{13}^{(3,0,3,3,0)}$
	\item $0_7^{(3,0,5,1,0)} \leftrightarrow 0_{14}^{(3,0,5,1,0)}$
\end{itemize}
\end{minipage}
\begin{minipage}{0.5\linewidth}
M-theory origins:
\begin{itemize}
	\item $3_7^{(1,5,0)} \rightarrow 4^{(1,5,0)}$
	\item $3_7^{(3,3,0)} \rightarrow 3^{(3,3,1)}$
	\item $3_7^{(5,1,0)} \rightarrow 3^{(5,2,0)}$
	\item $2_7^{(1,0,0,6,0)} \rightarrow 2^{(1,0,0,7,0)}$
	\item $2_7^{(1,0,2,4,0)} \rightarrow 2^{(1,0,3,4,0)}$
	\item $2_7^{(1,0,4,2,0)} \rightarrow 2^{(1,1,4,2,0)}$
	\item $2_7^{(1,0,6,0,0)} \rightarrow 2^{(2,0,6,0,0)}$
	\item $1_7^{(2,0,1,5,0)} \rightarrow 1^{(3,0,1,5,0)}$
	\item $1_7^{(2,0,3,3,0)} \rightarrow 1^{(1,2,0,3,3,0)}$
	\item $1_7^{(2,0,5,1,0)} \rightarrow 1^{(1,0,2,0,5,1,0)}$
	\item $0_7^{(3,0,0,6,0)} \rightarrow 0^{(1,0,3,0,0,6,0)}$
	\item $0_7^{(3,0,2,4,0)} \rightarrow 0^{(1,0,0,3,0,2,4,0)}$
	\item $0_7^{(3,0,4,2,0)} \rightarrow 0^{(1,0,0,0,3,0,4,2,0)}$
	\item $0_7^{(3,0,6,0,0)} \rightarrow 0^{(1,0,0,0,0,3,0,6,0,0)}$
\end{itemize}
\end{minipage}
\clearpage
\subsection{M-Theory Origins of Type IIA branes}
Here we collate all the M-theory lifts of every exotic Type IIA brane that we have introduced up to this point as well as their reductions. Since every parent in M-theory may be reduced in multiple ways, the existence of any one brane in Type IIA indicates the existence of multiple `siblings' obtained in this manner. Every single brane down to $g_s^{-7}$ have been housed in one of the duality orbits and so any gaps in the figure references correspond to lower powers of $\alpha$.\par
The format should be self-explanatory: the left-most brane in each column is an M-theory brane whilst the branes to the right of each brace are all the possible reductions that one can obtain from that brane. We stress that it is only within EFT that one may `reduce' along a non-isometric direction since this corresponds only to a re-identification of section. One may still `reduce' transverse to a codimension-1 brane in M-theory to give a codimension-0 brane in ten dimensions without worrying about isometries, for example, without obtaining a trivial result; the non-trivial structure is encoded in the dependence on wrapping directions, which distinguishes between the space-filling branes in 10-dimensions.
\begin{multicols}{2}
\begin{itemize}
	\item $0=\text{WM}
		\begin{cases}
			{}^10_0\text{=P; See Figure \ref{fig:10Orbit}}\\
			0_1\text{=D0; See Figure \ref{fig:11Orbit}}
		\end{cases}$
	\item $0^{(1,7)}
		\begin{cases}
			0_3^7\text{; See Figure \ref{fig:532Orbit}}\\
			0_4^{(1,6)}\text{; See Figure \ref{fig:146Orbit}}\\
			0_6^{(1,7)}\text{; See Figure \ref{fig:0617Orbit}}
		\end{cases}$
	\item $0^{(2,1,6)}
		\begin{cases}
			0_4^{(1,1,6)}\text{; See Figure \ref{fig:146Orbit}}\\
			0_5^{(2,0,6)}\text{; See Figure \ref{fig:2515Orbit}}\\
			0_6^{(2,1,5)}\text{; See Figure \ref{fig:0617Orbit}}\\
			0_8^{(2,1,6)}
		\end{cases}$
	\item $0^{(3,2,4)}
		\begin{cases}
			0_5^{(2,2,4)}\text{; See Figure \ref{fig:2515Orbit}}\\
			0_6^{(3,1,4)}\text{; See Figure \ref{fig:0617Orbit}}\\
			0_7^{(3,2,3)}\text{; See Figure \ref{fig:0753Orbit}}\\
			0_9^{(3,2,4)}
		\end{cases}$
	\item $0^{(5,1,3)}
		\begin{cases}
			0_6^{(4,1,3)}\text{; See Figure \ref{fig:0617Orbit}}\\
			0_7^{(5,0,3)}\text{; See Figure \ref{fig:0753Orbit}}\\
			0_8^{(5,1,2)}\\
			0_{10}^{(5,1,3)}
		\end{cases}$
	\item $0^{(1,0,5,3)}
		\begin{cases}
			0_4^{(5,3)}\text{; See Figure \ref{fig:4413Orbit}}\\
			0_6^{(1,0,4,3)}\text{; See Figure \ref{fig:1643Orbit}}\\
			0_7^{(1,0,5,2)}\text{; See Figure \ref{fig:0753Orbit}}\\
			0_9^{(1,0,5,3)}
		\end{cases}$
	\item $0^{(1,2,4,2)}
		\begin{cases}
			0_5^{(2,4,2)}\text{; See Figure \ref{fig:2515Orbit}}\\
			0_6^{(1,1,4,2)}\text{; See Figure \ref{fig:1643Orbit}}\\
			0_7^{(1,2,3,2)}\text{; See Figure \ref{fig:0753Orbit}}\\
			0_8^{(1,2,4,1)}\\
			0_{10}^{(1,2,4,2)}
		\end{cases}$
	\item $0^{(1,5,1,2)}
		\begin{cases}
			0_6^{(5,1,2)}\text{; See Figure \ref{fig:0617Orbit}}\\
			0_7^{(1,4,1,2)}\text{; See Figure \ref{fig:0753Orbit}}\\
			0_8^{(1,5,0,2)}\\
			0_9^{(1,5,1,1)}\\
			0_{11}^{(1,5,1,2)}
		\end{cases}$
	\item $0^{(2,2,4,1)}
		\begin{cases}
			0_6^{(1,2,4,1)}\text{; See Figure \ref{fig:1643Orbit}}\\
			0_7^{(2,1,4,1)}\text{; See Figure \ref{fig:0753Orbit}}\\
			0_8^{(2,2,3,1)}\\
			0_9^{(2,2,4,0)}\\
			0_{11}^{(2,2,4,1)}
		\end{cases}$
	\item $0^{(3,3,2,1)}
		\begin{cases}
			0_7^{(2,3,2,1)}\text{; See Figure \ref{fig:0753Orbit}}\\
			0_8^{(3,2,2,1)}\\
			0_9^{(3,3,1,1)}\\
			0_{10}^{(3,3,2,0)}\\
			0_{12}^{(3,3,2,1)}
		\end{cases}$
	\item $0^{(4,0,5,0)}
		\begin{cases}
			0_7^{(3,0,5,0)}\text{; See Figure \ref{fig:0753Orbit}}\\
			0_9^{(4,0,4,0)}\\
			0_{12}^{(4,0,5,0)}
		\end{cases}$
	\item $0^{(1,0,0,2,7)}
		\begin{cases}
			0_3^{(2,7)}\text{; See Figure \ref{fig:532Orbit}}\\
			0_6^{(1,0,0,1,7)}\text{; See Figure \ref{fig:0617Orbit}}\\
			0_7^{(1,0,0,2,6)}\text{; See Figure \ref{fig:1726Orbit}}
		\end{cases}$
	\item $0^{(1,0,2,1,6)}
		\begin{cases}
			0_4^{(2,1,6)}\text{; See Figure \ref{fig:146Orbit}}\\
			0_6^{(1,0,1,1,6)}\text{; See Figure \ref{fig:0617Orbit}}\\
			0_7^{(1,0,2,0,6)}\text{; See Figure \ref{fig:1726Orbit}}\\
			0_8^{(1,0,2,1,5)}
		\end{cases}$
	\item $0^{(1,0,2,6,0)}
		\begin{cases}
			0_5^{(2,6,0)}\text{; See Figure \ref{fig:2515Orbit}}\\
			0_7^{(1,0,1,6,0)}\text{; See Figure \ref{fig:17160Orbit}}\\
			0_8^{(1,0,2,5,0)}\\
			0_{11}^{(1,0,2,6,0)}
		\end{cases}$
	\item $0^{(1,0,6,1,1)}
		\begin{cases}
			0_6^{(6,1,1)}\text{; See Figure \ref{fig:0617Orbit}}\\
			0_8^{(1,0,5,1,1)}\\
			0_9^{(1,0,5,0,1)}\\
			0_{10}^{(1,0,5,1,0)}\\
			0_{12}^{(1,0,6,1,1)}
		\end{cases}$
	\item $0^{(1,1,2,1,5)}
		\begin{cases}
			0_5^{(1,2,1,5)}\text{; See Figure \ref{fig:2515Orbit}}\\
			0_6^{(1,0,2,1,5)}\text{; See Figure \ref{fig:0617Orbit}}\\
			0_7^{(1,1,1,1,5)}\text{; See Figure \ref{fig:1726Orbit}}\\
			0_8^{(1,1,2,0,5)}\\
			0_9^{(1,1,2,1,4)}
		\end{cases}$
	\item $0^{(1,1,3,4,0)}
		\begin{cases}
			0_6^{(1,3,4,0)}\text{; See Figure \ref{fig:1643Orbit}}\\
			0_7^{(1,0,3,4,0)}\text{; See Figure \ref{fig:17160Orbit}}\\
			0_8^{(1,1,2,4,0)}\\
			0_9^{(1,1,3,3,0)}\\
			0_{12}^{(1,1,3,4,0)}
		\end{cases}$
	\item $0^{(1,2,5,0,1)}
		\begin{cases}
			0_7^{(2,5,0,1)}\text{; See Figure \ref{fig:0753Orbit}}\\
			0_8^{(1,1,5,0,1)}\\
			0_9^{(1,2,4,0,1)}\\
			0_{11}^{(1,2,5,0,0)}\\
			0_{13}^{(1,2,5,0,1)}
		\end{cases}$
	\item $0^{(1,3,0,2,4)}
		\begin{cases}
			0_6^{(3,0,2,4)}\text{; See Figure \ref{fig:3624Orbit}}\\
			0_7^{(1,2,0,2,4)}\text{; See Figure \ref{fig:1726Orbit}}\\
			0_9^{(1,3,0,1,4)}\\
			0_{10}^{(1,3,0,2,3)}
		\end{cases}$
	\item $0^{(1,3,2,3,0)}
		\begin{cases}
			0_7^{(3,2,3,0)}\text{; See Figure \ref{fig:0753Orbit}}\\
			0_8^{(1,2,2,3,0)}\\
			0_9^{(1,3,1,3,0)}\\
			0_{10}^{(1,3,2,2,0)}\\
			0_{13}^{(1,3,2,3,0)}
		\end{cases}$
	\item $0^{(2,0,3,1,4)}
		\begin{cases}
			0_6^{(1,0,3,1,4)}\text{; See Figure \ref{fig:0617Orbit}}\\
			0_8^{(2,0,2,1,4)}\\
			0_9^{(2,0,3,0,4)}\\
			0_{10}^{(2,0,3,1,1)}
		\end{cases}$
	\item $0^{(2,0,5,2,0)}
		\begin{cases}
			0_7^{(1,0,5,2,0)}\text{; See Figure \ref{fig:17160Orbit}}\\
			0_9^{(2,0,4,2,0)}\\
			0_{10}^{(2,0,5,1,0)}\\
			0_{13}^{(2,0,5,2,0)}
		\end{cases}$
	\item $0^{(2,1,0,4,3)}
		\begin{cases}
			0_6^{(1,1,0,4,3)}\text{; See Figure \ref{fig:1643Orbit}}\\
			0_7^{(2,0,0,4,3)}\text{; See Figure \ref{fig:27133Orbit}}\\
			0_9^{(2,1,0,3,3)}\\
			0_{10}^{(2,1,0,4,2)}
		\end{cases}$
	\item $0^{(2,2,2,0,4)}
		\begin{cases}
			0_7^{(1,2,2,0,4)}\text{; See Figure \ref{fig:1726Orbit}}\\
			0_8^{(2,1,2,0,4)}\\
			0_9^{(2,2,1,0,4)}\\
			0_{11}^{(2,2,2,0,3)}
		\end{cases}$
	\item $0^{(3,0,2,2,3)}
		\begin{cases}
			0_7^{(2,0,2,2,3)}\text{; See Figure \ref{fig:27133Orbit}}\\
			0_9^{(3,0,1,2,3)}\\
			0_{10}^{(3,0,2,1,3)}\\
			0_{11}^{(3,0,2,2,2)}
		\end{cases}$
	\item $0^{(1,0,0,1,5,3)}
		\begin{cases}
			0_4^{(1,5,3)}\text{; See Figure \ref{fig:4413Orbit}}\\
			0_7^{(1,0,0,0,5,3)}\text{; See Figure \ref{fig:0753Orbit}}\\
			0_8^{(1,0,0,1,4,3)}\\
			0_9^{(1,0,0,1,5,2)}
		\end{cases}$
	\item $0^{(1,0,0,5,0,4)}
		\begin{cases}
			0_5^{(5,0,4)}\text{; See Figure \ref{fig:5522Orbit}}\\
			0_8^{(1,0,0,4,0,4)}\\
			0_{10}^{(1,0,0,5,0,3)}
		\end{cases}$
	\item $0^{(1,0,0,7,1,0)}
		\begin{cases}
			0_6^{(7,1,0)}\text{; See Figure \ref{fig:0617Orbit}}\\
			0_9^{(1,0,0,6,1,0)}\\
			0_{10}^{(1,0,0,7,0,0)}\\
			0_{13}^{(1,0,0,7,1,0)}
		\end{cases}$
	\item $0^{(1,0,1,2,3,3)}
		\begin{cases}
			0_5^{(1,2,3,3)}\text{; See Figure \ref{fig:2515Orbit}}\\
			0_7^{(1,0,0,2,3,3)}\text{; See Figure \ref{fig:0753Orbit}}\\
			0_8^{(1,0,1,1,3,3)}\\
			0_9^{(1,0,1,2,2,3)}\\
			0_{10}^{(1,0,1,2,3,2)}
		\end{cases}$
	\item $0^{(1,0,3,1,2,3)}
		\begin{cases}
			0_6^{(3,1,2,3)}\text{; See Figure \ref{fig:3624Orbit}}\\
			0_8^{(1,0,2,1,2,3)}\\
			0_9^{(1,0,3,0,2,3)}\\
			0_{10}^{(1,0,3,1,1,3)}\\
			0_{11}^{(1,0,3,1,2,2)}
		\end{cases}$
	\item $0^{(1,0,3,4,1,0)}
		\begin{cases}
			0_7^{(3,4,1,0)}\text{; See Figure \ref{fig:0753Orbit}}\\
			0_9^{(1,0,2,4,1,0)}\\
			0_{10}^{(1,0,3,3,1,0)}\\
			0_{11}^{(1,0,3,4,0,0)}\\
			0_{14}^{(1,0,3,4,1,0)}
		\end{cases}$
	\item $0^{(1,1,0,4,1,3)}
		\begin{cases}
			0_6^{(1,0,4,1,3)}\text{; See Figure \ref{fig:0617Orbit}}\\
			0_7^{(1,0,0,4,1,3)}\text{; See Figure \ref{fig:0753Orbit}}\\
			0_9^{(1,1,0,3,1,3)}\\
			0_{10}^{(1,1,0,4,0,3)}\\
			0_{11}^{(1,1,0,4,1,2)}
		\end{cases}$
	\item $0^{(1,1,0,7,0,0)}
		\begin{cases}
			0_7^{(1,0,7,0,0)}\text{; See Figure \ref{fig:17160Orbit}}\\
			0_8^{(1,0,0,7,0,0)}\\
			0_{10}^{(1,1,0,6,0,0)}\\
			0_{14}^{(1,1,0,7,0,0)}
		\end{cases}$
	\item $0^{(1,1,1,1,4,2)}
		\begin{cases}
			0_6^{(1,1,1,4,2)}\text{; See Figure \ref{fig:1643Orbit}}\\
			0_7^{(1,0,1,1,4,2)}\text{; See Figure \ref{fig:0753Orbit}}\\
			0_8^{(1,1,0,1,4,2)}\\
			0_9^{(1,1,1,0,4,2)}\\
			0_{10}^{(1,1,1,1,3,2)}\\
			0_{11}^{(1,1,1,1,4,1)}
		\end{cases}$
	\item $0^{(1,1,3,1,1,3)}
		\begin{cases}
			0_7^{(1,3,1,1,3)}\text{; See Figure \ref{fig:1726Orbit}}\\
			0_8^{(1,0,3,1,1,3)}\\
			0_9^{(1,1,2,1,1,3)}\\
			0_{10}^{(1,1,3,0,1,3)}\\
			0_{11}^{(1,1,3,1,0,3)}\\
			0_{12}^{(1,1,3,1,1,2)}
		\end{cases}$
	\item $0^{(1,2,0,4,0,3)}
		\begin{cases}
			0_7^{(2,0,4,0,3)}\text{; See Figure \ref{fig:27133Orbit}}\\
			0_8^{(1,1,0,4,0,3)}\\
			0_{10}^{(1,2,0,3,0,3)}\\
			0_{12}^{(1,2,0,4,0,2)}
		\end{cases}$
	\item $0^{(1,2,1,1,3,2)}
		\begin{cases}
			0_7^{(2,1,1,3,2)}\text{; See Figure \ref{fig:27133Orbit}}\\
			0_8^{(1,1,1,1,3,2)}\\
			0_9^{(1,2,0,1,3,2)}\\
			0_{10}^{(1,2,1,0,3,2)}\\
			0_{11}^{(1,2,1,1,2,2)}\\
			0_{12}^{(1,2,1,1,3,1)}
		\end{cases}$
	\item $0^{(2,0,1,3,2,2)}
		\begin{cases}
			0_7^{(1,0,1,3,2,2)}\text{; See Figure \ref{fig:0753Orbit}}\\
			0_9^{(2,0,0,3,2,2)}\\
			0_{10}^{(2,0,1,2,2,2)}\\
			0_{11}^{(2,0,1,3,1,2)}\\
			0_{12}^{(2,0,1,3,2,1)}
		\end{cases}$
	\item $0^{(2,0,2,0,5,1)}
		\begin{cases}
			0_7^{(1,0,2,0,5,1)}\text{; See Figure \ref{fig:0753Orbit}}\\
			0_9^{(2,0,1,5,0,1)}\\
			0_{11}^{(2,0,2,0,4,1)}\\
			0_{12}^{(2,0,2,0,5,0)}
		\end{cases}$
	\item $0^{(2,1,0,0,7,0)}
		\begin{cases}
			0_7^{(1,1,0,0,7,0)}\text{; See Figure \ref{fig:17160Orbit}}\\
			0_8^{(2,0,0,0,7,0)}\\
			0_{11}^{(2,1,0,0,6,0)}
		\end{cases}$
	\item $0^{(1,0,0,0,0,9,0)}
		\begin{cases}
			0_4^{(9,0)}\text{; See Figure \ref{fig:7420Orbit}}\\
			0_9^{(1,0,0,0,0,8,0)}
		\end{cases}$
	\item $0^{(1,0,0,0,5,2,2)}
		\begin{cases}
			0_5^{(5,2,2)}\text{; See Figure \ref{fig:5522Orbit}}\\
			0_9^{(1,0,0,0,4,2,2)}\\
			0_{10}^{(1,0,0,0,5,1,2)}\\
			0_{11}^{(1,0,0,0,5,2,1)}
		\end{cases}$
	\item $0^{(1,0,0,1,2,5,1)}
		\begin{cases}
			0_5^{(1,2,5,1)}\text{; See Figure \ref{fig:2515Orbit}}\\
			0_8^{(1,0,0,0,2,5,1)}\\
			0_9^{(1,0,0,1,1,5,1)}\\
			0_{10}^{(1,0,0,1,2,4,1)}\\
			0_{11}^{(1,0,0,1,2,5,0)}
		\end{cases}$
	\item $0^{(1,0,0,3,2,2,2)}
		\begin{cases}
			0_6^{(3,2,2,2)}\text{; See Figure \ref{fig:3624Orbit}}\\
			0_9^{(1,0,0,2,2,2,2)}\\
			0_{10}^{(1,0,0,3,1,2,2)}\\
			0_{11}^{(1,0,0,3,2,1,2)}\\
			0_{12}^{(1,0,0,3,2,2,1)}
		\end{cases}$
	\item $0^{(1,0,1,0,5,1,2)}
		\begin{cases}
			0_6^{(1,0,5,1,2)}\text{; See Figure \ref{fig:0617Orbit}}\\
			0_8^{(1,0,0,0,5,1,2)}\\
			0_{10}^{(1,0,1,0,4,1,2)}\\
			0_{11}^{(1,0,1,0,5,0,2)}\\
			0_{12}^{(1,0,1,0,5,1,1)}
		\end{cases}$
	\item $0^{(1,0,1,1,2,4,1)}
		\begin{cases}
			0_6^{(1,1,2,4,1)}\text{; See Figure \ref{fig:1643Orbit}}\\
			0_8^{(1,0,0,1,2,4,1)}\\
			0_9^{(1,0,1,0,2,4,1)}\\
			0_{10}^{(1,0,1,1,1,4,1)}\\
			0_{11}^{(1,0,1,1,2,3,1)}\\
			0_{12}^{(1,0,1,1,2,4,0)}
		\end{cases}$
	\item $0^{(1,0,1,4,0,2,2)}
		\begin{cases}
			0_7^{(1,4,0,2,2)}\text{; See Figure \ref{fig:1726Orbit}}\\
			0_9^{(1,0,0,4,0,2,2)}\\
			0_{10}^{(1,0,1,3,0,2,2)}\\
			0_{12}^{(1,0,1,4,0,1,2)}\\
			0_{13}^{(1,0,1,4,0,2,1)}
		\end{cases}$
	\item $0^{(1,0,2,1,3,1,2)}
		\begin{cases}
			0_7^{(2,1,3,1,2)}\text{; See Figure \ref{fig:27133Orbit}}\\
			0_9^{(1,0,1,1,3,1,2)}\\
			0_{10}^{(1,0,2,0,3,1,2)}\\
			0_{11}^{(1,0,2,1,2,1,2)}\\
			0_{12}^{(1,0,2,1,3,0,2)}\\
			0_{13}^{(1,0,2,1,3,1,1)}
		\end{cases}$
	\item $0^{(1,0,2,2,0,4,1)}
		\begin{cases}
			0_7^{(2,2,0,4,1)}\text{; See Figure \ref{fig:27133Orbit}}\\
			0_9^{(1,0,1,2,0,4,1)}\\
			0_{10}^{(1,0,2,1,0,4,1)}\\
			0_{12}^{(1,0,2,2,0,3,1)}\\
			0_{13}^{(1,0,2,2,0,4,0)}
		\end{cases}$
	\item $0^{(1,0,3,0,0,6,0)}
		\begin{cases}
			0_7^{(3,0,0,6,0)}\text{; See Figure \ref{fig:3760Orbit}}\\
			0_9^{(1,0,2,0,0,6,0)}\\
			0_{12}^{(1,0,3,0,0,5,0)}
		\end{cases}$
	\item $0^{(1,1,0,1,5,0,2)}
		\begin{cases}
			0_7^{(1,0,1,5,0,2)}\text{; See Figure \ref{fig:0753Orbit}}\\
			0_8^{(1,0,0,1,5,0,2)}\\
			0_{10}^{(1,1,0,0,5,0,2)}\\
			0_{11}^{(1,1,0,1,4,0,2)}\\
			0_{13}^{(1,1,0,1,5,0,1)}
		\end{cases}$
	\item $0^{(1,1,0,2,2,3,1)}
		\begin{cases}
			0_7^{(1,0,2,2,3,1)}\text{; See Figure \ref{fig:0753Orbit}}\\
			0_8^{(1,0,0,2,2,3,1)}\\
			0_{10}^{(1,1,0,1,2,3,1)}\\
			0_{11}^{(1,1,0,2,1,3,1)}\\
			0_{12}^{(1,1,0,2,2,2,1)}\\
			0_{13}^{(1,1,0,2,2,3,0)}
		\end{cases}$
	\item $0^{(1,1,1,0,2,5,0)}
		\begin{cases}
			0_7^{(1,1,0,2,5,0)}\text{; See Figure \ref{fig:17160Orbit}}\\
			0_8^{(1,0,1,0,2,5,0)}\\
			0_9^{(1,1,0,0,2,5,0)}\\
			0_{11}^{(1,1,1,0,1,5,0)}\\
			0_{12}^{(1,1,1,0,2,4,0)}
		\end{cases}$
	\item $0^{(1,0,0,0,0,5,4,0)}
		\begin{cases}
			0_5^{(5,4,0)}\text{; See Figure \ref{fig:5522Orbit}}\\
			0_{10}^{(1,0,0,0,0,4,4,0)}\\
			0_{11}^{(1,0,0,0,0,5,3,0)}
		\end{cases}$
	\item $0^{(1,0,0,0,3,3,2,1)}
		\begin{cases}
			0_6^{(3,3,2,1)}\text{; See Figure \ref{fig:3624Orbit}}\\
			0_{10}^{(1,0,0,0,2,3,2,1)}\\
			0_{11}^{(1,0,0,0,3,2,2,1)}\\
			0_{12}^{(1,0,0,0,3,3,1,1)}\\
			0_{13}^{(1,0,0,0,3,3,2,0)}
		\end{cases}$
	\item $0^{(1,0,0,1,0,6,1,1)}
		\begin{cases}
			0_6^{(1,0,6,1,1)}\text{; See Figure \ref{fig:0617Orbit}}\\
			0_9^{(1,0,0,0,0,6,1,1)}\\
			0_{11}^{(1,0,0,1,0,5,1,1)}\\
			0_{12}^{(1,0,0,1,0,6,0,1)}\\
			0_{13}^{(1,0,0,1,0,6,1,0)}
		\end{cases}$
	\item $0^{(1,0,0,1,1,3,4,0)}
		\begin{cases}
			0_6^{(1,1,3,4,0)}\text{; See Figure \ref{fig:1643Orbit}}\\
			0_9^{(1,0,0,0,1,3,4,0)}\\
			0_{10}^{(1,0,0,1,0,3,4,0)}\\
			0_{11}^{(1,0,0,1,1,2,4,0)}\\
			0_{12}^{(1,0,0,1,1,3,3,0)}
		\end{cases}$
	\item $0^{(1,0,0,1,4,2,0,2)}
		\begin{cases}
			0_7^{(1,4,2,0,2)}\text{; See Figure \ref{fig:1726Orbit}}\\
			0_{10}^{(1,0,0,0,4,2,0,2)}\\
			0_{11}^{(1,0,0,1,3,2,0,2)}\\
			0_{12}^{(1,0,0,1,4,1,0,2)}\\
			0_{14}^{(1,0,0,1,4,2,0,1)}
		\end{cases}$
	\item $0^{(1,0,0,2,2,2,2,1)}
		\begin{cases}
			0_7^{(2,2,2,2,1)}\text{; See Figure \ref{fig:27133Orbit}}\\
			0_{10}^{(1,0,0,1,2,2,2,1)}\\
			0_{11}^{(1,0,0,2,1,2,2,1)}\\
			0_{12}^{(1,0,0,2,2,1,2,1)}\\
			0_{13}^{(1,0,0,2,2,2,1,1)}\\
			0_{14}^{(1,0,0,2,2,2,2,0)}
		\end{cases}$
	\item $0^{(1,0,0,3,0,2,4,0)}
		\begin{cases}
			0_7^{(3,0,2,4,0)}\text{; See Figure \ref{fig:3760Orbit}}\\
			0_{10}^{(1,0,0,2,0,2,4,0)}\\
			0_{12}^{(1,0,0,3,0,1,4,0)}\\
			0_{13}^{(1,0,0,3,0,2,3,0}
		\end{cases}$
	\item $0^{(1,0,1,0,2,4,1,1)}
		\begin{cases}
			0_7^{(1,0,2,4,1,1)}\text{; See Figure \ref{fig:0753Orbit}}\\
			0_9^{(1,0,0,0,2,4,1,1)}\\
			0_{11}^{(1,0,1,0,1,4,1,1)}\\
			0_{12}^{(1,0,1,0,2,3,1,1)}\\
			0_{13}^{(1,0,1,0,2,4,0,1)}\\
			0_{14}^{(1,0,1,0,2,4,1,0)}
		\end{cases}$
	\item $0^{(1,0,1,0,3,1,4,0)}
		\begin{cases}
			0_7^{(1,0,3,1,4,0)}\text{; See Figure \ref{fig:0753Orbit}}\\
			0_9^{(1,0,0,0,0,3,1,4,0)}\\
			0_{11}^{(1,0,1,0,2,1,4,0)}\\
			0_{12}^{(1,0,1,0,3,0,4,0)}\\
			0_{13}^{(1,0,1,0,3,1,3,0)}
		\end{cases}$
	\item $0^{(1,0,1,1,0,4,3,0)}
		\begin{cases}
			0_7^{(1,1,0,4,3,0)}\text{; See Figure \ref{fig:17160Orbit}}\\
			0_9^{(1,0,0,1,0,4,3,0)}\\
			0_{10}^{(1,0,1,0,0,4,3,0)}\\
			0_{12}^{(1,0,1,1,0,3,3,0)}\\
			0_{13}^{(1,0,1,1,0,4,2,0)}
		\end{cases}$
	\item $0^{(1,0,0,0,0,3,4,2,0)}
		\begin{cases}
			0_6^{(3,4,2,0)}\text{; See Figure \ref{fig:3624Orbit}}\\
			0_{10}^{(1,0,0,0,0,2,4,2,0)}\\
			0_{11}^{(1,0,0,0,0,3,3,2,0)}\\
			0_{12}^{(1,0,0,0,0,3,4,1,0)}
		\end{cases}$
	\item $0^{(1,0,0,0,1,0,7,1,0)}
		\begin{cases}
			0_6^{(1,0,7,1,0)}\text{; See Figure \ref{fig:0617Orbit}}\\
			0_{10}^{(1,0,0,0,0,0,7,1,0)}\\
			0_{12}^{(1,0,0,0,1,0,6,1,0)}\\
			0_{13}^{(1,0,0,0,1,0,7,0,0)}
		\end{cases}$
	\item $0^{(1,0,0,0,1,5,1,1,1)}
		\begin{cases}
			0_7^{(1,5,1,1,1)}\text{; See Figure \ref{fig:1726Orbit}}\\
			0_{11}^{(1,0,0,0,0,5,1,1,1)}\\
			0_{12}^{(1,0,0,0,1,4,1,1,1)}\\
			0_{13}^{(1,0,0,0,1,5,0,1,1)}\\
			0_{14}^{(1,0,0,0,1,5,1,0,1)}\\
			0_{15}^{(1,0,0,0,1,5,1,1,0)}
		\end{cases}$
	\item $0^{(1,0,0,0,2,2,4,0,1)}
		\begin{cases}
			0_7^{(2,2,4,0,1)}\text{; See Figure \ref{fig:27133Orbit}}\\
			0_{11}^{(1,0,0,0,1,2,4,0,1)}\\
			0_{12}^{(1,0,0,0,2,1,4,0,1)}\\
			0_{13}^{(1,0,0,0,2,2,3,0,1)}\\
			0_{15}^{(1,0,0,0,2,2,4,0,0)}
		\end{cases}$
	\item $0^{(1,0,0,0,3,0,4,2,0)}
		\begin{cases}
			0_7^{(3,0,4,2,0)}\text{; See Figure \ref{fig:3760Orbit}}\\
			0_{11}^{(1,0,0,0,2,0,4,2,0)}\\
			0_{13}^{(1,0,0,0,3,0,3,2,0)}\\
			0_{14}^{(1,0,0,0,3,0,4,1,0)}
		\end{cases}$
	\item $0^{(1,0,0,1,0,3,3,2,0)}
		\begin{cases}
			0_7^{(1,0,3,3,2,0)}\text{; See Figure \ref{fig:0753Orbit}}\\
			0_{10}^{(1,0,0,0,0,3,3,2,0)}\\
			0_{12}^{(1,0,0,1,0,2,3,2,0)}\\
			0_{13}^{(1,0,0,1,0,3,2,2,0)}\\
			0_{14}^{(1,0,0,1,0,3,3,1,0)}
		\end{cases}$
	\item $0^{(1,0,0,1,1,0,6,1,0)}
		\begin{cases}
			0_7^{(1,1,0,6,1,0)}\text{; See Figure \ref{fig:17160Orbit}}\\
			0_{10}^{(1,0,0,0,1,0,6,1,0)}\\
			0_{11}^{(1,0,0,1,0,0,6,1,0)}\\
			0_{13}^{(1,0,0,1,1,0,5,1,0)}\\
			0_{14}^{(1,0,0,1,1,0,6,0,0)}
		\end{cases}$
	\item $0^{(1,0,0,0,2,3,1,3,0)}
		\begin{cases}
			0_7^{(2,3,1,3,0)}\text{; See Figure \ref{fig:27133Orbit}}\\
			0_{11}^{(1,0,0,0,1,3,1,3,0)}\\
			0_{12}^{(1,0,0,0,2,2,1,3,0)}\\
			0_{13}^{(1,0,0,0,2,3,0,3,0)}\\
			0_{14}^{(1,0,0,0,2,3,1,2,0)}
		\end{cases}$
	\item $0^{(1,0,0,0,1,0,3,5,0,0)}
		\begin{cases}
			0_7^{(1,0,3,5,0,0)}\text{; See Figure \ref{fig:0753Orbit}}\\
			0_{11}^{(1,0,0,0,0,0,3,5,0,0)}\\
			0_{13}^{(1,0,0,0,1,0,2,5,0,0)}\\
			0_{14}^{(1,0,0,0,1,0,3,4,0,0)}
		\end{cases}$
	\item $0^{(1,0,0,0,0,1,6,0,2,0)}
		\begin{cases}
			0_7^{(1,6,0,2,0)}\text{; See Figure \ref{fig:1726Orbit}}\\
			0_{12}^{(1,0,0,0,0,0,6,0,2,0)}\\
			0_{13}^{(1,0,0,0,0,1,5,0,2,0)}\\
			0_{15}^{(1,0,0,0,0,1,6,0,1,0)}
		\end{cases}$
	\item $0^{(1,0,0,0,0,2,3,3,1,0)}\\
		\begin{cases}
			0_7^{(2,3,3,1,0)}\text{; See Figure \ref{fig:27133Orbit}}\\
			0_{12}^{(1,0,0,0,0,1,3,3,1,0)}\\
			0_{13}^{(1,0,0,0,0,2,2,3,1,0)}\\
			0_{14}^{(1,0,0,0,0,2,3,2,1,0)}\\
			0_{15}^{(1,0,0,0,0,2,3,3,0,0)}
		\end{cases}$
	\item $0^{(1,0,0,0,0,3,0,6,0,0)}\\
		\begin{cases}
			0_7^{(3,0,6,0,0)}\text{; See Figure \ref{fig:3760Orbit}}\\
			0_{12}^{(1,0,0,0,0,2,0,6,0,0)}\\
			0_{14}^{(1,0,0,0,0,3,0,5,0,0)}
		\end{cases}$
	\item $0^{(1,0,0,0,0,0,1,6,2,0,0)}
		\begin{cases}
			0_7^{(1,6,2,0,0)}\text{; See Figure \ref{fig:1726Orbit}}\\
			0_{13}^{(1,0,0,0,0,0,0,6,2,0,0)}\\
			0_{14}^{(1,0,0,0,0,0,1,5,2,0,0)}\\
			0_{15}^{(1,0,0,0,0,0,1,6,1,0,0)}
		\end{cases}$
	\item $1^{(1,1,6)}
		\begin{cases}
			1_3^{(1,6)}\text{; See Figure \ref{fig:532Orbit}}\\
			1_4^{(1,0,6)}\text{; See Figure \ref{fig:146Orbit}}\\
			1_5^{(1,1,5)}\text{; See Figure \ref{fig:2515Orbit}}\\
			0_6^{(1,1,6)}\text{; See Figure \ref{fig:0617Orbit}}\\
			1_7^{(1,1,6)}\text{; See Figure \ref{fig:1726Orbit}}
		\end{cases}$
	\item $1^{(1,4,3)}
		\begin{cases}
			1_4^{(4,3)}\text{; See Figure \ref{fig:4413Orbit}}\\
			1_5^{(1,3,3)}\text{; See Figure \ref{fig:2515Orbit}}\\
			1_6^{(1,4,2)}\text{; See Figure \ref{fig:1643Orbit}}\\
			0_7^{(1,4,3)}\text{; See Figure \ref{fig:0753Orbit}}\\
			1_8^{(1,4,3)}
		\end{cases}$
	\item $1^{(2,5,1)}
		\begin{cases}
			1_5^{(1,5,1)}\text{; See Figure \ref{fig:2515Orbit}}\\
			1_6^{(2,4,1)}\text{; See Figure \ref{fig:1643Orbit}}\\
			1_7^{(2,5,0)}\text{; See Figure \ref{fig:17160Orbit}}\\
			0_8^{(2,5,1)}\\
			1_9^{(2,5,1)}
		\end{cases}$
	\item $1^{(4,4,0)}
		\begin{cases}
			1_6^{(3,4,0)}\text{; See Figure \ref{fig:1643Orbit}}\\
			1_7^{(4,3,0)}\text{; See Figure \ref{fig:17160Orbit}}\\
			0_9^{(4,4,0)}\\
			1_{10}^{(4,4,0)}
		\end{cases}$
	\item $1^{(7,1,0)}
		\begin{cases}
			1_7^{(6,1,0)}\text{; See Figure \ref{fig:17160Orbit}}\\
			1_8^{(7,0,0)}\\
			0_{10}^{(7,1,0)}\\
			1_{11}^{(7,1,0)}
		\end{cases}$
	\item $1^{(1,2,0,6)}
		\begin{cases}
			1_4^{(2,0,6)}\text{; See Figure \ref{fig:146Orbit}}\\
			1_5^{(1,1,0,6)}\text{; See Figure \ref{fig:2515Orbit}}\\
			1_7^{(1,2,0,5)}\text{; See Figure \ref{fig:1726Orbit}}\\
			0_8^{(1,2,0,6)}
		\end{cases}$
	\item $1^{(2,1,2,4)}
		\begin{cases}
			1_5^{(1,1,2,4)}\text{; See Figure \ref{fig:2515Orbit}}\\
			1_6^{(2,0,2,4)}\text{; See Figure \ref{fig:3624Orbit}}\\
			1_7^{(2,1,1,4)}\text{; See Figure \ref{fig:1726Orbit}}\\
			1_8^{(2,1,2,3)}\\
			0_9^{(2,1,2,4)}
		\end{cases}$
	\item $1^{(3,1,2,3)}
		\begin{cases}
			1_6^{(2,1,2,3)}\text{; See Figure \ref{fig:3624Orbit}}\\
			1_7^{(3,0,2,3)}\text{; See Figure \ref{fig:1726Orbit}}\\
			1_8^{(3,1,1,3)}\\
			1_9^{(3,1,2,2)}\\
			0_{10}^{(3,1,2,3)}
		\end{cases}$
	\item $1^{(4,2,0,3)}
		\begin{cases}
			1_7^{(3,2,0,3)}\text{; See Figure \ref{fig:1726Orbit}}\\
			1_8^{(4,1,0,3)}\\
			1_{10}^{(4,2,0,2)}\\
			0_{11}^{(4,2,0,3)}
		\end{cases}$
	\item $1^{(1,0,1,4,3)}
		\begin{cases}
			1_4^{(1,4,3)}\text{; See Figure \ref{fig:4413Orbit}}\\
			1_6^{(1,0,0,4,3)}\text{; See Figure \ref{fig:1643Orbit}}\\
			1_7^{(1,0,1,3,3)}\text{; See Figure \ref{fig:27133Orbit}}\\
			1_8^{(1,0,1,4,2)}\\
			1_9^{(1,0,1,4,3)}
		\end{cases}$
	\item $1^{(2,0,2,4,1)}
		\begin{cases}
			1_6^{(1,0,2,4,1)}\text{; See Figure \ref{fig:1643Orbit}}\\
			1_8^{(2,0,1,4,1)}\\
			1_9^{(2,0,2,3,1)}\\
			1_{10}^{(2,0,2,4,0)}\\
			0_{12}^{(2,0,2,4,1)}
		\end{cases}$
	\item $1^{(1,0,4,1,3)}
		\begin{cases}
			1_5^{(4,1,3)}\text{; See Figure \ref{fig:5522Orbit}}\\
			1_7^{(1,0,3,1,3)}\text{; See Figure \ref{fig:27133Orbit}}\\
			1_8^{(1,0,4,0,3)}\\
			1_9^{(1,0,4,1,2)}\\
			0_{10}^{(1,0,4,1,3)}
		\end{cases}$
	\item $1^{(1,1,1,4,2)}
		\begin{cases}
			1_5^{(1,1,4,2)}\text{; See Figure \ref{fig:2515Orbit}}\\
			1_6^{(1,0,1,4,2)}\text{; See Figure \ref{fig:1643Orbit}}\\
			1_7^{(1,1,0,4,2)}\text{; See Figure \ref{fig:27133Orbit}}\\
			1_8^{(1,1,1,3,2)}\\
			1_9^{(1,1,1,4,1)}\\
			0_{10}^{(1,1,1,4,2)}
		\end{cases}$
	\item $1^{(1,2,2,2,2)}
		\begin{cases}
			1_6^{(2,2,2,2)}\text{; See Figure \ref{fig:3624Orbit}}\\
			1_7^{(1,1,2,2,2)}\text{; See Figure \ref{fig:27133Orbit}}\\
			1_8^{(1,2,1,2,2)}\\
			1_9^{(1,2,2,1,2)}\\
			1_{10}^{(1,2,2,2,1)}\\
			0_{11}^{(1,2,2,2,2)}
		\end{cases}$
	\item $1^{(1,4,1,1,2)}
		\begin{cases}
			1_7^{(4,1,1,2)}\text{; See Figure \ref{fig:1726Orbit}}\\
			1_8^{(1,3,1,1,2)}\\
			1_9^{(1,4,0,1,2)}\\
			1_{10}^{(1,4,1,0,2)}\\
			1_{11}^{(1,4,1,1,1)}\\
			0_{12}^{(1,4,1,1,2)}
		\end{cases}$
	\item $1^{(2,1,4,0,2)}
		\begin{cases}
			1_7^{(1,1,4,0,2)}\text{; See Figure \ref{fig:27133Orbit}}\\
			1_8^{(2,0,4,0,2)}\\
			1_9^{(2,1,3,0,2)}\\
			1_{11}^{(2,1,4,0,1)}\\
			0_{12}^{(2,1,4,0,2)}
		\end{cases}$
	\item $1^{(2,2,1,3,1)}
		\begin{cases}
			1_7^{(1,2,1,3,1)}\text{; See Figure \ref{fig:27133Orbit}}\\
			1_8^{(2,1,1,3,1)}\\
			1_9^{(2,2,0,3,1)}\\
			1_{10}^{(2,2,1,2,1)}\\
			1_{11}^{(2,2,1,3,0)}\\
			0_{12}^{(2,2,1,3,1)}
		\end{cases}$
	\item $1^{(3,0,1,5,0)}
		\begin{cases}
			1_7^{(2,0,1,5,0)}\text{; See Figure \ref{fig:3760Orbit}}\\
			1_9^{(3,0,0,5,0)};\\
			1_{10}^{(3,0,1,4,0)};\\
			0_{12}^{(3,0,1,5,0)}
		\end{cases}$
	\item $1^{(1,0,0,4,3,1)}
		\begin{cases}
			1_5^{(4,1,3)}\text{; See Figure \ref{fig:5522Orbit}}\\
			1_8^{(1,0,0,3,3,1)}\\
			1_9^{(1,0,0,4,2,1)}\\
			1_{10}^{(1,0,0,4,3,0)}\\
			0_{11}^{(1,0,0,4,3,1)}
		\end{cases}$
	\item $1^{(1,0,1,1,6,0)}
		\begin{cases}
			1_5^{(1,1,6,0)}\text{; See Figure \ref{fig:2515Orbit}}\\
			1_7^{(1,0,0,1,6,0)}\text{; See Figure \ref{fig:17160Orbit}}\\
			1_8^{(1,0,1,0,6,0)}\\
			1_9^{(1,0,1,1,5,0)}\\
			1_{11}^{(1,0,1,1,6,0)}
		\end{cases}$
	\item $1^{(1,0,2,3,2,1)}
		\begin{cases}
			1_6^{(2,3,2,1)}\text{; See Figure \ref{fig:3624Orbit}}\\
			1_8^{(1,0,1,3,2,1)}\\
			1_9^{(1,0,2,2,2,1)}\\
			1_{10}^{(1,0,2,3,1,1)}\\
			1_{11}^{(1,0,2,3,2,0)}\\
			0_{12}^{(1,0,2,3,2,1)}
		\end{cases}$
	\item $1^{(1,0,5,0,2,1)}
		\begin{cases}
			1_7^{(5,0,2,1)}\text{: See Figure \ref{fig:1726Orbit}}\\
			1_9^{(1,0,4,0,2,1)}\\
			1_{11}^{(1,0,5,0,1,1)}\\
			1_{12}^{(1,0,5,0,2,0)}\\
			0_{13}^{(1,0,5,0,2,1)}
		\end{cases}$
	\item $1^{(1,1,0,3,4,0)}
		\begin{cases}
			1_6^{(1,0,3,4,0)}\text{; See Figure \ref{fig:1643Orbit}}\\
			1_7^{(1,0,0,3,4,0)}\text{; See Figure \ref{fig:17160Orbit}}\\
			1_9^{(1,1,0,2,4,0)}\\
			1_{10}^{(1,1,0,3,3,0)}\\
			0_{12}^{(1,1,0,3,4,0)}
		\end{cases}$
	\item $1^{(1,1,2,3,1,1)}
		\begin{cases}
			1_7^{(1,2,3,1,1)}\text{; See Figure \ref{fig:27133Orbit}}\\
			1_8^{(1,0,2,3,1,1)}\\
			1_9^{(1,1,1,3,1,1)}\\
			1_{10}^{(1,1,2,2,1,1)}\\
			1_{11}^{(1,1,2,3,0,1)}\\
			1_{12}^{(1,1,2,3,1,0)}\\
			0_{13}^{(1,1,2,3,1,1)}
		\end{cases}$
	\item $1^{(1,1,3,0,4,0)}
		\begin{cases}
			1_7^{(1,3,0,4,0)}\text{; See Figure \ref{fig:27133Orbit}}\\
			1_8^{(1,0,3,0,4,0)}\\
			1_9^{(1,1,2,0,4,0)}\\
			1_{11}^{(1,1,3,0,3,0)}\\
			0_{13}^{(1,1,3,0,4,0)}
		\end{cases}$
	\item $1^{(1,2,0,3,3,0)}
		\begin{cases}
			1_7^{(2,0,3,3,0)}\text{; See Figure \ref{fig:3760Orbit}}\\
			1_8^{(1,1,0,3,3,0)}\\
			1_{10}^{(1,2,0,2,3,0)}\\
			1_{11}^{(1,2,0,3,2,0)}\\
			0_{13}^{(1,2,0,3,3,0)}
		\end{cases}$
	\item $1^{(2,0,0,5,2,0)}
		\begin{cases}
			1_7^{(1,0,0,5,2,0)}\text{; See Figure \ref{fig:17160Orbit}}\\
			1_{10}^{(2,0,0,4,2,0)}\\
			1_{11}^{(2,0,0,5,1,0)}\\
			0_{13}^{(2,0,0,5,2,0)}
		\end{cases}$
	\item $1^{(1,0,0,2,4,2,0)}
		\begin{cases}
			1_6^{(2,4,2,0)}\text{; See Figure \ref{fig:3624Orbit}}\\
			1_9^{(1,0,0,1,4,2,0)}\\
			1_{10}^{(1,0,0,2,3,2,0)}\\
			1_{11}^{(1,0,0,2,4,1,0)}\\
			0_{13}^{(1,0,0,2,4,2,0)}
		\end{cases}$
	\item $1^{(1,0,0,5,2,0,1)}
		\begin{cases}
			1_7^{(5,2,0,1)}\text{; See Figure \ref{fig:1726Orbit}}\\
			1_{10}^{(1,0,0,4,2,0,1)}\\
			1_{11}^{(1,0,0,5,1,0,1)}\\
			1_{13}^{(1,0,0,5,2,0,0)}\\
			0_{14}^{(1,0,0,5,2,0,1)}
		\end{cases}$
	\item $1^{(1,0,1,3,2,2,0)}
		\begin{cases}
			1_7^{(1,3,2,2,0)}\text{; See Figure \ref{fig:27133Orbit}}\\
			1_9^{(1,0,0,3,2,2,0)}\\
			1_{10}^{(1,0,1,2,2,2,0)}\\
			1_{11}^{(1,0,1,3,1,2,0)}\\
			1_{12}^{(1,0,1,3,2,1,0)}\\
			0_{14}^{(1,0,1,3,2,2,0)}
		\end{cases}$
	\item $1^{(1,0,2,0,5,1,0)}
		\begin{cases}
			1_7^{(2,0,5,1,0)}\text{; See Figure \ref{fig:3760Orbit}}\\
			1_9^{(1,0,1,0,5,1,0)}\\
			1_{11}^{(1,0,2,0,4,1,0)}\\
			1_{12}^{(1,0,2,0,5,0,0)}\\
			0_{14}^{(1,0,2,0,5,1,0)}
		\end{cases}$
	\item $1^{(1,0,0,0,6,1,1,0)}
		\begin{cases}
			1_7^{(6,1,1,0)}\text{; See Figure \ref{fig:1726Orbit}}\\
			1_{11}^{(1,0,0,0,5,1,1,0)}\\
			1_{12}^{(1,0,0,0,6,0,1,0)}\\
			1_{13}^{(1,0,0,0,6,1,0,0)}\\
			0_{15}^{(1,0,0,0,6,1,1,0}
		\end{cases}$
	\item $1^{(1,1,0,0,7,0,0)}
		\begin{cases}
			1_7^{(1,0,0,7,0,0)}\text{; See Figure \ref{fig:17160Orbit}}\\
			1_8^{(1,0,0,0,7,0,0)}\\
			1_{11}^{(1,1,0,0,6,0,0)}\\
			0_{14}^{(1,1,0,0,7,0,0)}
		\end{cases}$
	\item $1^{(1,0,0,1,3,4,0,0)}
		\begin{cases}
			1_7^{(1,3,4,0,0)}\text{; See Figure \ref{fig:27133Orbit}}\\
			1_{10}^{(1,0,0,0,3,4,0,0)}\\
			1_{11}^{(1,0,0,1,2,4,0,0)}\\
			1_{12}^{(1,0,0,1,3,3,0,0)}\\
			1_{15}^{(1,0,0,1,3,4,0,0)}
		\end{cases}$
	\item $2=\text{M2}
		\begin{cases}
			1_0=\text{F1; See Figure \ref{fig:10Orbit}}\\
			2_1=\text{D2; See Figure \ref{fig:11Orbit}}
		\end{cases}$
	\item $2^6
		\begin{cases}
			2_3^5\text{; See Figure \ref{fig:532Orbit}}\\
			1_4^6\text{; See Figure \ref{fig:146Orbit}}\\
			2_5^6\text{; See Figure \ref{fig:2515Orbit}}
		\end{cases}$
	\item $2^{(4,3)}
		\begin{cases}
			2_4^{(3,3)}\text{; See Figure \ref{fig:4413Orbit}}\\
			2_5^{(4,2)}\text{; See Figure \ref{fig:2515Orbit}}\\
			1_6^{(4,3)}\text{; See Figure \ref{fig:1643Orbit}}\\
			2_7^{(4,3)}\text{; See Figure \ref{fig:27133Orbit}}
		\end{cases}$
	\item $2^{(7,0)}
		\begin{cases}
			2_5^{(6,0)}\text{; See Figure \ref{fig:2515Orbit}}\\
			1_7^{(7,0)}\text{; See Figure \ref{fig:17160Orbit}}\\
			2_8^{(7,0)}
		\end{cases}$
	\item $2^{(4,0,4)}
		\begin{cases}
			2_5^{(3,0,4)}\text{; See Figure \ref{fig:5522Orbit}}\\
			2_7^{(4,0,3)}\text{; See Figure \ref{fig:27133Orbit}}\\
			1_8^{(4,0,4)}
		\end{cases}$
	\item $2^{(1,0,2,5)}
		\begin{cases}
			2_3^{(2,5)}\text{; See Figure \ref{fig:532Orbit}}\\
			2_5^{(1,0,1,5)}\text{; See Figure \ref{fig:2515Orbit}}\\
			2_6^{(1,0,2,4)}\text{; See Figure \ref{fig:3624Orbit}}\\
			1_7^{(1,0,2,5)}\text{; See Figure \ref{fig:1726Orbit}}\\
		\end{cases}$
	\item $2^{(1,1,3,3)}
		\begin{cases}
			2_4^{(1,3,3)}\text{; See Figure \ref{fig:4413Orbit}}\\
			2_5^{(1,0,3,3)}\text{; See Figure \ref{fig:2515Orbit}}\\
			2_6^{(1,1,2,3)}\text{; See Figure \ref{fig:3624Orbit}}\\
			2_7^{(1,1,3,2)}\text{; See Figure \ref{fig:27133Orbit}}\\
			1_8^{(1,1,3,3)}
		\end{cases}$
	\item $2^{(1,3,2,2)}
		\begin{cases}
			2_5^{(3,2,2)}\text{; See Figure \ref{fig:5522Orbit}}\\
			2_6^{(1,2,2,2)}\text{; See Figure \ref{fig:3624Orbit}}\\
			2_7^{(1,3,1,2)}\text{; See Figure \ref{fig:27133Orbit}}\\
			2_8^{(1,3,2,1)}\\
			1_9^{(1,3,2,2)}
		\end{cases}$
	\item $2^{(2,0,5,1)}
		\begin{cases}
			2_5^{(1,0,5,1)}\text{; See Figure \ref{fig:2515Orbit}}\\
			2_7^{(2,0,4,1)}\text{; See Figure \ref{fig:27133Orbit}}\\
			2_8^{(2,0,5,0)}\\
			1_9^{(2,0,5,1)}
		\end{cases}$
	\item $2^{(2,3,2,1)}
		\begin{cases}
			2_6^{(1,3,2,1)}\text{; See Figure \ref{fig:3624Orbit}}\\
			2_7^{(2,2,2,1)}\text{; See Figure \ref{fig:27133Orbit}}\\
			2_8^{(2,3,1,1)}\\
			2_9^{(2,3,2,0)}\\
			1_{10}^{(2,3,2,1)}
		\end{cases}$
	\item $2^{(3,4,0,1)}
		\begin{cases}
			2_7^{(2,4,0,1)}\text{; See Figure \ref{fig:27133Orbit}}\\
			2_8^{(3,3,0,1)}\\
			2_{10}^{(3,4,0,0)}\\
			1_{11}^{(3,4,0,1)}
		\end{cases}$
	\item $2^{(4,1,3,0)}
		\begin{cases}
			2_7^{(3,1,3,0)}\text{; See Figure \ref{fig:27133Orbit}}\\
			2_8^{(4,0,3,0)}\\
			2_9^{(4,1,2,0)}\\
			1_{11}^{(4,1,3,0)}
		\end{cases}$
	\item $2^{(1,0,0,7,0)}
		\begin{cases}
			2_4^{(7,0)}\text{; See Figure \ref{fig:7420Orbit}}\\
			2_7^{(1,0,0,6,0)}\text{; See Figure \ref{fig:3760Orbit}}\\
			1_9^{(1,0,0,7,0)}
		\end{cases}$
	\item $2^{(1,0,3,4,0)}
		\begin{cases}
			2_5^{(3,4,0)}\text{; See Figure \ref{fig:5522Orbit}}\\
			2_7^{(1,0,2,4,0)}\text{; See Figure \ref{fig:3760Orbit}}\\
			2_8^{(1,0,3,3,0)}\\
			1_{10}^{(1,0,3,4,0)}
		\end{cases}$
	\item $2^{(1,1,4,2,0)}
		\begin{cases}
			2_6^{(1,4,2,0)}\text{; See Figure \ref{fig:3624Orbit}}\\
			2_7^{(1,0,4,2,0)}\text{; See Figure \ref{fig:3760Orbit}}\\
			2_9^{(1,1,4,1,0)}\\
			1_{11}^{(1,1,4,2,0)}
		\end{cases}$
	\item $2^{(1,3,3,1,0)}
		\begin{cases}
			2_7^{(3,3,1,0)}\text{; See Figure \ref{fig:27133Orbit}}\\
			2_8^{(1,2,3,1,0)}\\
			2_9^{(1,3,2,1,0)}\\
			2_{10}^{(1,3,3,0,0)}\\
			1_{12}^{(1,3,3,1,0)}
		\end{cases}$
	\item $2^{(2,0,6,0,0)}
		\begin{cases}
			2_7^{(1,0,6,0,0)}\text{; See Figure \ref{fig:3760Orbit}}\\
			2_9^{(2,0,5,0,0)};\\
			1_{12}^{(2,0,6,0,0)}
		\end{cases}$
	\item $3^{(2,4)}
		\begin{cases}
			3_3^{(1,4)}\text{; See Figure \ref{fig:532Orbit}}\\
			3_4^{(2,3)}\text{; See Figure \ref{fig:4413Orbit}}\\
			2_5^{(2,4)}\text{; See Figure \ref{fig:2515Orbit}}\\
			3_6^{(2,4)}\text{; See Figure \ref{fig:3624Orbit}}
		\end{cases}$
	\item $3^{(2,2,3)}
		\begin{cases}
			3_4^{(1,2,3)}\text{; See Figure \ref{fig:4413Orbit}}\\
			3_5^{(2,1,3)}\text{; See Figure \ref{fig:5522Orbit}}\\
			3_6^{(2,2,2)}\text{; See Figure \ref{fig:3624Orbit}}\\
			2_7^{(2,2,3)}\text{; See Figure \ref{fig:27133Orbit}}
		\end{cases}$
	\item $3^{(3,3,1)}
		\begin{cases}
			3_5^{(2,3,1)}\text{; See Figure \ref{fig:5522Orbit}}\\
			3_6^{(3,2,1)}\text{; See Figure \ref{fig:3624Orbit}}\\
			3_7^{(3,3,0)}\text{; See Figure \ref{fig:3760Orbit}}\\
			2_8^{(3,3,1)}
		\end{cases}$
	\item $3^{(5,2,0)}
		\begin{cases}
			3_6^{(4,2,0)}\text{; See Figure \ref{fig:3624Orbit}}\\
			3_7^{(5,1,0)}\text{; See Figure \ref{fig:3760Orbit}}\\
			2_9^{(5,2,0)}
		\end{cases}$
	\item $4^{(1,2,3)}
		\begin{cases}
			4_3^{(2,3)}\text{; See Figure \ref{fig:532Orbit}}\\
			4_4^{(1,1,3)}\text{; See Figure \ref{fig:4413Orbit}}\\
			4_5^{(1,2,2)}\text{; See Figure \ref{fig:5522Orbit}}\\
			3_6^{(1,2,3)}\text{; See Figure \ref{fig:3624Orbit}}
		\end{cases}$
	\item $4^{(1,5,0)}
		\begin{cases}
			4_4^{(5,0)}\text{; See Figure \ref{fig:7420Orbit}}\\
			4_5^{(1,4,0)}\text{; See Figure \ref{fig:5522Orbit}}\\
			3_7^{(1,5,0)}\text{; See Figure \ref{fig:3760Orbit}}
		\end{cases}$
	\item $5=\text{M5}
		\begin{cases}
			4_1=\text{D4; See Figure \ref{fig:11Orbit}}\\
			5_2=\text{NS5; See Figure \ref{fig:52Orbit}}
		\end{cases}$
	\item $5^3
		\begin{cases}
			5_2^2\text{; See Figure \ref{fig:52Orbit}}\\
			4_3^3\text{; See Figure \ref{fig:532Orbit}}\\
			5_4^3\text{; See Figure \ref{fig:4413Orbit}}
		\end{cases}$
	\item $5^{(1,3)}
		\begin{cases}
			5_2^3=\text{R-monopole; See Figure \ref{fig:52Orbit}}\\
			5_3^{(1,2)}\text{; See Figure \ref{fig:532Orbit}}\\
			4_4^{(1,3)}\text{; See Figure \ref{fig:4413Orbit}}\\
			5_5^{(1,3)}\text{; See Figure \ref{fig:5522Orbit}}
		\end{cases}$
	\item $5^{(1,0,4)}
		\begin{cases}
			5_2^4\text{; See Figure \ref{fig:52Orbit}}\\
			5_4^{(1,0,3)}\text{; See Figure \ref{fig:4413Orbit}}\\
			4_5^{(1,0,4)}\text{; See Figure \ref{fig:5522Orbit}}
		\end{cases}$
	\item $6^1=\text{KK6M}
		\begin{cases}
			6_1=\text{D6; See Figure \ref{fig:11Orbit}}\\
			5_2^1=\text{KK5A; See Figure \ref{fig:52Orbit}}\\
			6_3^1 \text{; See Figure \ref{fig:532Orbit}}
		\end{cases}$
	\item $6^{(3,1)}
		\begin{cases}
			6_3^{(2,1)}\text{; See Figure \ref{fig:532Orbit}}\\
			6_4^{(3,0)}\text{; See Figure \ref{fig:7420Orbit}}\\
			5_5^{(3,1)}\text{; See Figure \ref{fig:5522Orbit}}
		\end{cases}$
	\item $8^{(1,0)}=\text{KK8M}
		\begin{cases}
			8_1=\text{D8; See Figure \ref{fig:11Orbit}}\\
			7_3^{(1,0)}=\text{KK7A; See Figure \ref{fig:532Orbit}}\\
			8_4^{(1,0)}=\text{KK8A; See Figure \ref{fig:7420Orbit}}
		\end{cases}$
\end{itemize}
\end{multicols}
\clearpage
\newgeometry{hmargin=2cm,vmargin=2cm}
\subsection{Discussion}
\subsubsection{Known Exotic Branes in the Literature}
A small subset of the exotic branes presented here have appeared in the literature before. The starting point are, of course, the `standard' branes appearing at $g_s^0$ (P and F1), $g_s^{-1}$ (D$p$-branes) and $g_s^{-2}$ (NS5 and KK5). Additionally, the existence of the $7_3$-brane as the S-dual of the D7 with much work being conducted in the context of the $(p,q)$ 7-branes of F-theory. This was included amongst the codimension-2 exotic branes of \cite{Tellechea2001,deBoer:2012ma} (note that the former uses an alternative notation to what we use, e.g. the $5_2^2$ here is called an NS$5_2$ there). The latter also gives a detailed exposition of the T-duality chain $\text{NS5}=5_2 \xrightarrow{\text{T}} \text{KK5}=5_2^1 \xrightarrow{\text{T}} 5_2^2$. This prototypical chain was extended to include the $5_2^3$ in \cite{Bakhmatov:2016kfn} and then, more recently, a novel $5_2^4$-brane from DFT considerations in \cite{Kimura:2018}. This whole five-brane chain matches what is found in the work presented here, specifically Figure \ref{fig:52Orbit}.\par
Lower codimension objects are even less well-studied and understood and there is limited literature on the subject. However, it has been known since, at least, \cite{Hull1998,Obers:1998} (and references therein) that a massive deformation of 11-dimensional supergravity admits a domain wall solution in M-theory which has since appeared under various names such as the M9 in \cite{Bergshoeff1999} or KK9M in \cite{Ortin:2015hya}. However, as remarked in \cite{Obers:1998}, it should more properly be called an M8-brane or perhaps KK8 following its mass formula designation $8^{(1,0)}$. It is, perhaps, to be understood as an object that exists only as a lift of the D8-brane of Type IIA. The remaining reductions of the $8^{(1,0)}$ are the $7_3^{(1,0)}$ and $8_4^{(1,0)}$ which were also tabulated in \cite{Ortin:2015hya} (though named as KK8A and KK9A respectively, with the same caveat as above).\par
Finally, much like the D7-brane, the D9-brane also has an exotic S-dual (previously called an S9 or an NS9) but which we designate as a $9_4\text{B}$, as was done in \cite{Obers:1998}.
\subsubsection{Mixed-Symmetry Potentials in the Literature}\label{sec:MSP}
Recently, much work has been done on the mixed-symmetry potentials that these exotic branes couple to \cite{Bergshoeff:2017gpw,Lombardo:2016swq}. These have focused on trying to classify the T-duality orbits starting from the highest weight representations of the Lie algebra but miss out on the T-duality orbits that require S-dualities and/or lifts to M-theory to obtain. Nonetheless, there is significant overlap between their work and the work presented here and we summarise this in Table \ref{tab:BranePotentialCorrespondence}.\par
Another piece of work worth mentioning is \cite{Kleinschmidt:2011vu} in which a similar set of potentials were derived from $E_{11}$ and the tensor hierarchy associated to it. One may verify that the majority of the potentials that they obtain for Type II coincide with ours. Those that they are missing are, again expected to be those that appear in the $d=2,1,0$ duality groups whilst those that we are missing are expected to turn up at lower $g_s$ scaling (note that they have not organised their results in powers of $g_s$, complicating the comparison of results).
\begin{landscape}
\thispagestyle{empty}
\begin{table}[!ht]
\centering
\captionsetup{width=1.05\linewidth}
\begin{tabulary}{1.2\linewidth}{CCCLp{5.5cm}p{5cm}}
\toprule
\multirow{2}{*}{$\alpha$} & \multicolumn{2}{c}{Potentials} & \multirow{2}{*}{Figure} & \multirow{2}{*}{Conditions} & \multirow{2}{*}{Notes}\\
\cmidrule{2-3}
& IIA & IIB & &\\
\midrule
0 & \multicolumn{2}{c}{$B_2$} & Fig. \ref{fig:10Orbit} & & NS-NS potentials\\
-1 & $C_{2n+1}$ & $C_{2n}$ & Fig. \ref{fig:11Orbit} & $n \in \{0,1,2,3,4\}$ & R-R potentials\\
-2 &\multicolumn{2}{c}{$D_{6+n,n}$} & Fig. \ref{fig:52Orbit} & $n\in \{0,1,2,3,4\}$ & 5-brane duality chain\\
-3 & $E_{8+n, 2m+1, n}$ & $E_{8+n,2m,n}$ & Fig. \ref{fig:532Orbit} & $n \in \{0,1,2\}, m \in \{1,2,3,4\}$ &\\
-4 & \multicolumn{2}{c}{$F_{8+n,6+m,m,n}$} & Fig. \ref{fig:146Orbit} & $n \in \{0,1,2\}, m \in \{n, n+1\}$ &\\
& \multicolumn{2}{c}{$F_{9+n,3+m,m,n}$} & Fig. \ref{fig:4413Orbit} & $n \in \{0,1\}$ \newline $m \in \{n, n+1, \ldots, n+5\}$ &\\
& $F_{10,2n+1,2n+1}$ & $F_{10,2n,2n}$ & Fig. \ref{fig:7420Orbit} & $n \in \{0, \ldots, 4\}$ &\\
-5 & $G_{9+p,6+n,2m,n,p}$ & $G_{9+p, 6+n, 2m+1, n,p}$ & Fig. \ref{fig:2515Orbit} & $p \in \{0,1\}, n \in \{ p, p+1, p+2\}$ \newline $n \leq 2m,2m+1 \leq n+6$ & \\
& $G_{10,4+n,2m+1,n}$ & $G_{10,4+n,2m,n}$ & Fig. \ref{fig:5522Orbit} & $n \in \{0, 1, \ldots, 5\}$ \newline $n \leq 2m, 2m+1 \leq n+4$ &\\
-6 & \multicolumn{2}{c}{$H_{9+n,8+n,m+n,m+n-1,n,n}$} & Fig. \ref{fig:0617Orbit} & $m\in \{1,2, \ldots, 8\}, n \in \{0,1\}$ & Not included in the analyses of \cite{Bergshoeff:2017gpw,Fernandez-Melgarejo:2018yxq}\\
& \multicolumn{2}{c}{$H_{10, 6+n, 2+m, m,n}$} & Fig. \ref{fig:3624Orbit} & $n\in\{0,1,2,3\}$ \newline $m \in \{n, n+1, \ldots, n+4\}$ &\\
& \multicolumn{2}{c}{$H_{9+p, 7+n, 4+m, m,n,p}$} & Fig. \ref{fig:1643Orbit} & $p \in \{0,1\}, n \in \{p,p+1\}$ \newline $m \in \{n, n+1, n+2, n+3\}$ & Not included in the analysis of \cite{Bergshoeff:2017gpw}\\
-7 & $I_{10,8+p, n+2, 2m+1, n, p}$ & $I_{10, 8+p, 2+n, 2m, n, p}$ & Fig. \ref{fig:1726Orbit} & $p\in\{0,1\}, n \in \{p, p+1, \ldots, p+6\}$ \newline $n \leq 2m, 2m+1 \leq 2+n$ & Not included in the analyses of \cite{Bergshoeff:2017gpw,Fernandez-Melgarejo:2018yxq}\\
& $I_{9+p, 8+p, 5+n+p, 2m+1, n+p, p, p}$ & $I_{9+p, 8+p, 5+n+p, 2m, n+p, p, p}$ & Fig. \ref{fig:0753Orbit} & $p \in \{0,1\}, n \in \{0,1,2,3\}$ \newline $n+p \leq 2m,2m+1 \leq n+p+5$ & Not included in the analyses of \cite{Bergshoeff:2017gpw,Fernandez-Melgarejo:2018yxq}\\
& $I_{9+p,p+n+7,p+n+7,2m,n+p,n+p,p}$ & $I_{9+p,p+n+7,p+n+7,2m+1,n+p,n+p,p}$ & Fig. \ref{fig:17160Orbit} & $n \in \{0,1\}, p \in \{0,1\}$, \newline $n+p \leq 2m, 2m+1 \leq n+p+7$ & Not included in the analysis of \cite{Bergshoeff:2017gpw}\\
& $I_{10,7+p,4+n+p,2m,n+p,p}$ & $I_{10,7+p,4+n+p,2m+1,n+p,p}$ & Fig. \ref{fig:27133Orbit} & $p\in \{0,1,2\}, n \in \{0,1,2,3\}$\newline $n+p \leq 2m, 2m+1, \leq, 4 +n +p$ & Not included in the analysis of \cite{Bergshoeff:2017gpw}\\
& $I_{10, 6+n, 6+n, 2m+1, n,n}$ & $I_{10, 6+n, 6+n, 2m, n,n}$ & Fig. \ref{fig:3760Orbit} & $n \in \{0,1,2,3\}$ \newline $n \leq 2m,  2m+1 \leq, 6+n$ &\\
\bottomrule
\end{tabulary}
\caption{The mixed symmetry potentials that the exotic branes couple to. Each of the branes in the listed T-duality orbit couple to one type of potential, listed in the second and third columns. Those potentials that straddle the two columns appear common to both. Most of the lower $g_s^\alpha$ potentials have not been found yet.
}
\label{tab:BranePotentialCorrespondence}
\end{table}
\end{landscape}
Note that \cite{Fernandez-Melgarejo:2018yxq} obtains many of the branes that we have, through studying U-duality multiplets\footnote{We would like to thank the authors of \cite{Fernandez-Melgarejo:2018yxq} for pointing out an issue with the NS9-brane in an earlier version of this paper which we have now rectified. We have removed three small spurious orbits (one each at $g_s^{-2}, g_s{-4}$ and $g_s^{-6}$ respectively) following their comments.} and we find good agreement for the portions that overlap, specifically to $g_s^{-7}$. Here, we spell out the correspondence between their potentials (right-hand side) and ours (left-hand side) for $g_s^{-6}$ and $g_s^{-7}$ potentials only since the other potentials should hopefully be self-evident.
\begin{align}
\begin{aligned}
{\left. H_{10,6+n,2+m,m,n} \right|}_{m \rightarrow m+n} & \rightarrow E_{10,6+n,2+m+n,m+n,n}^{(6)}\\
{\left. H_{9+p,7+n,4+m,m,n,p} \right|}_{p=n=0} & \rightarrow E_{9,7,4+n,n}^{(6)}\\
{\left. \left( I_{9+p,p+n+7,p+n+7,,2m,n+p,n+p,p}, I_{9+p,p+n+7,p+n+7,2m+1,n+p,n+p,p} \right) \right|}_{p=n=0} & \rightarrow E_{9,7,7,q}^{(7)}\\
{\left. \left( I_{10,7+p,4+n+p,2m,n+p,p}, I_{10,7+p,4+n+p,2m+1,n+p,p} \right) \right|}_{p=0} & \rightarrow E_{10,7,4+n,q,n}^{(7)}\\
{\left. \left( I_{10,6+n,6+n,2m+1,n,n}, I_{10,6+n,6+n,2m,n,n} \right) \right|}_{p=0} & \rightarrow E_{10,6+n,6+n,q,n,n}^{(7)}
\end{aligned}
\end{align}
\subsubsection{Unification at Larger Duality Groups}
The proliferation of extoic branes is self-evident from the figures above and it is currently difficult to tell if this procedure will even terminate at all. However, the growing number of DFT and EFT solutions found to date, including the DFT monopole \cite{Berman:2014hna,Bakhmatov:2016kfn,Kimura:2018}, the $E_{7(7)}$ geometric solution \cite{Berman:2014hna} and the non-geometric solution presented here, all point to the over-arching theme of unification of branes in higher dimensions. Just as the possible wrappings of the M2 were found to give a unifying description of the F1 and D2 in one dimension higher, multiple branes have lifted to single solutions in DFT and EFT. That they only unify a small fraction of the branes that we have described is not a problem and, indeed, is probably to be expected given the awkward split between internal and external spaces that is inherent in EFT which puts a restriction on which branes can be lifted to the same solution in ExFT. More exciting is the possibility that every single brane presented in this section should all lift to one unified solution in ExFT at higher duality groups.\par
The rationale behind this claim is as follows. We have already mentioned that many of the novel branes that we have found at codimension-1 and 0 have not been found in the literature simply because previous efforts such as \cite{Bergshoeff:2017gpw,Fernandez-Melgarejo:2018yxq} have always classified them under U-duality representations under reductions down to $d=3$. It is thus natural to expect that the novel branes presented here will only appear when one considers reductions down to $d=2,1$ and even $d=0$. Put differently, if one were to consider the largest duality groups, one should be able to accomodate more and more of them until every single brane presented here is accommodated for. This also consistent with the observation that the procedure does not appear to have any clear termination point---there still remains the possibility that there are indeed an infinite number of exotic branes whose wrapping modes are then used to construct the extended spaces of the highest ExFTs.\par
We further note that since every one of the figures are inter-related by S- and T-dualities, the lift to M-theory means that every one of those figures are part of a single U-duality orbit\footnote{This is, of course, assuming that U-duality is strictly generated from combinations of S- and T-dualities only}. From the discussion before, it is tempting to call it a single U-dualtiy orbit of $E_{11}$ that fragments to smaller U-duality orbits only when one descends down the $E_n$-series. See \cite{West:2001as,West:2003fc,Tumanov:2015yjd,Tumanov:2016abm,Kleinschmidt:2003jf,Cook:2008bi,West:2004kb,West:2004iz,Cook:2009ri,Cook:2011ir} for a discussion on both standard and exotic branes in the context of the $E_{11}$ program. According to this conjecture one may thus only construct `truly' non-geometric objects (in the sense that there is no U-duality transformation that transforms it to a geometric solution) within the smaller duality groups; what appear to be distinct orbits in those groups should successively merge into fewer and fewer orbits of the higher duality groups until one is left with only a single U-duality orbit at $E_{11}$. Thus whilst we now have two distinct solutions in $E_{7(7)}$ EFT, covering different sets of branes and with no apparent way to transition between the two (see Figure \ref{fig:Brane}), one should expect that these two EFT solutions can be unified into a single solution of a larger EFT (along with a much wider class of exotic branes).
\section{Conclusion}\label{sec:Conclusion}
In this paper, we have argued that extended field theories (ExFT) provide an ideal laboratory in which to study exotic branes. Starting in Section \ref{sec:NonGeomSoln}, we have tried to emphasise the power of ExFT in unifying the known solutions of M-theory----a fact that is built into EFT by construction since the Type IIB and M-theory sections (hence, also, the Type IIA section) are derived from a single extended space. This is reminiscent of the remarkable result in M-theory that the numerous Type IIA branes were lifted to just a handful of M-theory branes in the conventional story. We demonstrated this fact by constructing a novel solution of $E_{7(7)}$ EFT that gives rise to all of the codimension-2 exotic states listed in \cite{deBoer:2012ma} upon taking either of the two sections (see Figure \ref{fig:Brane}).\par
Having demonstrated the utility of ExFT in studying these exotic branes, we moved onto a broader discussion of other types of exotic branes that one may hope to discover in Section \ref{sec:Map}. We have raised the possibility that the exotic branes are far more numerous than previously thought, and that there may even be an infinite number of them. In the course of this analysis it was found that the vast majority of these novel exotic states lie at codimension-1 or 0 which are particuly diffcult to interpret in conventional supergravity and may even require a DFT or EFT description to make sense at all. Having revealed the structure of the intricately woven dualities and lifts/reductions, we hope that they will pave the way to a more complete understanding of these exotic states. In particular, it is hoped that they will aid in constucting more solutions in ExFT that unify these objects---an exercise which has previously been complicated by the awkward split between internal and external spaces.
\section{Note Added}
Whilst this work was in production, the paper \cite{Fernandez-Melgarejo:2018yxq} appeared on the e-Print arXiv which contains overlapping material with Section \ref{sec:Map} presented here. We have expanded on the relation of their work with ours in Section \ref{sec:MSP}.
\section{Acknowledgements}
We thank the folowing for discussions: Ilya Bakhmatov, Chris Hull, Axel Kleinschmidt and Masaki Shigemori. DSB is supported by the STFC grant ST/L000415/1, ``String Theory, Gauge Theory and Duality'' and also acknowledges hospitality from DAMTP, Cambridge, during the completion of this work. The work of ETM was supported by the Russian state grant Goszadanie 3.9904.2017/8.9 and by the Alexander von Humboldt return fellowship and partially by the program of competitive growth of Kazan Federal University. ETM is grateful to Queen Mary University of London for hospitality during the initial stage of this project. RO is supported by an STFC studentship.
\clearpage
\begin{appendix}
\section{Exotic Brane Metrics}
We collect the backgrounds of the exotic branes appearing in the non-geometric solution of Section \ref{sec:NonGeomSoln} for convenience. For the Type II branes, we adopt the Einstein frame which is related to the string frame via $\textrm{d} s^2_{\textrm{s}} = e^{\frac{\phi}{2}} \textrm{d} s^2_{\text{E}}$ where $\phi$ is the dilaton. Here, we have obtained the metrics through a sequence of conventional T- and S-duality transformations, starting from a smeared NS5-brane. As such, all harmonic functions here $H$ are harmonic in $(r,\theta)$ only, with
\begin{align}
H(r) = h_0 + \sigma \ln \frac{\mu}{r}
\end{align}
where $h_0$ is a diverging bare quantity, $\mu$ a renormalisation scale and $\sigma$ a dimensionless constant (which is different for each solution but irrelevant for the discussion here). We also define $K \coloneqq H^2 + \sigma^2 \theta^2$.
\subsection{M-Theory Branes}
\begin{itemize}
	\item $5^3$
\end{itemize}
\begin{gather}
\begin{gathered}
\textrm{d} s^2 = {(HK^{-1})}^{-\frac{1}{3}} ( - \textrm{d} t^2 + \textrm{d} {\vec{x}}^2_{(5)} ) + {(HK^{-1})}^{\frac{2}{3}} \textrm{d} {\vec{y}}^2_{(3)} +  H^{\frac{2}{3}} K^{\frac{1}{3}} (\textrm{d} r^2 + r^2 \textrm{d} \theta^2)\\
A_{(3)} = - K^{-1} \sigma \theta \textrm{d} y^1 \wedge \textrm{d} y^2 \wedge \textrm{d} y^3, \qquad A_{(6)} = - H^{-1} K \textrm{d} t \wedge \textrm{d} x^1 \wedge \ldots \wedge \textrm{d} x^5
\end{gathered}
\end{gather}
\begin{itemize}
	\item $2^6$
\end{itemize}
\begin{gather}
\begin{gathered}
\textrm{d} s^2 = {(HK^{-1})}^{-\frac{2}{3}} ( - \textrm{d} t^2 + \textrm{d} {\vec{x}}^2_{(2)} ) + {(HK^{-1})}^{\frac{1}{3}} \textrm{d} {\vec{y}}^2_{(6)} +  H^{\frac{1}{3}} K^{\frac{2}{3}} (\textrm{d} r^2 + r^2 \textrm{d} \theta^2)\\
A_{(3)} = - H^{-1} K \textrm{d} t \wedge \textrm{d} x^1 \wedge \textrm{d} x^2, \qquad A_{(6)} = - K^{-1} \sigma \theta \textrm{d} y^1 \wedge \ldots \wedge \textrm{d} y^6
\end{gathered}
\end{gather}
\begin{itemize}
	\item $0^{(1,7)}$
\end{itemize}
\begin{gather}
\begin{gathered}
\textrm{d} s^2 = - H^{-1} K \textrm{d} t^2 + \textrm{d} {\vec{x}}^2_{(7)} + HK^{-1} {( \textrm{d} z - H^{-1}K \textrm{d} t)}^2 + K (\textrm{d} r^2 + r^2 \textrm{d} \theta^2)
\end{gathered}
\end{gather}
\subsection{Type II Branes}
Both Type IIA and IIB have $5_2^2-, 1_4^6$- and $0_4^{(1,6)}$-branes. Type IIA has $p_3^{7-p}$ for $p = 1, 3, 5,7$ whilst Type IIB has $p_3^{7-p}$ for $p = 0,2,4,6$. 
\begin{itemize}
	\item $5^2_2$---the only exotic brane with scaling $g_s^{-2}$
\end{itemize}
\begin{gather}
\begin{gathered}
\textrm{d} s^2_{\text{E}} = {(HK^{-1})}^{-\frac{1}{4}} \left( - \textrm{d} t^2 + \textrm{d} {\vec{x}}^2_{(5)} \right) + {(HK^{-1})}^{\frac{3}{4}} \textrm{d} {\vec{y}}^2_{(2)} + H^{\frac{3}{4}} K^{\frac{1}{4}} (\textrm{d} r^2 + r^2 \textrm{d} \theta^2)\\
B_{(2)} = - K^{-1} \sigma \theta \textrm{d} y^1 \wedge \textrm{d} y^2, \qquad B_{(6)} = - H^{-1} K \textrm{d} t \wedge \textrm{d} x^1 \wedge \ldots \wedge \textrm{d} x^5\\
e^{2(\phi - \phi_0)} = HK^{-1}
\end{gathered}
\end{gather}
\begin{itemize}
	\item For $p=0, \ldots, 6$, the $p_3^{7-p}$ exotic branes (with mass scaling $g_s^{-3}$) are given by
\end{itemize}
\begin{gather}
\begin{gathered}
\textrm{d} s^2_{\text{E}} = {(HK^{-1})}^{\frac{p-7}{8}} \left( - \textrm{d} t^2 + \textrm{d} {\vec{x}}^2_{(p)} \right) + {(HK^{-1})}^{\frac{p+1}{8}} \textrm{d} {\vec{y}}^2_{(7-p)} + H^{\frac{p+1}{8}} K^{\frac{7-p}{8}} \left( \textrm{d} r^2 + r^2 \textrm{d} \theta^2\right)\\
C_{(7-p)} = - K^{-1} \sigma \theta \textrm{d} y^1 \wedge \ldots \wedge \textrm{d} y^{7-p},\qquad  C_{(p+1)} = - H^{-1} K \textrm{d} t \wedge \textrm{d} x^1 \wedge \ldots \wedge \textrm{d} x^p\\ 
e^{2(\phi- \phi_0)} = {(HK^{-1})}^{-\frac{p-3}{2}}
\end{gathered}
\end{gather}
	The $7_3$, as with the D7, appears slightly anomalous.
\begin{itemize}
	\item There are two $g_s^{-4}$ branes. The $1_4^6$ has background
\end{itemize}
\begin{gather}
\begin{gathered}
\textrm{d} s^2_{\text{E}} = {(HK^{-1})}^{-\frac{3}{4}} \left( - \textrm{d} t^2 + \textrm{d} x^2 \right) + {(HK^{-1})}^{\frac{1}{4}} \textrm{d} {\vec{y}}^2_{(6)} + H^{\frac{1}{4}} K^{\frac{3}{4}} \left( \textrm{d} r^2 + r^2 \textrm{d} \theta^2\right)\\
B_{(6)} = - K^{-1} \sigma \theta \textrm{d} y^1 \wedge \ldots \wedge \textrm{d} y^{6},\qquad  B_{(2)} = - H^{-1} K \textrm{d} t \wedge \textrm{d} x^1 \wedge\textrm{d} x^2\\ 
e^{2(\phi- \phi_0)} = H^{-1} K
\end{gathered}
\end{gather}
	whilst the $0_4^{(1,6)}$ has a purely metric background
\begin{gather}
\begin{gathered}
\textrm{d} s^2_{\text{E}} = - {(HK^{-1})}^{-1} \textrm{d} t^2 + HK^{-1} {(\textrm{d} z - H^{-1} K \textrm{d} t)}^2 + \textrm{d} {\vec{x}}^2_{(6)} + K (\textrm{d} r^2 + r^2 \textrm{d} \theta^2)\\
e^{2(\phi - \phi_0)} = 1
\end{gathered}
\end{gather}
Note that the $(1_4^6, 1_3^6)$ and $(5_3^2, 5_2^2)$ each from S-duality doublets and thus share the same metric in the Einstein frame, exchanging only $B_{(p)} \xleftrightarrow{\text{S}} C_{(p)}$ and inverting the dilaton.
\section{Anti Self-Duality}\label{sec:AntiSelfDuality}
The anti self-duality relation is given by
\begin{align}
\mathcal{F}_{\mu \nu}{}^M & = - \frac{|g_{(4)}|}{2} \varepsilon_{\mu \nu \lambda \tau} g^{\lambda \rho} g^{\tau \sigma} \Omega^{MN} \mathcal{M}_{NK} \mathcal{F}_{\rho \sigma}{}^K
\end{align}
For the present solution, we shall apply the simplifications $\mathcal{B}_{\mu \nu, \bullet}= 0$ and $\partial_N \mathcal{A}_\mu{}^M = 0$ i.e. that the generalised vectors do not depend on the internal coordinates. Then, the covariantised generalised field strength reduces to the Abelian Field strength
\begin{align}
\mathcal{F}_{\mu \nu}{}^M \rightarrow F_{\mu \nu}{}^M = 2 \partial_{[\mu} \mathcal{A}_{\nu]}{}^M
\end{align}
For the ansatz given above, the only non-vanishing components of the field strength in the $5^3$ frame are
\begin{align}
\mathcal{F}_{\mu \nu}{}^M & = ( \mathcal{F}_{tr}{}^{\xi \chi}, \mathcal{F}_{t \theta}{}^{\xi \chi}, \mathcal{F}_{rz, \xi \chi}, \mathcal{F}_{\theta z, \xi \chi})\\
	& = \left( \frac{\sigma(\sigma^2 \theta^2 - H^2)}{rH^2} , \frac{2 \sigma^2 \theta}{H}, - \frac{2 \sigma^2 \theta H}{r K^2}, \frac{\sigma ( \sigma^2 \theta^2 - H^2)}{K^2}\right)
\end{align}
We begin by considering $\mathcal{F}_{tr}{}^{\xi \chi}$ component:
\begin{align}
\mathcal{F}_{tr}{}^{\xi \chi} & = - {\left| g_{(4)}\right|}^{\frac{1}{2}} \varepsilon_{tr \theta z} g^{\theta \theta} g^{zz} \Omega^{\xi \chi}{}_{\xi \chi} \mathcal{M}^{\xi \chi, \xi \chi} \mathcal{F}_{\theta z}{}^{\xi \chi}
\end{align}
We begin by substituting $\Omega^{\xi \chi}{}_{\xi \chi} = - e^{\Delta} \delta^{\xi \chi}_{\xi \chi} = - {|g_{(4)}|}^{\frac{1}{4}} \delta^{\xi \chi}_{\xi \chi}$ and $g^{\theta \theta},g^{zz}$ read off from \eqref{eq:Ext} to obtain
\begin{align}
\mathcal{F}_{tr}{}^{\xi \chi} & = {\left| g_{(4)}\right|}^{\frac{3}{4}} \varepsilon_{tr \theta z} \frac{\sigma ( \sigma^2 \theta^2 - H^2)}{r^2HK^2} \mathcal{M}^{\xi \chi, \xi \chi} 
\end{align}
Finally, using $\mathcal{M}^{\xi \chi, \xi \chi} = {|g_{(4)}|}^{-\frac{1}{4}} {(HK^{-1})}^{\frac{3}{2}}$ (read-off from \eqref{eq:53GenMetric}) and $|g_{(4)}| = r^2 HK$, we obtain
\begin{align}
\mathcal{F}_{tr}{}^{\xi \chi} & = {(r^2 HK)}^{\frac{1}{2}} \frac{\sigma ( \sigma^2 \theta^2 - H^2)}{r^2HK^2} {(HK^{-1})}^{-\frac{3}{2}}\\
	& = \frac{\sigma(\sigma^2 \theta^2 - H^2)}{rH^2}
\end{align}
and thus satisfies the duality relation. Likewise, the remaining relations all follow, with the only subtlety begin that $\Omega^{MN} \mathcal{M}_{NK} = {\tilde{\Omega}}^{MN} {\tilde{\mathcal{M}}}_{NK}$ i.e. $e^{-\Delta} = {|g_{(4)}|}^{-\frac{1}{4}}$ the scaling factors cancel:
\begin{align}
\mathcal{F}_{\theta z, \xi \chi} & = - \frac{{|g_{(4)}|}^{\frac{1}{2}}}{2} \times 2 \epsilon_{\theta z t r} g^{tt} g^{rr} {\tilde{\Omega}}_{\xi \chi}{}^{\xi \chi} {\tilde{\mathcal{M}}}^{\xi \chi, \xi \chi} \mathcal{F}_{tr}{}^{\xi \chi}\\
	& = - {(r^2 HK)}^{\frac{1}{2}} \epsilon _{tr \theta z} \left[(-1) \cdot {(HK)}^{-\frac{1}{2}} \right] {(HK^{-1})}^{-\frac{1}{2}} {(HK^{-1})}^{-\frac{3}{2}} \cdot \frac{\sigma}{r} \left( \frac{\sigma^2 \theta^2 - H^2}{H^2} \right)\\
	& = \frac{\sigma ( \sigma^2 \theta^2 - H^2)}{K^2}\\
\mathcal{F}_{t \theta}{}^{\xi \chi} & = - \frac{{|g_{(4)}|}^{\frac{1}{2}}}{2} \times 2 \epsilon_{t \theta r z} g^{rr} g^{zz} {\tilde{\Omega}}^{\xi \chi}{}_{\xi \chi} {\tilde{\mathcal{M}}}_{\xi \chi, \xi \chi} \mathcal{F}_{tr,\xi \chi}\\
	& = - {(r^2 HK)}^{\frac{1}{2}} \left[ -\epsilon _{tr \theta z} \right]{(HK)}^{-\frac{1}{2}} {(HK^{-1})}^{-\frac{1}{2}}  \cdot \left[ - \delta^{\xi \chi}_{\xi \chi} \right] \cdot {(HK^{-1})}^{-\frac{3}{2}} \cdot  \left( - \frac{2 \sigma^2 \theta H}{r K^2} \right)\\
	& = \frac{2 \sigma^2 \theta}{H}\\
\mathcal{F}_{rz, \xi \chi} & = - \frac{{|g_{(4)}|}^{\frac{1}{2}}}{2} \times 2 \epsilon_{r z t \theta} g^{tt} g^{\theta \theta} {\tilde{\Omega}}_{\xi \chi}{}^{\xi \chi} {\tilde{\mathcal{M}}}_{\xi \chi, \xi \chi} \mathcal{F}_{t\theta}{}^{\xi \chi}\\
	& = - {(r^2 HK)}^{\frac{1}{2}} \left[ -\epsilon _{tr \theta z} \right]\cdot \left[ (-1) \cdot {(HK)}^{-\frac{1}{2}} \right] {(HK^{-1})}^{-\frac{1}{2}} {(HK^{-1})}^{\frac{3}{2}} \cdot  \left( \frac{2 \sigma^2 \theta}{H} \right)\\
	& = - \frac{2 \sigma^2 \theta H}{K^2 r}
\end{align}
\section{Bianchi Identity}\label{sec:Bianchi}
For the solution discussed here, the Bianchi identity \eqref{eq:Bianchi} reduces to
\begin{align}
\mathbb{L}_{\mathcal{A}_{[\mu}} \mathcal{F}_{\nu \rho]}{}^\mu & = 0
\end{align}
but each term vanishes independently since $\partial_M \mathcal{A}_\nu{}^N = 0$.
\section{Generators of \texorpdfstring{$E_{7(7)}\times \mathbb{R}^+$}{E7(7)R+}}
Throughout the text, we adopt the convention that $\alpha = 1, \ldots, 133$ indexes the adjoint of the $E_{7(7)}$ algebra. With the conventions that we adopt, the adjoint is symmetric ${(t_\alpha)}_{(MN)}$. We also have
\begin{align}
{(t_\alpha)}_K{}^{(P} c^{QRS)K} & = 0\\
{(t_\beta)}_M{}^K {(t^\alpha)}_K{}^N & = \frac{19}{8} \delta_M{}^N\\
{(t^\alpha)}^{MK} {(t_\beta)}_{KL} {(t_\alpha)}^{LN} & = - \frac{7}{8} {(t_\beta)}^{MN}
\end{align}
The $\mathbb{R}^+$ generator is ${(t_0)}_M{}^N = - \delta_M^N$
\end{appendix}
\newgeometry{left=2cm,right=4cm,vmargin=2cm}
\appto{\bibsetup}{\raggedright}
\bibliographystyle{unsrt}
\bibliography{\jobname}
\end{document}